\pdfoutput=1
\documentclass[12pt]{article}
\usepackage[lmargin=1in, rmargin=1in, tmargin=1in, bmargin=1in]{geometry}
\usepackage{array}
\usepackage{amsfonts}
\usepackage{float}
\usepackage{empheq}
\usepackage{xcolor}
\usepackage{extarrows}
\usepackage[authoryear]{natbib}
\usepackage{bibunits}
\defaultbibliography{refs}
\defaultbibliographystyle{myabbrvnat}
\usepackage{graphicx,setspace,rotating}
\usepackage{amsmath, amssymb, amsthm, mathtools}
\usepackage{bm}
\usepackage{pdflscape}
\usepackage{placeins}
\usepackage{caption}
\usepackage{subcaption}
\usepackage{setspace}
\usepackage{changepage}
\usepackage{booktabs, siunitx, threeparttable, adjustbox}
\usepackage{longtable}
\usepackage{supertabular}
\usepackage{multirow}

\usepackage{hyperref}
\hypersetup{
    colorlinks=true,
     linkcolor=blue,
     filecolor=blue,
     citecolor = blue,
     urlcolor=cyan,
}

\usepackage[nottoc,numbib]{tocbibind}
\usepackage[toc,page,header]{appendix}
\usepackage{minitoc}

\newtheorem{theorem}{Theorem}
\newtheorem{lemma}[theorem]{Lemma}
\newtheorem{proposition}[theorem]{Proposition}

\DeclareMathOperator*{\argmin}{arg\,min}
\newcommand{\1}{\mathbf{1}}

\newcommand{\ATT}{\mathrm{ATT}}

\DeclareMathOperator{\Var}{Var}

\newcommand{\gls}{\mathrm{gls}}

\newcommand{\tfgls}{\mathrm{tfgls}}
\newcommand{\ols}{\mathrm{ols}}

\newcommand{\Kappa}{\mathcal{K}}
\newcommand{\Pspace}{\mathcal{P}}
\newcommand{\wh}[1]{\widehat{#1}}
\newcommand{\norm}[1]{\lVert #1 \rVert}

\newcommand{\bSig}{\bar{\Sigma}}

\newcommand{\rhostar}{\rho^*_I}
\newcommand{\rhohat}{\widehat{\rho}^*_I}

\newcommand{\tauhat}{\hat{\tau}_I}
\newcommand{\tauw}{\tau_w}

\newcommand{\tS}{\widetilde{\Sigma}}
\newcommand{\bY}{\bar{Y}}

\newcommand{\Sigbar}{\bar{\Sigma}}
\newcommand{\Pix}{\Pi}
\newcommand{\PI}{\Pi_I}
\newcommand{\tm}{\tilde{m}}

\newcommand{\zo}{[0]}
\newcommand{\zone}{[1]}
\newcommand{\zoz}{[00]}

\newcommand{\zto}{[10]}

\newcommand{\R}{\mathbb{R}}
\newcommand{\one}{\mathbf{1}}
\newcommand{\kdot}{\dot\kappa_{0,i}}
\newcommand{\kddot}{\ddot\kappa_{0,i}}
\newcommand{\xt}{\widetilde x_{0,i}}
\newcommand{\bhat}{\widehat\beta}
\newcommand{\that}{\widehat\tau_I(0)}
\newcommand{\rhat}{\widehat{\rho}^*_I}
\newcommand{\rstar}{\rho^*_I}
\newcommand{\Ghat}{\widehat G}
\newcommand{\Gtil}{\widetilde G}
\newcommand{\Hhat}{\widehat H}

\newcommand{\diam}{\operatorname{diam}}

\newcommand{\calP}{\mathcal{P}}
\newcommand{\tauhattfgls}{\hat{\tau}_{w,\tfgls}}
\newcommand{\rhohatstar}{\widehat{\rho}^*_I}

\newcommand{\HI}{H_I}

\newcommand{\phiI}{\Phi_I}
\newcommand{\KI}{K_I}

\newcommand{\Blead}{\mathcal{B}_{\mathrm{lead}}}

\newcommand{\DI}{D_I}
\newcommand{\LambdaI}{\Lambda_I}
\newcommand{\rem}{\mathrm{rem}}

\usepackage{titling}
\makeatletter\newif\iftightdisp
\g@addto@macro\normalsize{\iftightdisp\setlength\abovedisplayskip{\tdabove}\setlength\belowdisplayskip{\tdbelow}\fi}
\newcommand\setdisp[2]{\gdef\tdabove{#1}\gdef\tdbelow{#2}\global\tightdisptrue\global\abovedisplayskip=#1\relax\global\belowdisplayskip=#2\relax\normalsize}
\makeatother
\makeatletter
\newcommand{\longpage}[1]{\enlargethispage{#1\baselineskip}\thispagestyle{longpage}\def\longpage@amt{#1}}
\def\longpage@amt{0}
\def\ps@longpage{\let\@oddhead\@empty\let\@evenhead\@empty
  \def\@oddfoot{\reset@font\hfil\raisebox{-\dimexpr\longpage@amt\baselineskip-10pt\relax}[0pt][0pt]{\thepage}\hfil}\let\@evenfoot\@oddfoot}
\makeatother
\begin{document}
\begin{bibunit}

\title{Targeted Feasible Generalized Least Squares}
\author{Saraswata Chaudhuri\thanks{Department of Economics, McGill University \& CIREQ. Email: saraswata.chaudhuri@mcgill.ca. I thank B. Honor\'{e}, C. Muris, D. Benatia, E. Renault, G. Kuersteiner, G. Pouliot, J. Chao, J-M Dufour, J. Galbraith, J. MacKinnon, J. Wooldridge, K. Nagasawa, K. Xue, M. Carrasco, M. Kejriwal, M. Koles\'{a}r, M. Webb, P. Guggenberger, P. Hull, P. Sant'Anna, S. Bonhomme, S. Gon\c{c}alves, T. Christensen, V. Zinde-Walsh and X. Chen for helpful discussions. Y. Ning helped with earlier version's coding. I used AI for checking derivations, literature review, proofreading, and coding. I also thank SSHRC for financial support.}} 

\date{November 25, 2026}
\maketitle

\vspace{-1cm}
\begin{abstract}
\noindent Feasible generalized least squares (FGLS) usually fits a parsimonious parametric working model to the regression error covariance. When that working model is misspecified, FGLS can be less precise than ordinary least squares (OLS), even if the working model nests spherical errors. Focusing on the imputation approach of Borusyak, Jaravel and Spiess (2024) to difference-in-differences estimation of heterogeneous treatment effects under staggered adoption of a binary treatment, we develop a Targeted Feasible Generalized Least Squares (TFGLS) estimator that chooses the parameters of the same, possibly misspecified, working model to minimize directly the asymptotic variance of the GLS estimator of target parameters such as individual event-study coefficients or their aggregates. Hence, TFGLS is asymptotically at least as precise as OLS and FGLS based on the same working model. A parsimonious working model may be misspecified, precluding semiparametric efficiency of TFGLS. However, simulations show that, in small samples, TFGLS can still be more precise than semiparametrically efficient estimators, even when OLS and FGLS are not. Simulations also indicate that parsimony can reduce the underestimation of standard errors, and the resulting over-rejection of true null hypotheses by conventional asymptotic inference, in small samples.\end{abstract}



\newpage
\setdisp{7pt plus 3pt minus 3pt}{7pt plus 3pt minus 3pt}
\section{Introduction}

Recent difference-in-differences (DiD) estimators in staggered adoption of a binary treatment allow treatment effects to vary across units and periods and then aggregate them into scalar parameters of interest, such as an event-study average. Among these, the estimator of \citet{BJS2024}, \citet{Gardner2021} and \citet{Wooldridge2025}, which we call OLS, is efficient when errors are homoskedastic and serially uncorrelated. However, serial correlation is common in panel data \citep{HansenGLS2007}, and GLS can exploit it to improve precision.

Semiparametric FGLS uses nonparametric working models for error covariance, i.e., the nuisance parameters \citep{Kiefer1980, CuiWooldridge2026}. It is asymptotically efficient under homoskedasticity across entities $i=1,\ldots,I$. However, if $I$ is not very large then it can be imprecise, even as its standard errors may wrongly suggest otherwise, since the number of nuisance parameters to be estimated in the working covariance model grows with the number of periods $T$. Parametric FGLS, referred to simply as FGLS, instead fits a parsimonious working model, e.g., autoregressive of order one (AR(1)), for the error covariance \citep{HansenGLS2007,BCJ2018,BJS2024}. Parsimony can make FGLS stable in small samples, but it also makes the working model wrong in general. So FGLS may be inefficient; it can also be less precise than OLS even when the working model nests spherical errors.

Our paper proposes an alternative, Targeted FGLS (TFGLS), that keeps the parsimony of FGLS but changes what the working model is fitted for. Let $\widehat{\tau}(\rho)$ denote the GLS estimator of the parameter of interest $\tau$ when the working model has (nuisance) parameter $\rho$. Under standard assumptions of GLS, $\widehat{\tau}(\rho)$ is consistent and asymptotically normal for $\tau$ at every $\rho$, whether or not the working model is correct, but its asymptotic variance $\sigma^2(\rho)$ depends on $\rho$. Targeting chooses $\rho^*$ that minimizes $\sigma^2(\rho)$ with respect to $\rho$, and TFGLS obtains it by minimizing an estimator of $\sigma^2(\rho)$, and plugs it in for $\rho$ in $\widehat{\tau}(\rho)$. If the working model happens to be correct, the same $\rho^*$ works for all targets $\tau$, and TFGLS is efficient. If the model is wrong, $\rho^*$ depends on the target $\tau$, and TFGLS is still the most precise GLS estimator $\widehat{\tau}(\rho)$ that the working model can deliver. We develop TFGLS under \citet{BJS2024}'s framework for analysis of staggered adoption of binary treatment, complementing the related work on efficient difference-in-differences estimation; see, e.g., \citet{RothSantanna2023}, \citet{BJS2024}, \citet{Harmon2024}, \citet{CSX2025}, \citet{CuiWooldridge2026}.

Our theoretical contributions are as follows. First, with unrestricted heterogeneous treatment effects and with $T$ fixed and $I\to\infty$, we show that even if $\rho^*$ is estimated at a slow rate, TFGLS is asymptotically equivalent to the infeasible GLS estimator (oracle) that knows $\rho^*$. Hence it is asymptotically at least as precise as OLS and as FGLS based on the same working model. Second, we show that estimation of $\rho^*$ does not suffer from the Nickell-Solon bias that afflicts FGLS in short panels, so the oracle efficiency of TFGLS needs no bias correction, cf. \citet{HansenGLS2007}. We also establish the unbiasedness of TFGLS under \citet{Kakwani1967}-type classical conditions and then, without them, we characterize its higher-order bias, contrasting it with \citet{HansenGLS2007}'s result under homogeneous effects. Third, as a result of independent interest, we show that with the true error covariance the infeasible version of \citet{CSX2025}'s estimator is numerically identical to the infeasible efficient GLS estimator when the covariance is homogeneous within cohorts. This places OLS, FGLS, TFGLS, semiparametric FGLS and \citet{CSX2025} in one framework with a common efficiency benchmark. 

Extensive simulations show that TFGLS can perform very well relative to the extant estimators in terms of root mean squared error and empirical size of tests. As empirical illustrations of the precision of TFGLS relative to the extant methods under contrasting panel dimensions ($I,T$), we revisit \citet{Wolfers2006}'s study ($I \approx 50, T = 33$) of the effect of adoption of unilateral divorce laws on divorce rates, and \citet{Dobkin2018}'s study ($I = 656, T = 4$) of the effect of hospital admissions on out-of-pocket medical expenses. We follow the adaptation of these studies in the related econometric literature:  \citet{dCdH2023-survey} and \citet{SunAbraham2021,CSX2025} respectively.

Related literature: Finding nuisance parameters to minimize the asymptotic risk for estimation of the parameters of interest is a very familiar idea; e.g., cross validation or the optimal weighting matrix in estimating equations. Applications include: parametric Empirical Bayes for incidental parameters in the conditional score function \citep{Lindsay1985}; weighted least squares (WLS) \citep{Cragg1992}\footnote{\citet{Cragg1992} minimizes the trace or determinant of the asymptotic variance of the WLS estimator of the regression coefficients to define the $\rho^*$ in a possibly incorrect model of heteroskedasticity. \citet{SC-WLS19} generalizes its estimation of $\rho^*$. \citet{SCER22b} extend it to other estimators and emphasize obtaining $\rho^*$ for scalar targets (e.g., C-optimality) instead of \citet{Cragg1992}'s objective of A or D optimality.}, estimation with missing data \citep{RubinvdL2008,CTD09}, shrinkage estimation of heteroskedastic hierarchical models and fixed effects \citep{Xieetal2012,Kwon2025}. Variance targeting by nuisance parameters has also been applied to grouped (panel/longitudinal) data settings like ours in several largely independently developed strands of work. \citet{FanHuangLi2007} minimize the generalized variance \citep[like][]{Cragg1992} of the time-invariant coefficients' estimators in a semiparametric partially linear model (with time-varying coefficients) over the parameters of a parametric working correlation. With homogeneous coefficients/effects, \citet{Westgate2014,Westgate2016} select working correlations for generalized estimating equations using criteria aimed at coefficient precision, \citet{YoungShah2024,YoungShahSR2024} use sandwich (robust-variance) criteria to select weights or estimators within specified classes for estimation in partially linear models and generalized linear models, while \citet{SCYN2024}, building on \citet{SCER22b}, develop related targeting for classical fixed effects panels. For staggered adoption with heterogeneous effects, \citet{KennedyShaffer2026} chooses the variance-minimizing weights for combining unbiased $2\times 2$ DiD comparisons under a working covariance whose parameters are fixed in advance, and \citet{LyuWang2026} combine candidate correlation structures by quadratic inference functions \citep{QuLindsayLi2000} in randomized stepped-wedge trials. Relative to these, TFGLS estimates the working model parameters from the data to minimize the variance of estimation of a given \citet{BJS2024}-type aggregate of unrestricted heterogeneous effects, and comes with asymptotic theory, including its higher-order bias.

Outline: Section~\ref{sec:background} is the background. Section~\ref{sec:tfgls} presents TFGLS and its asymptotic properties. Section~\ref{sec:mc} studies small-sample behavior by simulations. Section~\ref{sec:empirical} presents two empirical illustrations. Section~\ref{sec:conclusion} concludes. {\hyperref[sec:proofs]{Appendix~A}} proves the results from the main text. Appendix~\ref{app:aux} (supplemental) contains further theoretical and simulation results.

\section{Background}\label{sec:background}

\subsection{Motivation using a canonical example}\label{sec:motivation}

Let $y_{i} = (y_{i,1},\ldots, y_{i,T})'$ be independent and identically distributed (i.i.d.) for $i=1,\ldots,I$ with $E[y_i] = \tau 1_{T}$ and $Var(y_i) = \Sigma$, both unknown. $1_T$ denotes a $T \times 1$ vector of ones. The infeasible efficient GLS estimator of $\tau$ is the best linear unbiased estimator (BLUE):

\begin{equation*}
\widehat{\tau}_{\text{blue}} := \frac{1_T'\Sigma^{-1}\bar{y}}{1_T'\Sigma^{-1}1_T} \text{~~with~~} Var(\widehat{\tau}_{\text{blue}}) = V_{\text{blue}} := \frac{1}{I} \frac{1}{1_T'\Sigma^{-1}1_T},
\end{equation*}   
where $\bar{y} := \frac{1}{I}\sum_{i=1}^I y_i$. The OLS estimator $\widehat{\tau}_{ols}  = \frac{1}{T}1'_T\bar{y}$ is not $\widehat{\tau}_{\text{blue}}$ in general unless $1_T$ is an eigenvector of $\Sigma$, e.g., $\Sigma$ is spherical. However, one can estimate $\Sigma$ nonparametrically using its sample analog $\widehat{\Sigma} := \frac{1}{I-1}\sum_i (y_i - \bar{y})y'_i$ and subsequently obtain a semiparametric FGLS estimator:   
\begin{equation*}
\widehat{\tau}_{\text{KFGLS}} := \frac{1_T'\widehat{\Sigma}^{-1}\bar{y}}{1_T'\widehat{\Sigma}^{-1}1_T}.
\end{equation*}   
We denote estimators of this kind as KFGLS in reference to \citet{Kiefer1980}. The variance of $\widehat{\tau}_{\text{KFGLS}}$ can be severely affected by the estimation of $T(T+1)/2$ unrestricted parameters in $\Sigma$ when $T$ is not too small relative to $I$. Consider normality to see this cleanly.

\begin{lemma}
\label{lemma:KFGLSmse}
Let $y_{i} \sim \text{Normal}(\tau 1_T, \Sigma)$ i.i.d. for $i=1,\ldots,I$. Then $E[\widehat{\tau}_{\text{KFGLS}}] = \tau$ and 
\begin{equation*}
Var(\widehat{\tau}_{\text{KFGLS}}) = V_{\text{blue}}\left[1 + \frac{T-1}{I - 1 - T}\right] \text{~~for~~} I > T+1. 
\end{equation*}
\end{lemma}

\noindent \textbf{Remark:} Lemma \ref{lemma:KFGLSmse} suggests that the promise of precision like BLUE may remain largely unrealized in smaller samples, e.g., $I = 50 $ (US states) and $T=10,20$ time periods where the finite sample variance becomes approximately 23.1\% and 65.5\% bigger than $V_{\text{blue}}$. 

\citet{HansenGLS2007} presents Monte Carlo evidence of similar problems with KFGLS (he calls it ``FGLS-U'') in DiD for both underestimation of true variability (in his Table 2) and its consequence on reliability of inference (in his Tables 4 and 5) with $I=51$ and $ T = 6, 12, 23$ and notes ``While this is unsurprising for moderate T, the poor performance with $T= 6$ suggests that even with a reasonably short time series dimension the added variability induced by estimating an unconstrained variance matrix poses a serious problem for inference''. \citet{HausmanKuersteiner2008} reach similar conclusions for inference in DiD: with an unrestricted $\widehat{\Sigma}$, the first-order (K)FGLS $t$-test ignores the estimation of $\Sigma$, and the second-order size correction they derive grows with $T$ relative to $I$; their simulations use $I=50$ and $T=5,10, 15, 20$ and note that unrestricted ``feasible GLS does suffer from a many parameter problem and exhibits severe small sample size distortions when N [$I$] is not large enough''. Thus, despite its promise of asymptotic efficiency, one should be careful with KFGLS if $I$ is not too large.  

Parametric FGLS reduces the dimension of the serial correlation problem from $T$ to, e.g., $d_\rho$ for some small $d_\rho$ by using a parsimonious and very likely misspecified working model for $\Sigma$. Compared to KFGLS, this reduction in dimension helps with more reliable inference, and in some cases can help with the precision. However, such FGLS can be even less precise than OLS when the working model is incorrect. To address this and also increase the likelihood of small-sample precision gains over KFGLS, while maintaining the parsimony of the working model to help with reliable inference, we will work with the targeting approach in our paper.

Parsimony comes at a cost: unless the working model is rich enough --- as for KFGLS here or, more generally, by letting $d_\rho$ grow with $I$ --- FGLS, targeted or not, is generally not semiparametrically efficient. Our paper is therefore complementary to those that pursue semiparametric efficiency, since our approach of targeting to extract all available benefits from given parsimonious working models (chosen by user) provides a useful alternative when $I$ is not very large, e.g., $I=50$ (states), i.e., cases where the asymptotic guarantees of good precision and reliable inference are, ceteris paribus, more difficult (Lemma \ref{lemma:KFGLSmse}).

\setdisp{7pt plus 3pt minus 3pt}{8pt plus 3pt minus 3pt}
\subsection{Framework and preliminaries}\label{sec:framework}

\textbf{Model:} We follow the framework of \citet{BJS2024}, \citet{Gardner2021}, \citet{GTTY2024}, and \citet{Wooldridge2025}, and the notation of \citet{BJS2024}; see also \citet{Xu2017}, \citet{AtheyMatCompl2021}, \citet{Harmon2024}, etc. for related and in some cases more general setups. Let $S_i \ge 1$ and $T_i \le T$ be the start and end times for entity $i=1,\ldots,I$ in the data, and $E_i \in \{S_{i}+1,\ldots, T_i, \infty\}$ be its event date, with $E_i := \infty$ if $i$ is never treated. Let $\tau_{i,t}$ be the $(i,t)$-specific treatment effect. The outcome model for entity $i$ at time $t = S_i,\ldots,T_i$ is:
\begin{equation}
\label{eq:model}
y_{i,t} ~=~ a'_{i,t}\lambda_i + x'_{i,t}\beta + 1(t \ge E_i)\tau_{i,t} + u_{i,t} ~\equiv~ r'_{i,t}\eta + 1(t \ge E_i)\tau_{i,t} + u_{i,t}.
\end{equation}
$a_{i,t}$ can be, e.g., $a_t = 1$ or $[1, t]'$, and $x_{i,t}$ can include time dummies; $r_{i,t}$ stacks them and $\eta$ stacks their coefficients (entity effects, time effects, etc.), subject to the usual normalization.

\textbf{Notation:} For a generic $z_{i,t}$ representing $y_{i,t}$, $u_{i,t}$, $r_{i,t}$, etc., stack the $N_{[0],i} := E_i - S_i$ and $N_{[1],i} := T_i - E_i + 1$ pre and post treatment observations for $i$ to obtain:
\begin{equation}
\label{eq:grouping}
z_i := \left(z'_{[0],i},z'_{[1],i}\right)'\text{~~where~~} z_{[0],i} := \left(z'_{i,S_i},\ldots, z'_{i,E_i-1}\right)' \text{~~and~~} z_{[1],i} := \left(z'_{i,E_i},\ldots, z'_{i,T_i}\right)'.
\end{equation}
Of course if $E_i = \infty$, i.e., $i$ is never treated, then $N_{[0],i} = T_i - S_i +1$, $N_{[1],i} = 0$, $z_i = z_{[0],i}$ and $z_{[1],i}$ is null. We sometimes also group observations by entity nested inside treatment status:
\begin{equation}
\label{eq:grouping2}
Z_{[0]} := \left(z'_{[0],1},\ldots,z'_{[0],I}\right)' \text{~~and~~} Z_{[1]} := \left(z'_{[1],1},\ldots,z'_{[1],I}\right)' \text{~~giving~~} Z := \left(Z'_{[0]}, Z'_{[1]}\right)'.
\end{equation} 
$N_{[j]} := \sum_{i=1}^I N_{[j],i}$ for $j=0,1$ are the total number of pre and post treatment observations. 
\medskip

\textbf{GLS construction:} Let $D := \left(0_{N_{[1]} \times N_{[0]}}, I_{N_{[1]}}\right)'$. GLS estimator (see Appendix~\ref{app:glsderiv} for details) solves:
\begin{equation*}
\min_{\eta, \tau} \left(Y - R\eta - D\tau\right)'\Omega \left(Y - R\eta - D\tau\right)
\end{equation*} 
for some $N \times N$ weight matrix $\Omega =(\Omega_{[j,k]})_{j,k=0}^1$, where $N := N_{[0]}+ N_{[1]}$. $\Omega = I_N$ gives OLS:
\begin{equation*}
\widehat{\eta}_{\ols} ~:=~ \widehat{\eta}(I_N)~=~ \left(R'_{[0]}R_{[0]} \right)^{-1}R'_{[0]}Y_{[0]}, \text{~~and~~}\widehat{\tau}_{\ols} ~:=~ \widehat{\tau}(I_N) ~=~  Y_{[1]} - R_{[1]}\widehat{\eta}_{\ols}
\end{equation*}
that is studied in \citet{BJS2024}, \citet{GTTY2024}, \citet{Wooldridge2025}, etc. $\Omega = \Sigma^{-1}$ with $\Sigma := \Var(U)$ gives BLUE GLS (see Section A.5, \citet{BJS2024}):
\begin{equation*}
\begin{aligned}
\widehat{\eta}_{\text{blue}} ~:=~& \widehat{\eta}(\Sigma^{-1}) = \left(R'_{[0]}\Sigma^{-1}_{[00]}R_{[0]} \right)^{-1}R'_{[0]}\Sigma^{-1}_{[00]}Y_{[0]}\\
\widehat{\tau}_{\text{blue}} ~:=~& \widehat{\tau}(\Sigma^{-1}) = \left(Y_{[1]} - R_{[1]}\widehat{\eta}_{\text{blue}}\right) - \Sigma_{[10]}\Sigma^{-1}_{[00]}\left(Y_{[0]} - R_{[0]}\widehat{\eta}_{\text{blue}}\right).
\end{aligned}
\end{equation*} 
While $\widehat{\tau}_{\ols}$ is trivial to implement, $\widehat{\tau}_{\text{blue}}$ requires knowledge of $\Sigma$, which is typically unknown.

\textbf{Parsimonious working model:} It is evident that there are two blocks of $\Sigma$, namely, $\Sigma_{[00]}$ and $\Sigma_{[10]}$, for which one would posit working models. As noted in Section~\ref{sec:motivation}, too general working models may be harmful. \citet{BJS2024} warn ``A model which allows the variances to vary arbitrarily across cohort–periods may have too many parameters to allow for an effective estimation of the variance as soon as $T$ is moderately large, in which case the theoretical efficiency gain may not materialize in smaller samples'' and recommend ``In particular, one could model the errors as an autoregressive process with homoskedastic innovations (e.g., AR(1)), estimate the autoregression parameters, and achieve an efficiency improvement.'' Therefore, utilizing the independence across $i$'s, consider working models known up to a small number of parameters, $\rho_0$ for $\bar\Sigma_{[00]}$ and $\rho_{10}$ for $\bar\Sigma_{[10]}$:
\begin{equation*}
\underset{N_{[0]} \times N_{[0]}}{\underbrace{\bar{\Sigma}_{[00]}(\rho_0)}} =  \text{diag}\left(\underset{N_{[0],1} \times N_{[0],1}}{\underbrace{\bar{\Sigma}_{[00],1}(\rho_0)}},\ldots, \underset{N_{[0],I} \times N_{[0],I}}{\underbrace{\bar{\Sigma}_{[00],I}(\rho_0)}} \right),~ 
\underset{N_{[1]} \times N_{[0]}}{\underbrace{\bar{\Sigma}_{[10]}(\rho_{10})}} =  \text{diag}\left(\underset{N_{[1],1} \times N_{[0],1}}{\underbrace{\bar{\Sigma}_{[10],1}(\rho_{10})}},\ldots, \underset{N_{[1],I} \times N_{[0],I}}{\underbrace{\bar{\Sigma}_{[10],I}(\rho_{10})}} \right).
\end{equation*}      
We maintain that $\bar{\Sigma}_{[00]}(0) = I_{N_{[0]}}$ and $\bar{\Sigma}_{[10]}(0) = 0_{N_{[1]}\times N_{[0]}}$, i.e., OLS is a special case.

\smallskip

\textbf{Target estimand:} Following \citet{BJS2024}, our scalar parameter of interest is:
\begin{equation}
\label{eq:target}
\tau_w \equiv \tau_{w,I} := \sum_{i=1}^I \sum_{t=E_i}^{T_i} w_{i,t}\tau_{i,t} = \sum_{i=1}^I w'_{[1],i}\tau_{[1],i} =  w'_{[1]}\tau_{[1]} 
\end{equation} 
where $w_{i,t}$'s are some user-specified weights. We introduced compact representation in \eqref{eq:target} using the notation $\tau_{[1],i} := \left(\tau_{i,E_i},\ldots, \tau_{i,T_i}\right)'$ for each $i$ and stack them as the $N_{[1]} \times 1$ vector $\tau_{[1]} := \left(\tau'_{[1],1},\ldots, \tau'_{[1],I}\right)'$. Similarly we defined $w_{[1],i}$ and $w_{[1]}$ for the user-specified weights.

\smallskip

\textbf{Generic estimator:} Based on our above discussion of: (i) the construction of  least squares estimators and (ii) the parsimonious working model indexed by $\rho$ consisting of the distinct elements in $\rho_0,\rho_{10}$, we will focus on the generic GLS estimator of the target $\tau_w$:
\begin{equation}
\label{eq:gls}
\begin{aligned}
\widehat{\tau}_{I}(\rho) ~~=~~& w'_{[1]}\left\{\left(Y_{[1]} - R_{[1]}\widehat{\eta}(\rho_0)\right) ~-~ \bar{\Sigma}_{[10]}(\rho_{10})\bar{\Sigma}^{-1}_{[00]}(\rho_0) \left(Y_{[0]} - R_{[0]}\widehat{\eta}(\rho_0)\right)\right\},\\
\widehat{\eta}(\rho_0) ~~=~~& \left(R'_{[0]}\bar{\Sigma}^{-1}_{[00]}(\rho_0)R_{[0]} \right)^{-1}R'_{[0]}\bar{\Sigma}^{-1}_{[00]}(\rho_0)Y_{[0]}.
\end{aligned}
\end{equation}
An important special case of \eqref{eq:gls} is the aforementioned OLS estimator:\footnote{\label{fn:ar1}See from \eqref{eq:gls} that an AR(1) working model with $\rho_0$ as the AR(1) coefficient for $u_{i,t}$'s until $t= E_i-1$ (untreated periods) and $\rho_1$ that from $t=E_i-1$ onward (so $\rho_{10} = \rho = (\rho_0,\rho_1)'$), gives (see Appendix~\ref{app:ar1wm}): \begin{equation*}
\widehat{\tau}_{I}(\rho) = \sum\nolimits_{i=1}^I\sum\nolimits_{t=E_i}^{T_i} w_{i,t}\left\{\widehat{u}_{i,t}(\rho_0) ~-~ \rho_1^{\,t-E_i+1}\,\widehat{u}_{i,E_i-1}(\rho_0)\right\}, \quad \widehat{u}_{i,t}(\rho_0) := y_{i,t} - r'_{i,t}\widehat{\eta}(\rho_0)
\end{equation*}
$\rho_0$ determines how $\eta$ is fitted from the untreated cells, and $\rho_1$ how much of the last untreated residual of entity $i$ (since AR(1) is Markov) is carried forward to impute the untreated outcome of its treated cells. Special cases. (i) $\rho = 0$ gives the OLS imputation estimator of \citet{BJS2024}: $\eta$ is fitted by OLS and nothing is carried forward. (ii) $\rho_1 = 0$ fits $\eta$ by GLS but carries nothing forward, whereas $\rho_0 = 0 \neq \rho_1$ does the reverse. (iii) $\rho_0 = \rho_1 = \rho$ is the stationary AR(1) model used for most of our numerical results; for $|\rho| < 1$ the weight $\rho^{t-E_i+1}$ on the entity's own last residual decays with the horizon $t - E_i + 1$, so the imputation relies more on the fitted $\eta$ at longer horizons. (iv) With $a_{i,t}=1$ and $\rho_1=1$, $\widehat{\lambda}_i(\rho_0)$ cancels for every $\rho_0$. If also $\rho_0\to1$, the random walk limit, $\widehat{\beta}(\rho_0)$ tends to the first-difference fit on consecutive untreated periods, and $\widehat{\tau}_{I}(\rho)$ tends to $\sum_{i,t} w_{i,t}\{(y_{i,t} - y_{i,E_i-1}) - (x_{i,t} - x_{i,E_i-1})'\widehat{\beta}_{\text{FD}}\}$. It is BLUE under random walk errors with i.i.d.\ shocks. If $x_{i,t}$ consists of time dummies, it is the stepwise DiD (SWDD) estimator of \citet{Harmon2024}. His subgroup DiD estimator coincides with it at horizon 0 and in balanced panels without staggered adoption. Unlike Harmon's, the limit also allows missing periods: a first difference over a gap of $k$ periods gets weight $1/k$. $\rho_0 = 1$ is excluded to keep $\bar{\Sigma}_{[00]}(\rho_0)$ nonsingular. With population weights like Section~\ref{sec:emp-divorce}, the limit differs (end of {\hyperref[sec:proofs]{Appendix~A}}). Finally, $\bar{\Sigma}_{[11]}(.)$ never enters \eqref{eq:gls}, so $\rho_1$ need not lie in $(-1,1)$.}
\begin{equation}
\label{eq:ols special gls}
\widehat{\tau}_{w,\ols} \equiv \widehat{\tau}_{I}(0), ~~ \widehat{\eta}_{\ols} \equiv \widehat{\eta}(0).
\end{equation}

Since any parsimonious working model is very likely incorrect, there is no guarantee that the standard FGLS --- i.e., one that estimates its $\rho$ by, e.g., least-squares fitting the working model to the error covariance --- is more precise than $\widehat{\tau}_{w,\ols}$. By contrast, we take a direct route to optimality by proposing to use the $\rho$-value that minimizes the asymptotic variance of $\widehat{\tau}_I(\rho)$ over $\rho \in \mathcal{P}$. Since this $\rho$-value is unknown, the feasible version of our estimator, the targeted feasible GLS (TFGLS) estimator of Section~\ref{sec:tfgls}, uses an estimate of it.

TFGLS builds on three background results, developed under standard assumptions in Section~\ref{sec:bgresults}: (i) the asymptotic distribution of $\widehat{\tau}_I(\rho)$, (ii) an oracle property, and (iii) the uniform convergence of the estimator of its asymptotic variance, i.e., of the TFGLS objective.

\subsection{Background results and assumptions}\label{sec:bgresults}

By the outcome model \eqref{eq:model}, $\widehat{\tau}_{I}(\rho)$ is linear in the errors:
\begin{equation*}
\widehat{\tau}_{I}(\rho) ~=~ \Kappa(\rho) Y ~=~ \sum_{i=1}^I \kappa'_{i,I}(\rho)y_i  ~=~ \tau_w + \sum_{i=1}^I \kappa'_{i,I}(\rho)u_i \equiv \tau_w + \sum_{i=1}^I \left(\kappa'_{[0],i,I}(\rho)u_{[0],i} + \kappa'_{[1],i,I}u_{[1],i}\right),
\end{equation*}
where the row vector $\Kappa(\rho)$ collects the entity-level weights $\kappa_{i,I}(\rho) := \left(\kappa'_{[0],i,I}(\rho),\kappa'_{[1],i,I}\right)'$ in the ordering of $Y$, $\kappa_{[1],i,I} := w_{[1],i}$ are the weights in the target $\tau_w = \sum_{i=1}^I w'_{[1],i}\tau_{[1],i}$, and: \begin{equation*}
\begin{aligned}
\kappa_{[0],i,I}(\rho) ~:=~ & - \bar{\Sigma}^{-1}_{[00],i}(\rho_0)\bar{\Sigma}_{[01],i}(\rho_{10})w_{[1],i} \\
~~ &  -  \bar{\Sigma}^{-1}_{[00],i}(\rho_0)r_{[0],i}\left(\sum_{j=1}^I r'_{[0],j} \bar{\Sigma}^{-1}_{[00],j}(\rho_0)r_{[0],j}\right)^{-1}  \left(\sum_{j=1}^I \left[r_{[1],j} -  \bar{\Sigma}_{[10],j}(\rho_{10})\bar{\Sigma}^{-1}_{[00],j}(\rho_0)r_{[0],j}\right]'w_{[1],j}\right).  
\end{aligned}
\end{equation*}

Consider the class of estimators $\{\widehat{\tau}_{I}(\rho):\rho \in \mathcal{P}\}$ where the compact $\mathcal{P} \subset \mathbb{R}^{d_\rho}$, specified with the working model by the user, contains $\rho = 0_{d_\rho}$, i.e., OLS is a special case. Under the assumptions below, the estimators $\widehat{\tau}_{I}(\rho)$'s are unbiased for $\tau_w$ with variances:
\begin{equation}
\label{eq:2ndmom}
\sigma^2_I(\rho) := V\left(\widehat{\tau}_{I}(\rho)\right) = \Kappa(\rho) \Sigma \Kappa'(\rho) = \sum_{i=1}^I \kappa'_{i,I}(\rho)\Sigma_i\kappa_{i,I}(\rho), \qquad \Sigma_i := \Var(u_i).
\end{equation}
As in \citet{BJS2024}, we treat the regressors as fixed (equivalently, condition on them) and state the assumptions at the entity level. $T$ is considered fixed.

\medskip
\noindent {\textbf{Assumptions:}}
\begin{itemize}
\item[(A1)] $u_i$'s are independent with $E[u_i] = 0$ for all $i$ and $\sup_{i} E \left[\|u_i\|^{2+\delta}\right] < \infty$ for some $\delta >0$.

\item[(A2)] $K_I := \sup_{\rho \in \mathcal{P}} \sum_{i=1}^I \|\kappa_{i,I}(\rho)\|^2 = O(1/I)$. $\sup_{\rho \in \mathcal{P}} \max_{1 \le i \le I} \|\kappa_{i,I}(\rho)\|^2 = o(K_I)$.

\item[(A3)] $\sigma^2_I(\rho) = \sum_{i=1}^I \kappa'_{i,I}(\rho)\Sigma_i\kappa_{i,I}(\rho)$ is non-degenerate: $\liminf_{I \rightarrow \infty} \inf_{\rho \in \mathcal{P}} I  \sigma^2_I(\rho) > 0$.

\item[(A4)] For all $i, I$, the entity-level weights are Lipschitz on $\mathcal{P}$: there exist constants $c_{i,I} \geq 0$ such that $\left\|\kappa_{i,I}(\rho) - \kappa_{i,I}(\rho')\right\| \leq c_{i,I}\left\|\rho - \rho'\right\|$ for all $\rho, \rho' \in \mathcal{P}$, and $\sum_{i=1}^I c^2_{i,I} = O(1/I)$.
\end{itemize}

\noindent \textbf{Remark:} (A1)--(A3) are very similar to those in \citet{BJS2024} up to and including their Proposition 8; see their extensive discussion. We additionally impose uniformity over the \textit{user-specified} parameter space $\mathcal{P}$ because TFGLS is based on the uniform convergence of the variance estimator of $\widehat{\tau}_{I}(\rho)$. (A4) is the Lipschitz analogue of the first part of (A2), and guarantees the asymptotic properties of TFGLS even when the estimated $\rho$ converges very slowly. Primitive sufficient conditions on the working model and the target weights for (A2) and (A4) are given in Proposition \ref{prop:primitive} in Appendix~\ref{app:primitive} along with further discussion.

\begin{proposition}
\label{prop:cons}
The model \eqref{eq:model}, and assumptions (A1)--(A2) imply that for every sequence $\{\rho_I\} \in \mathcal{P}$: $ \widehat{\tau}_{I}(\rho_I) - \tau_w ~\xrightarrow{p}~ 0$ as $I \to \infty$.
\end{proposition}

\begin{proposition}
\label{prop:clt}
The model \eqref{eq:model}, and assumptions (A1)--(A3) imply that for every sequence $\{\rho_I\} \in \mathcal{P}$:
\[
  \frac{\widehat{\tau}_{I}(\rho_I) - \tau_w}{\sigma_I(\rho_I)} ~\xrightarrow{d}~ \mathrm{Normal}(0, 1)
  \quad \text{as } I \to \infty.
\]
\end{proposition}

\begin{proposition}
\label{prop:oracle}
The model \eqref{eq:model}, and assumptions (A1)--(A4) imply that for any estimator $\widehat{\rho}_I$ such that $\widehat{\rho}_I - \rho_I = o_P(1)$ for some sequence $\{\rho_I\} \in \mathcal{P}$:
\[
  \frac{\widehat{\tau}_{I}(\widehat{\rho}_I) - \tau_w}{\sigma_I(\rho_I)} ~=~ \frac{\widehat{\tau}_{I}(\rho_I) - \tau_w}{\sigma_I(\rho_I)} + o_P(1) ~\xrightarrow{d}~ \mathrm{Normal}(0, 1)
  \quad \text{as } I \to \infty.
\]
\end{proposition}

\medskip

\noindent \textbf{Remark:} $\widehat{\tau}_{w,\ols} \equiv \widehat{\tau}_{I}(0)$; see \eqref{eq:ols special gls}. If the working model is correct up to scale, i.e., $(\Sigma_{[00]},\Sigma_{[10]}) =\allowbreak c\,(\bar{\Sigma}_{[00]}(\rho^\dag_{0,I}), \bar{\Sigma}_{[10]}(\rho^\dag_{10,I}))$ for some $c>0$ and $\rho^{\dag}_I$, then $\widehat{\tau}_{I}(\rho^{\dag}_I) = w'_{[1]}\widehat{\tau}_{\text{blue}}$, the BLUE, and, by Proposition~\ref{prop:oracle}, $\widehat{\tau}_{I}(\widehat{\rho}_I)$ is asymptotically equivalent to it for any $\widehat{\rho}_I$ with $\widehat{\rho}_I - \rho^{\dag}_I = o_P(1)$. 

\smallskip

Finally, for inference and as the key building block of TFGLS, we estimate $\sigma^2_I(\rho)$ in \eqref{eq:2ndmom}, possibly conservatively, following \citet{BJS2024}. For uniformity of presentation we use OLS estimates for this purpose but, thanks to Proposition \ref{prop:cons}, other estimates can also be used. The variance estimator is:
\begin{equation}
\label{eq:var-est}
\widehat{\sigma}^2_I(\rho) ~:=~ \sum_{i=1}^I \left(\kappa'_{i,I}(\rho)\widehat{u}_{i,I}\right)^2 \equiv \sum_{i=1}^I \left(\kappa'_{[0],i,I}(\rho)\widehat{u}_{[0],i,I} + \kappa'_{[1],i,I}\widehat{u}_{[1],i,I}\right)^2
\end{equation} 
imposing independence across entities (see (A1)). Its residuals use both groupings in \eqref{eq:grouping}--\eqref{eq:grouping2}, by entity ($z_i$) and by treatment status ($Z_{[j]}$), which order the same observations differently:  
\begin{equation*}
\widehat{u}_{[0],i,I} := y_{[0],i} - r_{[0],i}\left(R'_{[0]}R_{[0]}\right)^{-1}R'_{[0]}Y_{[0]} = u_{[0],i} - r_{[0],i}\left(R'_{[0]}R_{[0]}\right)^{-1}R'_{[0]}U_{[0]} 
\end{equation*}
is the OLS residual for $i$ in its untreated periods; and in the spirit of \citet{BJS2024}:
\begin{equation*}
\begin{aligned}
\widehat{u}_{[1],i,I} &:=& y_{[1],i} - r_{[1],i}\left(R'_{[0]}R_{[0]}\right)^{-1}R'_{[0]}Y_{[0]} - 1_{N_{[1],i}}\widehat{\tau}_{I}(0) ~\equiv~   \widehat{\tau}_{[1],i,I}(0)  - 1_{N_{[1],i}}\widehat{\tau}_{I}(0)\\
&=& u_{[1],i} - r_{[1],i}\left(R'_{[0]}R_{[0]}\right)^{-1}R'_{[0]}U_{[0]} - 1_{N_{[1],i}}\left(\widehat{\tau}_{I}(0)-\tau_w\right)  +  \left(\tau_{[1],i}  - 1_{N_{[1],i}}\tau_w\right)
\end{aligned}
\end{equation*}
are the treated residuals. As in Theorem 3 of \citet{BJS2024}, uniform consistency of $\widehat{\sigma}^2_I(\rho)$ to a conservative limit requires the following additional assumptions.

\noindent \textbf{Assumptions} (continued)\textbf{:}
\begin{itemize}

\item[(A5)] $\sup_i E[\|u_i\|^4] < \infty$.

\item[(A6)]  $\sup_{i}\,\|x_{i}\| < \infty$. And, $ \frac{1}{I}\sum_{i=1}^{I}
  \tilde{x}_{[0],i}'\,\tilde{x}_{[0],i}$ is finite and positive definite uniformly for all $I$ large, where
$\tilde{x}_{[0],i} := x_{[0],i} - a_{[0],i}(a_{[0],i}'a_{[0],i})^{-1}a_{[0],i}'x_{[0],i}$.

\item[(A7)] $\sup_i \|\tau_{[1],i}\| < \infty$.

\item[(A8)] $\sum_{i=1}^I \|w_{[1],i}\|^2 = O(1/I)$ and $\max_{1 \leq i \leq I} \|w_{[1],i}\|^2 = o\!\left(\sum_{j=1}^I \|w_{[1],j}\|^2\right)$.

\end{itemize}

\begin{proposition}
\label{prop:cons_var}
Consider model \eqref{eq:model}, the target parameter $\tau_w = w'_{[1]}\tau_{[1]}$, and the family of estimators $\wh{\tau}_I(\rho)$ defined in \eqref{eq:gls}, indexed by $\rho \in \Pspace$. Define $\sigma^2_{\tau,I} := \sum_{i} \bigl(w'_{[1],i}(\tau_{[1],i} - 1_{N_{[1],i}}\tau_w)\bigr)^2$. Let $\wh{\sigma}^2_I(\rho)$ be the variance estimator defined in \eqref{eq:var-est}. Under assumptions (A1)--(A8):
\begin{equation*}
\sup_{\rho \in \Pspace} \frac{1}{K_I}\left|\wh{\sigma}^2_I(\rho) - \left(\sigma^2_I(\rho) + \sigma^2_{\tau,I}\right)\right|, \text{~equivalently,~}  \sup_{\rho \in \Pspace} ~I \left|\wh{\sigma}^2_I(\rho) - \left(\sigma^2_I(\rho) + \sigma^2_{\tau,I}\right)\right| ~\xrightarrow{p}~ 0.
\end{equation*}
\end{proposition}

\noindent \textbf{Remark:} The conservativeness of $\wh{\sigma}^2_I(\rho)$ is due to the construction of the treated ``residuals'' $\widehat{u}_{[1],i,I}$, and cannot be avoided without further assumptions \citep{BJS2024}.\footnote{\label{fn:bjs0res}The actual treated residuals are identically 0 by construction. So $\widehat{u}_{[1],i,I}$ uses the aggregate $\widehat{\tau}_{I}(0)$, and the conservativeness reflects the heterogeneity of $\tau_{i,t}$ around it. Proposition~\ref{prop:cons_var} holds if, cell by cell, $\widehat{\tau}_{I}(0)$ is replaced by the aggregate $\tilde\tau_g$ of a fixed group $g$ of treated cells, provided $\sup_{\rho\in\mathcal P}(\tilde\tau_g - c_g)^2 = O_P(K_I)$ for nonrandom bounded $c_g$, which replaces $\tau_w$ in $\sigma^2_{\tau,I}$. This fails if a few entities dominate a group, e.g., a singleton cohort, whose treated residuals are then 0. Sections~\ref{sec:mc} and \ref{sec:emp-hrs} use \citet{BJS2024}'s optimal cohort-period grouping, weighting cells by their squared target weights, for all methods --- including TFGLS --- related to theirs. Section~\ref{sec:emp-divorce} has singleton treated cohorts and hence it instead pools all treated cells.} Finally, given a consistent estimator $\widehat{\rho}_I$ for some $\rho_I$, one can perform conservative inference asymptotically equivalently to the ``oracle'' that knows $\rho_I$: by Propositions \ref{prop:clt}--\ref{prop:cons_var}, assumption (A3), and the equicontinuity of $I\sigma^2_I(\cdot)$ along $\rho_I$ (by (A1), (A2) and (A4)) (see, e.g., \eqref{eq:equicont}), 
\begin{equation}
\label{eq:equivconsinfer}
\frac{\widehat{\tau}_{I}(\widehat{\rho}_I) - \tau_w}{\widehat{\sigma}_I(\widehat{\rho}_I)} ~~=~~ \frac{\widehat{\tau}_{I}(\rho_I) - \tau_w}{\widehat{\sigma}_I(\rho_I)} + o_P(1) ~~=~~ \frac{\widehat{\tau}_{I}(\rho_I) - \tau_w}{\sigma_I(\rho_I)}\sqrt{\frac{\sigma^2_I(\rho_I)}{\sigma^2_I(\rho_I) + \sigma^2_{\tau,I}}} ~+~ o_P(1).\end{equation}
Hence, inference based on $\widehat{\sigma}_I(\widehat{\rho}_I)$ and normal critical values is asymptotically conservative, and exact if $I\sigma^2_{\tau,I} \to 0$, e.g., if all $\tau_{i,t}$ equal $\tau_w$; Proposition \ref{prop:tfgls} below applies this to TFGLS.

\setdisp{12pt plus 3pt minus 7pt}{12pt plus 3pt minus 7pt}\section{TFGLS: Theory}
\label{sec:tfgls}

\subsection{Definition}\label{sec:tfgls-def}

We define the target $\tau_w$-specific optimal sequence $\rho^*_I \in \mathcal{P}$ for TFGLS by using \eqref{eq:2ndmom} as:
\begin{equation}
\label{eq:optrho}
\rho^*_I := \arg\min_{\rho \in \mathcal{P}} I\sigma^2_I(\rho).
\end{equation}
Since $\mathcal{P}$ is compact, the existence of $\rho^*_I$ in \eqref{eq:optrho} follows for all $I$ by continuity of $\sigma^2_I(\rho)$ following from (A4). Targeting this optimal $\rho^*_I$ for plug-in, we define the TFGLS estimator as:
\begin{equation}
\label{eq:tfgls}
\widehat{\tau}_{w,\tfgls} := \widehat{\tau}_I\left(\widehat{\rho}^*_I\right)
\end{equation}
using \eqref{eq:gls}, and with the estimated plug-in $\widehat{\rho}^*_I$ defined by using \eqref{eq:var-est} as:
\begin{equation}
\label{eq:estrho}
\widehat{\rho}^*_I := \arg\min_{\rho \in \mathcal{P}} I\widehat{\sigma}^2_I(\rho).
\end{equation}

\noindent \textbf{Remark:} By Proposition \ref{prop:cons_var}, uniformly over $\rho \in \mathcal{P}$, the sample objective in \eqref{eq:estrho} differs from the population objective in \eqref{eq:optrho} asymptotically only by the additive term $I\sigma^2_{\tau,I}$, which does not depend on $\rho$ and so does not affect the minimizer of the probability limit of $I\widehat{\sigma}^2_I(\rho)$. Hence, with fixed $T$ and without bias correction, $\widehat{\rho}^*_I$ consistently estimates $\rho^*_I$, whereas conventional FGLS, which fits its working model to residuals contaminated by $\widehat{\lambda}_i - \lambda_i$, suffers from the Nickell-Solon bias; see, e.g., \citet[between his equations (8) and (9)]{HansenGLS2007}.\footnote{\label{fn:NickellBias}\lineskip=0.5pt Consider $y_{i,t} = \lambda_i + u_{i,t}$ with $i=1,\ldots,I$ and $t = 1,\ldots,T$ to focus on the problem that conventional FGLS faces in fitting, e.g., an AR(1) model using OLS residuals $\widehat{u}_{i,t} := y_{i,t} - \widehat{\lambda}_i =  u_{i,t} + (\lambda_i - \widehat{\lambda}_i)=  u_{i,t} - \bar{u}_i$ where $\bar{u}_i := \frac{1}{T}\sum_{s=1}^T u_{i,s}$. The FGLS sample objective $\hat{Q}(\rho) = \frac{1}{I(T-1)}\sum_{i=1}^I\sum_{t=2}^T(\widehat{u}_{i,t} - \rho\widehat{u}_{i,t-1})^2$ converges in probability to 
$\bar{Q}(\rho) = \frac{1}{T-1}\sum_{t=2}^T E[(\widehat{u}_{i,t} - \rho\widehat{u}_{i,t-1})^2]$ under standard conditions, 
whereas the population objective is $Q(\rho) = E[(u_{i,t} - \rho u_{i,t-1})^2]$. 
Since $\widehat{u}_{i,t} - \rho\widehat{u}_{i,t-1} = (u_{i,t} - \rho u_{i,t-1}) - (1-\rho)\bar{u}_i$, we obtain: $\bar{Q}(\rho) - Q(\rho) = -2(1-\rho)\frac{1}{T-1}\sum_{t=2}^T E[(u_{i,t} - \rho u_{i,t-1})\bar{u}_i] + (1-\rho)^2 E[\bar{u}_i^2]$ which is $O(1/T)$ but nonzero for fixed $T$ in general unless $\rho = 1$. This discrepancy depends on $\rho$, so the sample minimizer $\hat{\rho}:= \arg\min_\rho \hat{Q}(\rho)$ converges 
to $\bar{\rho} = \arg\min_\rho \bar{Q}(\rho)$ which is generally not equal to the population minimizer $\rho_0:= \arg\min_\rho Q(\rho)$. Coming back to our paper, as the proof of Proposition \ref{prop:cons_var} explains, the entity fixed effects cannot have the same adverse effect in our case because $\widehat{\tau}_I(\rho)$ is invariant to them, $\kappa'_{i,I}(\rho)a_i = 0$ for every $\rho$, so the estimation errors $\widehat{\lambda}_i(0) - \lambda_i$ ($\rho = 0$ for OLS) cancel exactly in \eqref{eq:var-est}, while those of $\widehat{\beta}(0)$ and $\widehat{\tau}_I(0)$ vanish as $I \to \infty$.}

\setdisp{7pt plus 3pt minus 3pt}{6pt plus 3pt minus 3pt}
\subsection{First-order asymptotic properties}\label{sec:tfgls-first}

\noindent \textbf{Assumptions} (continued)\textbf{:}
\begin{itemize}

\item[(A9)] $\rho^*_I$ in \eqref{eq:optrho} is well separated: $\liminf_I \inf_{\rho \in \mathcal{P}: \|\rho - \rho^*_I\| > \epsilon}  I(\sigma^2_I(\rho) - \sigma^2_I(\rho^*_I)) > 0$ for all $\epsilon >0$. 

\end{itemize}

\begin{lemma}
\label{lemma:rhocons}
Consider model \eqref{eq:model}, and let assumptions (A1)--(A9) hold. Then: $\widehat{\rho}^*_I - \rho^*_I = o_P(1)$. 
\end{lemma}

\noindent \textbf{Remark:} Given the drifting target $\rho^*_I$, Assumption (A9) requires the separation gap to hold uniformly in large $I$ in order to apply standard arguments, e.g., Theorem 5.7 of \citet{Aad98}, in the proof of Lemma \ref{lemma:rhocons}. Given $\widehat{\rho}^*_I - \rho^*_I = o_P(1)$, the oracle-efficiency of TFGLS in Proposition
\ref{prop:tfgls} follows by Proposition \ref{prop:oracle} and the steps used for \eqref{eq:equivconsinfer}.\footnote{(A9) requires that $\rho^*_I$ be unique. $\sigma^2_I(\rho)$ is a rational function of $\rho$ for standard AR working models, so it may not be convex and hence $\rho^*_I$ may not be unique. Since this may not be of first-order importance in practice, we relegate the treatment of non-unique $\rho^*_I$ to Appendix~\ref{app:multirho}. It shows that TFGLS continues to be $\sqrt{I}$-consistent without (A9) (Proposition \ref{prop:tfgls-not-unique-rho}). For the cases where non-uniqueness could possibly distort the asymptotic normality result below in Proposition \ref{prop:tfgls}, it restores asymptotic normality by drawing $\rho$ from a vanishing band around the minimized criterion, using randomness independent of $U$ (Proposition \ref{prop:bandTFGLS}). The novelty (to our knowledge) of this is to let this band vanish at a rate slower than the rate of convergence of $\widehat{\sigma}^2_I(\rho)$ so the randomization in effect acts on $\sigma^2_I(\rho)$, a non-random quantity, breaking the dependence on $U$.}

\begin{proposition}
\label{prop:tfgls}
Consider model \eqref{eq:model}, and let assumptions (A1)--(A9) hold. Then:
\begin{equation*}
\frac{\widehat{\tau}_{w,\tfgls} - \tau_w}{\widehat{\sigma}_I(\widehat{\rho}^*_I)}  ~~=~~  \frac{\widehat{\tau}_{I}(\rho^*_I) - \tau_w}{\widehat{\sigma}_I(\rho^*_I)} + o_P(1) ~~=~~  \frac{\widehat{\tau}_{I}(\rho^*_I) - \tau_w}{\sigma_I(\rho^*_I)}\sqrt{\frac{\sigma^2_I(\rho^*_I)}{\sigma^2_I(\rho^*_I) + \sigma^2_{\tau,I}}} ~+~ o_P(1) ~ \overset{d}{\preceq} \mathrm{Normal}(0, 1)
\end{equation*}
where the notation $X_{I} \overset{d}{\preceq} X$ denotes $\lim\sup_I P(|X_I| > c) \le P(|X| >c)$ for all $c \ge 0$.
\end{proposition}

\vspace{-.2cm}
\subsection{Higher-order bias}\label{sec:tfgls-hobias}
\vspace{-.2cm}

OLS estimation of $\tau_w$ is unbiased. Estimation of the weights $\kappa_i(\rho)$ causes finite-sample bias in FGLS-type estimators (FGLS, KFGLS, TFGLS) in general, although the \citet{Kakwani1967} argument gives unbiasedness under homogeneous $\tau_{i,t}$ and symmetric distribution of $U$.   

\begin{lemma}
\label{lemma:exactunb}
Consider model \eqref{eq:model}. Let (A1), (A2) and (A4) hold. If $\tau_{i,t}$ is homogeneous (more generally, $\delta_i := w'_{[1],i}(\tau_{[1],i} - 1_{N_{[1],i}}\tau_w) = 0$ for all $i=1,\ldots,I$), and the distribution of $U$ is symmetric, then $E[\widehat{\tau}_{w,\tfgls}] = \tau_w$ for every $I$ if the plugged-in $\widehat{\rho}^*_I$ from \eqref{eq:estrho} depends on $U$ only through the criterion function $\rho\mapsto\widehat{\sigma}^2_I(\rho)$.
\end{lemma}

Homogeneity of treatment effect $\tau_{i,t}$ and symmetry of $U$ are restrictive. Although bias does not appear under first-order asymptotics even when these two conditions are not met, it is useful to have a formal characterization of the higher-order bias caused by the estimation of weights. We present the result for TFGLS. Its qualitative implication holds in general for all estimators where estimation of the weights $\kappa_i(\rho)$ introduces dependence with the regression model error $u_i$. The novelty of our result is the contribution of heterogeneity of $\tau_{i,t}$ to this bias, contrasting it with the homogeneous $\tau_{i,t}$ setup  of \citet{HansenGLS2007}'s Proposition 3.

\medskip 

\noindent \textbf{Assumptions} (continued)\textbf{:}
\begin{itemize}

\item[(A5')] $\sup_i E\big[\|u_i\|^{\,2(d_\rho\vee 4)+\varepsilon}\big]<\infty$ for some $\varepsilon>0$ (strengthens $\sup_i E[\|u_i\|^{4}]<\infty$ from (A5)).

\item[(A8')] $\max_{1\le i \le I} \|w_{[1],i}\|=O(I^{-1})$ (strengthens $\max_{1\le i \le I} \|w_{[1],i}\|=o(I^{-1/2})$ from (A8)).

\item[(P6')] The maps $\rho_0\mapsto\bar\Sigma^{-1}_{[00],i}(\rho_0)$ and $\rho_{10}\mapsto\bar\Sigma_{[10],i}(\rho_{10})$ are three times continuously differentiable on $\mathcal{P}$, and these derivatives are bounded uniformly in $i$, $I$, and
$\rho\in\mathcal{P}$.

\item[(A10)] $H_I := \frac{\partial^2}{\partial \rho \partial \rho'} \left[\frac{I}{2}\sigma^2_I(\rho^*_I)\right]$ is finite and positive definite uniformly in $I$ for large $I$.

\end{itemize}

\noindent \textbf{Remarks:} The moment requirement in (A5') simplifies the proof by allowing integer-exponents in Rosenthal's inequality (at $L^4$ and $L^8$ for terms, respectively, quadratic and linear in $u$). For one part of the proof we need the order of finite moments to depend on $d_\rho$, i.e., the number of nuisance parameters estimated --- this estimation is causing the higher-order bias. (A8') holds for standard estimands such as the average treatment effect on the treated, event study and average event study estimands, etc. (P6') strengthens (P6) from Appendix~\ref{app:primitive}. The delicate part of (P6) or (P6') is the map $\rho_0\mapsto\bar\Sigma^{-1}_{[00],i}(\rho_0)$, and it does not cause problem in standard working models if the parameter space for $\rho_0$ is bounded away from non-stationarity. The curvature condition in (A10) is local and complements (A9).

\begin{proposition}
\label{prop:ho-bias}
Let the conditions of Proposition \ref{prop:tfgls} hold. Suppose every limit point of $\{\rho^*_I\}$ lies in  $\mathrm{interior}(\mathcal{P})$. Let $\Phi_I := \max_i\|\kappa_{i,I}(\rho^*_I)\|^2/K_I$, which is $o(1)$ under (A2). Let $\Blead := \mathcal{BA}_1 + \mathcal{BA}_2 + \mathcal{BB}_2$ where:
\begin{eqnarray*}
\mathcal{BA}_1 &:=& -I\sum_i E\left[u'_{i}\kappa_{i,I}(\rho^*_I)u'_{[0],i}\frac{\partial \kappa_{[0],i,I}(\rho^*_I)}{\partial\rho'}H^{-1}_I\frac{\partial \kappa'_{[0],i,I}(\rho^*_I)}{\partial\rho}u_{[0],i}\right]=O(\Phi^{1/2}_IK^{1/2}_I),\\
\mathcal{BA}_2  &:=& -I\sum_i \delta_i E\left[u'_{[0],i}\frac{\partial \kappa_{[0],i,I}(\rho^*_I)}{\partial\rho'}H^{-1}_I\frac{\partial \kappa'_{[0],i,I}(\rho^*_I)}{\partial\rho}u_{[0],i}\right]=O(\Phi^{1/2}_IK^{1/2}_I),\\
\mathcal{BB}_2 &:=& I \cdot \sum_i \sum_j \delta_i E\left[(\widehat{\beta}(0) - \beta)'x'_{[0],i}\frac{\partial \kappa_{[0],i,I}(\rho^*_I)}{\partial\rho'}H^{-1}_I \frac{\partial \kappa'_{[0],j,I}(\rho^*_I)}{\partial\rho}u_{[0],j}\right] = O(I^{-1}),
\end{eqnarray*}
with $\widehat{\beta}(0)$ the OLS estimator of $\beta$: $(\wh{\eta}_{\ols} - \eta) = ((\wh{\lambda}_1(0) - \lambda_1)', \ldots, (\wh{\lambda}_I(0) - \lambda_I)', (\wh{\beta}(0) - \beta)')'$.
\begin{itemize}
\item[(i)] If (A5'), (P6'), (A10) and the entity-level condition (A12) in Appendix~\ref{app:hobias-lemmas} hold then $\mathcal{BA}_1,\mathcal{BA}_2 =O(\Phi^{1/2}_IK^{1/2}_I) = o(I^{-1/2})$, and:~~ $E[\hat\tau_{w,\tfgls}-\tau_w] \;=\; \mathcal{B}_{\mathrm{lead}} \;+\; o\bigl(\Phi_I^{1/2}\,I^{-1/2}\bigr)$.
\item[(ii)] If additionally (A8') holds then $\Phi_I = O(I^{-1})$ by (A12), which gives $\mathcal{BA}_1,\mathcal{BA}_2 =O(\Phi^{1/2}_IK^{1/2}_I) = O(I^{-1})$, and:~~$  E[\hat\tau_{w,\tfgls}-\tau_w] \;=\; \mathcal{B}_{\mathrm{lead}} \;+\; O(I^{-3/2})$.
\end{itemize}
\end{proposition}

\noindent\textbf{Remarks:} First, the no-rate assumption for $\Phi_I$ in (A2) caused the awkward term $o\bigl(\Phi_I^{1/2}\,I^{-1/2}\bigr)$ that we cannot write as $o(I^{-1})$ to cleanly distinguish its order from that of $\mathcal{BB}_2$. The assumption (A8') solves this problem --- the orders of $\Blead$ and the next higher-order $O(I^{-3/2})$ in (ii) are the more familiar representation in the literature on higher-order bias.

Second, $\mathcal{BA}_1$ vanishes if $u_i$ is symmetric, e.g., $u_i \sim N(0,\Sigma_i)$ (like assumption HO1 in \citet{HansenGLS2007}); see Lemma \ref{lem:bias-order} (iii) in Appendix~\ref{app:hobias-lemmas}. Contrasting this higher-order result from \citet{HansenGLS2007}'s, the terms $\mathcal{BA}_2$ and $\mathcal{BB}_2$ are generally non-zero unless treatment effects $\tau_{i,t}$ are homogeneous in the sense that $\delta_i = 0$ for all $i$. We note that Lemma \ref{lemma:exactunb} already gave unbiasedness under such symmetry and homogeneity.

Third, the compact $\mathcal{P}$ chosen by the user, e.g., $\rho_0 \in [-1+c, 1-c]$ in the AR(1) working model to keep $\bar\Sigma_{[00]}(\rho_0)$ positive definite, can put $\rho^*_I$ on the boundary when the error is highly persistent. Let $d_\rho =1$. If the one-sided derivative of $I\sigma^2_I(\rho)$ at that boundary is bounded away from zero, then $P(\widehat{\rho}^*_I \ne \rho^*_I) = O(\Phi^2_I)$ and, since $E[\widehat{\tau}_I(\rho^*_I)] = \tau_w$, Cauchy-Schwarz bounds the bias by $2E^{1/2}[\sup_{\rho \in \mathcal{P}} (\widehat{\tau}_{I}(\rho)-\tau_w)^2]\,P^{1/2}(\widehat{\rho}^*_I \ne \rho^*_I) = O(\Phi_I I^{-1/2})$ under (A8), and $O(I^{-3/2})$ under (A8') (Lemma \ref{lemma:hobiasboundary}, Appendix~\ref{app:hobias-boundary}). If instead the derivative is zero, the unconstrained minimizer lies outside $\mathcal{P}$ (thus $\widehat{\rho}^*_I = \rho^*_I$) with probability approaching $1/2$ \citep{Andrews1999}. Let us give a heuristic. Assume (A8'), and with $\widehat{G}:=\partial[\tfrac I2\widehat{\sigma}^2_I(\rho^*_I)]/\partial\rho$ denoting the score, let $\sqrt{I}\bigl(\widehat{G},\partial\widehat{\tau}_I(\rho^*_I)/\partial\rho\bigr)$ have a mean-zero normal limit $(Z_1,Z_2)$ with convergent moments. At an upper (lower) boundary, the expansion behind Proposition \ref{prop:ho-bias} then holds with $\widehat{G}$ replaced by $\max(\widehat{G},0)$ ($\min(\widehat{G},0)$), i.e., by $\tfrac12(\widehat{G}\pm|\widehat{G}|)$. The $|\widehat{G}|$ part adds no bias of order $I^{-1}$ since $E[|Z_1|Z_2]=0$ (by symmetry), so the bias is $\tfrac12\Blead+o(I^{-1})$, i.e., the leading term is halved. Proposition \ref{prop:tfgls} is unaffected in all cases, as it needs only consistency of $\widehat{\rho}^*_I$, not a rate.

\section{Monte Carlo Experiment}\label{sec:mc}

\subsection{Theoretical underpinning of what we are comparing and why?}\label{sec:mc-theory}

\citet{CSX2025} propose a new estimator and demonstrate using simulations its big improvement in precision for estimating an event study average parameter over the estimators of \citet{dcdH2020}, \citet{CallawaySantanna2021}, \citet{SunAbraham2021} and, to a lesser extent, over what we call the OLS estimator, i.e., the \citet{BJS2024}, \citet{Wooldridge2025}, \citet{Gardner2021}, \citet{GTTY2024} estimator.

\citet{CSX2025}'s estimator, denoted CSX henceforth, is semiparametrically efficient. Hence, it is a feasible benchmark for assessing the small-sample performance of TFGLS. (An infeasible benchmark would use the unknown true error covariance structure.) To formalize the comparisons, we first establish a common framework that clarifies the relationship between the estimators and the efficiency benchmark against which they can be compared.

Focus on the average treatment effect at period $t$ on those that were treated in period $g \le t$:
\vspace{-.1cm}
\begin{equation*}
ATT(g,t) := \frac{1}{n_g}\sum_{i: E_i = g} \tau_{i,t} \text{~~where~~} n_g := \sum_{i=1}^I \1(E_i = g).
\vspace{-.1cm}
\end{equation*}
This is the building block for many aggregate causal parameters. Proposition \ref{prop:unif} concerns arbitrary $ATT(g,t)$'s and hence, under its assumptions, extends to their linear aggregates. 

\begin{proposition}
\label{prop:unif}
Let $\tau_w := ATT(g,t) = w'_{[1]}\tau_{[1]}$ where the elements of the weight vector are $w_{i,s} =1/n_g$ if $E_i = g$ and $s=t$; and $=0$ otherwise. Consider \eqref{eq:model} with $r'_{i,t}\eta = \lambda_i + \beta_t$, and balanced panel. Let $u_i$'s be independent, $E[u_{i}] = 0$ and $Var(u_i) = \Sigma_i$ positive definite for all $i=1,\ldots,I$. Let $\Sigma_i = \tilde{\Sigma}^{c}$ for all $i=1,\ldots,I$ such that $E_i = c$ for $c \in \mathcal{G} := \mathcal{G}_{\text{trt}} \cup \{\infty\}$, i.e., there is within cohort homoskedasticity for each treatment cohort $g \in \mathcal{G}_{\text{trt}}\subseteq \{2,\ldots,T\}$ and the never treated cohort $g = \infty$. Then, infeasible CSX, by which we mean CSX that uses the true variance conditional on $E_1,\ldots,E_I$, is numerically identical to $\widehat{\tau}_{w,\text{blue}}:= w'_{[1]}\widehat{\tau}(\Sigma^{-1})$.  
\end{proposition}

\noindent \textbf{Remarks:} Since infeasible CSX is a linear unbiased estimator, it cannot be more precise than $\widehat{\tau}_{w,\text{blue}}$. The interesting and, to our knowledge, novel part is that under within cohort homogeneity of $\Sigma_i$ we can show that infeasible CSX becomes numerically identical to $\widehat{\tau}_{w,\text{blue}}$. 

Proposition \ref{prop:unif} gives a framework for comparing feasible estimators. If we knew the true covariance structure, then we would use it and get numerically identical $\widehat{\tau}_{w,\text{blue}}$ and infeasible CSX. Because we do not know it, the FGLS approach would postulate a low-dimensional parametric working model for it, estimate that model, and use it for FGLS estimation. Or, as in KFGLS (also see \citet{CuiWooldridge2026}), one could postulate a nonparametric working covariance structure and estimate it using sample analogs after imposing homoskedasticity across $i$ (see Section~\ref{sec:motivation}). The feasible CSX approach also takes a route similar to KFGLS and hence, like it, could face problems when $T$ is not small. The cost could be more for CSX because its nonparametric working covariance structure imposes fewer working constraints --- roughly speaking, cf. Lemma \ref{lemma:KFGLSmse} with $n_c$'s instead of $I$.

\subsection{Experiment design}\label{sec:mc-design}

Our data generating processes (DGPs) build on Section 5.2 of \citet{CSX2025} who, in turn, design their experiment following \citet{BakerLarckerWang2022}. They use the decomposition of the return on assets (ROA) of U.S.-incorporated, non-financial Compustat firms over the years 1980--2015 into firm fixed effects (FE) $\tilde{\lambda}_i$'s, year (period) FE $\tilde{\beta}_t$'s and residuals $\tilde{\epsilon}_{i,t}$'s (source: https://github.com/andrewchbaker/JFE\_DID). In all the designs below, the untreated potential outcome of firm $i$ in period $t$ is $\tilde{\lambda}_i + \tilde{\beta}_t + \tilde{u}_{i,t}$, $t = 1,\ldots,T$, $i = 1,\ldots,I$. 

Treatment timing and effects are the same in all designs and follow \citet{CSX2025}. With $T = 10$, each firm is independently assigned, with probability 1/3 each, to the cohort treated at $t = 5$, to the cohort treated at $t = 8$, or to the never-treated cohort. \citet{CSX2025} impose homogeneity in $\tau_{i,t}$ within a treatment cohort: $\tau_{i,t}:= \tau_t(E_i) := \left\{1(E_i=5)\times .5 + 1(E_i=8)\times .3\right\}\times .309 \times (t- (E_i-1))$ for $t = E_i,\ldots, 10$, where $.309$ is the sample standard deviation of ROA reported by \citet{CSX2025}. In each simulated sample, $ES(e)$ is the average of $\tau_{i,t}$ over the treated firm-periods at event time $e$, so it reflects the realized cohort sizes; and the target parameter $\tau_w$ is their average $ES_{\text{avg}} = \sum_{e=0}^5 ES(e)/6$. 

\textbf{Design 1:} We randomly pick a block of $T$ consecutive years from the 36 years 1980--2015, and then randomly pick (without replacement) $I$ firms from those that are observed in all $T$ years of the block. Following \citet{CSX2025}, we take $\tilde{u}_{i,t}= \nu_{1}\tilde{u}_{i,t-1} + \tilde{\epsilon}_{i,t}$ for $t=1,\ldots, T$ with $\nu_1 \in \{0, \pm .5, \pm 1\}$, excluding their explosive cases $\nu_1 = \pm 1.1$. The initial value $\tilde{u}_{i,0}$ is one random draw from the entire pool of residuals $\{\tilde{\epsilon}_{j,s}\}$ ($j$: all firms, $s$: all years). Unlike \citet{CSX2025}, the innovations $\tilde{\epsilon}_{i,1},\ldots,\tilde{\epsilon}_{i,T}$ are firm $i$'s own estimated residuals in the $T$ years of the block, retaining the within-firm serial correlation of the Compustat residuals --- so, the model error $\tilde{u}_{i,t}$ is \emph{not} AR(1), and an AR(1) working model is misspecified.\footnote{\label{fn:CSXdes}\citet{CSX2025} draw each $\tilde{\epsilon}_{i,t}$ independently from the entire pool of residuals. Our corresponding results, which are comparatively more favorable to TFGLS, are in Appendix \ref{app:csxdes}.}

\textbf{Design 2:} Innovations $\epsilon_{i,t}$ are now drawn independently for all $i,t$ from $N(0, .114)$, where $.114$ is the variance of the pooled Compustat residuals $\tilde{\epsilon}_{j,s}$'s, and each stationary process below is started from its stationary distribution. This makes available the infeasible GLS estimator $\widehat{\tau}_{w,\text{blue}}$, which uses the true error covariance, as a benchmark for how much of the possible gain over OLS by $\widehat{\tau}_{w,\text{blue}}$ can be delivered by feasible estimators. $\tilde{\lambda}_i$'s and $\tilde{\beta}_t$'s are drawn with replacement from the estimated firm FE's and the estimated year FE's.
    \begin{itemize}\setlength{\itemsep}{0pt}
    \item Usual benchmarks (AR(1) working model is correct): \textbf{Design 2(a):} i.i.d.\ errors, $\tilde{u}_{i,t} = \epsilon_{i,t}$, for which OLS is $\widehat{\tau}_{w,\text{blue}}$; and \textbf{Design 2(b):} AR(1) errors, $\tilde{u}_{i,t} = .6\,\tilde{u}_{i,t-1} + \epsilon_{i,t}$, for which GLS with the AR(1) working model evaluated at $\rho = .6$ is $\widehat{\tau}_{w,\text{blue}}$.
    \item \textbf{Design 2(c):} $\tilde{u}_{i,t} =  -.9025\tilde{u}_{i,t-2} + \epsilon_{i,t}$, which is particularly unfavorable to any AR(1) working model because this AR(2) model gives autocorrelations at lags $1, 2, 3, 4, 5, 6, \ldots$ as $0, -.90, 0, .81, 0, -.74, \ldots$, which no AR(1) model can come close to representing.
            \item MA(1): $\tilde{u}_{i,t}= \epsilon_{i,t} + \theta_1\epsilon_{i,t-1}$ with \textbf{Design 2(d):} $\theta_1 = .6$ and \textbf{Design 2(e):} $\theta_1 = -.6$.
    \item ARMA(1,1): $\tilde{u}_{i,t} = \nu_1\tilde{u}_{i,t-1} + \epsilon_{i,t} + \theta_1\epsilon_{i,t-1}$ with \textbf{Design 2(f):} $(\nu_1,\theta_1) = (.6, .6)$ and \textbf{Design 2(g):} $(\nu_1,\theta_1) = (-.6, -.6)$.
    \item Designs 2(h)--2(k) combine a persistent stationary ARMA(1,1) process $e_{i,t} = .85\,e_{i,t-1} + \epsilon_{i,t} + \theta_1\epsilon_{i,t-1}$, with $\theta_1 = .2$ in Designs 2(h) and 2(j) and $\theta_1 = -.2$ in Designs 2(i) and 2(k), with a variance of $\tilde{u}_{i,t}$ that changes over time, so that no stationary AR(1) model represents the error covariance well. Let $c_t := (t-1)/(T-1)$. \textbf{Designs 2(h)--2(i):} U-shaped variance, $\tilde{u}_{i,t} = \{1+8(c_t-.5)^2\}e_{i,t}$ --- the standard deviation in the first and last periods is three times that in the middle. \textbf{Designs 2(j)--2(k):} falling variance, $\tilde{u}_{i,t} = \{1+3(1-c_t)^2\}e_{i,t}$ --- the standard deviation falls from 4 times to its final level.
    \end{itemize}

Besides $I = 400$ as in \citet{CSX2025}, we report results for the smaller $I=100, 50$, where gains in precision may be more useful. To see how the quality of approximation by the $I \rightarrow \infty$ asymptotics varies with different values of $T$ for different estimators, we also report results with $T=20$. We take the event dates to be $t = 10$ and $t=16$ when $T = 20$ (in proportion to the case with $T = 10$), with the same cohort probabilities and the same $\tau_{i,t}$ formula; so $ES_{\text{avg}}$ now averages $ES(e)$ over $e = 0,\ldots,10$.

\subsection{The five estimators for the experiment}\label{sec:mc-estimators}

\indent (1) TFGLS using an AR(1) working model following \citet{BJS2024}'s proposal ``... model the errors as an autoregressive process with homoskedastic innovations (e.g., AR(1))''. For simplicity, we impose $\rho_0 = \rho_1 =\rho \in \mathcal{P} :=[-.99, .99]$; see footnote \ref{fn:ar1} or Appendix~\ref{app:ar1wm}.

(2) The OLS estimator of \citet{BJS2024}, \citet{Gardner2021}, \citet{Wooldridge2025}. 

(3) FGLS has many versions --- Cochrane-Orcutt or Prais-Winsten, with or without \citet{HansenGLS2007}'s bias correction, based on the untreated cells or the never-treated units --- with no clear winner across our designs. As a parsimonious benchmark that is true to the AR(1) working model, we instead report the \emph{infeasible, idealized version of FGLS}, denoted FGLS$^*$, whose $\widehat\rho$ is the pooled first-order autoregression coefficient of the true errors $\tilde u_{i,t}$.

(4) KFGLS estimates the GLS weight matrix by the sample covariance of residuals, as in \citet{Kiefer1980}. Which residuals to use matters, and each choice has a drawback. With the residuals of the never-treated units only, KFGLS is asymptotically equivalent to $\widehat{\tau}_{w,\text{blue}}$ under homoskedasticity across firms, but the estimated $T\times T$ matrix is singular or nearly so unless the number of never-treated units is large relative to $T$ --- with $I=50$ there are only about 17 of them on average, against $T=10$ or $20$ --- and a Moore-Penrose inverse then gives unstable results. Using the residuals of all untreated cells need not give a positive semidefinite matrix, and in our designs often does not, even at $I=400$. Using the residuals of all units, including the treated cells, avoids both problems, but under the heterogeneity in $\tau_{i,t}$ allowed in our theory the treated residuals contain the deviations of $\tau_{i,t}$ from their cohort-period means, and the resulting KFGLS can have unbounded asymptotic risk.\footnote{\label{fn:KFGLSproblem}The sample covariance of all-units residuals then converges not to the error covariance but to it plus the covariance of these $\tau_{i,t}$ deviations. Such a KFGLS remains consistent but is no longer asymptotically efficient, and its asymptotic variance can be arbitrarily larger than even that of OLS. As an illustrative example of this asymptotic phenomenon, let $\tau_{i,t} = \tau_t(E_i) + b_i\,(t-E_i+1)^p$ for $t \ge E_i$, where $\tau_t(E_i)$ is the cohort-period effect of our designs and the $b_i$'s are bounded constants whose mean squared deviation from their cohort mean converges, in each treated cohort, to half the variance of the $\tilde{u}_{i,t}$'s. Let the $\tilde{u}_{i,t}$'s be i.i.d. as in Design 2(a), so that OLS is $\widehat{\tau}_{w,\text{blue}}$, and TFGLS is asymptotically equivalent to it ($\rho^*_I = 0$). Let $T = 10$, and let the target be $\tau_w = ES_{\text{avg}}$. Then, as $I\to\infty$, the asymptotic variance of KFGLS that uses the residuals of all units is 1.57, 6.61, 23.43, 77.65, 261.03, 899.47 times that of OLS for $p=1,\ldots,6$, whereas the asymptotic variance of KFGLS using only the never-treateds (and that of TFGLS) equals that of OLS.} Our designs impose homogeneity within cohort-periods, so this last problem does not arise here, which favors KFGLS. To favor it further, we report as KFGLS$^*$ whichever of the never-treated and all-units versions has the smaller standard error in each sample. (This selection has essentially no adverse effect on inference reliability in our simulations; see footnote \ref{fn:infrel}.)

(5) Our implementation of the estimator of \citet{CSX2025}, i.e., CSX. At their specification of sample size, $I = 400$ and $T = 10$, and under their design (see our footnote \ref{fn:CSXdes}), it reproduces their reported performance and is, if anything, more favorable to CSX.

TFGLS and KFGLS* also use the construction of the treated residuals for the standard errors of \citet{BJS2024}, from whose paper OLS and FGLS are drawn.

For all estimators, we compare: (i) precision, as measured by their root mean squared error (RMSE) and (ii) reliability in inference, as measured by the rejection probability of the truth by their (robust) Wald test. Both are estimated based on 10000 Monte Carlo trials. 

\subsection{Experiment results}\label{sec:mc-results}

Tables \ref{tab:des1T10}--\ref{tab:des2T20} report the RMSE of each estimator relative to that of TFGLS and the rejection probability of the truth by the nominal 5\% Wald test. Tables \ref{tab:des1stdbse} and \ref{tab:des2stdbse} in Appendix \ref{app:sdse} compare the Monte Carlo standard deviation of each estimator with the square root of the mean of its estimated asymptotic variance. Three findings on precision gains stand out.

\textbf{(i) TFGLS has a precision advantage over OLS and FGLS$^*$ even in small samples.} OLS is never materially more precise than TFGLS, and is up to 49\% (Design 1) and 53\% (Design 2) less precise. FGLS$^*$, although its $\widehat{\rho}$ uses the true errors, is never more precise than TFGLS in Design 1 (and up to 4\% less precise), and is within 2\% of TFGLS in the stationary Designs 2(d)--2(g). In Designs 2(h)--2(k), FGLS$^*$ is 5--14\% less precise than TFGLS. Even in the benchmark Designs 2(a) and 2(b), where OLS and FGLS$^*$, respectively, are (essentially) $\widehat{\tau}_{w,\text{blue}}$, the RMSE of TFGLS is at most 1.4\% larger than theirs. 

\textbf{(ii) The choice of working model matters.} Design 2(c) is an example where the working model is inadequate: no AR(1) model comes close to mimicking its error covariance. RMSE of $\widehat{\tau}_{w,\text{blue}}$ is about 30\% smaller than that of TFGLS. KFGLS$^*$ and (generally) CSX are much more precise than TFGLS. Targeting cannot make up for an inadequate working model. Still, OLS and FGLS$^*$ (whose $\widehat{\rho} \approx 0$ here) are up to 15\% less precise than TFGLS.

\textbf{(iii) Targeting recovers what misspecification costs standard FGLS, and can even beat the semiparametrically efficient estimators at moderate $I$.} In Designs 2(h)--2(k), the time-varying error variance makes the AR(1) working model misspecified: FGLS$^*$ is 5--14\% less precise than TFGLS which is within 4\% of $\widehat{\tau}_{w,\text{blue}}$. KFGLS$^*$ and CSX, which are meant to be asymptotically efficient, have RMSE similar to that of TFGLS at $I=400$ but become less precise as $I$ falls. CSX is less precise than TFGLS by 3--5\% ($T=10$) and 11--15\% ($T=20$) when $I=100$, and by 9--12\% ($T=10$) and 39--42\% ($T=20$) when $I=50$. KFGLS$^*$ is less precise than TFGLS by at most 5\% when $I=100$ and 9\% when $I=50$. The same pattern holds in Designs 2(a)--2(g) except 2(c). In Design 1 with $I=400$ and $T=10$, CSX and KFGLS$^*$ are more precise than TFGLS; the advantage can fall with smaller $I$. The relative performance of TFGLS generally improves when \emph{reading down each table, where $T$ is fixed and only $I$ falls}. A larger $T$ generally amplifies the precision loss of CSX and, to a lesser extent, of KFGLS$^*$ relative to TFGLS, consistent with the parameter proliferation that \citet{BJS2024} caution against; see our motivating Section \ref{sec:motivation}.

\textbf{Inference reliability:} The empirical size of TFGLS, OLS and FGLS$^*$ stays close to the nominal 5\% in all designs. KFGLS$^*$ and CSX over-reject when $I$ is small relative to $T$ because their standard errors heavily understate their variability  (Tables \ref{tab:des1stdbse} and \ref{tab:des2stdbse}).\footnote{\label{fn:infrel}For KFGLS$^*$ this is the finite-sample cost of estimating an unrestricted weight matrix, which its standard error treats as known; as noted in Section \ref{sec:motivation}, this was also pointed out in \citet{HansenGLS2007}, \citet{HausmanKuersteiner2008}, etc. In the canonical case of Lemma \ref{lemma:KFGLSmse} in Section \ref{sec:motivation}, the estimation of the unrestricted weight matrix inflates the variance by $1 + (T-1)/(I-1-T)$, and the ratio of the Monte Carlo standard deviation of KFGLS$^*$ to the root of its mean estimated variance tracks the square root of this factor: when $I=50$ it is 1.10--1.14 ($T=10$) and 1.23--1.29 ($T=20$) in Designs 1 and 2, against 1.11 and 1.29. Accordingly, its empirical size rises from 5.3--5.9\% when $I=400$ to 7.5--8.8\% ($T=10$) and 11.2--12.8\% ($T=20$) when $I=50$, close to the 7.7\% and 12.8\% at which a nominal 5\% test rejects when the standard error is too small by factors of 1.11 and 1.29. Choosing the version with the smaller standard error costs little: KFGLS$^*$'s rejection rate never exceeds the higher of those of the two versions by more than .41 percentage point.}$^,$\footnote{Appendix \ref{app:boot} shows that clustered bootstrap standard error, as in \citet{CSX2025}, helps to bring the empirical size to close to 5\% when $I=100$ and $T=10$ but is otherwise conservative. It is extremely conservative (empirical size is 0.0\%) when $I=50$ and $T=20$. The bootstrap also cannot address the high variability of CSX in small samples, and its validity under the allowed heterogeneity in $\tau_{i,t}$ is unclear to us.} Parsimony seems to aid the reliability of standard cluster-robust asymptotic inference.

\subsection{Extension 1: More treated cohorts}\label{sec:mc-morecoh}

What happens if there are more treated cohorts? To maintain reasonably sized cohorts for the experiment, we consider $I = 400$ and $I = 100$. Similarly, since we already saw that $T = 20$ affects KFGLS$^*$ and CSX more adversely than TFGLS (and OLS and FGLS$^*$), we set $T=10$ here. In addition to the two cohorts treated at $t=5,8$ as in \citet{CSX2025}, we include two more cohorts treated at $t=6,7$: so in total four treated cohorts and one never-treated cohort; each firm is independently assigned to each of these five cohorts with probability 1/5. Extending \citet{CSX2025}'s construction to this setup with more cohorts, we set $\tau_{i,t}:= \tau_t(E_i) := \left\{1(E_i=5)\times .6 + 1(E_i=6)\times .5 + 1(E_i=7)\times .4 + 1(E_i=8)\times .3\right\}\times .309\times(t- (E_i-1))$ for $t = E_i,\ldots, 10$. With $I=100$, each cohort now has about 20 firms on average, against about 33 in Tables \ref{tab:des1T10} and \ref{tab:des2T10}. The rest is exactly as in Sections \ref{sec:mc-design} and \ref{sec:mc-estimators}.

Table \ref{tab:morecoh} presents RMSE and empirical size, and Table \ref{tab:morecohsdse} (Appendix \ref{app:sdse}) the ratio of the Monte Carlo standard deviation to the square root of the average of estimated asymptotic variance. Compared with Tables \ref{tab:des1T10} and \ref{tab:des2T10}, the following observations stand out.

First, the precision advantage of TFGLS over OLS and FGLS$^*$ is not diminished by more cohorts. OLS is up to 37\% (Design 1) and 66\% (Design 2) less precise than TFGLS, and FGLS$^*$ is 4--15\% less precise in Designs 2(h)--2(k). 

Second, perhaps due to its cohort-specific operations, more cohorts hurt CSX the most. For example, with $I=100$, CSX is 10--13\% less precise than TFGLS in Design 2 except 2(c), against 2--5\% with two treated cohorts, while KFGLS$^*$ is 2--7\% less precise (against 0--2\%). 

Third, the cost of an inadequate working model is less in Design 2(c): KFGLS$^*$ and CSX are now 14--20\% and 10--19\% more precise than TFGLS (against 27--29\% in Table~\ref{tab:des2T10}).

Finally, consider inference reliability. The empirical size of TFGLS, OLS and FGLS$^*$ stays within 4.7--6.6\% thanks, perhaps, to parsimony. On the other hand, with $I=100$, the rejection rate of KFGLS$^*$ and CSX rises to 6.8--9.9\% and 10.6--14.3\%, respectively (against 5.8--7.0\% and 7.3--8.4\% respectively in Section~\ref{sec:mc-results} where there were only two treated cohorts).

\subsection{Extension 2: $(i,t)$-specific treatment effect $\tau_{i,t}$ heterogeneity}\label{sec:mc-hetit}

Our theory, like the modern DiD literature, lets $\tau_{i,t}$ vary across $(i,t)$, not just across cohort-periods, subject only to (A7) that $\sup_i \|\tau_{[1],i}\| < \infty$. Although simulations with such heterogeneity are rare in the literature, we wish to report our simulations allowing heterogeneity at the $(i,t)$ level because they reveal a weakness of TFGLS in small samples relative to standard FGLS, without requiring a comparison with the idealized FGLS$^*$ of Sections~\ref{sec:mc-estimators}--\ref{sec:mc-morecoh}.\footnote{\label{fn:hetit}TFGLS $\widehat{\rho}^*_I$ minimizes $\widehat{\sigma}^2_I(\rho)$. As in the proof of Proposition~\ref{prop:cons_var}, $\widehat{\sigma}^2_I(\rho)=\sum_i\xi^2_i(\rho)+2\sum_i\xi_i(\rho)\delta_i+\sigma^2_{\tau,I}+o_P(K_I)$ uniformly in $\rho$, with $\xi_i(\rho):=\kappa'_{i,I}(\rho)u_i$ and $\delta_i:=w'_{[1],i}(\tau_{[1],i}-1_{N_{[1],i}}\tau_w)$. The cross term $\sum_i\xi_i(\rho)\delta_i$ has mean zero and is $o_P(K_I)$ under (A7)--(A8), so $\widehat{\rho}^*_I$ remains consistent (Lemma \ref{lemma:rhocons}). But it varies with $\rho$, and its standard deviation relative to $\sigma^2_I(\rho^*_I)$ is of order $r_I/\sqrt{I}$ for diffuse weights, where $r^2_I:=\sigma^2_{\tau,I}/\sigma^2_I(\rho^*_I)$ is $O(1)$ by (A3), (A7)--(A8) but can be large. So, large and persistent entity-level heterogeneity can dominate the differences $\sigma^2_I(\rho)-\sigma^2_I(\rho^*_I)$ that identify $\rho^*_I$, making TFGLS noisier than the oracle $\widehat{\tau}_I(\rho^*_I)$ in finite samples. Unlike the Nickell-Solon bias of FGLS, this distortion vanishes as $I\to\infty$. FGLS and OLS, whose weights use no treated residuals, are immune to it. KFGLS using all units (footnote~\ref{fn:KFGLSproblem}) and CSX are affected more.}

We use Design~2 but set $\tau_{i,t}=\tau_t(E_i)+\sqrt{T-t}\,z_i$ for $t\ge E_i$. $z_i$'s are i.i.d.\ $N(0,1)$ truncated to $[-3,3]$, so that (A7) holds, and are independent of everything else. We take $T=20$ to amplify the problem, so $\tau_{i,t}$'s deviate persistently from their cohort-period means with standard deviation up to 3.12 against an error standard deviation of .34 in Design~2(a). We use $I=50, 100, 400, 1000, 2000, 4000$.\footnote{Design~1 is infeasible for $I\ge 1000$: with $T=20$, at most 593 firms are observed in all years of any block.} We replace FGLS$^*$ with the Prais-Winsten FGLS whose $\widehat{\rho}_I$ is the AR(1) coefficient of the OLS residuals over consecutive \emph{untreated} cells. 

Table~\ref{tab:hetT20} reports RMSEs relative to TFGLS. It also includes an infeasible TFGLS (TFGLS$^*$) that uses the trial-specific true $\rho^*_I$'s, and its comparison with TFGLS reflects the loss of the latter due to estimation of $\rho^*_I$ in this design. Three new observations now stand out.\footnote{\label{fn:hetitinf}Bias is negligible and inference is conservative (CSX still over-rejects at $I=50$); see Appendix~\ref{app:hetaux}.} 
 
First, such heterogeneity costs TFGLS precision when $I$ is small, but this cost vanishes as $I$ grows. Except in Design~2(c), where no AR(1) working model is adequate, the RMSE of $\widehat{\tau}_{w,\text{blue}}$ relative to TFGLS rises from .44--.98 when $I=50$ to .92--1.00 when $I=4000$.

Second, FGLS $\hat{\rho}$ using untreated cells is unaffected by the $z_i$'s. FGLS is more precise than TFGLS when $I$ is small in Designs~2(a)--(g), more than twice so in 2(g) when $I=50$, but this advantage disappears as $I$ grows. On the other hand, in Designs~2(h)--(k), FGLS is 14--27\% less precise than TFGLS at every $I$, due to the Nickell-Solon bias and the misspecified AR(1) working model, and OLS is 19--48\% less precise as it ignores serial correlation.

Third, KFGLS$^*$ and CSX, whose weights can also depend on the treated residuals, are up to 2.3 and 7.2 times less precise than TFGLS when $I=50$. The disadvantage shrinks slowly as $I$ grows: they are still up to 10\% and 35\% less precise respectively even when $I = 4000$.\footnote{TFGLS takes larger $I$ to recover against FGLS in Designs~2(a)--(g) if $\tau_{i,t}$ is $\tau_t(E_i)+(t-E_i+1)^pz_i$ (see footnote \ref{fn:KFGLSproblem}) instead of $\tau_t(E_i)+\sqrt{T-t}\,z_i$. KFGLS$^*$ and CSX need orders of magnitude larger $I$ for the same.}

\begin{table}[H]
\centering
\footnotesize
\renewcommand{\arraystretch}{1.04}
\setlength{\tabcolsep}{4pt}
\newcolumntype{C}{>{\centering\arraybackslash}p{1.2cm}}
\captionsetup{font=small,skip=4pt}
\begin{tabular}{rr@{\hspace{12pt}}|CCCC@{\hspace{10pt}}|CCCCC}
\hline
\multicolumn{2}{c|}{Specifications}
& \multicolumn{4}{c|}{\begin{tabular}[c]{@{}c@{}}RMSE relative\\to TFGLS\end{tabular}}
& \multicolumn{5}{c}{\begin{tabular}[c]{@{}c@{}}Rejection probability (\%) of truth\\by 5\% Wald test\end{tabular}} \\
\hline
$I$ & $\nu_1$ & OLS & FGLS$^*$ & KFGLS$^*$ & CSX & TFGLS & OLS & FGLS$^*$ & KFGLS$^*$ & CSX \\
\hline
&&&&&&&&&&\\[-1.6ex]
400 & $-1$ & 1.33 & 1.03 & 0.99 & 0.98 & 5.0 & 4.9 & 5.0 & 5.3 & 5.3 \\
400 & $-.5$ & 1.01 & 1.00 & 0.98 & 0.97 & 5.3 & 5.2 & 5.1 & 5.5 & 5.2 \\
400 & 0 & 1.02 & 1.00 & 0.99 & 0.99 & 5.3 & 5.2 & 5.2 & 5.3 & 5.5 \\
400 & .5 & 1.08 & 1.01 & 0.99 & 0.99 & 5.2 & 4.8 & 5.2 & 5.4 & 5.6 \\
400 & 1 & 1.33 & 1.01 & 0.93 & 0.92 & 5.3 & 5.1 & 5.3 & 5.3 & 6.0 \\[0.8ex]
\hline
&&&&&&&&&&\\[-1.6ex]
100 & $-1$ & 1.35 & 1.04 & 0.98 & 0.98 & 5.1 & 5.3 & 5.0 & 6.2 & 7.3 \\
100 & $-.5$ & 1.01 & 1.01 & 0.97 & 0.97 & 5.8 & 5.4 & 5.5 & 6.7 & 7.6 \\
100 & 0 & 1.02 & 1.01 & 0.98 & 0.98 & 5.9 & 5.3 & 5.5 & 6.7 & 8.0 \\
100 & .5 & 1.09 & 1.01 & 0.97 & 0.96 & 5.7 & 5.6 & 5.4 & 6.7 & 8.4 \\
100 & 1 & 1.32 & 1.01 & 0.92 & 0.89 & 5.3 & 5.5 & 5.3 & 5.8 & 7.8 \\[0.8ex]
\hline
&&&&&&&&&&\\[-1.6ex]
50 & $-1$ & 1.34 & 1.03 & 1.01 & 1.05 & 5.4 & 5.7 & 5.0 & 7.5 & 11.2 \\
50 & $-.5$ & 1.01 & 1.01 & 0.99 & 1.03 & 6.2 & 5.6 & 5.6 & 7.6 & 11.6 \\
50 & 0 & 1.03 & 1.02 & 1.00 & 1.03 & 6.1 & 5.5 & 5.5 & 7.5 & 11.9 \\
50 & .5 & 1.10 & 1.02 & 0.98 & 0.97 & 5.9 & 5.6 & 5.4 & 7.8 & 12.4 \\
50 & 1 & 1.33 & 1.01 & 0.91 & 0.86 & 5.8 & 5.6 & 5.6 & 7.6 & 12.7 \\[0.8ex]
\hline
\end{tabular}
\caption{Design 1, $\mathbf{T = 10}$, 10000 Monte Carlo trials. Left panel: RMSE of estimators relative to that of TFGLS ($> 1$ means estimator is less precise than TFGLS). Right panel: rejection probability (\%) of the truth by the nominal 5\% Wald test. $I = 400$ is \citet{CSX2025}'s sample size.}\label{tab:des1T10}
\end{table}

\begin{table}[H]
\centering
\footnotesize
\renewcommand{\arraystretch}{1.04}
\setlength{\tabcolsep}{4pt}
\newcolumntype{C}{>{\centering\arraybackslash}p{1.2cm}}
\captionsetup{font=small,skip=4pt}
\begin{tabular}{rr@{\hspace{12pt}}|CCCC@{\hspace{10pt}}|CCCCC}
\hline
\multicolumn{2}{c|}{Specifications}
& \multicolumn{4}{c|}{\begin{tabular}[c]{@{}c@{}}RMSE relative\\to TFGLS\end{tabular}}
& \multicolumn{5}{c}{\begin{tabular}[c]{@{}c@{}}Rejection probability (\%) of truth\\by 5\% Wald test\end{tabular}} \\
\hline
$I$ & $\nu_1$ & OLS & FGLS$^*$ & KFGLS$^*$ & CSX & TFGLS & OLS & FGLS$^*$ & KFGLS$^*$ & CSX \\
\hline
&&&&&&&&&&\\[-1.6ex]
400 & $-1$ & 1.26 & 1.02 & 0.96 & 0.97 & 5.2 & 5.5 & 5.3 & 5.8 & 5.3 \\
400 & $-.5$ & 1.00 & 1.00 & 0.96 & 0.96 & 5.4 & 5.3 & 5.3 & 5.9 & 5.4 \\
400 & 0 & 1.02 & 1.01 & 0.99 & 0.99 & 5.2 & 5.2 & 5.3 & 5.8 & 6.3 \\
400 & .5 & 1.10 & 1.01 & 0.98 & 0.98 & 5.2 & 5.2 & 5.1 & 5.8 & 6.6 \\
400 & 1 & 1.47 & 1.01 & 0.90 & 0.90 & 5.3 & 5.1 & 5.3 & 5.9 & 7.1 \\[0.8ex]
\hline
&&&&&&&&&&\\[-1.6ex]
100 & $-1$ & 1.25 & 1.02 & 0.99 & 1.05 & 5.4 & 5.1 & 5.1 & 8.2 & 10.6 \\
100 & $-.5$ & 1.01 & 1.01 & 0.97 & 1.03 & 5.5 & 5.1 & 5.2 & 8.2 & 11.8 \\
100 & 0 & 1.02 & 1.01 & 0.98 & 1.03 & 5.8 & 5.4 & 5.3 & 7.9 & 12.9 \\
100 & .5 & 1.10 & 1.02 & 0.97 & 0.98 & 5.8 & 5.4 & 5.4 & 7.8 & 13.9 \\
100 & 1 & 1.48 & 1.01 & 0.88 & 0.86 & 4.7 & 4.7 & 4.7 & 7.6 & 14.3 \\[0.8ex]
\hline
&&&&&&&&&&\\[-1.6ex]
50 & $-1$ & 1.21 & 1.01 & 1.01 & 1.20 & 5.6 & 5.6 & 5.1 & 12.5 & 23.7 \\
50 & $-.5$ & 1.01 & 1.01 & 0.98 & 1.18 & 6.3 & 5.7 & 5.7 & 11.6 & 25.8 \\
50 & 0 & 1.03 & 1.01 & 0.99 & 1.18 & 6.3 & 5.6 & 5.6 & 11.5 & 28.5 \\
50 & .5 & 1.11 & 1.02 & 0.94 & 1.05 & 6.1 & 5.4 & 5.5 & 11.2 & 30.3 \\
50 & 1 & 1.49 & 1.01 & 0.89 & 0.87 & 5.5 & 5.3 & 5.3 & 11.5 & 30.9 \\[0.8ex]
\hline
\end{tabular}
\caption{Design 1, $\mathbf{T = 20}$, 10000 Monte Carlo trials. Left panel: RMSE of estimators relative to that of TFGLS ($> 1$ means estimator is less precise than TFGLS). Right panel: rejection probability (\%) of the truth by the nominal 5\% Wald test.  $T=20$ is not considered by \citet{CSX2025}. In Tables~\ref{tab:des1T10}--\ref{tab:des1T20}, the non-stationary AR(1) working model of Appendix~\ref{app:ar1wm} with $\rho\in[-2,2]$ lowers the RMSE of TFGLS by 5--9\% when $\nu_1=1$ and leaves it similar otherwise; size also remains similar.}\label{tab:des1T20}
\end{table}

\begin{table}[H]
\centering
\footnotesize
\renewcommand{\arraystretch}{1.07}
\setlength{\tabcolsep}{4pt}
\newcolumntype{C}{>{\centering\arraybackslash}p{1.2cm}}
\captionsetup{font=small,skip=4pt}
\begin{tabular}{rc@{\hspace{12pt}}|CCCCC@{\hspace{10pt}}|CCCCC}
\hline
\multicolumn{2}{c|}{Specifications}
& \multicolumn{5}{c|}{\begin{tabular}[c]{@{}c@{}}RMSE relative\\to TFGLS\end{tabular}}
& \multicolumn{5}{c}{\begin{tabular}[c]{@{}c@{}}Rejection probability (\%) of truth\\by 5\% Wald test\end{tabular}} \\
\hline
$I$ & Design & $\widehat{\tau}_{w,\text{blue}}$ & OLS & FGLS$^*$ & KFGLS$^*$ & CSX & TFGLS & OLS & FGLS$^*$ & KFGLS$^*$ & CSX \\
\hline
&&&&&&&&&&&\\[-1.6ex]
400 & (a) & 1.00 & 1.00 & 1.00 & 1.01 & 1.01 & 5.2 & 5.1 & 5.1 & 5.5 & 5.7 \\
400 & (b) & 1.00 & 1.05 & 1.00 & 1.01 & 1.01 & 5.2 & 5.3 & 5.1 & 5.5 & 5.9 \\
400 & (c) & 0.71 & 1.02 & 1.02 & 0.72 & 0.72 & 5.1 & 5.1 & 5.1 & 5.6 & 5.6 \\
400 & (d) & 0.99 & 1.02 & 1.00 & 1.00 & 1.01 & 5.2 & 5.2 & 5.1 & 5.5 & 5.7 \\
400 & (e) & 0.99 & 1.07 & 1.00 & 1.00 & 1.00 & 5.0 & 5.2 & 5.0 & 5.3 & 5.4 \\
400 & (f) & 0.98 & 1.09 & 1.00 & 0.99 & 0.99 & 5.2 & 4.8 & 5.2 & 5.6 & 5.9 \\
400 & (g) & 0.97 & 1.32 & 1.00 & 0.98 & 0.98 & 5.1 & 4.6 & 5.1 & 5.4 & 5.4 \\
400 & (h) & 0.98 & 1.30 & 1.07 & 0.99 & 0.99 & 5.3 & 5.4 & 5.4 & 5.6 & 5.8 \\
400 & (i) & 1.00 & 1.25 & 1.11 & 1.01 & 1.01 & 5.2 & 5.4 & 5.4 & 5.5 & 5.9 \\
400 & (j) & 0.99 & 1.49 & 1.07 & 0.99 & 0.99 & 5.4 & 5.2 & 5.4 & 5.7 & 6.1 \\
400 & (k) & 0.99 & 1.36 & 1.11 & 1.00 & 1.00 & 5.5 & 5.0 & 5.3 & 5.8 & 6.0 \\[0.8ex]
\hline
&&&&&&&&&&&\\[-1.6ex]
100 & (a) & 1.00 & 1.00 & 0.99 & 1.02 & 1.05 & 5.9 & 5.6 & 5.5 & 7.0 & 8.4 \\
100 & (b) & 0.99 & 1.05 & 0.99 & 1.02 & 1.04 & 5.9 & 5.6 & 5.5 & 7.0 & 8.4 \\
100 & (c) & 0.69 & 1.01 & 1.01 & 0.71 & 0.73 & 6.0 & 5.5 & 5.6 & 6.5 & 7.8 \\
100 & (d) & 0.99 & 1.02 & 1.00 & 1.02 & 1.04 & 5.7 & 5.4 & 5.5 & 7.0 & 8.3 \\
100 & (e) & 0.98 & 1.06 & 1.00 & 1.01 & 1.03 & 5.7 & 5.1 & 5.4 & 6.8 & 7.8 \\
100 & (f) & 0.98 & 1.10 & 1.00 & 1.00 & 1.03 & 5.6 & 5.6 & 5.4 & 6.5 & 8.0 \\
100 & (g) & 0.96 & 1.32 & 1.00 & 1.00 & 1.02 & 5.6 & 5.5 & 5.5 & 6.7 & 8.2 \\
100 & (h) & 0.98 & 1.30 & 1.07 & 1.01 & 1.03 & 5.4 & 5.6 & 5.5 & 6.5 & 8.2 \\
100 & (i) & 0.99 & 1.24 & 1.10 & 1.02 & 1.05 & 5.5 & 5.7 & 5.5 & 6.6 & 8.3 \\
100 & (j) & 0.98 & 1.53 & 1.08 & 1.00 & 1.03 & 5.0 & 5.8 & 5.2 & 5.9 & 7.8 \\
100 & (k) & 0.98 & 1.38 & 1.11 & 1.01 & 1.04 & 5.1 & 5.6 & 5.3 & 6.1 & 7.9 \\[0.8ex]
\hline
&&&&&&&&&&&\\[-1.6ex]
50 & (a) & 0.99 & 0.99 & 0.99 & 1.04 & 1.13 & 6.6 & 6.0 & 6.1 & 8.6 & 13.4 \\
50 & (b) & 0.99 & 1.04 & 0.99 & 1.03 & 1.12 & 6.3 & 5.9 & 5.7 & 8.3 & 13.7 \\
50 & (c) & 0.69 & 1.00 & 1.00 & 0.74 & 0.78 & 7.0 & 6.0 & 6.1 & 8.6 & 12.8 \\
50 & (d) & 0.98 & 1.01 & 0.99 & 1.03 & 1.12 & 6.4 & 6.2 & 5.8 & 8.5 & 13.4 \\
50 & (e) & 0.98 & 1.05 & 0.99 & 1.03 & 1.11 & 6.7 & 6.1 & 5.8 & 8.8 & 13.6 \\
50 & (f) & 0.97 & 1.09 & 0.99 & 1.01 & 1.09 & 6.4 & 5.9 & 5.7 & 8.3 & 12.8 \\
50 & (g) & 0.95 & 1.32 & 0.98 & 1.02 & 1.09 & 6.0 & 5.8 & 5.1 & 8.6 & 13.0 \\
50 & (h) & 0.97 & 1.29 & 1.06 & 1.03 & 1.10 & 6.1 & 6.0 & 6.1 & 8.2 & 13.3 \\
50 & (i) & 0.98 & 1.22 & 1.09 & 1.04 & 1.12 & 6.2 & 6.1 & 6.1 & 8.5 & 13.3 \\
50 & (j) & 0.97 & 1.49 & 1.06 & 1.02 & 1.09 & 6.1 & 6.2 & 5.9 & 7.9 & 13.0 \\
50 & (k) & 0.98 & 1.35 & 1.09 & 1.02 & 1.10 & 6.4 & 6.3 & 6.0 & 8.0 & 13.2 \\[0.8ex]
\hline
\end{tabular}
\caption{Design 2, $\mathbf{T = 10}$, 10000 Monte Carlo trials. Left panel: RMSE of each estimator relative to that of TFGLS (a value larger than 1 means the estimator is less precise than TFGLS); $\widehat{\tau}_{w,\text{blue}}$ is the infeasible GLS estimator that uses the true error covariance. Right panel: rejection probability (\%) of the truth by the nominal 5\% Wald test. Designs (a): i.i.d.; (b): AR(1) with coefficient .6; (c): AR(2) with $(\nu_1,\nu_2) = (0,-.9025)$; (d)--(e): MA(1) with $\theta_1 = .6, -.6$; (f)--(g): ARMA(1,1) with $(\nu_1,\theta_1) = (.6,.6), (-.6,-.6)$; (h)--(k): ARMA(1,1) $e_{i,t}$ with $(\nu_1,\theta_1) = (.85,.2)$ in (h), (j) and $(.85,-.2)$ in (i), (k), with a U-shaped (h, i) or falling (j, k) variance over time. Design 2(c) is adversarial to the AR(1) working model.}\label{tab:des2T10}
\end{table}

\begin{table}[H]
\centering
\footnotesize
\renewcommand{\arraystretch}{1.07}
\setlength{\tabcolsep}{4pt}
\newcolumntype{C}{>{\centering\arraybackslash}p{1.2cm}}
\captionsetup{font=small,skip=4pt}
\begin{tabular}{rc@{\hspace{12pt}}|CCCCC@{\hspace{10pt}}|CCCCC}
\hline
\multicolumn{2}{c|}{Specifications}
& \multicolumn{5}{c|}{\begin{tabular}[c]{@{}c@{}}RMSE relative\\to TFGLS\end{tabular}}
& \multicolumn{5}{c}{\begin{tabular}[c]{@{}c@{}}Rejection probability (\%) of truth\\by 5\% Wald test\end{tabular}} \\
\hline
$I$ & Design & $\widehat{\tau}_{w,\text{blue}}$ & OLS & FGLS$^*$ & KFGLS$^*$ & CSX & TFGLS & OLS & FGLS$^*$ & KFGLS$^*$ & CSX \\
\hline
&&&&&&&&&&&\\[-1.6ex]
400 & (a) & 1.00 & 1.00 & 1.00 & 1.01 & 1.03 & 5.1 & 5.1 & 5.1 & 5.7 & 6.0 \\
400 & (b) & 1.00 & 1.04 & 1.00 & 1.01 & 1.02 & 5.1 & 5.0 & 5.1 & 5.6 & 6.2 \\
400 & (c) & 0.71 & 1.15 & 1.15 & 0.72 & 0.73 & 5.3 & 5.1 & 5.1 & 5.7 & 5.8 \\
400 & (d) & 1.00 & 1.01 & 1.00 & 1.01 & 1.02 & 5.2 & 5.1 & 5.0 & 5.9 & 6.1 \\
400 & (e) & 0.97 & 1.08 & 1.01 & 0.99 & 1.00 & 4.9 & 5.0 & 4.8 & 5.4 & 4.9 \\
400 & (f) & 0.99 & 1.07 & 1.01 & 1.01 & 1.02 & 5.2 & 5.2 & 5.1 & 5.8 & 6.8 \\
400 & (g) & 0.95 & 1.33 & 1.00 & 0.97 & 0.98 & 5.0 & 5.2 & 5.0 & 5.6 & 4.8 \\
400 & (h) & 1.00 & 1.30 & 1.08 & 1.01 & 1.03 & 4.9 & 4.8 & 4.8 & 5.5 & 6.4 \\
400 & (i) & 0.99 & 1.24 & 1.13 & 1.00 & 1.02 & 5.0 & 5.0 & 4.8 & 5.6 & 6.2 \\
400 & (j) & 0.99 & 1.46 & 1.07 & 1.01 & 1.02 & 5.2 & 5.2 & 5.2 & 5.8 & 6.5 \\
400 & (k) & 0.98 & 1.36 & 1.14 & 0.99 & 1.00 & 5.2 & 5.4 & 5.3 & 5.7 & 6.5 \\[0.8ex]
\hline
&&&&&&&&&&&\\[-1.6ex]
100 & (a) & 1.00 & 1.00 & 1.00 & 1.04 & 1.14 & 5.8 & 5.6 & 5.6 & 8.3 & 13.6 \\
100 & (b) & 1.00 & 1.04 & 1.00 & 1.04 & 1.14 & 5.9 & 5.6 & 5.8 & 8.3 & 13.7 \\
100 & (c) & 0.72 & 1.15 & 1.15 & 0.76 & 0.83 & 5.1 & 5.0 & 5.0 & 8.0 & 12.4 \\
100 & (d) & 0.99 & 1.01 & 1.00 & 1.04 & 1.14 & 5.9 & 5.7 & 5.5 & 8.4 & 14.0 \\
100 & (e) & 0.97 & 1.07 & 1.00 & 1.02 & 1.12 & 5.6 & 5.3 & 5.5 & 8.5 & 12.0 \\
100 & (f) & 0.98 & 1.07 & 1.00 & 1.03 & 1.13 & 5.7 & 5.6 & 5.3 & 8.1 & 14.0 \\
100 & (g) & 0.95 & 1.34 & 1.00 & 1.01 & 1.09 & 5.4 & 5.4 & 5.2 & 8.1 & 10.9 \\
100 & (h) & 0.99 & 1.30 & 1.07 & 1.05 & 1.15 & 5.3 & 5.5 & 5.2 & 8.0 & 14.2 \\
100 & (i) & 0.99 & 1.24 & 1.12 & 1.04 & 1.14 & 5.3 & 5.5 & 5.2 & 8.1 & 14.0 \\
100 & (j) & 0.99 & 1.46 & 1.06 & 1.04 & 1.13 & 5.3 & 5.5 & 5.5 & 7.9 & 13.7 \\
100 & (k) & 0.97 & 1.34 & 1.13 & 1.02 & 1.11 & 5.6 & 5.4 & 5.6 & 8.0 & 13.9 \\[0.8ex]
\hline
&&&&&&&&&&&\\[-1.6ex]
50 & (a) & 0.99 & 0.99 & 0.99 & 1.06 & 1.42 & 6.5 & 6.0 & 6.0 & 11.3 & 31.7 \\
50 & (b) & 0.99 & 1.04 & 0.99 & 1.06 & 1.43 & 6.4 & 5.9 & 5.9 & 11.3 & 32.6 \\
50 & (c) & 0.71 & 1.14 & 1.14 & 0.80 & 1.02 & 6.0 & 5.9 & 5.8 & 12.3 & 31.4 \\
50 & (d) & 0.99 & 1.00 & 1.00 & 1.06 & 1.41 & 6.3 & 6.0 & 5.8 & 11.4 & 32.2 \\
50 & (e) & 0.97 & 1.07 & 1.00 & 1.04 & 1.38 & 6.4 & 5.7 & 5.9 & 11.7 & 29.3 \\
50 & (f) & 0.98 & 1.06 & 1.00 & 1.06 & 1.43 & 6.0 & 5.8 & 5.5 & 11.3 & 33.5 \\
50 & (g) & 0.95 & 1.32 & 0.98 & 1.06 & 1.37 & 6.1 & 5.8 & 5.4 & 12.5 & 29.0 \\
50 & (h) & 0.99 & 1.28 & 1.07 & 1.09 & 1.42 & 6.6 & 6.5 & 6.2 & 12.8 & 34.6 \\
50 & (i) & 0.98 & 1.22 & 1.10 & 1.07 & 1.40 & 6.9 & 6.4 & 6.3 & 12.7 & 34.6 \\
50 & (j) & 0.98 & 1.45 & 1.05 & 1.08 & 1.42 & 6.0 & 6.3 & 5.8 & 11.6 & 33.1 \\
50 & (k) & 0.96 & 1.34 & 1.12 & 1.05 & 1.39 & 6.3 & 6.4 & 6.1 & 11.6 & 33.1 \\[0.8ex]
\hline
\end{tabular}
\caption{Design 2, $\mathbf{T = 20}$, 10000 Monte Carlo trials. Left panel: RMSE of each estimator relative to that of TFGLS (a value larger than 1 means the estimator is less precise than TFGLS); $\widehat{\tau}_{w,\text{blue}}$ is the infeasible GLS estimator that uses the true error covariance. Right panel: rejection probability (\%) of the truth by the nominal 5\% Wald test. Designs (a): i.i.d.; (b): AR(1) with coefficient .6; (c): AR(2) with $(\nu_1,\nu_2) = (0,-.9025)$; (d)--(e): MA(1) with $\theta_1 = .6, -.6$; (f)--(g): ARMA(1,1) with $(\nu_1,\theta_1) = (.6,.6), (-.6,-.6)$; (h)--(k): ARMA(1,1) $e_{i,t}$ with $(\nu_1,\theta_1) = (.85,.2)$ in (h), (j) and $(.85,-.2)$ in (i), (k), with a U-shaped (h, i) or falling (j, k) variance over time. Design 2(c) is adversarial to the AR(1) working model.}\label{tab:des2T20}
\end{table}

\clearpage

\begin{table}[H]
\centering
\footnotesize
\renewcommand{\arraystretch}{1.02}
\setlength{\tabcolsep}{4pt}
\newcolumntype{C}{>{\centering\arraybackslash}p{1.2cm}}
\captionsetup{font=small,skip=4pt}
\begin{tabular}{rc@{\hspace{12pt}}|CCCCC@{\hspace{10pt}}|CCCCC}
\hline
\multicolumn{2}{c|}{Specifications}
& \multicolumn{5}{c|}{\begin{tabular}[c]{@{}c@{}}RMSE relative\\to TFGLS\end{tabular}}
& \multicolumn{5}{c}{\begin{tabular}[c]{@{}c@{}}Rejection probability (\%) of truth\\by 5\% Wald test\end{tabular}} \\
\hline
\multicolumn{12}{l}{Design 1}\\
\hline
$I$ & $\nu_1$ & $\widehat{\tau}_{w,\text{blue}}$ & OLS & FGLS$^*$ & KFGLS$^*$ & CSX & TFGLS & OLS & FGLS$^*$ & KFGLS$^*$ & CSX \\
\hline
&&&&&&&&&&&\\[-1.6ex]
400 & $-1$ & -- & 1.32 & 1.05 & 0.98 & 0.95 & 5.4 & 5.5 & 5.4 & 6.1 & 6.1 \\
400 & $-.5$ & -- & 1.01 & 1.00 & 1.02 & 0.99 & 5.6 & 5.4 & 5.3 & 6.4 & 6.2 \\
400 & 0 & -- & 1.01 & 1.00 & 1.01 & 1.00 & 5.4 & 4.9 & 5.0 & 5.9 & 6.2 \\
400 & .5 & -- & 1.08 & 1.00 & 0.98 & 0.98 & 5.2 & 5.2 & 5.0 & 5.4 & 6.2 \\
400 & 1 & -- & 1.37 & 1.01 & 0.94 & 0.93 & 5.1 & 5.0 & 5.1 & 5.5 & 6.7 \\[0.8ex]
\hline
&&&&&&&&&&&\\[-1.6ex]
100 & $-1$ & -- & 1.31 & 1.03 & 0.97 & 1.04 & 5.2 & 5.7 & 4.8 & 6.8 & 10.6 \\
100 & $-.5$ & -- & 1.01 & 1.00 & 1.00 & 1.04 & 6.1 & 5.5 & 5.4 & 7.2 & 11.9 \\
100 & 0 & -- & 1.02 & 1.01 & 1.00 & 1.02 & 5.9 & 5.6 & 5.4 & 7.1 & 12.3 \\
100 & .5 & -- & 1.09 & 1.01 & 0.99 & 0.97 & 6.0 & 5.8 & 5.5 & 7.0 & 12.8 \\
100 & 1 & -- & 1.36 & 1.01 & 0.92 & 0.87 & 5.8 & 6.0 & 5.4 & 6.9 & 13.2 \\[0.8ex]
\hline
\multicolumn{12}{l}{Design 2}\\
\hline
$I$ & Design & $\widehat{\tau}_{w,\text{blue}}$ & OLS & FGLS$^*$ & KFGLS$^*$ & CSX & TFGLS & OLS & FGLS$^*$ & KFGLS$^*$ & CSX \\
\hline
&&&&&&&&&&&\\[-1.6ex]
400 & (a) & 1.00 & 1.00 & 1.00 & 1.02 & 1.02 & 4.8 & 4.7 & 4.7 & 5.9 & 6.2 \\
400 & (b) & 1.00 & 1.06 & 1.00 & 1.02 & 1.02 & 5.9 & 6.0 & 5.8 & 6.4 & 6.9 \\
400 & (c) & 0.79 & 1.11 & 1.11 & 0.80 & 0.81 & 5.1 & 5.2 & 5.2 & 5.6 & 5.9 \\
400 & (d) & 0.99 & 1.02 & 1.00 & 1.02 & 1.02 & 5.5 & 5.1 & 5.3 & 5.9 & 6.4 \\
400 & (e) & 0.99 & 1.05 & 1.00 & 1.01 & 1.02 & 5.1 & 5.1 & 5.1 & 6.0 & 6.1 \\
400 & (f) & 0.98 & 1.10 & 1.00 & 1.00 & 1.01 & 5.6 & 5.5 & 5.3 & 5.9 & 6.8 \\
400 & (g) & 0.97 & 1.20 & 1.01 & 1.01 & 1.00 & 5.3 & 5.2 & 5.2 & 6.3 & 6.2 \\
400 & (h) & 0.98 & 1.24 & 1.05 & 1.00 & 1.01 & 5.6 & 5.3 & 5.4 & 6.3 & 6.7 \\
400 & (i) & 1.00 & 1.18 & 1.07 & 1.02 & 1.03 & 5.1 & 5.4 & 5.4 & 6.1 & 6.4 \\
400 & (j) & 0.98 & 1.66 & 1.09 & 1.00 & 1.01 & 5.3 & 5.7 & 5.5 & 5.6 & 6.5 \\
400 & (k) & 0.99 & 1.48 & 1.15 & 1.00 & 1.02 & 5.5 & 5.5 & 5.6 & 5.7 & 6.5 \\[0.8ex]
\hline
&&&&&&&&&&&\\[-1.6ex]
100 & (a) & 0.99 & 0.99 & 0.99 & 1.06 & 1.13 & 6.2 & 5.9 & 5.9 & 8.8 & 13.5 \\
100 & (b) & 0.99 & 1.05 & 0.99 & 1.05 & 1.13 & 6.4 & 5.9 & 5.9 & 8.5 & 14.0 \\
100 & (c) & 0.79 & 1.11 & 1.11 & 0.86 & 0.90 & 5.9 & 5.8 & 5.8 & 8.7 & 13.2 \\
100 & (d) & 0.99 & 1.01 & 0.99 & 1.05 & 1.12 & 6.2 & 5.9 & 5.5 & 9.0 & 13.6 \\
100 & (e) & 0.98 & 1.04 & 0.99 & 1.05 & 1.12 & 6.6 & 6.1 & 6.0 & 9.3 & 13.3 \\
100 & (f) & 0.97 & 1.10 & 0.99 & 1.03 & 1.10 & 6.6 & 6.2 & 6.2 & 9.0 & 14.2 \\
100 & (g) & 0.97 & 1.19 & 1.00 & 1.07 & 1.11 & 6.2 & 5.7 & 5.7 & 9.9 & 12.8 \\
100 & (h) & 0.97 & 1.23 & 1.04 & 1.06 & 1.13 & 6.0 & 6.1 & 5.6 & 9.3 & 14.2 \\
100 & (i) & 0.99 & 1.17 & 1.06 & 1.06 & 1.13 & 6.6 & 6.4 & 6.2 & 9.4 & 14.3 \\
100 & (j) & 0.98 & 1.63 & 1.08 & 1.02 & 1.10 & 5.5 & 6.3 & 5.8 & 6.9 & 13.1 \\
100 & (k) & 0.99 & 1.47 & 1.13 & 1.03 & 1.12 & 5.3 & 5.6 & 5.3 & 7.1 & 12.7 \\[0.8ex]
\hline
\end{tabular}
\caption{More cohorts: four treated cohorts (treated in periods 5, 6, 7 and 8) and a never-treated cohort, each with probability 1/5, $\mathbf{T = 10}$, 10000 Monte Carlo trials. Left panel: RMSE of each estimator relative to that of TFGLS (a value larger than 1 means the estimator is less precise than TFGLS); $\widehat{\tau}_{w,\text{blue}}$ is the infeasible GLS estimator that uses the true error covariance (Design 2 only). Right panel: rejection probability (\%) of the truth by the nominal 5\% Wald test. The designs for the errors are those of Tables \ref{tab:des1T10} and \ref{tab:des2T10}.}
\label{tab:morecoh}
\end{table}

\begin{table}[H]
\centering
\footnotesize
\renewcommand{\arraystretch}{1.07}
\setlength{\tabcolsep}{3.5pt}
\captionsetup{font=small,skip=4pt}
\begin{tabular}{c@{\hspace{8pt}}|cccccc@{\hspace{7pt}}|cccccc}
\hline
& \multicolumn{12}{c}{RMSE relative to TFGLS} \\
\hline
Design & $\widehat{\tau}_{w,\text{blue}}$ & TFGLS$^*$ & OLS & FGLS & KFGLS$^*$ & CSX & $\widehat{\tau}_{w,\text{blue}}$ & TFGLS$^*$ & OLS & FGLS & KFGLS$^*$ & CSX \\
\hline
& \multicolumn{6}{c|}{$I=4000$} & \multicolumn{6}{c}{$I=400$} \\[0.3ex]
(a) & 0.99 & 0.99 & 0.99 & 0.99 & 1.04 & 1.11 & 0.95 & 0.95 & 0.95 & 0.95 & 1.30 & 1.82 \\
(b) & 1.00 & 1.00 & 1.04 & 1.00 & 1.01 & 1.03 & 0.98 & 0.98 & 1.03 & 0.99 & 1.09 & 1.29 \\
(c) & 0.71 & 1.00 & 1.15 & 1.14 & 0.77 & 0.86 & 0.68 & 0.95 & 1.10 & 1.10 & 1.07 & 1.63 \\
(d) & 0.99 & 1.00 & 1.01 & 1.00 & 1.01 & 1.05 & 0.97 & 0.97 & 0.98 & 0.97 & 1.13 & 1.43 \\
(e) & 0.96 & 0.98 & 1.06 & 0.99 & 1.09 & 1.29 & 0.86 & 0.88 & 0.95 & 0.88 & 1.69 & 2.71 \\
(f) & 0.99 & 1.00 & 1.07 & 1.00 & 1.00 & 1.01 & 0.98 & 0.99 & 1.07 & 0.99 & 1.03 & 1.13 \\
(g) & 0.92 & 0.97 & 1.31 & 0.97 & 1.10 & 1.35 & 0.81 & 0.85 & 1.12 & 0.85 & 1.81 & 2.98 \\
(h) & 1.00 & 1.00 & 1.30 & 1.18 & 1.00 & 1.00 & 0.99 & 1.00 & 1.28 & 1.17 & 1.01 & 1.05 \\
(i) & 0.99 & 1.00 & 1.23 & 1.17 & 0.99 & 1.00 & 0.99 & 1.00 & 1.23 & 1.18 & 1.02 & 1.06 \\
(j) & 1.00 & 1.00 & 1.47 & 1.26 & 1.00 & 1.00 & 0.99 & 1.00 & 1.45 & 1.24 & 1.02 & 1.05 \\
(k) & 0.98 & 1.00 & 1.36 & 1.26 & 0.98 & 0.98 & 0.97 & 0.99 & 1.35 & 1.25 & 1.01 & 1.05 \\[0.8ex]
\hline
& \multicolumn{6}{c|}{$I=2000$} & \multicolumn{6}{c}{$I=100$} \\[0.3ex]
(a) & 0.99 & 0.99 & 0.99 & 0.99 & 1.07 & 1.20 & 0.83 & 0.83 & 0.83 & 0.83 & 1.67 & 3.26 \\
(b) & 1.00 & 1.00 & 1.05 & 1.01 & 1.02 & 1.06 & 0.93 & 0.93 & 0.98 & 0.94 & 1.26 & 2.04 \\
(c) & 0.70 & 0.99 & 1.13 & 1.13 & 0.80 & 0.96 & 0.60 & 0.85 & 0.97 & 0.97 & 1.56 & 3.17 \\
(d) & 0.99 & 0.99 & 1.01 & 0.99 & 1.02 & 1.09 & 0.89 & 0.90 & 0.90 & 0.90 & 1.37 & 2.43 \\
(e) & 0.94 & 0.97 & 1.04 & 0.97 & 1.19 & 1.53 & 0.62 & 0.64 & 0.69 & 0.64 & 2.10 & 4.36 \\
(f) & 0.99 & 1.00 & 1.07 & 1.00 & 1.00 & 1.02 & 0.95 & 0.96 & 1.03 & 0.96 & 1.13 & 1.60 \\
(g) & 0.91 & 0.95 & 1.26 & 0.95 & 1.22 & 1.65 & 0.58 & 0.61 & 0.81 & 0.61 & 2.27 & 4.81 \\
(h) & 1.00 & 1.00 & 1.30 & 1.17 & 1.00 & 1.01 & 0.99 & 0.99 & 1.28 & 1.16 & 1.07 & 1.26 \\
(i) & 0.99 & 1.00 & 1.24 & 1.18 & 1.00 & 1.01 & 0.97 & 0.98 & 1.23 & 1.17 & 1.09 & 1.37 \\
(j) & 0.99 & 1.00 & 1.48 & 1.27 & 1.00 & 1.01 & 0.98 & 0.99 & 1.45 & 1.24 & 1.06 & 1.25 \\
(k) & 0.97 & 1.00 & 1.35 & 1.26 & 0.99 & 0.99 & 0.95 & 0.97 & 1.33 & 1.23 & 1.06 & 1.31 \\[0.8ex]
\hline
& \multicolumn{6}{c|}{$I=1000$} & \multicolumn{6}{c}{$I=50$} \\[0.3ex]
(a) & 0.98 & 0.98 & 0.98 & 0.98 & 1.13 & 1.38 & 0.77 & 0.77 & 0.77 & 0.77 & 1.95 & 5.56 \\
(b) & 0.99 & 0.99 & 1.04 & 1.00 & 1.03 & 1.12 & 0.90 & 0.90 & 0.95 & 0.91 & 1.43 & 3.51 \\
(c) & 0.70 & 0.98 & 1.13 & 1.13 & 0.89 & 1.17 & 0.53 & 0.75 & 0.85 & 0.85 & 1.74 & 5.13 \\
(d) & 0.98 & 0.99 & 1.00 & 0.99 & 1.06 & 1.19 & 0.85 & 0.86 & 0.87 & 0.86 & 1.58 & 4.27 \\
(e) & 0.93 & 0.95 & 1.02 & 0.95 & 1.37 & 1.94 & 0.52 & 0.53 & 0.57 & 0.53 & 2.26 & 6.83 \\
(f) & 0.99 & 1.00 & 1.07 & 1.00 & 1.01 & 1.05 & 0.93 & 0.94 & 1.01 & 0.94 & 1.23 & 2.57 \\
(g) & 0.87 & 0.92 & 1.22 & 0.92 & 1.43 & 2.12 & 0.44 & 0.46 & 0.61 & 0.46 & 2.29 & 7.18 \\
(h) & 1.00 & 1.00 & 1.30 & 1.17 & 1.01 & 1.02 & 0.98 & 0.98 & 1.26 & 1.14 & 1.12 & 1.72 \\
(i) & 0.99 & 1.00 & 1.23 & 1.17 & 1.00 & 1.02 & 0.95 & 0.96 & 1.19 & 1.14 & 1.13 & 1.97 \\
(j) & 1.00 & 1.00 & 1.47 & 1.26 & 1.01 & 1.02 & 0.96 & 0.97 & 1.41 & 1.22 & 1.11 & 1.75 \\
(k) & 0.97 & 1.00 & 1.33 & 1.24 & 0.99 & 1.01 & 0.93 & 0.95 & 1.28 & 1.19 & 1.12 & 1.93 \\[0.8ex]
\hline
\end{tabular}
\caption{Design 2 with entity-level heterogeneity $\tau_{i,t}=\tau_t(E_i)+\sqrt{T-t}\,z_i$ (Section~\ref{sec:mc-hetit}), $\mathbf{T = 20}$, 10000 Monte Carlo trials. RMSE of each estimator relative to that of TFGLS (a value larger than 1 means the estimator is less precise than TFGLS) for the realized target $ES_{\text{avg}}$. $\widehat{\tau}_{w,\text{blue}}$ is the infeasible GLS estimator that uses the true error covariance. TFGLS$^*$ is the infeasible oracle $\widehat{\tau}_I(\rho^*_I)$, with $\rho^*_I$ minimizing the true variance $\sigma^2_I(\rho)$ given the realized cohorts --- its smaller RMSE reflects the cost to TFGLS due to estimation of $\rho^*_I$ under this setup of heterogeneous $\tau_{i,t}$. FGLS is the feasible FGLS of Section~\ref{sec:mc-hetit}. Designs (a)--(k) are as in Table~\ref{tab:des2T10}.}
\label{tab:hetT20}
\end{table}
\clearpage

\section{Empirical illustrations}\label{sec:empirical}

We present two examples --- the effect of adoption of unilateral divorce laws on divorce rates \citep{Wolfers2006} and the effect of hospitalization on out-of-pocket medical expenses \citep{Dobkin2018}. The divorce law example revisited in \citet{dCdH2023-survey} involves $I=47$, $T=33$ and 10 treated cohorts (four singleton). We do not include KFGLS$^*$ and CSX in this study because our simulations in Section~\ref{sec:mc} suggest that such setups may not be particularly suitable for those estimators.\footnote{KFGLS$^*$ and CSX are not reported because $I=47$ is small relative to $T=33$, and there are many small cohorts (four are singletons). By Lemma \ref{lemma:KFGLSmse} and Section~\ref{sec:mc-results}, the standard errors of KFGLS$^*$ would likely understate its variability by a large factor --- about 1.9 if the balanced i.i.d. benchmark of the lemma is applied here as a back-of-the-envelope calculation. CSX would suffer from many-moments problem since its estimates use $I=47$ (40 with balanced panel) while its PT-All condition gives 173, i.e., about four times as many, moment restrictions because the 10 treated cohorts change the law in periods 14, 15, $\ldots$, 22 and 30.} The hospitalization example revisited in \citet{SunAbraham2021} involves $I=656$, $T = 4$ and only 3 treated cohorts, and all cohorts are large. \citet{CSX2025} use this example as the empirical application of their CSX estimator, so we include all estimators for comparison here. The DiD literature does not rule out the unconditional parallel trend and no-anticipation in these two examples, so we maintain them to abstract from identification and instead focus on the matter of precision.

\subsection{Unilateral divorce laws: \citet{Wolfers2006}}
\label{sec:emp-divorce}

We follow the empirical analysis in the survey \citet{dCdH2023-survey}, which revisits \citet{Wolfers2006}'s study of the effect of the staggered adoption of unilateral divorce laws in 29 US states between 1968 and 1988 on post-adoption divorce rates. 

We focus on the event study plot in Figure 3 of \citet{dCdH2023-survey}. Like them, (i) we estimate the effects 0--15 years after the law change, with 10 placebo leads and endpoint binning; and (ii) state-specific time trends are not used since the leads estimates using all methods were already quite small without them. In addition to the five estimators --- the standard two-way fixed effects estimator (TWFE), \citet{SunAbraham2021}'s estimator (SA), \citet{CallawaySantanna2021}'s estimator (CS), \citet{dcdH2020}'s estimator (dCDH), and BJS (OLS) --- considered by them, we also consider FGLS and TFGLS, constructed as in Section~\ref{sec:mc}.\footnote{\label{fn:empconstr}TFGLS uses the AR(1) working model with $\rho_0=\rho_1=\rho \in \mathcal{P} = [-.99,.99]$ (Section~\ref{sec:mc-estimators}). FGLS uses Prais-Winsten-type AR(1) weights with $\widehat{\rho}_I$ the AR(1) coefficient of the BJS (OLS) residuals over all pairs of consecutive untreated cells (Section~\ref{sec:mc-hetit}). As in \citet{dCdH2023-survey}, all estimators weight by state population $p_{i,t}$: with $P_i:=\mathrm{diag}(p_{i,t})$ and $R_i(\rho)$ the AR(1) correlation matrix of state $i$'s observed years, FGLS and TFGLS use $\bar\Sigma_i(\rho)=P_i^{-1/2}R_i(\rho)P_i^{-1/2}$, so $\rho=0$ gives BJS (OLS), and our theory covers this (end of {\hyperref[sec:proofs]{Appendix~A}}). The panel has 37 missing untreated outcomes (25 for Louisiana).} Since four of the ten treated cohorts are singletons, whose treated residuals would be identically zero after subtracting their cohort-period averages, and three have only two states, the standard errors of BJS (OLS), FGLS and TFGLS constructed as in Section~\ref{sec:mc} would be artificially small. This might be a reason behind the reported high precision of BJS (OLS) estimator of instantaneous effect in \citet{dCdH2023-survey}. Hence, we use the conservative construction in Section~\ref{sec:bgresults} (footnote~\ref{fn:bjs0res}), with the treated residuals of BJS (OLS) subtracting its estimate and those of FGLS and TFGLS subtracting their own estimates of $\tau_w$.

\medskip

\begin{figure}[H]\centering
\includegraphics[width=.95\textwidth]{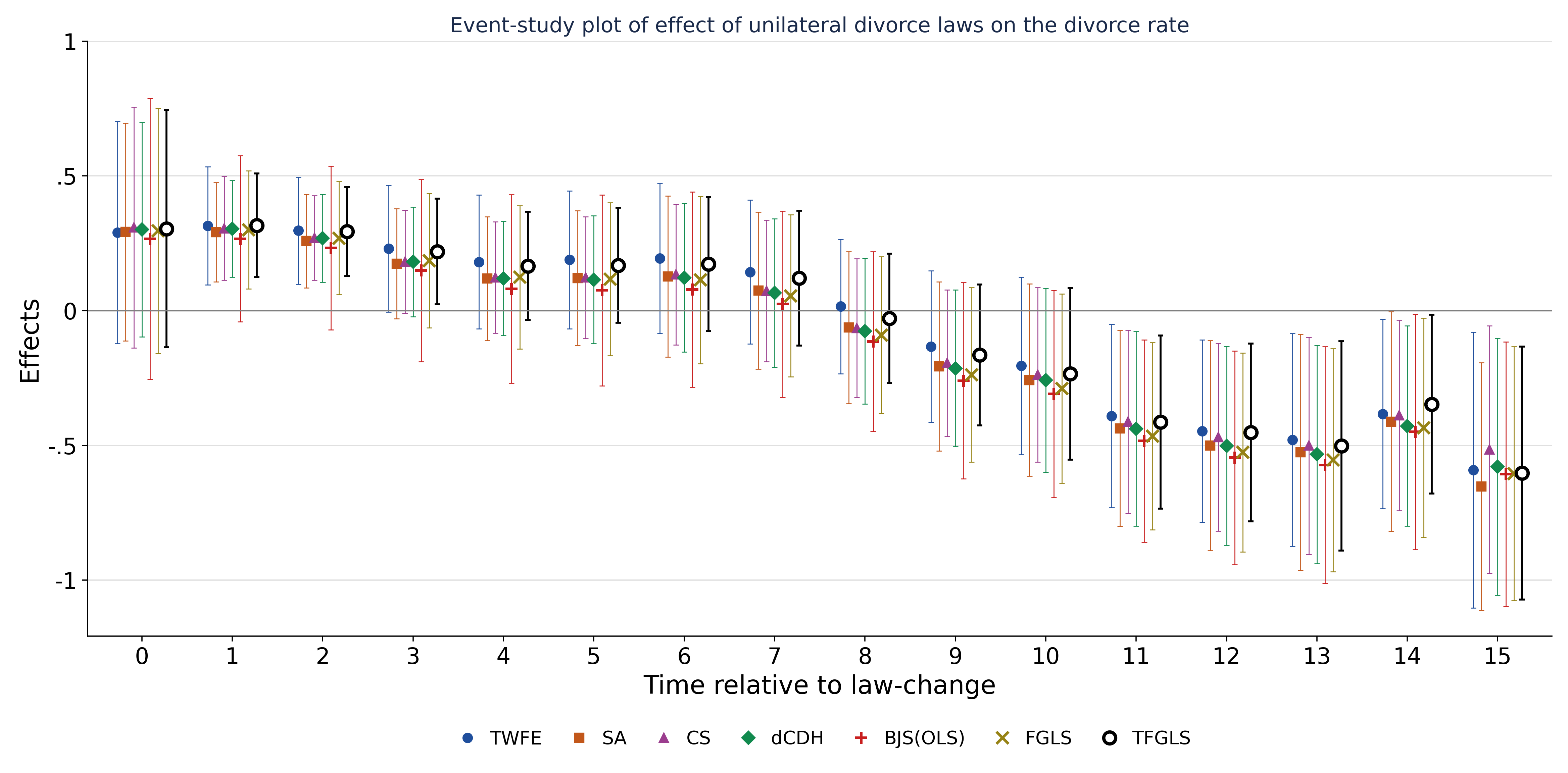}
\captionsetup{skip=2pt}
\caption{\footnotesize{Event study plot with point estimates and 95\% confidence intervals, $\pm t_{.975}(46)$ standard errors, from seven estimators. TWFE: two-way fixed effects estimator; SA: \citet{SunAbraham2021}; CS: \citet{CallawaySantanna2021}; dCDH: \citet{dcdH2020}; all four from the replication package of \citet{dCdH2023-survey}. BJS (OLS), FGLS and TFGLS are computed by our code with the same weighting by state population, etc., as in \citet{dCdH2023-survey}; see footnote~\ref{fn:empconstr}. For comparability, we drop the two always treated states AK and OK, and the two states IN and NM with missing post-treatment outcomes. The treated residuals in the standard errors subtract the BJS (OLS) estimate for BJS (OLS) and their own estimates of $\tau_w$ for FGLS and TFGLS (footnote~\ref{fn:bjs0res}). The $\widehat{\rho}_I$ for the FGLS plug-in is .80. The $\widehat{\rho}^*_I$ for the TFGLS plug-in is target-specific: TFGLS searches over $\mathcal{P} = [-.99,.99]$, and $\widehat{\rho}^*_I$ is between .96 and .99 for event times 0--14 and is .87 for the binned end time 15.}}
\label{fig:es}
\end{figure}

Point estimates are similar across methods; see Figure~\ref{fig:es}. Also evident from Figure~\ref{fig:es}, and the ratio of the length of those confidence intervals reported in Table~\ref{tab:ci_ratio_common_2dp}, utilizing the serial correlation by FGLS can be somewhat useful. Further and, in this case, much bigger precision gain follows for TFGLS that utilizes the same serial correlation by means of targeting. 

\begin{table}[H]
\centering
\captionsetup{skip=3pt}
\renewcommand{\arraystretch}{0.90}
\caption{Ratio with respect to TFGLS of the length of 95\% confidence interval for effects}
\label{tab:ci_ratio_common_2dp}
\resizebox{\textwidth}{!}{\small
\begin{tabular}{lrrrrrrrrrrrrrrrr}
\toprule
& \multicolumn{16}{c}{Years relative to law change} \\
\cmidrule(l){2-17}
Estimator & 0 & 1 & 2 & 3 & 4 & 5 & 6 & 7 & 8 & 9 & 10 & 11 & 12 & 13 & 14 & 15 \\
\midrule
TWFE & \textbf{0.94} & 1.14 & 1.20 & 1.20 & 1.24 & 1.20 & 1.12 & 1.07 & 1.04 & 1.08 & 1.03 & 1.06 & 1.03 & 1.02 & 1.06 & 1.09 \\
SA & \textbf{0.92} & \textbf{0.96} & 1.05 & 1.04 & 1.14 & 1.17 & 1.20 & 1.16 & 1.17 & 1.20 & 1.12 & 1.13 & 1.18 & 1.13 & 1.23 & \textbf{0.98} \\
CS & 1.01 & 1.00 & \textbf{0.95} & \textbf{0.97} & 1.03 & 1.06 & 1.05 & 1.05 & 1.07 & 1.04 & 1.02 & 1.06 & 1.06 & 1.04 & 1.06 & \textbf{0.98} \\
dCDH & \textbf{0.90} & \textbf{0.93} & \textbf{0.98} & 1.04 & 1.05 & 1.11 & 1.11 & 1.11 & 1.12 & 1.11 & 1.07 & 1.12 & 1.12 & 1.04 & 1.12 & 1.02 \\
BJS(OLS) & 1.18 & 1.60 & 1.83 & 1.73 & 1.74 & 1.66 & 1.46 & 1.38 & 1.39 & 1.39 & 1.21 & 1.17 & 1.20 & 1.13 & 1.31 & 1.04 \\
FGLS & 1.03 & 1.14 & 1.26 & 1.28 & 1.32 & 1.33 & 1.25 & 1.20 & 1.21 & 1.24 & 1.10 & 1.08 & 1.12 & 1.07 & 1.23 & 1.00 \\
\bottomrule
\end{tabular}}
\vspace{-0.3925em}
\begin{minipage}{\textwidth}
\footnotesize \textit{Notes.} Intervals shorter than that of TFGLS are bolded. At event time 15, CS and dCDH estimate the effect exactly 15 years after the law change, the others the binned effect 15 or more years after it. The standard errors of dCDH are from 200 bootstrap replications, as in \citet{dCdH2023-survey}. Population-weighted SWDD \citep{Harmon2024}: ratios are .98--1.02, except 1.08 at 15, where $\widehat{\rho}^*_I=.87$.
\end{minipage}
\end{table}

\subsection{Hospital admissions: \citet{Dobkin2018}}
\label{sec:emp-hrs}

Following \citet{SunAbraham2021} and, in particular, Section 6 of \citet{CSX2025}, we estimate the effect of hospital admission on out-of-pocket medical spending using \citet{Dobkin2018}'s Health and Retirement Study panel. The sample consists of 656 individuals first hospitalized at ages 50--59 in one of the biennial waves 8--11. We use waves 7--10, so $T=4$, and the 65 individuals hospitalized in wave 11 serve as the never-treated; the three treated cohorts have 252, 176 and 163 individuals. Unlike in Section~\ref{sec:emp-divorce}, $I$ is large relative to $T$ and the cohorts are few and large, so KFGLS$^*$ and CSX can also be computed. In fact, \citet{CSX2025} show that in this example their CSX estimator is uniformly more precise than \citet{SunAbraham2021}'s estimator (SA), \citet{CallawaySantanna2021}'s estimator (CS), \citet{dcdH2020}'s estimator (dCDH), and BJS (OLS) for various target parameters. \citet{CSX2025} note that CS using never treated units as comparison group is SA and denote it as CS-SA, while CS using not-yet-treated units as comparison group is dCDH and denote it by CS-dCDH. We follow their notation. 

How does TFGLS perform relative to all these estimators? 

We will explore this for the following four targets: the event study targets $ES(0), ES(1), ES(2)$ like in Section~\ref{sec:emp-divorce}, and the $ES_{\text{avg}}$ as in Sections~\ref{sec:mc-design}--\ref{sec:mc-results}. All these targets are considered in \citet{CSX2025}. Just like Section~\ref{sec:emp-divorce}, we also include TWFE, along with BJS (OLS), FGLS and TFGLS. Additionally we include KFGLS$^*$ as constructed in Section~\ref{sec:mc}.

{\looseness=-1 Table~\ref{tab:hrs} shows that the point estimates are similar, except that TWFE and CS-SA give smaller estimates of $ES(1)$. TFGLS gives shorter intervals than TWFE, CS-SA, CS-dCDH, BJS (OLS) and FGLS for almost all targets, by up to 16\%. FGLS, whose $\widehat{\rho}_I = -.52$ possibly reflects the Nickell-Solon bias with $T=4$, is even less precise than BJS (OLS) for all but $ES(1)$. KFGLS$^*$ is about as precise as TFGLS, while CSX gives the shortest intervals: 4--10\% shorter than those of TFGLS. This is in line with Section~\ref{sec:mc}: with $I$ large relative to $T$, the estimators based on unrestricted covariance structures can deliver their asymptotic efficiency.\par}

So far TFGLS imposed $\rho_0 = \rho_1 \in [-.99,.99]$. Our theory also allows them to differ, and $\rho_1$ to lie outside $(-1,1)$ (footnote~\ref{fn:ar1} and Appendix~\ref{app:ar1wm}). The last row of Table~\ref{tab:hrs} reports TFGLS$(\rho_0,\rho_1)$ with $\rho_1 \in [-2,2]$. This simple relaxation brings TFGLS close to this example's most precise estimator, CSX --- CSX intervals now are 1--4\% shorter for $ES(0)$--$ES(2)$ and equally long for $ES_{\text{avg}}$. Although TFGLS$(\rho_0,\rho_1)$ is only exploratory, even TFGLS remains competitive: both TFGLS and TFGLS$(\rho_0,\rho_1)$ are more precise than all the new DiD estimators if we compare \citet{CSX2025}-type bootstrap standard errors (Appendix~\ref{app:hrsboot}).

\medskip

\begin{table}[H]
\centering
\captionsetup{skip=3pt}
\renewcommand{\arraystretch}{0.97}
\caption{Hospital admissions: estimates of effects on out-of-pocket medical spending, and ratio with respect to TFGLS of the length of 95\% confidence interval}
\label{tab:hrs}
\resizebox{\textwidth}{!}{\small
\begin{tabular}{lrrrrcrrrr}
\toprule
& \multicolumn{4}{c}{Estimate (standard error)} & & \multicolumn{4}{c}{Ratio of confidence-interval length} \\
\cmidrule(lr){2-5}\cmidrule(l){7-10}
Estimator & $ES(0)$ & $ES(1)$ & $ES(2)$ & $ES_{\text{avg}}$ & & $ES(0)$ & $ES(1)$ & $ES(2)$ & $ES_{\text{avg}}$ \\
\midrule
TWFE & 2978 (500) & 511 (608) & 861 (932) & 1450 (609) & & \textbf{0.95} & 1.18 & 1.10 & 1.05 \\
CS-SA & 2960 (539) & 530 (585) & 800 (1008) & 1430 (647) & & 1.02 & 1.13 & 1.19 & 1.12 \\
CS-dCDH & 3134 (536) & 779 (570) & 800 (1008) & 1571 (566) & & 1.02 & 1.10 & 1.19 & \textbf{0.98} \\
BJS(OLS) & 3017 (554) & 788 (586) & 1021 (849) & 1609 (580) & & 1.05 & 1.13 & 1.00 & 1.00 \\
FGLS & 2951 (623) & 730 (562) & 895 (873) & 1525 (596) & & 1.18 & 1.09 & 1.03 & 1.03 \\
KFGLS$^*$ & 3049 (532) & 846 (553) & 1074 (855) & 1656 (552) & & 1.01 & 1.07 & 1.01 & \textbf{0.95} \\
CSX & 3024 (486) & 692 (471) & 1038 (816) & 1585 (521) & & \textbf{0.92} & \textbf{0.91} & \textbf{0.96} & \textbf{0.90} \\
\midrule
TFGLS & 3090 (526) & 692 (517) & 1030 (849) & 1630 (580) & & 1.00 & 1.00 & 1.00 & 1.00 \\
TFGLS ($\rho_0,\rho_1$) & 3046 (492) & 719 (485) & 1030 (849) & 1577 (522) & & \textbf{0.93} & \textbf{0.94} & 1.00 & \textbf{0.90} \\
\bottomrule
\end{tabular}}
\vspace{0.2em}
\begin{minipage}{\textwidth}
\footnotesize \textit{Notes.} Intervals shorter than that of TFGLS are bolded. TFGLS: $\rho_0=\rho_1\in [-.99,.99]$; TFGLS ($\rho_0,\rho_1$): $\rho_0 \in [-.99,.99]$, $\rho_1 \in [-2,2]$ (footnote~\ref{fn:ar1} and Appendix~\ref{app:ar1wm}). SWDD \citep{Harmon2024}: ratios 1.02,\,1.33,\,1.10,\,1.10.
\end{minipage}
\end{table}

\FloatBarrier
\vspace{-1cm}
\section{Conclusion}\label{sec:conclusion}

Choosing nuisance parameters to minimize the variance of the estimator of interest is a familiar principle, applied largely independently in several literatures, whenever misspecification of the nuisance does not bias the estimator of interest. Such robustness can happen naturally or by explicit orthogonalization. We develop this principle for GLS estimation of aggregates of unrestricted heterogeneous treatment effects under staggered roll-out of a binary treatment. The targeted estimator has exactly the form of its untargeted counterpart; only the plugged-in value of the nuisance parameter changes to the variance-minimizing one. 

We establish an equivalence result for infeasible estimators, which helps to show that the feasible estimators we compare seek to approximate the same BLUE benchmark. Parsimony of a working model generally prevents TFGLS from attaining this benchmark. But parsimony can be advantageous when the sample is not large. In our simulations, TFGLS is often more precise in small samples than the existing estimators, including the semiparametrically efficient ones. There will be cases where the working model used by TFGLS is inadequate for such gains against the semiparametrically efficient estimators. Even then, TFGLS remains competitive --- targeting makes it asymptotically at least as precise as OLS and FGLS based on the same working model, while parsimony helps make its inference more reliable than that of the semiparametrically efficient estimators in small samples. 

Our paper held $T$ fixed and let the asymptotics be driven by $I \rightarrow \infty$. The asymptotic approximation for even $I=50$ seemed to work well for TFGLS for small and not so small fixed values of $T$, except under large and persistent entity-level heterogeneity in $\tau_{i,t}$ (Section~\ref{sec:mc-hetit}). Letting $T \rightarrow \infty$ is nontrivial without imposing some sort of homogeneity of treatment effects, because, e.g., the standard error of even OLS can then be biased by a term that is diverging due to heterogeneity, and result in inference with no power. Alternatives under unrestricted heterogeneity while letting $T,I \rightarrow \infty$ are a topic of our ongoing research.  

{\setlength{\bibsep}{0.45\bibsep}\putbib}

\vspace{-.2cm}
\section*{Appendix A: Proofs of results in the main text}\label{sec:proofs}
\begingroup \setlength{\abovedisplayskip}{7pt plus 3pt minus 3pt}\setlength{\belowdisplayskip}{6pt plus 3pt minus 3pt}
\setlength{\abovedisplayshortskip}{4pt plus 2pt}\setlength{\belowdisplayshortskip}{5pt plus 2pt minus 2pt}
\setlength{\emergencystretch}{3em}

\noindent \textbf{Proof of Lemma \ref{lemma:KFGLSmse}:} $\widehat\Sigma= \frac{1}{I-1}\sum_i (y_i - \bar{y})y'_i$. Hence, by normality, $(I-1)\widehat\Sigma \sim \text{Wishart}_T(I-1,\Sigma)$ and $\widehat\Sigma \perp \bar y$. (For the independence, apply Basu's theorem with $\Sigma$ held fixed: $\bar y$ is then complete sufficient for the mean and $\widehat\Sigma$ is ancillary.) Write $\widehat\tau_{\text{KFGLS}} = w(\widehat\Sigma)'\bar y$ with $w(\Sigma) := \Sigma^{-1}1_T/(1_T'\Sigma^{-1}1_T)$, so that $w(\widehat\Sigma)$ is exactly the sample global minimum-variance portfolio weight vector, with ``assets'' relabeled as ``time periods'' and ``portfolio weight'' as ``GLS weight.'' Since $w(\widehat\Sigma)'1_T = 1$, conditioning on $\widehat\Sigma$ gives $V(\widehat\tau_{\text{KFGLS}}) = \frac{1}{I}E\left[w(\widehat\Sigma)'\Sigma w(\widehat\Sigma)\right]$. The result then follows from the unbiasedness and the exact variance formula in Propositions 2 and 5 of \citet{KempfMemmel2005} (equivalently, Proposition 1 and Corollary 2 of \citet{OkhrinSchmid2006}), applied with their $T$ and $N$ equal to our $I$ and $T$. \qed

\medskip

\noindent \textbf{Proof of Proposition \ref{prop:cons}:} $\widehat{\tau}_{I}(\rho_I) - \tau_w = \sum_{i=1}^I \kappa'_{i,I}(\rho_I)u_i$. $E[u_i]=0$ in (A1) implies unbiasedness $E[\widehat{\tau}_I(\rho_I)] = \tau_w$ holds for
any $\rho_I$. (A1) also gives, for some generic constant $C$ not dependent on $\{\rho_I\}$,
\begin{equation*}
\sigma^2_I(\rho_I) = \sum_{i=1}^I \kappa'_{i,I}(\rho_I)\Sigma_i \kappa_{i,I}(\rho_I) \le \sup_i \|\Sigma_i\| \sup_{\rho \in \mathcal{P}} \sum_{i=1}^I\|\kappa_{i,I}(\rho)\|^2 \le C K_I.
\end{equation*}
So $\sigma^2_I(\rho_I) \rightarrow 0$, and $\widehat{\tau}_{I}(\rho_I)$ is consistent for any $\{\rho_I\}\in \mathcal{P}$ as $IK_I = O(1)$ by (A2). \qed

\medskip

\noindent \textbf{Proof of Proposition \ref{prop:clt}:} The proof follows by verifying the Lyapounov condition --- for the independent mean zero triangular arrays $\kappa'_{i,I}(\rho_I)u_i$ --- that for the $\delta > 0$ in (A1):
\begin{equation*}
\lim_{I \rightarrow \infty} \frac{1}{\sigma^{2+\delta}_I(\rho_I)}\sum_{i=1}^I E \left[|\kappa'_{i,I}(\rho_I)u_i|^{2+\delta}\right] = 0. 
\end{equation*} 
Since $\sum_i \|\kappa_{i,I}(\rho_I)\|^{2+\delta} \le \bigl(\max_i \|\kappa_{i,I}(\rho_I)\|^2\bigr)^{\delta/2}\sum_i \|\kappa_{i,I}(\rho_I)\|^2$, (A1) gives, for some generic constant $C$ not dependent on $\{\rho_I\}$:
\begin{equation*}
\begin{aligned}
\frac{1}{\sigma^{2+\delta}_I(\rho_I)}\sum_{i=1}^I E \left[|\kappa'_{i,I}(\rho_I)u_i|^{2+\delta}\right] &\le \frac{I^{1+\delta/2}\sup_i E\left[\|u_i\|^{2+\delta}\right]}{\left(I\sigma^2_I(\rho_I)\right)^{1 + \delta/2}}\Bigl(\sup_{\rho \in \mathcal{P}} \max_i \|\kappa_{i,I}(\rho)\|^2 \Bigr)^{\delta/2}\sup_{\rho \in \mathcal{P}}\sum_{i=1}^I \|\kappa_{i,I}(\rho)\|^{2}\\
&\le C I^{1+\delta/2}\left(o(K_I)\right)^{\delta/2}K_I = C \left(o(IK_I)\right)^{\delta/2}(IK_I),
\end{aligned}
\end{equation*}
which is $o(1)$ not depending on $\{\rho_I\}$ by (A2) and (A3). This completes the proof. \qed

\medskip

\noindent \textbf{Remark:} We repeatedly use in the sequel that $IK_I = O(1)$ by (A2) and $\liminf_I IK_I > 0$ by (A3). Also, the proof of Proposition \ref{prop:oracle} uses Lemma \ref{lemma:maxineq}. See Appendix~\ref{app:maxineq} for its proof and related standard references that we modified slightly to obtain the required result.

\medskip

\begin{lemma}
\label{lemma:maxineq}
Let (A1) and (A4) hold. Let $A_I := (\sum_{i=1}^I c^2_{i,I})^{1/2}$ with $c_{i,I}$ from (A4), $D_{\mathcal{P}} := \diam(\mathcal{P}) := \sup_{\rho,\rho' \in \mathcal{P}}\|\rho - \rho'\|$, and $\omega(\mu) := \mu\,(1 + \log(D_{\mathcal{P}}/\mu))^{1/2}$ for $\mu \in (0, D_{\mathcal{P}}]$. Then there is a finite constant $C$, depending only on $d_\rho$ and $\sup_i E\left[\|u_i\|^2\right]$, such that for all $I$ and all $\mu \in (0, D_{\mathcal{P}}]$:
\vspace{-.1cm}
\begin{equation*}
E\Bigl[\sup_{\rho,\rho' \in \mathcal{P}:\, \|\rho - \rho'\| \leq \mu}\Bigl|\sum_{i=1}^I \left(\kappa_{i,I}(\rho) - \kappa_{i,I}(\rho')\right)'u_i\Bigr|\Bigr] ~\leq~ C\,\omega(\mu)\,A_I.
\end{equation*}
\end{lemma}

\medskip

\noindent \textbf{Proof of Proposition \ref{prop:oracle}:} Proposition \ref{prop:clt} (CLT) implies that the result follows if: 
\vspace{-.1cm}
\begin{equation}
\label{eq:gap}
\Psi_I(\widehat{\rho}_I) ~:=~ \frac{\widehat{\tau}_I(\widehat{\rho}_I) - \widehat{\tau}_I(\rho_I)}{\sigma_I(\rho_I)} ~=~ o_P(1).
\end{equation}
Note that:
\vspace{-.1cm}
\begin{equation*}
\Psi_I(\rho) ~=~ \sum_{i=1}^I \psi_{i,I}(\rho) \text{~~for~~} \psi_{i,I}(\rho) ~:=~ \frac{\left(\kappa_{i,I}(\rho) - \kappa_{i,I}(\rho_I)\right)' u_i}{\sigma_I(\rho_I)}
\end{equation*}
where $\psi_{i,I}(\rho_I) = 0$, and by (A1) $E[\psi_{i,I}(\rho)] = 0$ and $\{\psi_{i,I}(\rho)\}_{i=1}^I$ are independent across $i$. 

With $A_I$, $D_{\mathcal{P}}$ and $\omega(\mu)$ as in Lemma \ref{lemma:maxineq}, the lemma (with its constant $C$) gives, for every $\mu \in (0, D_{\mathcal{P}}]$:
\vspace{-.1cm}
\begin{equation*}
E\Bigl[\sup_{\rho \in \mathcal{P}:\, \|\rho - \rho_I\| \leq \mu}|\Psi_I(\rho)|\Bigr] ~\leq~ \frac{C\,\omega(\mu)\,A_I}{\sigma_I(\rho_I)} ~=~ C\,\omega(\mu)\left(\frac{I A_I^2}{I\sigma^2_I(\rho_I)}\right)^{1/2} ~\leq~ C'\,\omega(\mu)
\end{equation*}
for all large $I$ and a finite $C'$ not depending on $\mu$ or $\{\rho_I\}$, because $IA_I^2 = O(1)$ by (A4) and $\liminf_I \inf_{\rho} I\sigma^2_I(\rho) > 0$ by (A3). Hence, for every $\eta > 0$, Markov's inequality gives:
\vspace{-.1cm}
\begin{equation*}
P\left(|\Psi_I(\widehat\rho_I)| > \eta\right) ~\leq~ P\left(\|\widehat\rho_I - \rho_I\| > \mu\right) + \frac{C'\,\omega(\mu)}{\eta}.
\end{equation*}
Since $\widehat{\rho}_I - \rho_I = o_P(1)$, $\limsup_I P(|\Psi_I(\widehat\rho_I)| > \eta) \leq C'\omega(\mu)/\eta$ for every $\mu \in (0, D_{\mathcal{P}}]$, and $\omega(\mu) \to 0$ as $\mu \downarrow 0$. This gives \eqref{eq:gap}.\qed

\medskip

\noindent \textbf{Proof of Proposition \ref{prop:cons_var}:} From the definitions of $\wh{u}_{[0],i,I}$ and $\wh{u}_{[1],i,I}$:
\vspace{-.1cm}
\begin{equation}
\label{eq:decomp}
\kappa'_{i,I}(\rho)\wh{u}_{i,I} 
= \underbrace{\kappa'_{i,I}(\rho)\,u_i}_{=:\,\xi_i(\rho)} 
+ \underbrace{w'_{[1],i}(\tau_{[1],i} - 1_{N_{[1],i}}\tau_w)}_{=:\,\delta_i} 
- \underbrace{\kappa'_{i,I}(\rho)\,r_i\,(\wh{\eta}_{\ols} - \eta)}_{=:\,\phi_i(\rho)} 
- \underbrace{w'_{[1],i}1_{N_{[1],i}}(\wh{\tau}_I(0) - \tau_w)}_{=:\,\psi_i}.
\vspace{-.1cm}
\end{equation}
$\xi_i(\rho)$ is the true error contribution. $\delta_i$ is the nonrandom treatment effect heterogeneity and does not depend on $\rho$. $\phi_i(\rho)$ arises from estimating $\eta$. $\psi_i$ arises from estimating $\tau_w$. 

Now, note that by analogy with the fact that regression residuals are invariant to shift in intercept:\footnote{\label{fn:fe_inv}\noindent The parameter vector is $\eta = (\lambda'_1, \ldots, \lambda'_I, \beta')'$. Let $J_i$ be the selector matrix such that $r_{[0],i}\,J_i = a_{[0],i}$ and $r_{[1],i}\,J_i = a_{[1],i}$, i.e., $J_i$ picks out the $\lambda_i$-component from~$\eta$. Consider the perturbation $\lambda_i \to \lambda^*_i := \lambda_i + \upsilon_i$ for all $i$ simultaneously with arbitrary $\upsilon_i$. This replaces $Y_{[0]}$ by $Y_{[0]} + R_{[0]}J\upsilon$ and $Y_{[1]}$ by $Y_{[1]} + R_{[1]}J\upsilon$, where $J\upsilon := (J_1\upsilon_1, \ldots, J_I\upsilon_I, 0_\beta)'$. The GLS estimator $\wh{\eta}^*(\rho_0)$ under the perturbed data satisfies:
\vspace{-.1cm}
\begin{equation*}
\wh{\eta}^*(\rho_0) 
= \wh{\eta}(\rho_0) + \left[R'_{[0]}\bar{\Sigma}^{-1}_{[00]}(\rho_0)\,R_{[0]}\right]^{-1}R'_{[0]}\bar{\Sigma}^{-1}_{[00]}(\rho_0)\,R_{[0]}\,J\upsilon 
= \wh{\eta}(\rho_0) + J\upsilon
\vspace{-.1cm}
\end{equation*}
since $R_{[0]}\,J\upsilon$ lies in the column space of~$R_{[0]}$. Hence, $Y^*_{[0]} - R_{[0]}\wh{\eta}^*(\rho_0) = Y_{[0]} - R_{[0]}\wh{\eta}(\rho_0)$ and $Y^*_{[1]} - R_{[1]}\wh{\eta}^*(\rho_0) = Y_{[1]} - R_{[1]}\wh{\eta}(\rho_0)$. Since $\wh{\tau}_I(\rho)$ depends on the data only through these two residual vectors (see \eqref{eq:gls}), $\wh{\tau}_I(\rho)$ is invariant. The equation below follows by writing $\wh{\tau}_I(\rho) = \sum_i \kappa'_{i,I}(\rho)\,y_i$ and using that the $\upsilon_i$'s are arbitrary.}
\vspace{-.1cm}
\begin{equation*}
\kappa'_{i,I}(\rho)\,a_i = 0 \quad \text{for all } i \text{ and all } \rho \in \Pspace.
\end{equation*}
Therefore, partitioning $(\wh{\eta}_{\ols} - \eta) = ((\wh{\lambda}_1(0) - \lambda_1)', \ldots, (\wh{\lambda}_I(0) - \lambda_I)', (\wh{\beta}(0) - \beta)')'$, it follows:
\vspace{-.1cm}
\begin{equation*}
\phi_i(\rho) 
= \kappa'_{i,I}(\rho)\,r_i\,(\wh{\eta}_{\ols} - \eta) 
= \kappa'_{i,I}(\rho)\,a_i\,(\wh{\lambda}_i(0) - \lambda_i) + \kappa'_{i,I}(\rho)\,x_i\,(\wh{\beta}(0) - \beta) 
= \kappa'_{i,I}(\rho)\,x_i\,(\wh{\beta}(0) - \beta).
\end{equation*}

\textbf{Step 1:} Define $\bar{\sigma}^2_I(\rho) := \sum_i (\xi_i(\rho) + \delta_i)^2$. We show that $\phi_i(\rho)$ and $\psi_i$ in \eqref{eq:decomp} are asymptotically negligible in the sense that $\sup_\rho |\wh{\sigma}^2_I(\rho) - \bar{\sigma}^2_I(\rho)|/K_I \xrightarrow{p} 0$. Let $c_i := \xi_i(\rho) + \delta_i - \phi_i(\rho) - \psi_i$ and $d_i := \xi_i(\rho) + \delta_i$. Then $|c^2_i - d^2_i| \equiv |c_i-d_i|\,|c_i+d_i| = |\phi_i(\rho) + \psi_i||2d_i-(\phi_i(\rho) + \psi_i)|$ gives: 
\[
|\wh{\sigma}^2_I(\rho) - \bar{\sigma}^2_I(\rho)| 
\leq \sum_i |c_i^2 - d_i^2| 
\leq 2\!\left(\sum_i (\xi_i(\rho) + \delta_i)^2\right)^{1/2}\!\left(\sum_i (\phi_i(\rho) + \psi_i)^2\right)^{1/2} + \sum_i (\phi_i(\rho) + \psi_i)^2.
\]
by Cauchy-Schwarz. Hence, $\sup_\rho |\wh{\sigma}^2_I(\rho) - \bar{\sigma}^2_I(\rho)|/K_I \xrightarrow{p} 0$ follows by showing (a)--(c) below.

(a) \textit{$\sup_\rho \sum_i \phi^2_i(\rho) = \sum_i \left[\kappa'_{i,I}(\rho)\,x_i (\wh{\beta}(0) - \beta)\right]^2 =o_P(K_I)$.} Indeed:
\begin{equation*}
\sup_\rho \sum_i \phi^2_i(\rho) \le \|\widehat{\beta}(0) - \beta\|^2\sup_i \|x_i\|^2 \sup_\rho  \sum_i \|\kappa_{i,I}(\rho)\|^2
= O_p(I^{-1})O(1)O(I^{-1}) = O_P(I^{-2})
\end{equation*}
using (A6) and (A1) for $\widehat{\beta}(0) - \beta = O_p(I^{-1/2})$, and (A6) and (A2) for the other terms.

(b) \textit{$\sum_i \psi_i^2 = (\wh{\tau}_I(0) - \tau_w)^2 \sum_i (w'_{[1],i}1_{N_{[1],i}})^2  \le (\wh{\tau}_I(0) - \tau_w)^2 T\sum_i \|w_{[1],i}\|^2 = o_P(K_I)$.} By~(A1)--(A3), $(\wh{\tau}_I(0) - \tau_w)^2 = O_P(\sigma^2_I(0)) = O_P(K_I)$. By~(A8), $\sum_i \|w_{[1],i}\|^2 = O(1/I)$. Hence $\sum_i \psi_i^2 = O_P(K_I/I) = o_P(K_I)$.

(c) \textit{ $\sup_\rho \sum_i (\xi_i(\rho) + \delta_i)^2 = O_P(K_I)$.} Indeed, $\sup_\rho\sum_i (\xi_i(\rho) + \delta_i)^2 \leq 2\sup_\rho\sum_i \xi^2_i(\rho) + 2\sum_i \delta_i^2 = O_P(K_I)$
because (i) for a fixed $\rho^\circ\in\mathcal P$, (A4) gives $\sup_{\rho}\|\kappa_{i,I}(\rho)\|\le\|\kappa_{i,I}(\rho^\circ)\|+c_{i,I}\diam(\mathcal P)$, so that $E[\sup_\rho\sum_i \xi^2_i(\rho)] \le \sum_i E\|u_i\|^2\sup_\rho\|\kappa_{i,I}(\rho)\|^2 \le 2\sup_i E\|u_i\|^2\,\bigl(K_I + \diam^2(\mathcal P)\sum_i c^2_{i,I}\bigr) = O(K_I)$ by (A1)--(A4) (recall $\liminf_I IK_I>0$ by (A3)), and Markov's inequality applies; and (ii) $\sum_i \delta_i^2 = \sum_i (w'_{[1],i}(\tau_{[1],i} - 1_{N_{[1],i}}\tau_w))^2 \leq \sup_i \|\tau_{[1],i} - 1_{N_{[1],i}}\tau_w\|^2 \cdot \sum_i \|w_{[1],i}\|^2 = O(1/I) = O(K_I)$ by (A7)--(A8) and (A2).

\textbf{Step 2:} We show that $\sup_\rho |\bar{\sigma}^2_I(\rho)- (\sigma^2_I(\rho) + \sigma^2_{\tau,I})|/K_I \xrightarrow{p} 0$. Since $\bar{\sigma}^2_I(\rho) = \sum_i \xi^2_i(\rho) + 2\sum_i \xi_i(\rho)\,\delta_i + \sigma^2_{\tau,I}$, as $\sum_i \delta^2_i =: \sigma^2_{\tau,I}$, this current step simply involves showing that:
\begin{equation*}
(i)~~ \sup_\rho \left|\sum_i \xi^2_i(\rho) - \sigma^2_I(\rho)\right| / K_I \xrightarrow{p} 0 \text{~~and~~} (ii)~~ \sup_\rho \left|\sum_i \xi_i(\rho)\,\delta_i\right| / K_I \xrightarrow{p} 0.
\end{equation*}
We will now establish $(i)$ and $(ii)$ by verifying pointwise convergence and a stochastic Lipschitz condition that jointly imply the uniform convergence by standard $\epsilon$-net arguments.
    
\textit{(i)} holds pointwise because $E[\sum_i \xi^2_i(\rho)] = \sigma^2_I(\rho)$ and $V(\sum_i \xi^2_i(\rho)) = o(K_I^2)$ since $V\left(\sum_i \xi^2_i(\rho)\right)  \leq \sum_i E[\xi^4_i(\rho)] \leq \sup_i E[\|u_i\|^4] \cdot \sum_i \|\kappa_{i,I}(\rho)\|^4$ by (A1) and (A5), and since $\sum_i \|\kappa_{i,I}(\rho)\|^4 \leq \max_i \|\kappa_{i,I}(\rho)\|^2 \cdot K_I = o(K_I^2)$ by (A2).

\textit{(ii)} holds pointwise because $E[\sum_i \xi_i(\rho)\,\delta_i] = 0$ and $V(\sum_i \xi_i(\rho)\,\delta_i) = o(K_I^2)$ since:
\[
V\left(\sum_i \xi_i(\rho)\,\delta_i\right) 
= \sum_i \delta_i^2 \cdot \kappa'_{i,I}(\rho)\,\Sigma_i\,\kappa_{i,I}(\rho) 
\leq \sup_i \|\Sigma_i\| \cdot \max_i \delta_i^2 \cdot K_I
\]
where $\max_i \delta_i^2 \leq \sup_i \|\tau_{[1],i} - 1_{N_{[1],i}}\tau_w\|^2 \cdot \max_i \|w_{[1],i}\|^2 = O(1) \cdot o(1/I) = o(K_I)$ by (A7)--(A8).

\textit{(i)} satisfies a stochastic Lipschitz bound condition. This follows by first noting that: 
\[
\left|\sum_i \xi^2_i(\rho) - \sum_i \xi^2_i(\rho')\right| 
\leq 
\left(\sum_i (\xi_i(\rho)-\xi_i(\rho'))^2\right)^{1/2}\left(\sum_i (\xi_i(\rho)+\xi_i(\rho'))^2\right)^{1/2}\!.
\]
The first factor on the right hand side is $\le \|\rho - \rho'\| \cdot (\sum_i c^2_{i,I}\|u_i\|^2)^{1/2} \leq C \|\rho - \rho'\| \cdot O_p(\sqrt{K_I})$, with $c_{i,I}$ from (A4), by using (A4) first and then (A1) and (A3) (recall $\liminf_I IK_I > 0$), for some finite $C$. In the same way as in Step 1(c), the second factor is $O_P(\sqrt{K_I})$ uniformly in $\rho$. Finally, note that, for some finite $C$:
\[
|\sigma^2_I(\rho) - \sigma^2_I(\rho')| \leq C\,\|\rho - \rho'\| \cdot K_I.
\]
since $|\sigma^2_I(\rho) - \sigma^2_I(\rho')| \leq 2\sup_i\|\Sigma_i\| \sum_i (\|\kappa_{i,I}(\rho)\| + \|\kappa_{i,I}(\rho')\|)\|\kappa_{i,I}(\rho) - \kappa_{i,I}(\rho')\| \leq C\|\rho - \rho'\| \sum_i (\|\kappa_{i,I}(\rho)\| + \|\kappa_{i,I}(\rho')\|)c_{i,I}$ by (A4), and $\sum_i \|\kappa_{i,I}(\rho)\| c_{i,I} \leq K_I^{1/2}(\sum_i c^2_{i,I})^{1/2} = O(K_I)$ by Cauchy-Schwarz and (A2)--(A4).
Hence $\left|\left(\sum_i \xi^2_i(\rho) - \sigma^2_I(\rho)\right) - \left(\sum_i \xi^2_i(\rho') - \sigma^2_I(\rho')\right) \right| \leq L_I \cdot \|\rho - \rho'\|$ with $L_I/K_I = O_P(1)$.

\textit{(ii)} satisfies a stochastic Lipschitz bound condition. This follows by noting that:
\begin{equation*}
\begin{aligned}
\left|\sum_i \xi_i(\rho)\delta_i - \sum_i \xi_i(\rho')\delta_i\right| 
&\leq  \left(\sum_i \left(\xi_i(\rho) - \xi_i(\rho')\right)^2\right)^{1/2}\left(\sum_i \delta^2_i\right)^{1/2}\\
&\le \left[C \|\rho - \rho'\| \cdot O_p(\sqrt{K_I})\right]O(\sqrt{K_I})
\end{aligned}
\end{equation*}
where the term inside $[.]$ is the bound for the first factor in \textit{(i)}, and the term outside follows in the same way as in Step 1(c). Hence, the product $\leq L'_I \cdot \|\rho - \rho'\|$ with $L'_I/K_I = O_P(1)$. 

Then the required uniform convergence follows from pointwise convergence and Lipschitz bound by standard $\epsilon$-net arguments. \qed

\medskip

\noindent \textbf{Proof of Lemma \ref{lemma:rhocons}:} The proof is the same as that of Theorem 5.7 of \citet{Aad98}. Define $\wh{M}_I(\rho) := I\hat{\sigma}^2_I(\rho)$ and $M_I(\rho) := I(\sigma^2_I(\rho) + \sigma^2_{\tau,I})$ (equivalently, divide by $K_I$). Since $\sigma^2_{\tau,I}$ does not depend on $\rho$, the minimizers of $M_I(\rho)$ over $\mathcal{P}$ and of $I\sigma^2_I(\rho)$ over $\mathcal{P}$ coincide, i.e., $\rho^*_I = \arg\min_\rho M_I(\rho)$. Similarly, $\widehat{\rho}^*_I = \arg\min_\rho \wh{M}_I(\rho)$. Fix $\epsilon > 0$. Define the separation gap $\Delta_I(\epsilon) := \inf_{\rho \in \mathcal{P}:\, \|\rho - \rho^*_I\| > \epsilon} \left[M_I(\rho) - M_I(\rho^*_I)\right]$. Since $\widehat{\rho}^*_I$ minimizes $\wh{M}_I(\rho)$, $\wh{M}_I(\widehat{\rho}^*_I) \le \wh{M}_I(\rho^*_I)$ and hence $M_I(\widehat{\rho}^*_I) - M_I(\rho^*_I) \leq 2\sup_{\rho \in \mathcal{P}} |\wh{M}_I(\rho) - M_I(\rho)|$.
Then, denoting $\underline{\Delta}(\epsilon) := \liminf_I \Delta_I(\epsilon) > 0$ by assumption (A9), $\Delta_I(\epsilon) \geq \underline{\Delta}(\epsilon)/2$ for all large $I$, and hence:
\[
P\big(\|\widehat{\rho}^*_I - \rho^*_I\| > \epsilon\big) \leq P\Big(\sup_{\rho} |\wh{M}_I(\rho) - M_I(\rho)| \geq \underline{\Delta}(\epsilon)/4\Big).
\]
The right hand side goes to 0 as $I \rightarrow \infty$ by Proposition \ref{prop:cons_var} that holds under (A1)--(A8). \qed

\medskip 

\noindent \textbf{Proof of Proposition \ref{prop:tfgls}:} The proof follows exactly in the way described in the remark below Proposition \ref{prop:cons_var} once we note that: (i) $\widehat{\rho}^*_I - \rho^*_I = o_P(1)$ by Lemma \ref{lemma:rhocons} that holds under (A1)--(A9), and (ii) (A1), (A2) and (A4) give the required continuity of $\sigma^2_I(\rho)$ along the drifting sequence $\rho^*_I$, i.e., for every $\epsilon > 0$ there exists $\varsigma > 0$ such that for all large $I$,
\begin{equation}\label{eq:equicont}
\|\rho - \rho^*_I\| < \varsigma \implies I |\sigma^2_I(\rho) - \sigma^2_I(\rho^*_I)| < \epsilon
\end{equation}
that will allow the application of the continuous mapping theorem. \eqref{eq:equicont} follows from the bound $|\sigma^2_I(\rho) - \sigma^2_I(\rho')| \leq C\,\|\rho - \rho'\| \cdot K_I$ for all $\rho, \rho' \in \mathcal{P}$, derived in Step 2 of the proof of Proposition \ref{prop:cons_var}, together with $K_I = O(I^{-1})$ in (A2). \qed

\medskip

\noindent \textbf{Proof of Lemma \ref{lemma:exactunb}:} The argument is similar to that of \citet{Kakwani1967}. By \eqref{eq:decomp} with $\delta_i = 0$, $\kappa'_{i,I}(\rho)\widehat{u}_{i,I} = \xi_i(\rho) - \phi_i(\rho) - \psi_i$, where $\xi_i(\rho) = \kappa'_{i,I}(\rho)u_i$, $\phi_i(\rho) = \kappa'_{i,I}(\rho)x_i(\widehat{\beta}(0)-\beta)$ and $\psi_i = w'_{[1],i}1_{N_{[1],i}}\sum_j \kappa'_{j,I}(0)u_j$ are all linear in $U$. Hence $\widehat{\sigma}^2_I(\rho) = \sum_i (\kappa'_{i,I}(\rho)\widehat{u}_{i,I})^2$ is an even function of $U$ for every $\rho$, so the set of its minimizers is the same at $U$ and at $-U$. If this set is not a singleton, $\widehat{\rho}^*_I$ is selected by any rule that depends on $U$ only through the function $\rho\mapsto\widehat{\sigma}^2_I(\rho)$, which is even in $U$. Therefore $\widehat{\rho}^*_I(U) = \widehat{\rho}^*_I(-U)$. Consequently, $g(U) := \widehat{\tau}_I(\widehat{\rho}^*_I(U)) - \tau_w = \sum_i \kappa'_{i,I}(\widehat{\rho}^*_I(U))u_i$ satisfies $g(-U) = -g(U)$. Since $|g(U)| \le \sum_i \sup_{\rho \in \mathcal{P}}\|\kappa_{i,I}(\rho)\|\,\|u_i\|$, which is integrable for every $I$ by (A1) and (A2), it follows that $E[g(U)]$ exists. Then, $U \overset{d}{=} -U$ gives $E[g(U)] = E[g(-U)] = -E[g(U)]$, i.e., $E[g(U)] = 0$. \qed

\medskip

\noindent \textbf{Proof of Proposition~\ref{prop:ho-bias}:} Write $\rhohatstar - \rhostar = -\HI^{-1}\,\Ghat + \rem_1$ where $\Ghat := \frac{\partial}{\partial \rho} \frac{I}{2} \widehat{\sigma}^2_I(\rho^*_I) =  I\sum_i\bigl(\dot{\kappa}'_{0,i}\,\hat{u}_{0,i}\bigr)\bigl(\kappa'_i\,\hat{u}_i\bigr)$, $\kappa_i := \kappa_{i,I}(\rhostar)$,  $\dot{\kappa}_{0,i}:=\partial\kappa_{[0],i,I}(\rhostar)/\partial\rho'$, $u_{0,i} := u_{[0],i}$, $\hat{u}_{0,i} := \hat{u}_{[0],i}$, and $\rem_1$ is the second-order remainder from the expansion of the first-order condition $\partial\bigl[\frac{I}{2} \widehat{\sigma}^2_I(\widehat{\rho}^*_I)\bigr]/\partial\rho = 0$ around $\rho^*_I$. Expand $\tauhat(\rhohatstar)-\tauhat(\rhostar)$ around
$\rhostar$ to obtain: $\tauhat(\rhohatstar)-\tauhat(\rhostar) = (\rhohatstar-\rhostar)'\sum_i\dot{\kappa}'_{0,i}\,u_{0,i}
  + \rem_2$ where $\rem_2$ is the second order remainder. Then:
\[
  \tauhat(\rhohatstar)-\tauhat(\rhostar)  = -\Ghat'\HI^{-1}\sum_i\dot{\kappa}'_{0,i}\,u_{0,i} + \rem_3,
\]
where $\rem_3:=\rem_1'\sum_i\dot{\kappa}'_{0,i}u_{0,i}+\rem_2$. Recalling that $E[\tauhat(\rhostar)]=\tauw$, it follows now that:
\begin{equation}
\label{eq:bias-expand}
  E[\tauhattfgls-\tauw] 
  = -E\left[\Ghat'\HI^{-1}\sum_i\dot{\kappa}'_{0,i}\,u_{0,i}\right] + E[\rem_3].
\end{equation}
Lemma \ref{lem:bias-order} in Appendix~\ref{app:hobias-lemmas} gives: $-E\left[\Ghat'\HI^{-1}\sum_i\dot{\kappa}'_{0,i}\,u_{0,i}\right] = \Blead + o(I^{-1})$ where:
\begin{eqnarray*}
\Blead &:=& \mathcal{BA}_1 + \mathcal{BA}_2 + \mathcal{BB}_2,\\
\mathcal{BA}_1 &:=& -I\sum_i E[u'_i\kappa_iu'_{0,i}\dot\kappa_{0,i}H^{-1}_I\dot\kappa'_{0,i}u_{0,i}] = o(I^{-1/2}),\\
\mathcal{BA}_2  &:=& -I\sum_i \delta_i E[u'_{0,i}\dot\kappa_{0,i}H^{-1}_I\dot\kappa'_{0,i}u_{0,i}] = o(I^{-1/2}),\\
\mathcal{BB}_2 &:=& I \cdot \sum_i \sum_j \delta_i E\left[(\widehat{\beta}(0) - \beta)'x'_{0,i}\dot\kappa_{0,i}H^{-1}_I \dot\kappa'_{0,j}u_{0,j}\right] = O(I^{-1}).
\end{eqnarray*}
If (A8') also holds, the same expansion holds with $o(I^{-1})$ replaced by $O(I^{-2})$, and then $\mathcal{BA}_1, \mathcal{BA}_2, \mathcal{BB}_2 = O(I^{-1})$.

Lemma \ref{lemma:rem3bound} in Appendix~\ref{app:hobias-lemmas} gives $E[\rem_3] = O(\Phi_I I^{-1/2})$ where $\Phi_I := \max_i\|\kappa_i\|^2/K_I$. Under (A2), $\Phi_I = o(1)$, so $E[\rem_3] = o(\Phi_I^{1/2} I^{-1/2})$; under (A8'), $\Phi_I = O(I^{-1})$, so $E[\rem_3] = O(I^{-3/2})$. As $\Phi_I \ge c/I$ by (A1)--(A3), these and Lemma \ref{lem:bias-order} give: $ E[\tauhattfgls - \tauw] \;=\; \Blead \;+\; o(\Phi_I^{1/2} I^{-1/2})$ and, under (A8'), $E[\tauhattfgls - \tauw] \;=\; \Blead \;+\; O(I^{-3/2})$. \qed

\medskip

\noindent \textbf{Proof of Proposition \ref{prop:unif}:} We first need to introduce the following notation using the treatment cohorts $c \in \mathcal{G} := \mathcal{G}_{\text{trt}}\cup \{\infty\}$ with $c = \infty$ signifying never treated. For unit $i \in c$, i.e., $i$ such that $E_i = c$, under the conditions of the proposition, the outcome model \eqref{eq:model} is:
\begin{equation*}
  y_{i,s} = \lambda_i + \beta_s + \1(s \ge c)\,\tau_{i,s} + u_{i,s}.
\end{equation*}
$n_c := \sum_{i=1}^I \mathbf{1}(E_i=c) \equiv \sum_{i \in c}1$, i.e., the cohort size. The cohort-period mean is $\bY^c_s = n_c^{-1}\sum_{i\in c} y_{i,s}$,
with
\[
  E[\bY^c_s] = \lambda^c + \beta_s + \1(s \ge c)\,\bar{\tau}^c_s,
\]
where $\lambda^c := n_c^{-1}\sum_{i \in c}\lambda_i$ and $\bar{\tau}^c_s := n_c^{-1}\sum_{i \in c}\tau_{i,s}$. The object of interest $ATT(g,t) := \bar{\tau}^g_t$.

Based on the fact that $c= \infty$ is never treated and $s=1$ is a pre-treatment period for all cohorts (there is no always treated), we know, using $q$ for the baseline period and $p$ for the intermediate period, that the following two families of estimators are unbiased for $\ATT(g,t)$:
\begin{eqnarray}
\label{eq:fam1}
  \hat\theta^{\infty}_{q}
    &:=& \bigl(\bY^{g}_{t} - \bY^{g}_{q}\bigr)
        - \bigl(\bY^{\infty}_{t} - \bY^{\infty}_{q}\bigr),
        \qquad q < g, \\
\label{eq:fam2}
  \hat\theta^{g'}_{p,\,q}
    &:= &\bigl(\bY^{g}_{t} - \bY^{g}_{q}\bigr)
        - \bigl(\bY^{\infty}_{t} - \bY^{\infty}_{p}\bigr)
        - \bigl(\bY^{g'}_{p} - \bY^{g'}_{q}\bigr),
        \quad q < g,\; p < g',\; q < g'. 
\end{eqnarray}
Every $(c,s)$ cell in \eqref{eq:fam2} other than $(g,t)$ is untreated --- \eqref{eq:fam2} augments the complete cases \eqref{eq:fam1} with the available cases (in the terminology of the missing data literature). 

The proof follows in two steps. Step 1 establishes that every cohort-mean-based unbiased estimator of $ATT(g,t)$ is a linear combination of these contrasts. Step 2 establishes that the infeasible CSX estimator is equal to the minimum variance unbiased estimator among the estimators generated by linear combination of those contrasts, and hence equal to $\widehat{\tau}_{w,\text{blue}}$.

\textbf{Step 1:} Consider a cohort-period-mean-based generic estimator $\widehat{\mu} := \sum_{c,s} \gamma^c_s\bY^c_s$ for $\bar{\tau}^g_t$. Following the same steps as in, e.g., Appendix A.5 of \citet{BJS2024}, $E[\widehat{\mu}] = \bar{\tau}^g_t$ if and only if the weights $\gamma^c_s$ satisfies: (R1) $\sum_s \gamma^c_s = 0$ for all $c$, (R2) $\sum_c \gamma^c_s = 0$ for all $s$, and (R3) $\gamma^c_s = \mathbf{1}((c,s) = (g,t))$ at every treated cell, i.e., $s \ge c$. Let $\gamma(\widehat{\mu})$ denote the matrix $\Gamma := (\gamma^c_s)_{c,s}$, (R1) is a restriction on its rows, (R2) on its columns, and (R3) is a condition imposed by the target $\bar{\tau}^g_t$ on the upper diagonal of $\Gamma$. Let $\mathcal{U} := \{\Gamma: \text{(R1), (R2), (R3) hold}\}$ be the subspace generated due to this unbiasedness. For convenience of dealing with the entire upper diagonal in the same way define the linear subspace generated by unbiasedness for 0 as: 
\begin{equation*}
\mathcal{U}_0 := \{\Gamma: \text{(R1) and (R2) hold, and $\gamma^c_s = 0$ for all } s \ge c\}.
\end{equation*}
$\mathcal{U}$ is an affine subspace, i.e., $\mathcal{U} = \Gamma_0 + \mathcal{U}_0$ for any $\Gamma_0 \in \mathcal{U}$.

Let the estimators/contrasts in the two families \eqref{eq:fam1}--\eqref{eq:fam2} be generically denoted by $\hat{\theta}_j$ and let $\mathcal{J}$ denote the largest class such that the estimators are non-redundant, i.e., their joint variance covariance matrix is positive definite. It is easy to see that $\Gamma_j :=\gamma(\hat{\theta}_j)$, i.e., matrix of the cohort-period weights in $\hat{\theta}_j$ belongs in $\mathcal{U}$. Then, if we consider the matrix of cohort-period weights for a linear combination $\sum_{j \in \mathcal{J}} h_j \hat{\theta}_j$ with $h_j \in \mathbb{R}$, that will satisfy (R1) and (R2) by linearity and satisfy (R3) iff $\sum_{j \in \mathcal{J}} h_j =1$. In other words, $\gamma\left(\sum_{j \in \mathcal{J}} h_j \hat{\theta}_j\right) \in \mathcal{U}$ iff $\sum_{j \in \mathcal{J}} h_j =1$. ($h_j< 0$ is fine since each $E[\hat{\theta}_j] = \bar{\tau}^g_t$ for all $j$; what matters is $\sum_j h_j =1$.)  

Now we establish the key result that any unbiased estimator of $\bar{\tau}^g_t$ is generated by these contrasts $\{\hat{\theta}_j: j \in \mathcal{J}\}$, i.e., any $\widehat{\mu}$ with $\gamma(\widehat{\mu}) \in \mathcal{U}$ can be expressed as $\sum_{j \in \mathcal{J}} h_j \hat{\theta}_j$ for some $h_j$'s with $h_j \in\mathbb{R}$ and $\sum_j h_j =1$. Since $\gamma(\hat{\theta}^{\infty}_1) \in \mathcal{U}$, we will use $\hat{\theta}^{\infty}_1$ as the reference, and note that every $\Gamma \in \mathcal{U}$ can be written as $\Gamma = \gamma(\hat{\theta}^{\infty}_1) + V$ where $V:=\Gamma - \gamma(\hat{\theta}^{\infty}_1)\in \mathcal{U}_0$. So, for our purpose it suffices to show that: 
\begin{equation}
\label{eq:temp1-unitprop}
\text{any $V  \in \mathcal{U}_0$ is a linear combination of differences $\{\gamma(\hat{\theta}_j) - \gamma(\hat{\theta}_{j'})\}_{j,j'\in \mathcal{J}}$.}
\end{equation} 
For treated cohorts $c \ne \infty$, in their untreated periods $s \in \{2,\ldots, c-1\}$, define:
\begin{equation*}
D_{c,s} := (\bY^c_s - \bY^c_1) - (\bY^\infty_s - \bY^\infty_1)
\end{equation*} 
a cohort-period-mean-based estimator that is unbiased for 0. Hence $\gamma(D_{c,s}) \in \mathcal{U}_0$. $D_{c,s}$'s are differences of the contrasts from \eqref{eq:fam1}--\eqref{eq:fam2}; in particular, for all $c \ne \infty$ and $s \in \{2,\ldots, c-1\}$:
\begin{equation*}
\begin{aligned}
D_{g,s} ~=~& \hat{\theta}^{\infty}_1 - \hat{\theta}^{\infty}_s  \text{~~for~~} 2 \le s \le g-1\\
D_{g',s} ~=~& \hat{\theta}^{\infty}_1 - \hat{\theta}^{g'}_{p=s,q=1}  \text{~~for~~} g \ne g', 2 \le s < g'.
\end{aligned}
\end{equation*} 
To prove \eqref{eq:temp1-unitprop} it is enough to show that any $V  \in \mathcal{U}_0$ is a linear combination of these $\{\gamma(D_{c,s})\}$.

Let $\hat{\mu} = \sum_{c,s} v^c_s\bY^c_s$ with $V:= \gamma(\hat{\mu}) \in \mathcal{U}_0$. Hence, $v^c_s = 0$ at every treated cell, so the only potentially nonzero entries are at untreated cells: $(c,s)$ with $c \neq \infty$, $s < c$,
and $(\infty, s)$ for any $s$. The proof follows by verifying that $\hat{\mu}$ is equal to the linear combination $\sum_{\substack{c \neq \infty \\ 2 \leq s < c}} v^c_s \, D_{c,s}$.

Substituting $D_{c,s} = \bar{Y}^c_s - \bar{Y}^c_1 - \bar{Y}^\infty_s + \bar{Y}^\infty_1$ into $\hat{\mu} - \sum_{c \neq \infty,\, 2 \leq s < c} v^c_s \, D_{c,s}$, the terms $v^c_s\bar{Y}^c_s$ with $c \neq \infty$, $2 \leq s < c$ cancel, leaving:
\begin{equation*}
\begin{aligned}
\text{rem} := \hat{\mu} - \sum_{\substack{c \neq \infty \\ 2 \leq s < c}} v^c_s \, D_{c,s} &= \sum_{c \neq \infty} v^c_1 \bar{Y}^c_1
       + \sum_s v^\infty_s \bar{Y}^\infty_s
       + \sum_{\substack{c \neq \infty \\ 2 \leq s < c}} v^c_s \bar{Y}^c_1
       + \sum_{\substack{c \neq \infty \\ 2 \leq s < c}} v^c_s \bar{Y}^\infty_s
       - \sum_{\substack{c \neq \infty \\ 2 \leq s < c}} v^c_s \bar{Y}^\infty_1.
\end{aligned}
\end{equation*}
Grouping by $c \neq \infty$, the coefficient on $\bar{Y}^c_1$ is: $v^c_1 + \sum_{2 \leq s < c} v^c_s = \sum_{s < c} v^c_s = 0$ by (R1), using $v^c_s = 0$ at treated cells. On the other hand, the coefficient on $\bar{Y}^\infty_1$ is: $ v^\infty_1 - \sum_{\substack{c \neq \infty \\ 2 \leq s < c}} v^c_s  = v^\infty_1 + \sum_{c \neq \infty} v^c_1 = \sum_c v^c_1 = 0$ where we use (R1) (giving $\sum_{2 \leq s < c} v^c_s = -v^c_1$ for each $c \neq \infty$) and then (R2) at $s = 1$. Finally, grouping by period $s$, the
coefficient on $\bar{Y}^\infty_s$ is: $  v^\infty_s + \sum_{\substack{c \neq \infty \\ c > s}} v^c_s
  = \sum_c v^c_s = 0$ where we use $v^c_s = 0$ for treated cohorts $c \leq s$, and then (R2). Hence, all coefficients vanish. So $\text{rem} = 0$, meaning: $V  = \sum_{\substack{c \neq \infty \\ 2 \leq s < c}} v^c_s \, \gamma(D_{c,s})$.
Since each $\gamma(D_{c,s})$ is a difference of two contrast coefficient arrays, this proves \eqref{eq:temp1-unitprop}. Hence, every cohort-mean-based unbiased estimator of $ATT(g,t)$ is a linear combination of the contrasts in \eqref{eq:fam1}--\eqref{eq:fam2}.

\textbf{Step 2:} We show that $\widehat{\tau}_{w,\text{blue}}$ is itself a cohort-mean-based unbiased estimator, and hence covered by Step 1, and then that it coincides with the infeasible CSX estimator.

Consider a generic linear estimator $\hat\mu = \sum_{i,s} d_{i,s}\,y_{i,s}$. As before, its unbiasedness for $\bar{\tau}^g_t$ follows if and only if: (R1') $\sum_s d_{i,s} = 0$ for every entity $i$; (R2') $\sum_i d_{i,s} = 0$ for every period $s$; and (R3') $d_{i,s} = \tfrac{1}{n_g}\1(E_i = g,\, s = t)$ at every treated cell $(i,s)$. Letting $d_i := (d_{i,s})_s$ for each $i$, by independence of the $u_{i}$'s and homoskedasticity within cohorts, $\Var(\hat\mu) = \sum_i d_i'\tS^{E_i}d_i$. $\widehat{\tau}_{w,\text{blue}}$ has coefficients $d_i$'s that minimize this variance subject to the unbiasedness constraints. 

Define the cohort column-sums $d^c_s := \sum_{i\in c} d_{i,s}$ for $c \in \mathcal{G}$. Fix $(d^c_s)$ and consider the within-cohort variance minimization subproblem for cohort $c$:
\begin{equation*}
  \min_{(d_i)_{i \in c}}\ \sum_{i \in c} d_i'\tS^{c}d_i
  \quad\text{s.t.}\quad
  \sum_{i \in c} d_{i,s} = d^c_s\ \ \forall s,
  \qquad
  \sum_s d_{i,s} = 0\ \ \forall i \in c.
\end{equation*}
The objective is strictly convex (since $\tS^{c}$ is positive definite), so the minimizer is unique. The objective and the feasible set are both unchanged by permuting the units of cohort $c$: the objective because $\Sigma_i = \tS^{c}$ for all $i \in c$, and the feasible set because the constraints treat every $i$. A unique minimizer of a strictly convex problem whose data are permutation-invariant must
itself be permutation-invariant. Hence $d_i = d_{i'}$ for all $i, i' \in c$, giving $d_{i,s} = d^c_s / n_c$.  Therefore:
\begin{equation*}
  \widehat{\tau}_{w,\text{blue}}
  = \sum_c \sum_{i \in c} \sum_s d_{i,s}\,y_{i,s}
  = \sum_{c,s} d^c_s \Bigl(\frac{1}{n_c}\sum_{i \in c} y_{i,s}\Bigr)
  = \sum_{c,s} d^c_s\,\bY^c_s,
\end{equation*}
with the $d^c_s$'s being the variance minimizing ones. Importantly, $\widehat{\tau}_{w,\text{blue}}$ is an unbiased estimator based on linear combinations of the cohort-period means. Therefore, by Step 1, $\widehat{\tau}_{w,\text{blue}} = \sum_{j \in \mathcal{J}} \tilde{h}_j \hat{\theta}_j$ for some $\tilde{h}_j \in \mathbb{R}$ with $\sum_{j \in \mathcal{J}} \tilde{h}_j = 1$. Therefore, letting $\Omega := Var((\hat{\theta}_j)_{j\in \mathcal{J}})$, by construction $Var\left(\widehat{\tau}_{w,\text{blue}}\right) \ge Var(\sum_{j \in \mathcal{J}} h_{j*} \hat{\theta}_j) = h'_*\Omega h_*$ where $h_* = (h_{j})_{j\in \mathcal{J}} := \arg\min_h h'\Omega h$ such that $h'\mathbf{1} = 1$, i.e., $h_* = \frac{\Omega^{-1}\mathbf{1}}{\mathbf{1}'\Omega^{-1}\mathbf{1}}$. This is the infeasible CSX estimator (see Section 4.1 of \citet{CSX2025} for this no-covariate case), since $\Omega$ is the variance given the realized cohort sizes $n_c$, i.e., $\Omega^*$ evaluated at $\tilde\Sigma^c$ and the realized shares $n_c/I$, as in the statement. Hence the variance of the infeasible CSX estimator cannot exceed that of $\widehat{\tau}_{w,\text{blue}}$; and, being linear and unbiased, it cannot be smaller either. So both attain the minimum variance, and since the minimum-variance linear unbiased estimator is unique (the variance is strictly convex in the weights), $\widehat{\tau}_{w,\text{blue}}$ is numerically identical to the infeasible CSX estimator. \qed 

\medskip

\noindent \textbf{Remark:} The within-cohort homoskedasticity is obviously not needed for the efficiency of BLUE $\widehat{\tau}_{w,\text{blue}}$. It is maintained to express $\widehat{\tau}_{w,\text{blue}}$ as a linear estimator based on cohort-period means, that helped us show that the infeasible CSX is at least as precise as $\widehat{\tau}_{w,\text{blue}}$.

\medskip

\noindent \textbf{Population weights in Section~\ref{sec:emp-divorce} (continuing footnotes~\ref{fn:empconstr} and~\ref{fn:ar1}):} There $w_{i,t}\propto p_{i,t}$ on the treated cells of each event time, and $\bar\Sigma_i(\rho)=P_i^{-1/2}R_i(\rho)P_i^{-1/2}$ (footnote~\ref{fn:empconstr}), so BJS (OLS), at $\rho=0$, fits $\eta$ by WLS with weights $p_{i,t}$. Thus $\bar\Sigma_i(0)=P_i^{-1}$ replaces the normalization $\bar\Sigma_{[00]}(0)=I$ of Section~\ref{sec:framework}, which only serves to nest BJS (OLS). As $p_{i,t}$ is fixed, our theory (Sections~\ref{sec:background}--\ref{sec:tfgls}) covers this entity-specific working model. In footnote~\ref{fn:ar1}, $\rho_1^{\,t-E_i+1}$ becomes $\rho_1^{\,t-E_i+1}(p_{i,E_i-1}/p_{i,t})^{1/2}$, so in (iv) $\widehat{\lambda}_i(\rho_0)$ cancels only if $p_{i,t}=p_{i,E_i-1}$, $t\ge E_i$.
 
\endgroup

\end{bibunit}

\begin{bibunit}

\newpage
\clearpage
\begin{center}
{\large Supplemental Materials for}
\vspace{0.5cm}

{\Large \textbf{Targeted Feasible Generalized Least Squares}}

\vspace{0.5cm}

{\large Saraswata Chaudhuri}
\end{center}

\doparttoc \faketableofcontents 

\part{~} \vspace{-2.0cm}
\parttoc

\appendix
\setcounter{section}{1}
\renewcommand{\topfraction}{.9}\renewcommand{\bottomfraction}{.9}\renewcommand{\textfraction}{.07}\renewcommand{\floatpagefraction}{.7}
\allowdisplaybreaks[1] \setstretch{1.6} 

\section{Appendix~\ref{app:aux}: Auxiliary material}\label{app:aux}

\subsection{GLS estimator derived from the GLS principle}\label{app:glsderiv}

We derive the GLS estimator from first principles under the restrictions $\tau_{[1]} = \Gamma \theta$ considered in \citet{BJS2024}, where $\Gamma$ is a known $N_{[1]} \times p$ matrix of full column rank and $\theta$ is an unknown $p \times 1$ vector. $\Gamma = I_{N_{[1]}}$ implies no restriction, $\Gamma = 1_{N_{[1]}\times 1}$ imposes fully homogeneous treatment effect, e.g., \citet{Kiefer1980}, \citet{ImAhnSchmidtWooldridge1999}, \citet{HansenGLS2007}, etc. Accordingly, define $D := \left(0_{N_{[1]} \times N_{[0]}}, I_{N_{[1]}}\right)'$ and consider the solution $\widehat{\tau} = \Gamma \widehat{\theta}_{(\Gamma,\Omega)}$ for the problem:
\begin{equation*}
\min_{\eta, \theta} \left(Y - R\eta - D\Gamma\theta\right)'\Omega \left(Y - R\eta - D\Gamma\theta\right)
\end{equation*} 
for some symmetric GLS-weighting matrix $\Omega$. Under suitable full-rank condition, the solution is $\Gamma \widehat{\theta}_{(\Gamma,\Omega)}$ where:
\begin{equation*}
\widehat{\theta}_{(\Gamma,\Omega)} ~:=~ \left(\Gamma'D'\Omega Q_{R:\Omega} D\Gamma\right)^{-1}\left(\Gamma'D'\Omega Q_{R:\Omega}Y\right) ~ \equiv ~ \left(\Gamma'D'\Omega D\Gamma\right)^{-1}\Gamma'D'\Omega \left(Y - R\widehat{\eta}_{(\Gamma,\Omega)}\right)
\end{equation*}
where $\widehat{\eta}_{(\Gamma,\Omega)} := \left(R'\Omega Q_{D\Gamma:\Omega} R \right)^{-1}\left(R'\Omega Q_{D\Gamma:\Omega}Y\right)
$ and $Q_{A:B} := I - A(A'B A)^{-1}A'B$, with conformable matrices, the GLS-annihilator. Taking $\Gamma = I_{N_{[1]}}$ and $\Omega = I_N$ where $N:= N_{[0]} + N_{[1]}$, the OLS estimator is:
\begin{equation*}
\widehat{\tau}_{w,\ols} :=   
w'_{[1]} \left(I_{N_{[1]}} \widehat{\theta}_{(I_{N_{[1]}},I_N)}\right)
=  w'_{[1]}\left(D'D\right)^{-1}D'\left(Y - R\widehat{\eta}_{(I_{N_{[1]}},I_N)}\right)  =  w'_{[1]}\left(Y_{[1]} - R_{[1]}\widehat{\eta}_{(I_{N_{[1]}},I_N)}\right)
\end{equation*}
where $ \widehat{\eta}_{(I_{N_{[1]}},I_N)} = \left(R'_{[0]}R_{[0]}\right)^{-1}R'_{[0]}Y_{[0]}$. Similarly, taking $\Gamma = I_{N_{[1]}}$ and writing $\Omega$ with the usual partitioned matrix notation as $(\Omega_{[jk]})$ for $j,k=0,1$, the GLS estimator is (writing $\widehat{\eta}_{I_{N_{[1]}},\Omega}$ as $\widehat{\eta}$):
\begin{eqnarray*}
\widehat{\tau}_{w,\gls}(\Omega) &:=&   
w'_{[1]} \left(I_{N_{[1]}} \widehat{\theta}_{(I_{N_{[1]}},\Omega)}\right)
=  w'_{[1]}\left(D'\Omega D\right)^{-1}D'\Omega\left(Y - R\widehat{\eta}\right) \\
 &=&  w'_{[1]}\left\{\left(Y_{[1]} - R_{[1]}\widehat{\eta}\right) + \Omega^{-1}_{[11]}\Omega_{[10]}\left(Y_{[0]} - R_{[0]}\widehat{\eta}\right)\right\}
\end{eqnarray*}  
where $\widehat{\eta} \equiv \widehat{\eta}_{I_{N_{[1]}},\Omega} := (R'_{[0]}\Omega_{[00.1]}R_{[0]})^{-1} R'_{[0]}\Omega_{[00.1]}Y_{[0]}$, using the notation $A_{[jj.k]} := A_{[jj]} - A_{[jk]}A^{-1}_{[kk]}A_{[kj]}$ for $j,k = 0,1$. Setting $\Omega = \Sigma^{-1}$ gives the optimal GLS estimator. \citet{BJS2024} caution against such restrictions, as the gain in precision may not be worth the bias if they are incorrect. Hence we take $\Gamma = I_{N_{[1]}}$ (no restriction) throughout.

\subsection{The AR(1) working model: TFGLS$(\rho)$ and TFGLS$(\rho_0,\rho_1)$}\label{app:ar1wm}

Consider entity $i$ with untreated periods $s \in \{S_i,\ldots,E_i-1\}$ and treated periods $t\in\{E_i,\ldots,T_i\}$, and let $\ell_i := E_i-1$ be its last untreated period. (For a never-treated entity only $\bar\Sigma_{[00],i}(\rho_0)$ is needed.) The AR(1) working model sets
\begin{equation}\label{eq:ar1wm}
\bigl[\bar\Sigma_{[00],i}(\rho_0)\bigr]_{s,s'} = \rho_0^{|s-s'|},
\qquad
\bigl[\bar\Sigma_{[10],i}(\rho_{10})\bigr]_{t,s} = \rho_1^{\,t-\ell_i}\,\rho_0^{\,\ell_i-s},
\end{equation}
i.e., with $N_{[0],i}$ and $N_{[1],i}$ untreated and treated periods,
\begin{equation*}
\bar\Sigma_{[00],i}(\rho_0)=
\begin{pmatrix}
1 & \rho_0 & \cdots & \rho_0^{N_{[0],i}-1}\\
\rho_0 & 1 & \cdots & \rho_0^{N_{[0],i}-2}\\
\vdots & & \ddots & \vdots\\
\rho_0^{N_{[0],i}-1} & \rho_0^{N_{[0],i}-2} & \cdots & 1
\end{pmatrix},
\qquad
\bar\Sigma_{[10],i}(\rho_{10}) =
\underbrace{\begin{pmatrix}\rho_1\\ \rho_1^2\\ \vdots\\ \rho_1^{N_{[1],i}}\end{pmatrix}}_{=:\,g_{i}(\rho_1)}
\bigl(\rho_0^{N_{[0],i}-1},\ldots,\rho_0,1\bigr).
\end{equation*}
$\bar\Sigma_{[00],i}(\rho_0)$ is positive definite for $|\rho_0|<1$, and when $N_{[0],i} \ge 2$, its inverse is tridiagonal: $(1-\rho_0^2)^{-1}$ times the matrix with diagonal $(1,1+\rho_0^2,\ldots,1+\rho_0^2,1)$ and off-diagonal $-\rho_0$. With missing untreated periods, $|s-s'|$ is the calendar distance (Section~\ref{sec:emp-divorce}) and the inverse is tridiagonal with gap-dependent entries. At $\rho_0=\rho_1=0$, $\bar\Sigma_{[00],i}(\rho_{0})=I$ and $\bar\Sigma_{[10],i}(\rho_{10})=0$, so OLS is a special case.

For most TFGLS results we impose $\rho_0=\rho_1=\rho$, so $[\bar\Sigma_{[10],i}(\rho_{10})]_{t,s}=\rho^{t-s}$ and \eqref{eq:ar1wm} gives the $[00]$ and $[10]$ blocks of the stationary AR(1) correlation matrix of $(u_{i,S_i},\ldots,u_{i,T_i})'$; here $d_\rho=1$ and $\mathcal P=[-.99,.99]$. TFGLS$(\rho_0,\rho_1)$, used in Section~\ref{sec:emp-hrs}, lets $\rho_0$ and $\rho_1$ differ, so $d_\rho=2$ and we take $\mathcal P=[-.99,.99]\times[-2,2]$.

\textbf{What $\rho_1$ does.} Write $\bar\Sigma_{[10],i}(\rho_{10}) = g_i(\rho_1)\,h_i(\rho_0)'$ with $h_i(\rho_0) := (\rho_0^{N_{[0],i}-1},\ldots,\rho_0,1)'$. Since $[\bar\Sigma_{[00],i}(\rho_0)]_{s,\ell_i} = \rho_0^{\ell_i-s}$, $h_i(\rho_0)$ is the last column of $\bar\Sigma_{[00],i}(\rho_0)$, i.e., $h_i(\rho_0) = \bar\Sigma_{[00],i}(\rho_0)\,e_{N_{[0],i}}$ with $e_{N_{[0],i}} := (0,\ldots,0,1)'$. By symmetry of $\bar\Sigma_{[00],i}(\rho_0)$, $h_i(\rho_0)'\bar\Sigma^{-1}_{[00],i}(\rho_0) = e'_{N_{[0],i}}$, so
\begin{equation*}
\bar\Sigma_{[10],i}(\rho_{10})\,\bar\Sigma^{-1}_{[00],i}(\rho_0) = g_{i}(\rho_1)\,e'_{N_{[0],i}},
\end{equation*}
which does not depend on $\rho_0$. This is the Markov property of the AR(1) working model: the working best linear predictor of a treated-period error given the untreated errors uses only the last one. (This holds with missing untreated periods if $\ell_i$ is observed, as in Section~\ref{sec:empirical}; otherwise the coefficient is $\rho_0^{\ell_i-s_m}g_i(\rho_1)$, $s_m$ being the last observed untreated period.) In the notation of Section~\ref{sec:framework}, $\rho_{10}=(\rho_0,\rho_1)'$ here, and \eqref{eq:gls} and $\kappa_{i,I}(\rho)$ depend on $\bar\Sigma_{[10],i}(\rho_{10})$ only through the product above, which depends on $\rho_1$ alone. Substituting into \eqref{eq:gls}, with $\widehat u_{i,t}(\rho_0) := y_{i,t}-r'_{i,t}\widehat\eta(\rho_0)$,
\begin{equation*}
\widehat\tau_I(\rho_0,\rho_1) = \sum_{i=1}^I\sum_{t=E_i}^{T_i} w_{i,t}\Bigl\{\widehat u_{i,t}(\rho_0) - \rho_1^{\,t-\ell_i}\,\widehat u_{i,\ell_i}(\rho_0)\Bigr\}.
\end{equation*}
Thus $\rho_0$ governs the GLS fit $\widehat\eta(\rho_0)$ on the untreated cells, and $\rho_1$ governs how much of the last untreated residual is carried forward, $t-\ell_i$ periods ahead, when imputing the untreated outcome of a treated cell. Three remarks are in order.

(i) $\Sigma_{[11]}$ never enters $\widehat{\tau}_{w,\text{blue}}$, and so $\bar\Sigma_{[11]}(\rho)$ never enters $\widehat{\tau}_{I}(\rho)$. Hence $\rho_1$ need not lie in $(-1,1)$. Only $\rho_0$ must be bounded away from $\pm1$. (P2) and (P6) hold for fixed $T$ and compact $\mathcal{P}$ with $\rho_0$-projection in $(-1,1)$: $\bar\Sigma^{\pm1}_{[00],i}(\rho_0)$ are smooth and bounded as $N_{[0],i}\le T$, and $\|h_i(\rho_0)\|\le\sqrt{T}$, $\|g_i(\rho_1)\|\le \sqrt{T}\max(1,|\rho_1|)^{T}$.

(ii) If $N_{[0],i}=1$ and the regressors include entity intercepts, then $\widehat u_{i,\ell_i}(\rho_0)=0$, since the intercept fits the single untreated observation exactly. If all the weight $w_{[1]}$ falls on such entities, neither $\widehat\tau_I(\rho)$ nor $\widehat\sigma^2_I(\rho)$ depends on $\rho_1$, and $\rho_1$ is not identified (nor needed). Likewise, if $t-\ell_i$ is even for every weighted cell of every entity with $N_{[0],i}\ge2$, then only $\rho_1^2$ enters and the sign of $\rho_1$ is not identified. Both cases are instances of the non-unique minimizers of Appendix~\ref{app:multirho}, and in both $\widehat\tau_I$ is the same at every minimizer, i.e., where RTFGLS is not needed.

(iii) Example (Section~\ref{sec:emp-hrs}, hospital admissions; waves 7--10). The cohort first hospitalized in wave 8 has $N_{[0],i}=1$. $ES(2)$ puts weight only on this cohort (wave 10), so $\rho_1$ drops out and $\widehat\rho^*_{0}=.03$. $ES(1)$ puts weight on this cohort (wave 9) and on the wave-9 cohort at $t-\ell_i=2$ (wave 10), so only $\rho_1^2$ enters: $\widehat\rho^*_{0}=-.99$, $|\widehat\rho^*_{1}|=1.53$. For $ES(0)$, $t-\ell_i=1$ throughout: $(\widehat\rho^*_{0},\widehat\rho^*_{1})=(-.91,1.93)$. For $ES_{\text{avg}}$: $(-.61,2.00)$; freeing $\rho_1$ beyond $[-2,2]$ gives $\widehat\rho^*_{1}=2.03$ and changes the standard error from 521.72 to 521.69.

\textbf{Non-stationary AR(1) working model.} The working model can instead start the AR(1) process from zero just before the entity's first period: $u_{i,S_i-1}=0$ and $u_{i,t}=\rho_0u_{i,t-1}+e_{i,t}$ for $t\ge S_i$. Then $\bar\Sigma^{-1}_{[00],i}(\rho_0)$ is tridiagonal with off-diagonal $-\rho_0$ and diagonal $(1+\rho_0^2,\ldots,1+\rho_0^2,1)$. Apart from the factor $(1-\rho_0^2)^{-1}$, it differs from the stationary inverse above only in its first diagonal entry. It is positive definite for every real $\rho_0$, so for fixed $T$ (P2) and (P6) hold on any compact $\mathcal{P}$, e.g., $[-2,2]$. GLS then keeps the first untreated period untransformed and quasi-differences the rest. In the trend model of \citet{CanjelsWatson1997}, this is their ``conditional case'' GLS estimator, which sets the pre-sample error to zero ($\kappa=0$ in their notation); the stationary model gives their Prais-Winsten estimator, the limit as $\kappa\to\infty$. \citet{CanjelsWatson1997} show that the conditional-case estimator is efficient when $\kappa=0$ but loses this advantage quickly as $\kappa$ grows, whereas Prais-Winsten is nearly efficient unless $\kappa$ is very small. In Design 1 with $\nu_1=1$, the initial error is a single draw while the shocks accumulate, close to $\kappa=0$; this is consistent with the gains reported in Table~\ref{tab:des1T20}. $\rho_0=0$ still gives OLS, and $\rho_0=1$ is the random walk, whose GLS fit is the first-difference fit of footnote~\ref{fn:ar1}. The model is Markov, so the display above for $\widehat\tau_I(\rho_0,\rho_1)$ is unchanged. For $|\rho_0|<1$ it treats the first period differently from the stationary model, so neither model nests the other.

\subsection{Primitive conditions for (A2) and (A4)}\label{app:primitive}

The main text maintains (A2) and (A4) directly on $\kappa_{[0],i,I}(\rho)$ for simplicity; Proposition \ref{prop:primitive} derives them from primitive conditions on more interpretable quantities. $C$ denotes a generic constant. Recall that $\kappa_{[1],i,I} = w_{[1],i}$, and write $\kappa_{[0],i,I}(\rho)$ as:
\begin{equation}
\label{eq:kappa0}
\kappa_{[0],i,I}(\rho) = -\bSig_{[00],i}^{-1}(\rho_0)\,\bSig_{[01],i}(\rho_{10})\,w_{[1],i} - \bSig_{[00],i}^{-1}(\rho_0)\,r_{[0],i}\,\Pi^{-1}_I(\rho)\,v_I(\rho)
\end{equation}
with $\Pi_I(\rho) := \sum_{j=1}^{I} r_{[0],j}'\,\bSig_{[00],j}^{-1}(\rho_0)\,r_{[0],j}$ and $v_I(\rho) := \sum_{j=1}^{I} \bigl[r_{[1],j} - \bSig_{[10],j}(\rho_{10})\,\bSig^{-1}_{[00],j}(\rho_0)\,r_{[0],j}\bigr]'\,w_{[1],j}$.

\begin{proposition}
\label{prop:primitive}
Suppose the following conditions hold:
\begin{enumerate}
\item[\textup{(P1)}] $\mathcal{P} \subset \mathbb{R}^{d_\rho}$ for some finite $d_\rho$, and $\mathcal{P}_0 \subseteq \{\rho_0 : \bSig_{[00],i}(\rho_0) \text{ is positive definite for all } i\}$.

\item[\textup{(P2)}] The working model matrices are uniformly bounded, and the eigenvalues of $\bSig_{[00],i}(\rho_0)$ are bounded away from zero and infinity:
\[
\sup_{\rho_0 \in \mathcal{P}_0}\sup_i \left(\norm{\bSig_{[00],i}^{-1}(\rho_0)} + \norm{\bSig_{[00],i}(\rho_0)}\right) \le C \quad\text{and}\quad \sup_{\rho \in \mathcal{P}}\sup_i \norm{\bSig_{[10],i}(\rho_{10})} \le C.
\]

\item[\textup{(P3)}]  $ \frac{1}{I}\sum_{i=1}^{I} \tilde{x}_{[0],i}'\,\tilde{x}_{[0],i}$ is finite and positive definite uniformly for all $I$ large, where
$\tilde{x}_{[0],i} := x_{[0],i} - a_{[0],i}(a_{[0],i}'a_{[0],i})^{-1}a_{[0],i}'x_{[0],i}$ is the within-entity OLS residuals of $x_{[0],i}$ on $a_{[0],i}$.

\item[\textup{(P4)}] The regressors are uniformly bounded: $\sup_i \norm{r_{[0],i}} \le C$ and $\sup_i \norm{r_{[1],i}} \le C$.

\item[\textup{(P5)}] The target weights satisfy the Herfindahl condition (A8):
\[
\sum_{i=1}^{I} \norm{w_{[1],i}}^2 = O(1/I) \quad\text{and}\quad \max_{1 \le i \le I} \norm{w_{[1],i}}^2 = o\!\left(\sum_{j=1}^{I} \norm{w_{[1],j}}^2\right).
\]

\item[\textup{(P6)}] The maps $\rho_0 \mapsto \bSig_{[00],i}^{-1}(\rho_0)$ and $\rho_{10} \mapsto \bSig_{[10],i}(\rho_{10})$ are Lipschitz with constants uniform in $i$:
\[
\sup_i \sup_{\rho_0 \neq \rho_0'} \frac{\norm{\bSig_{[00],i}^{-1}(\rho_0) - \bSig_{[00],i}^{-1}(\rho_0')}}{\norm{\rho_0 - \rho_0'}} \le C, \qquad
\sup_i \sup_{\rho_{10} \neq \rho_{10}'} \frac{\norm{\bSig_{[10],i}(\rho_{10}) - \bSig_{[10],i}(\rho_{10}')}}{\norm{\rho_{10} - \rho_{10}'}} \le C.
\]
\item[\textup{(P7)}] The untreated entity-specific designs are uniformly well conditioned: $\inf_i\mu_{\min}(a_{[0],i}'a_{[0],i})\ge c>0$, where $\mu_{\min}(\cdot)$ denotes the smallest eigenvalue.
\end{enumerate}
Assumption \textup{(A2)} follows from (P1)--(P5) and (P7). Assumption \textup{(A4)} follows from (P1)--(P7), with $c_{i,I} = C_L(\|w_{[1],i}\| \vee I^{-1})$ for a finite constant $C_L$.
\end{proposition}

\medskip

\noindent \textbf{Remark:} Write $r_{[0],i} = [a_{[0],i}',\, x_{[0],i}']'$ (keeping implicit the usual normalization, e.g. drop a time dummy, for less messy notation) and partition $\Pi_I(\rho)$ conformably:
\[
  \Pi_I(\rho)
  = \begin{bmatrix}
      \Pi_{aa}(\rho) & \Pi_{ax}(\rho) \\
      \Pi_{xa}(\rho) & \Pi_{xx}(\rho)
    \end{bmatrix},
\]
where:
\begin{align*}
  \Pi_{aa}(\rho)
  &= \sum_{i=1}^{I} a_{[0],i}'\bSig_{[00],i}^{-1}(\rho_0)a_{[0],i}   = \mathrm{block\text{-}diag}
    \bigl(a_{[0],i}'\bSig_{[00],i}^{-1}(\rho_0)a_{[0],i}\bigr)_{i=1}^{I},\\
  \Pi_{xx}(\rho)
  &= \sum_{i=1}^{I} x_{[0],i}'\bSig_{[00],i}^{-1}(\rho_0)x_{[0],i},\\
 \Pi_{ax}(\rho)
  &= \sum_{i=1}^{I} a_{[0],i}'\bSig_{[00],i}^{-1}(\rho_0)x_{[0],i}
  = \begin{pmatrix}
      a_{[0],1}'\bSig_{[00],1}^{-1}(\rho_0)x_{[0],1} \\
      \vdots \\
      a_{[0],I}'\bSig_{[00],I}^{-1}(\rho_0)x_{[0],I}
    \end{pmatrix},
\end{align*}
The block-diagonal structure follows because $a_{i,t}$ contributes entity-specific columns to $r_{i,t}$: e.g., $a_{i,t} = 1$ contributes the entity dummy $1(j=i)$ for each unit $j=1,\ldots,I$, while $a_{i,t} = [1,t]'$ contributes both $1(j=i)$
and $1(j=i)\cdot t$. Consequently, $a_{[0],i}$ and $a_{[0],j}$
have disjoint column supports for $i \neq j$ in the full parameter space
of $(\lambda_1,\ldots,\lambda_I)$. With this in mind, it follows by using (P1), (P2) and (P7) that: $\Pi^{-1}_{aa}(\rho) = \mathrm{block\text{-}diag}
\bigl((a_{[0],i}'\bSig_{[00],i}^{-1}(\rho_0)a_{[0],i})^{-1}\bigr)_{i=1}^{I}$. Finally, in the same spirit partition
$v_I(\rho) = [v_{I,\lambda}'(\rho),\, v_{I,\beta}'(\rho)]'$ where:
\begin{align*}
  v_{I,\lambda_i}(\rho)
  &:= \bigl(a_{[1],i}
    - \bSig_{[10],i}(\rho_{10})\bSig^{-1}_{[00],i}(\rho_0)a_{[0],i}
    \bigr)'w_{[1],i},\\
  v_{I,\beta}(\rho)
  &:= \sum_{i=1}^{I}\bigl(x_{[1],i}
    - \bSig_{[10],i}(\rho_{10})\bSig^{-1}_{[00],i}(\rho_0)x_{[0],i}
    \bigr)'w_{[1],i}.
\end{align*}
Because $\Pix_{aa}(\rho)$ is block-diagonal, the quantity
\[
g_I(\rho) := v_{I,\beta}(\rho) - \Pix_{xa}(\rho)\Pix_{aa}^{-1}(\rho)v_{I,\lambda}(\rho),
\]
which will appear in the partitioned-inverse formula \eqref{eq:beta_comp}, admits the sum over entities representation:
\[
g_I(\rho) = \sum_{i=1}^I \tm_{\zo,i}'(\rho)\, w_{\zone,i}
\]
where:
\begin{align*}
\tm_{\zo,i}(\rho) &:= \bigl(x_{\zone,i} - \Sigbar_{\zto,i}(\rho_{10})\Sigbar_{\zoz,i}^{-1}(\rho_0)x_{\zo,i}\bigr) \\
&\quad - \bigl(a_{\zone,i} - \Sigbar_{\zto,i}(\rho_{10})\Sigbar_{\zoz,i}^{-1}(\rho_0)a_{\zo,i}\bigr)
\bigl(a_{\zo,i}'\Sigbar_{\zoz,i}^{-1}(\rho_0)a_{\zo,i}\bigr)^{-1}
a_{\zo,i}'\Sigbar_{\zoz,i}^{-1}(\rho_0)x_{\zo,i}.
\end{align*}
By (P2), (P4), (P7), and finite $T$ (hence of all matrix dimensions finite),
$\sup_i\sup_\rho \|\tm_{\zo,i}(\rho)\| \leq C$. That $\|g_I(\rho)\| = O(1)$
uniformly in $\rho$ then follows from its definition, since $\|v_{I,\beta}(\rho)\| = O(1)$ by \eqref{eq:vbeta_bound} and
$\|\Pix_{xa}(\rho)\Pix_{aa}^{-1}(\rho)v_{I,\lambda}(\rho)\|
\leq C\sum_i\|w_{\zone,i}\| = O(1)$ by (P2), (P4), (P7), and \eqref{eq:vlambda_bound}. Note that, by (P2) and (P4):
\begin{equation}
\label{eq:vlambda_bound}
  \norm{v_{I,\lambda_i}(\rho)} \leq C\norm{w_{[1],i}},
  \quad\text{uniformly in } \rho.
\end{equation}
On the other hand, by (P2), (P4), (P5), and Cauchy-Schwarz:
\begin{equation}
\label{eq:vbeta_bound}
  \norm{v_{I,\beta}(\rho)}
  \leq C\sum_{j=1}^{I}\norm{w_{[1],j}}
  \leq CI^{1/2}\Bigl(\sum_{j=1}^{I}\norm{w_{[1],j}}^2\Bigr)^{1/2}
  = O(1),
  \quad\text{uniformly in } \rho.
\end{equation}

\medskip

\noindent \textbf{Proof of Proposition \ref{prop:primitive}:}
\textbf{Part 1: (A2) follows from (P1)--(P5) and (P7).}

\eqref{eq:kappa0} implies that:
\[
\norm{\kappa_{[0],i,I}(\rho)} \le \underbrace{\norm{\bSig_{[00],i}^{-1}(\rho_0)}\cdot\norm{\bSig_{[01],i}(\rho_{10})}\cdot\norm{w_{[1],i}}}_{=:\,A_i(\rho)}
\;+\; \underbrace{\norm{\bSig_{[00],i}^{-1}(\rho_0)}\cdot\norm{r_{[0],i}}\cdot\norm{\Pi^{-1}_I(\rho)v_I(\rho)}}_{=:\,B_i(\rho)}.
\]

\noindent\emph{Bounding $A_i(\rho)$.}  (P2) implies that: $A_i(\rho) \le C^2\,\norm{w_{[1],i}}$ uniformly over $\rho$ and~$i$.

\noindent\emph{Bounding $\norm{\Pi^{-1}_I(\rho)v_I(\rho)}$.} We bound its $\beta$ and $\lambda_i$ components in turn. Using the partitioned inverse formula we first write the $\beta$ and $\lambda_i$ components of $\Pi^{-1}_I(\rho)v_I(\rho)$ as:
\begin{eqnarray}
\label{eq:beta_comp}
\hspace{-.5cm} [\Pi^{-1}_I(\rho)v_I(\rho)]_\beta
  &=& S^{-1}_I(\rho)
    \bigl(v_{I,\beta}(\rho)
    - \Pi_{xa}(\rho)\Pi^{-1}_{aa}(\rho)v_{I,\lambda}(\rho)\bigr)\\
\label{eq:lambda_comp}
\hspace{-1cm} [\Pi^{-1}_I(\rho)v_I(\rho)]_{\lambda_i}
  &=& \bigl(a_{[0],i}'\bSig_{[00],i}^{-1}(\rho_0)a_{[0],i}\bigr)^{-1}
    \Bigl(v_{I,\lambda_i}(\rho)
    - a_{[0],i}'\bSig_{[00],i}^{-1}(\rho_0)x_{[0],i}\,
    [\Pi_I^{-1}(\rho)v_I(\rho)]_\beta\Bigr)
\end{eqnarray}
where the Schur complement is:
\begin{equation}
\label{eq:schur}
  S_I(\rho)
  := \Pi_{xx}(\rho) - \Pi_{xa}(\rho)\Pi^{-1}_{aa}(\rho)\Pi_{ax}(\rho)
  = \sum_{i=1}^{I}
    \tilde{x}_{[0],i}'(\rho_0)\,\bSig_{[00],i}^{-1}(\rho_0)\,
    \tilde{x}_{[0],i}(\rho_0),
\end{equation}
with the last equality following from the standard partitioned-regression
identity using the block-diagonal structure of $\Pi_{aa}(\rho)$, and where:
\[
  \tilde{x}_{[0],i}(\rho_0)
  := x_{[0],i}
  - a_{[0],i}\bigl(a_{[0],i}'\bSig_{[00],i}^{-1}(\rho_0)a_{[0],i}\bigr)^{-1}
    a_{[0],i}'\bSig_{[00],i}^{-1}(\rho_0)\,x_{[0],i}
\]
is the GLS within-entity residual of $x_{[0],i}$ on $a_{[0],i}$ under the working model indexed by $\rho_0$. Lemma \ref{lemma:s-prop} (i) gives:
\begin{equation}
\label{eq:schur_inv}
  \norm{S^{-1}_I(\rho)} = O(I^{-1}) \quad\text{uniformly in }\rho_0
  \in \mathcal{P}_0.
\end{equation}

\noindent For the $\beta$-component in \eqref{eq:beta_comp}, $\|\Pi_{xa}(\rho)\Pi_{aa}^{-1}(\rho)v_{I,\lambda}(\rho)\| \le C\sum_{i}\norm{w_{[1],i}} = O(1)$, as shown in the Remark following Proposition \ref{prop:primitive}. Hence, by \eqref{eq:vbeta_bound}
and \eqref{eq:schur_inv}, uniformly in $\rho$:
\begin{equation}
\label{eq:beta_order}
  \norm{[\Pi_I^{-1}(\rho)v_I(\rho)]_\beta}
  \leq O(I^{-1}) \cdot O(1) = O(I^{-1}).
\end{equation}
Now, consider the $\lambda_i$-component from \eqref{eq:lambda_comp}. Noting that 
$\left(a_{[0],i}'\bSig_{[00],i}^{-1}(\rho_0)a_{[0],i}\right)^{-1} = O(1)$ by (P2) and (P7), we use \eqref{eq:vlambda_bound} and \eqref{eq:beta_order} to obtain that uniformly in $\rho$:
\begin{equation}
\label{eq:lambda_order}
  \norm{[\Pi_I^{-1}(\rho)v_I(\rho)]_{\lambda_i}}
  \leq O(1)\cdot\bigl(C\norm{w_{[1],i}} + C\cdot O(I^{-1})\bigr)
  = O\bigl(\norm{w_{[1],i}} \vee I^{-1}\bigr).
\end{equation}

\noindent\emph{Bounding $B_i(\rho)$.} Writing
$r_{[0],i}\Pi_I^{-1}(\rho)v_I(\rho)
= a_{[0],i}[\Pi_I^{-1}(\rho)v_I(\rho)]_{\lambda_i}
+ x_{[0],i}[\Pi_I^{-1}(\rho)v_I(\rho)]_\beta$
and applying~(P2), (P4), \eqref{eq:beta_order}, and~\eqref{eq:lambda_order}:
\begin{eqnarray*}
  B_i(\rho)
  &:=& \norm{\bSig_{[00],i}^{-1}(\rho_0)\,r_{[0],i}\,\Pi_I^{-1}(\rho)v_I(\rho)}
  \leq C\bigl(\norm{a_{[0],i}}\cdot O(\norm{w_{[1],i}}\vee I^{-1})
  + \norm{x_{[0],i}}\cdot O(I^{-1})\bigr)\\
  &=& O\bigl(\norm{w_{[1],i}}\vee I^{-1}\bigr).
\end{eqnarray*}

\noindent\emph{Establish (A2).} Since $\|\kappa_{[0],i,I}(\rho)\| \leq A_i(\rho) + B_i(\rho)
\leq C(\norm{w_{[1],i}}\vee I^{-1})$ and $\kappa_{[1],i,I} = w_{[1],i}$:
\[
  \norm{\kappa_{i,I}(\rho)}^2
  \leq C\bigl(\norm{w_{[1],i}}^2 + I^{-2}\bigr).
\]
Summing and taking the supremum over $\rho$:
\[
  K_I
  := \sup_{\rho \in \mathcal{P}}\sum_{i=1}^{I}\norm{\kappa_{i,I}(\rho)}^2
  \leq C\Bigl(\sum_{i=1}^{I}\norm{w_{[1],i}}^2 + I^{-1}\Bigr)
  = O(I^{-1}),
\]
by~(P5). For the no-dominance condition, note that the $I^{-1}$ terms above arise only from $\|g_I(\rho)\|\le C\sum_j\|w_{[1],j}\|$ (see \eqref{eq:vbeta_bound} and the Remark following Proposition \ref{prop:primitive}), so in fact $\|\kappa_{i,I}(\rho)\|\le C(\|w_{[1],i}\|+\bar w_I)$ with $\bar w_I:=I^{-1}\sum_j\|w_{[1],j}\|$, and $\bar w^2_I\le I^{-1}\varpi_I$ with $\varpi_I:=\sum_j\|w_{[1],j}\|^2$ by Cauchy-Schwarz. Since $\kappa_{[1],i,I}=w_{[1],i}$, $K_I\ge\varpi_I$. Hence, by (P5):
\[
  \sup_{\rho \in \mathcal{P}}\max_{1\leq i\leq I}\norm{\kappa_{i,I}(\rho)}^2
  \leq C\bigl(\max_{1\leq i\leq I}\norm{w_{[1],i}}^2 + I^{-1}\varpi_I\bigr)
  = o(\varpi_I) = o(K_I).
\]

\bigskip
\noindent\textbf{Part 2: (A4) follows from (P1)--(P7).}

\medskip
Write $\kappa_{[0],i,I}(\rho) - \kappa_{[0],i,I}(\rho') = T_{1,i}(\rho,\rho') + T_{2,i}(\rho,\rho') + T_{3,i}(\rho,\rho')$ where:
\begin{align*}
T_{1,i}(\rho,\rho') &:= -\bigl[\bSig_{[00],i}^{-1}(\rho_0)\,\bSig_{[01],i}(\rho_{10}) - \bSig_{[00],i}^{-1}(\rho_0')\,\bSig_{[01],i}(\rho_{10}')\bigr]\,w_{[1],i}, \\
T_{2,i}(\rho,\rho') &:= -\bigl[\bSig_{[00],i}^{-1}(\rho_0) - \bSig_{[00],i}^{-1}(\rho_0')\bigr]\,r_{[0],i}\,\Pi_I^{-1}(\rho)\,v_I(\rho), \\
T_{3,i}(\rho,\rho') &:= -\bSig_{[00],i}^{-1}(\rho_0')\,r_{[0],i}\bigl[\Pi^{-1}_I(\rho)\,v_I(\rho) - \Pi^{-1}_I(\rho')\,v_I(\rho')\bigr].
\end{align*}

\smallskip

\noindent\textit{Bounding $T_{1,i}(\rho,\rho')$.} Decompose
$\bSig_{[00],i}^{-1}(\rho_0)\bSig_{[01],i}(\rho_{10})
- \bSig_{[00],i}^{-1}(\rho_0')\bSig_{[01],i}(\rho_{10}')
= \bigl(\bSig_{[00],i}^{-1}(\rho_0)
- \bSig_{[00],i}^{-1}(\rho_0')\bigr)\bSig_{[01],i}(\rho_{10})
+ \bSig_{[00],i}^{-1}(\rho_0')
\bigl(\bSig_{[01],i}(\rho_{10}) - \bSig_{[01],i}(\rho_{10}')\bigr)$.
By~(P2) and~(P6):
\[
  \norm{T_{1,i}(\rho,\rho')} \leq C\norm{\rho-\rho'}\cdot\norm{w_{[1],i}}.
\]

\noindent\textit{Bounding $T_{2,i}(\rho,\rho')$.} As in the step \emph{Bounding $B_i(\rho)$} of Part 1, $r_{[0],i}\Pi_I^{-1}(\rho)v_I(\rho) = a_{[0],i}[\Pi_I^{-1}(\rho)v_I(\rho)]_{\lambda_i} + x_{[0],i}[\Pi_I^{-1}(\rho)v_I(\rho)]_\beta$ has norm $O(\norm{w_{[1],i}}\vee I^{-1})$ by (P4), \eqref{eq:beta_order} and \eqref{eq:lambda_order}. Hence, by (P6):
\[
  \norm{T_{2,i}(\rho,\rho')}
  \leq C\norm{\rho_0-\rho_0'}\cdot\norm{r_{[0],i}\Pi^{-1}_I(\rho)v_I(\rho)}
  = O\bigl(\norm{\rho-\rho'}(\norm{w_{[1],i}}\vee I^{-1})\bigr).
\]

\noindent \emph{Bounding $T_{3,i}(\rho,\rho')$.} It suffices to bound
$\|\PI^{-1}(\rho)v_I(\rho) - \PI^{-1}(\rho')v_I(\rho')\|$ component by
component using \eqref{eq:beta_comp}--\eqref{eq:lambda_comp}.

\medskip

\noindent First, consider the $\beta$-component. By \eqref{eq:beta_comp}, $[\PI^{-1}(\rho)v_I(\rho)]_\beta
= S^{-1}_I(\rho)g_I(\rho)$, where $g_I$ is defined in the Remark following Proposition \ref{prop:primitive}. Write:
\[
[\PI^{-1}v_I]_\beta(\rho) - [\PI^{-1}v_I]_\beta(\rho')
= \bigl(S^{-1}_I(\rho) - S^{-1}_I(\rho')\bigr)g_I(\rho)
+ S^{-1}_I(\rho')\bigl(g_I(\rho) - g_I(\rho')\bigr).
\]
For the first term, the identity
$S^{-1}_I(\rho) - S^{-1}_I(\rho') = -S^{-1}_I(\rho)(S_I(\rho) - S_I(\rho'))S^{-1}_I(\rho')$
along with Lemma \ref{lemma:s-prop} (i),(ii) and $\|g_I(\rho)\|=O(1)$ gives:
\[
\bigl\|(S^{-1}_I(\rho) - S^{-1}_I(\rho'))g_I(\rho)\bigr\|
\leq O(I^{-1})\cdot CI\|\rho-\rho'\|\cdot O(I^{-1})\cdot O(1)
= O(I^{-1})\|\rho-\rho'\|.
\]
For the second term, using the summed over entity representation
$g_I(\rho) = \sum_i \tm_{\zo,i}'(\rho)w_{\zone,i}$,
\[
\|g_I(\rho) - g_I(\rho')\|
\leq \sup_i\|\tm_{\zo,i}(\rho)-\tm_{\zo,i}(\rho')\| \cdot \sum_i \|w_{\zone,i}\|
\leq C\|\rho-\rho'\|\cdot O(1) = O(\|\rho-\rho'\|),
\]
by Lemma \ref{lemma:s-prop} (iii), Cauchy-Schwarz, and (P5). Hence by Lemma \ref{lemma:s-prop} (i),
$\|S_I^{-1}(\rho')(g_I(\rho)-g_I(\rho'))\| = O(I^{-1})\|\rho-\rho'\|$. Combining the two terms, we obtain:
\begin{equation}
\label{eq:piinv_v}
\bigl\|[\PI^{-1}v_I]_\beta(\rho) - [\PI^{-1}v_I]_\beta(\rho')\bigr\|
= O(I^{-1})\|\rho-\rho'\|.
\end{equation}

\medskip

\noindent Now, consider the $\lambda_i$-component. By \eqref{eq:lambda_comp}:
\[
[\PI^{-1}(\rho)v_I(\rho)]_{\lambda_i}
= \left(a_{\zo,i}'\Sigbar_{\zoz,i}^{-1}(\rho_0)a_{\zo,i}\right)^{-1}\Bigl(v_{I,\lambda_i}(\rho) - a_{\zo,i}'\Sigbar_{\zoz,i}^{-1}(\rho_0)x_{\zo,i}\,[\PI^{-1}v_I]_\beta(\rho)\Bigr).
\]
By (P2), (P4), (P6), and (P7), both $(a_{\zo,i}'\Sigbar_{\zoz,i}^{-1}(\rho_0)a_{\zo,i})^{-1}$ and $a_{\zo,i}'\Sigbar_{\zoz,i}^{-1}(\rho_0)x_{\zo,i}$ are uniformly bounded and uniformly
Lipschitz in $\rho$ across $i$. Moreover, by (P2), (P4), and (P6),
$\|v_{I,\lambda_i}(\rho)\| \leq C\|w_{\zone,i}\|$ and
$\|v_{I,\lambda_i}(\rho) - v_{I,\lambda_i}(\rho')\| \leq C\|w_{\zone,i}\|\,\|\rho-\rho'\|$. Combining these with \eqref{eq:beta_order}, \eqref{eq:piinv_v}, and the uniform boundedness and Lipschitz
property of $a_{[0],i}'\bar{\Sigma}_{[00],i}^{-1}(\rho_0)x_{[0],i}$
established above, via the standard product rule for Lipschitz functions,
\begin{equation*}
\bigl\|[\PI^{-1}v_I]_{\lambda_i}(\rho) - [\PI^{-1}v_I]_{\lambda_i}(\rho')\bigr\|
\leq C\bigl(\|w_{\zone,i}\| \vee I^{-1}\bigr)\|\rho-\rho'\|.
\end{equation*}

\medskip

Finally, writing:
$r_{\zo,i}\PI^{-1}v_I(\rho) = a_{\zo,i}[\PI^{-1}v_I]_{\lambda_i}(\rho)
+ x_{\zo,i}[\PI^{-1}v_I]_\beta(\rho)$ and differencing, then premultiplying by
$\Sigbar_{\zoz,i}^{-1}(\rho_0')$ which is $O(1)$ by (P2), and using
$\|a_{\zo,i}\|, \|x_{\zo,i}\| \leq C$ from (P4) together with the two displayed
bounds above,
\[
\|T_{3,i}(\rho,\rho')\| \leq C\bigl(\|w_{\zone,i}\| \vee I^{-1}\bigr)\|\rho-\rho'\|.
\]
Combining the bounds on $T_{1,i}, T_{2,i}, T_{3,i}$ gives, for a finite constant $C_L$, $\|\kappa_{i,I}(\rho) - \kappa_{i,I}(\rho')\| \leq C_L(\|w_{[1],i}\| \vee I^{-1})\|\rho - \rho'\|$ for all $\rho,\rho' \in \mathcal{P}$ and all $i, I$. This is (A4) with $c_{i,I} := C_L(\|w_{[1],i}\| \vee I^{-1})$, since $\sum_i c^2_{i,I} \leq C^2_L(\sum_i \|w_{[1],i}\|^2 + I^{-1}) = O(1/I)$ by (P5).\qed

\bigskip

\noindent \textbf{Remark:} It is useful to verify the primitive conditions in Proposition \ref{prop:primitive} specialized to the AR(1) working model $\bSig_{[00],i}(\rho_0)_{st} = \rho_0^{|s-t|}$ with $\rho_0 \in \mathcal{P}_0 \subseteq [-1+c, 1-c]$ for some $c > 0$:

\begin{itemize}
\item[(P1)] Positive definiteness holds since the AR(1) covariance matrix is positive definite if and only if $|\rho_0| < 1$.

\item[(P2)] The precision matrix $\bSig_{[00],i}^{-1}(\rho_0)$ is tridiagonal with diagonal entries bounded by $(1+\rho_0^2)/(1-\rho_0^2)$ and off-diagonal entries equal to $-\rho_0/(1-\rho_0^2)$.  Let $\norm{A}_{\mathrm{op}}:=\sup_{\norm{v}=1}\norm{Av}$ denote the operator (spectral) norm, as distinct from the Frobenius norm $\norm{\cdot}$ used for matrices elsewhere; recall $\norm{A}_{\mathrm{op}}\le\norm{A}\le\sqrt{n}\,\norm{A}_{\mathrm{op}}$ for an $n\times n$ matrix $A$. Since the spectral norm of a symmetric matrix is at most its maximum absolute row sum, $\norm{\bSig_{[00],i}^{-1}(\rho_0)}_{\mathrm{op}} \le (1 + 2|\rho_0| + \rho_0^2)/(1-\rho_0^2) = (1+|\rho_0|)/(1-|\rho_0|)$, and hence $\norm{\bSig_{[00],i}^{-1}(\rho_0)} \le \sqrt{T}\,(1+|\rho_0|)/(1-|\rho_0|)$. Both bounds hold uniformly over $|\rho_0| \le 1 - c$ and all panel lengths $N_{[0],i} \le T$. Also, being a correlation matrix, $\norm{\bSig_{[00],i}(\rho_0)} \le N_{[0],i} \le T$.

\item[(P3)] Since $T$ is finite and the regressors include entity-specific intercepts, $\Pi_I(\rho)/I$ inherits the structure of a within-entity Gram matrix averaged over $I$ units.  Condition (P3) is the standard full-rank condition on the within-entity regressors; by Lemma \ref{lemma:s-prop} (i) it implies that the smallest eigenvalue of the Schur complement $S_I(\rho)/I$ in \eqref{eq:schur} is bounded away from zero.

\item[(P4)] Since $T$ is finite this assumption is standard for the same reason as above. 

\item[(P5)] The weights in the target estimand have nothing to do with the working model.

\item[(P6)] The map $\rho_0 \mapsto \bSig_{[00],i}^{-1}(\rho_0)$ is analytic on $|\rho_0| < 1$, so it is Lipschitz on the compact set $|\rho_0| \le 1-c$ with a constant depending only on $c$ and the maximal panel length $\max_i N_{[0],i} \le T$.

\item[(P7)] It holds with entity intercepts, since $a_{[0],i}'a_{[0],i}=N_{[0],i}\ge1$, and, with fixed $T$, with entity-specific linear trends $a_{i,t}=[1,t]'$ whenever every entity has at least two untreated periods (finitely many configurations, each of full rank).

\item The cross-block parameter $\rho_1$ need not satisfy $|\rho_1| < 1$ since $\bSig_{[10],i}(\rho_{10})$, $\rho_{10}=(\rho_0,\rho_1)'$, is a cross-period covariance matrix not required to be positive definite. For $\bSig_{[10],i}(\rho_{10}) = g_i(\rho_1)h_i(\rho_0)'$ of Appendix~\ref{app:ar1wm}, (P2) holds on any compact $\mathcal{P}$ with $|\rho_0|\le 1-c$, and (P6) holds with Lipschitz constant depending on $\sup_{\rho \in \mathcal{P}}|\rho_1|$, $c$ and~$T$.

\item (P1)--(P7) do not require $\mathcal{P}$ to be compact. Compactness would be needed to deduce the uniform bounds in (P2), (P3), and (P6) from continuity alone, but since these bounds are assumed directly, compactness is unnecessary. The $\epsilon$-net arguments elsewhere in the paper (e.g., Lemma \ref{lemma:maxineq} and the proof of Proposition \ref{prop:tfgls-not-unique-rho}) rely on the compactness, hence boundedness, of $\mathcal{P}$ maintained in Section~\ref{sec:bgresults}, not on (P1): finite dimension alone does not give finite $\epsilon$-nets on an unbounded set.
    
\end{itemize}

\bigskip

\begin{lemma}
\label{lemma:s-prop}
Let the conditions of Proposition \ref{prop:primitive} hold. Then with $S_I(\rho)$ defined in \eqref{eq:schur} and $\tm_{\zo,i}(\rho)$ defined in the Remark following Proposition \ref{prop:primitive}:
\begin{itemize}
\item[(i)] $\|S_I^{-1}(\rho)\| = O(I^{-1})$ uniformly in $\rho_0\in \mathcal{P}_0$;
\item[(ii)] $\|S_I(\rho) - S_I(\rho')\| \leq I C\|\rho_0-\rho_0'\|$;
\item[(iii)] there exists $C<\infty$ uniform in $i$ such that, for all
$\rho,\rho'\in\mathcal{P}$, we have: $\sup_i\sup_\rho \|\tm_{\zo,i}(\rho)\| \leq C$ and $\sup_i\|\tm_{\zo,i}(\rho) - \tm_{\zo,i}(\rho')\| \leq C\|\rho-\rho'\|$.
\end{itemize}
\end{lemma}

\bigskip

\noindent \textbf{Proof of Lemma \ref{lemma:s-prop}:} (i) Recall \eqref{eq:schur} and the description of the involved quantities from below it. By (P2), the eigenvalues of $\bSig_{[00],i}^{-1}(\rho_0)$ are uniformly bounded above and away from zero over $\rho_0 \in \mathcal{P}_0$ and $i$, and hence for some constants $0 < c \leq C < \infty$ uniform in $i$, $\rho_0$, $I$:
\begin{equation}
\label{eq:gls_ols_norm}
  c\,\bigl\|\tilde{x}_{[0],i}(\rho_0)\,b\bigr\|^2
  ~\leq~
  b'\,\tilde{x}_{[0],i}'(\rho_0)\,\bSig_{[00],i}^{-1}(\rho_0)\,
  \tilde{x}_{[0],i}(\rho_0)\,b
  ~\leq~
  C\,\bigl\|\tilde{x}_{[0],i}(\rho_0)\,b\bigr\|^2
  \quad\text{for every }b.
\end{equation}
Moreover, write $\tilde{x}_{[0],i}(\rho_0) = M_{a,i}(\rho_0)\,x_{[0],i}$ and
$\widetilde{x}_{[0],i} = M^{\mathrm{ols}}_{a,i}\,x_{[0],i}$, where $M_{a,i}(\rho_0)$ is the GLS
within-entity residual maker implicit in the definition of $\tilde{x}_{[0],i}(\rho_0)$ below
\eqref{eq:schur} and written out in part (ii) below, and
$M^{\mathrm{ols}}_{a,i} := I_{N_{[0],i}} - a_{[0],i}(a_{[0],i}'a_{[0],i})^{-1}a_{[0],i}'$ is its OLS
counterpart from (A6). The two share the null space $\mathrm{col}(a_{[0],i})$: since
$M^{\mathrm{ols}}_{a,i}a_{[0],i} = 0$ we get $M^{\mathrm{ols}}_{a,i}M_{a,i}(\rho_0) = M^{\mathrm{ols}}_{a,i}$
and hence $\widetilde{x}_{[0],i} = M^{\mathrm{ols}}_{a,i}\,\tilde{x}_{[0],i}(\rho_0)$. As
$M^{\mathrm{ols}}_{a,i}$ is an orthogonal projector, this gives
\begin{equation}
\label{eq:ols_dominated}
  \bigl\|\widetilde{x}_{[0],i}\,b\bigr\|
  ~=~ \bigl\|M^{\mathrm{ols}}_{a,i}\,\tilde{x}_{[0],i}(\rho_0)\,b\bigr\|
  ~\leq~ \bigl\|\tilde{x}_{[0],i}(\rho_0)\,b\bigr\|
  \quad\text{for every }b,
\end{equation}
uniformly in $i$ and $\rho_0$, and with no norm-equivalence constant.
Combining \eqref{eq:gls_ols_norm} and \eqref{eq:ols_dominated} and summing over $i$, for every $b$:
\[
  b'\bigl(I^{-1}S_I(\rho)\bigr)b
  ~\geq~ c\,I^{-1}\sum_{i=1}^{I}\bigl\|\tilde{x}_{[0],i}(\rho_0)\,b\bigr\|^2
  ~\geq~ c\,I^{-1}\sum_{i=1}^{I}\bigl\|\widetilde{x}_{[0],i}\,b\bigr\|^2
  ~=~ c\,b'\Bigl(I^{-1}\sum_{i=1}^{I}\widetilde{x}_{[0],i}'\,\widetilde{x}_{[0],i}\Bigr)b,
\]
so that $\mu_{\min}\left(I^{-1}S_I(\rho)\right) \;\geq\; c\,\mu_{\min}\!\left(I^{-1}\sum_{i=1}^{I}
  \widetilde{x}_{[0],i}'\,\widetilde{x}_{[0],i}\right)
  \;>\; 0$ uniformly in $\rho_0 \in \mathcal{P}_0$ and $I$ by (P3). Therefore $\norm{S^{-1}_I(\rho)} = O(I^{-1})$ uniformly in $\rho_0 \in \mathcal{P}_0$.

\noindent (ii) Note that $S_I(\rho)$ depends on $\rho$ only through $\rho_0$. The GLS within-entity residual $\tilde{x}_{[0],i}(\rho_0) = M_{a,i}(\rho_0)x_{[0],i}$ depends on $\rho_0$
through:
\[
  M_{a,i}(\rho_0)
  = I_{N_{[0],i}}
  - a_{[0],i}
    \bigl(a_{[0],i}'\bSig_{[00],i}^{-1}(\rho_0)a_{[0],i}\bigr)^{-1}
    a_{[0],i}'\bSig_{[00],i}^{-1}(\rho_0).
\]
Since $\bSig_{[00],i}^{-1}(\rho_0)$ is Lipschitz by~(P6) and uniformly
bounded by~(P2), and since $a_{[0],i}'\bSig_{[00],i}^{-1}(\rho_0)a_{[0],i}$ is uniformly bounded
away from zero by (P2) and (P7) (since $a_{[0],i}'\bSig_{[00],i}^{-1}(\rho_0)a_{[0],i}\ge C^{-1}a_{[0],i}'a_{[0],i}$), the map $\rho_0 \mapsto M_{a,i}(\rho_0)$ is Lipschitz with a constant uniform in $i$, as it involves only
products and inverses of uniformly bounded Lipschitz maps. Combined with $\norm{x_{[0],i}} \leq C$ from (P4), we obtain uniformly in $i$:
\begin{equation}
\label{eq:xtilde_lip}
  \norm{\tilde{x}_{[0],i}(\rho_0) - \tilde{x}_{[0],i}(\rho_0')}
  \leq C\norm{\rho_0 - \rho_0'}.
\end{equation}
Now recall from \eqref{eq:schur} that $S_I(\rho) = \sum_{i=1}^{I}
\tilde{x}_{[0],i}'(\rho_0)\bSig_{[00],i}^{-1}(\rho_0)
\tilde{x}_{[0],i}(\rho_0)$. Therefore:
\begin{eqnarray*}
S_I(\rho) - S_I(\rho')  &=& \sum_{i=1}^I \tilde{x}_{[0],i}'(\rho_0)\bSig_{[00],i}^{-1}(\rho_0)
    \bigl(\tilde{x}_{[0],i}(\rho_0) - \tilde{x}_{[0],i}(\rho_0')\bigr)\\
  &&~~
    + \sum_{i=1}^I  \bigl(\tilde{x}_{[0],i}(\rho_0)
    - \tilde{x}_{[0],i}(\rho_0')\bigr)'
    \bSig_{[00],i}^{-1}(\rho_0)\tilde{x}_{[0],i}(\rho_0')\\
  &&~~
    + \sum_{i=1}^I \tilde{x}_{[0],i}'(\rho_0')
    \bigl(\bSig_{[00],i}^{-1}(\rho_0)
    - \bSig_{[00],i}^{-1}(\rho_0')\bigr)
    \tilde{x}_{[0],i}(\rho_0')\\
    && \le I C\norm{\rho_0 - \rho_0'}
\end{eqnarray*}
because each term inside each sum has norm bounded by $C\norm{\rho_0 - \rho_0'}$, using
$\norm{\tilde{x}_{[0],i}(\rho_0)} \leq C\norm{x_{[0],i}} \leq C$
from~(P2), (P4) and (P7) (the oblique projector $M_{a,i}(\rho_0)$ is uniformly bounded, though its norm may exceed one), $\norm{\bSig_{[00],i}^{-1}(\rho_0)} \leq C$ from~(P2),
\eqref{eq:xtilde_lip}, and (P6).

\noindent (iii) Boundedness was established in the Remark following Proposition \ref{prop:primitive}. For the Lipschitz part, write $\tm_{\zo,i}(\rho) = m^x_i(\rho) - m^a_i(\rho)G_i(\rho_0)$, where $m^x_i(\rho) := x_{\zone,i} - \Sigbar_{\zto,i}(\rho_{10})\Sigbar_{\zoz,i}^{-1}(\rho_0)x_{\zo,i}$, $m^a_i(\rho) := a_{\zone,i} - \Sigbar_{\zto,i}(\rho_{10})\Sigbar_{\zoz,i}^{-1}(\rho_0)a_{\zo,i}$ and $G_i(\rho_0) := (a_{\zo,i}'\Sigbar_{\zoz,i}^{-1}(\rho_0)a_{\zo,i})^{-1}a_{\zo,i}'\Sigbar_{\zoz,i}^{-1}(\rho_0)x_{\zo,i}$. By (P2), (P4) and (P6), $m^x_i$ and $m^a_i$ are bounded and Lipschitz in $\rho$, uniformly in $i$; e.g., $\|m^x_i(\rho) - m^x_i(\rho')\| \leq \bigl(\|\Sigbar_{\zto,i}(\rho_{10})\|\,\|\Sigbar_{\zoz,i}^{-1}(\rho_0) - \Sigbar_{\zoz,i}^{-1}(\rho_0')\| + \|\Sigbar_{\zto,i}(\rho_{10}) - \Sigbar_{\zto,i}(\rho_{10}')\|\,\|\Sigbar_{\zoz,i}^{-1}(\rho_0')\|\bigr)\|x_{\zo,i}\| \leq C\|\rho - \rho'\|$. Let $Q_i(\rho_0) := a_{\zo,i}'\Sigbar_{\zoz,i}^{-1}(\rho_0)a_{\zo,i}$. By (P2) and (P7), $Q_i(\rho_0) \geq C^{-1}a_{\zo,i}'a_{\zo,i} \geq c$, so $\|Q^{-1}_i(\rho_0)\| \leq C$; and, by (P4) and (P6), $\|Q^{-1}_i(\rho_0) - Q^{-1}_i(\rho_0')\| = \|Q^{-1}_i(\rho_0)\{Q_i(\rho_0') - Q_i(\rho_0)\}Q^{-1}_i(\rho_0')\| \leq C\|\rho_0 - \rho_0'\|$. Hence, with (P2), (P4) and (P6), $G_i$ is also bounded and Lipschitz in $\rho_0$, uniformly in $i$. Since sums and products of uniformly bounded, uniformly Lipschitz maps are uniformly bounded and uniformly Lipschitz, $\|\tm_{\zo,i}(\rho) - \tm_{\zo,i}(\rho')\| \leq C\|\rho-\rho'\|$ uniformly in $i$.\qed

\subsection{Maximal inequality for Propositions \ref{prop:oracle} and \ref{prop:tfgls-not-unique-rho}}\label{app:maxineq}

\begin{trivlist}\item[\hskip\labelsep{\bfseries Lemma \ref{lemma:maxineq}} (restated from {\hyperref[sec:proofs]{Appendix~A}})\textbf{.}]\itshape
Let (A1) and (A4) hold. Let $A_I := (\sum_{i=1}^I c^2_{i,I})^{1/2}$ with $c_{i,I}$ from (A4), $D_{\mathcal{P}} := \diam(\mathcal{P}) := \sup_{\rho,\rho' \in \mathcal{P}}\|\rho - \rho'\|$, and $\omega(\mu) := \mu\,(1 + \log(D_{\mathcal{P}}/\mu))^{1/2}$ for $\mu \in (0, D_{\mathcal{P}}]$. Then there is a finite constant $C$, depending only on $d_\rho$ and $\sup_i E\left[\|u_i\|^2\right]$, such that for all $I$ and all $\mu \in (0, D_{\mathcal{P}}]$:
\vspace{-.1cm}
\begin{equation*}
E\Bigl[\sup_{\rho,\rho' \in \mathcal{P}:\, \|\rho - \rho'\| \leq \mu}\Bigl|\sum_{i=1}^I \left(\kappa_{i,I}(\rho) - \kappa_{i,I}(\rho')\right)'u_i\Bigr|\Bigr] ~\leq~ C\,\omega(\mu)\,A_I.
\end{equation*}
\end{trivlist}

\noindent \textbf{Proof of Lemma \ref{lemma:maxineq}:} We adapt standard stochastic equicontinuity arguments for Lipschitz-in-parameter processes \citep{Pollard1990,Andrews94} to make $\omega(\mu)$ and $A_I$ explicit. Let $Z_i(\rho,\rho') := (\kappa_{i,I}(\rho) - \kappa_{i,I}(\rho'))'u_i$, indexed by $T_\mu := \{(\rho,\rho') \in \mathcal{P}^2: \|\rho - \rho'\| \leq \mu\}$. By (A1), $Z_1,\ldots,Z_I$ are independent mean zero processes, and by (A4) they are continuous in $(\rho,\rho')$, so all suprema below can be taken over a countable dense subset. Let $\varepsilon_1,\ldots,\varepsilon_I$ be i.i.d.\ Rademacher variables independent of $U$. The symmetrization inequality \citep[Lemma 2.3.6]{VW96} gives:
\vspace{-.1cm}
\begin{equation*}
E\Bigl[\sup_{T_\mu}\Bigl|\sum_i Z_i(\rho,\rho')\Bigr|\Bigr] ~\leq~ 2\,E\Bigl[E_\varepsilon\Bigl[\sup_{T_\mu}\bigl|X^\varepsilon(\rho) - X^\varepsilon(\rho')\bigr|\Bigr]\Bigr], \text{~~where~~} X^\varepsilon(\rho) := \sum_i \varepsilon_i\,\kappa'_{i,I}(\rho)u_i,
\end{equation*}
and $E_\varepsilon$ denotes expectation over $\varepsilon$ given $U$. Given $U$, by Hoeffding's inequality \citep[Lemma 2.2.8]{VW96}, $\{X^\varepsilon(\rho): \rho \in \mathcal{P}\}$ is sub-Gaussian with respect to $d_U(\rho,\rho') := \bigl(\sum_i ((\kappa_{i,I}(\rho) - \kappa_{i,I}(\rho'))'u_i)^2\bigr)^{1/2}$, and by (A4), $d_U(\rho,\rho') \leq B_U\|\rho - \rho'\|$ with $B_U := (\sum_i c^2_{i,I}\|u_i\|^2)^{1/2}$. Hence $T_\mu \subseteq \{d_U(\rho,\rho') \leq B_U\mu\}$, and the $d_U$-packing numbers of $\mathcal{P}$ satisfy $D(\epsilon,\mathcal{P},d_U) \leq D(\epsilon/B_U,\mathcal{P},\|\cdot\|) \leq (3D_{\mathcal{P}}B_U/\epsilon)^{d_\rho}$ for $\epsilon \leq D_{\mathcal{P}}B_U$. So, by \citet[Corollary 2.2.9]{VW96}, for a universal constant $K$:
\vspace{-.1cm}
\begin{equation*}
\begin{aligned}
E_\varepsilon\Bigl[\sup_{T_\mu}\bigl|X^\varepsilon(\rho) - X^\varepsilon(\rho')\bigr|\Bigr] ~&\leq~ K\int_0^{B_U\mu}\sqrt{d_\rho\log(3D_{\mathcal{P}}B_U/\epsilon)}\,d\epsilon\\
~&=~ K\sqrt{d_\rho}\,B_U\int_0^{\mu}\sqrt{\log(3D_{\mathcal{P}}/t)}\,dt ~\leq~ C_1\,\omega(\mu)\,B_U,
\end{aligned}
\end{equation*}
with $C_1$ depending only on $d_\rho$, using $\sqrt{\log(3D_{\mathcal{P}}/t)} \leq \sqrt{\log(3D_{\mathcal{P}}/\mu)} + \sqrt{\log(\mu/t)}$ and $\int_0^\mu\sqrt{\log(\mu/t)}\,dt = \mu\sqrt{\pi}/2$. Finally, $E[B_U] \leq (E[B_U^2])^{1/2} \leq (\sup_i E\|u_i\|^2)^{1/2}A_I$ by Jensen's inequality. \qed

\subsection{TFGLS under multiple optimal nuisance $\rho^*_I$}\label{app:multirho}

Uniqueness of $\rho^*_I$ is necessary for (A9). But generic conditions for it are hard to give since the presence of $\widehat\eta(\rho_0)$ in $\widehat\tau_I(\rho)$ makes $\sigma^2_I(\rho)$ a rational function of $\rho$ with standard
AR working models. While Lemma \ref{lemma:rhocons} can be easily extended to a set-valued $\rho^*_I$ with consistency holding in terms of the distance of $\widehat{\rho}^*_I$ from the set \citep{Redner1981}, the proof of Proposition \ref{prop:tfgls} exploits the convergence of $\widehat{\rho}^*_I$ to a singleton $\rho^*_I$. This appendix considers the case where $\rho^*_I$ is not unique, i.e., the set of minimizers in the optimization problem \eqref{eq:optrho} is not a singleton.

We first show what properties of TFGLS are preserved under non-uniqueness of $\rho^*_I$.

\begin{proposition}
\label{prop:tfgls-not-unique-rho}
Consider model \eqref{eq:model}. Let $\mathcal P$ be compact. Let (A1), (A2) and (A4) hold. Then:
\[
\sup_{\rho\in\mathcal P}\bigl|\widehat\tau_I(\rho)-\tau_w\bigr| \;=\; O_P\bigl(I^{-1/2}\bigr).
\]
Hence, $\widehat\tau_I(\widehat\rho_I)-\tau_w = O_P(I^{-1/2})$ for every measurable
$\widehat\rho_I =\widehat\rho_I(U)\in \mathcal{P}$, regardless of $\widehat\rho_I$'s convergence in probability and dependence on $U$. The result holds in particular for $\widehat\rho_I = \widehat \rho^*_I$.
\end{proposition}

\noindent \textbf{Remark:} TFGLS remains $\sqrt{I}$-consistent for $\tau_w$. However, asymptotic normality, asymptotic unbiasedness and, in some cases, minimum variance cannot be guaranteed without further conditions. The non-uniqueness issue can be visually ``checked'' by plotting $\widehat\sigma^2_I(\rho)$ against $\rho$, and can sometimes be addressed by using a bigger, e.g., AR(2) as opposed to AR(1), working model, albeit at the cost of parsimony. The discussion below concerns cases where a bigger working model does not help or where any loss of parsimony is not acceptable to the user.   

Non-uniqueness of $\rho^*_I$, or of the sample minimizer $\widehat\rho^*_I$, does not necessarily mean non-uniqueness of the TFGLS estimator: $\rho$ enters $\widehat\sigma^2_I(\rho)$ and $\widehat\tau_I(\rho)$ only through the weights $\Kappa(\rho)$, so wherever that map is not injective --- redundant parameters, common-factor cancellation in ARMA working models, root-flipping in MA models --- distinct minimizers deliver exactly the same TFGLS estimator. The user can check this by plotting $\widehat\sigma^2_I(\rho)$ and $\widehat\tau_I(\rho)$ against $\rho$. 

The leading example where $\widehat\tau_I(\cdot)$ may not move with different minimizers of $\sigma^2_I(\rho)$ is perhaps the case where those minimizers form a continuum. This is because a connected set along which $\widehat\tau_I(\rho)$ \emph{does} move would instead require $\Kappa(\rho)$ to trace, as $\rho$ ranges over the curve, infinitely many distinct points lying exactly on the ellipsoid $\Kappa(.)\Sigma \Kappa'(.) =\min_{\rho} \sigma^2_I(\rho)$. That would perhaps only be a coincidence between the true $\Sigma$ and its working model $\bar{\Sigma}(\rho)$. Hence, the case we will consider is that of a finite number $J_I$ of minimizers such that $\Kappa(\rho^*_{I,j})$ not necessarily equal to $\Kappa(\rho^*_{I,k})$ and hence $\widehat{\tau}_I(\rho^*_{I,j})$ not necessarily equal to $\widehat{\tau}_I(\rho^*_{I,k})$ for $j\neq k$.

\medskip
\noindent\textbf{Assumptions (continued):}
\begin{itemize}
\item[(A9')] $\mathcal{P}^*_I:=\arg\min_{\rho\in\mathcal P} I\sigma^2_I(\rho) =\{\rho^*_{I,1},\dots,\rho^*_{I,J_I}\}$ with $\limsup_I J_I<\infty$ and $\liminf_I\min_{j\neq k}\|\rho^*_{I,j}-\rho^*_{I,k}\|>0$. (A9) holds uniformly over the basins: $\liminf_I\min_{j\le J_I}\inf\{I(\sigma^2_I(\rho)-\sigma^2_I(\rho^*_{I,j})):\rho\in\mathcal P,\ \min_{k\le J_I}\|\rho-\rho^*_{I,k}\|=\|\rho-\rho^*_{I,j}\|>\epsilon\}>0$ for all $\epsilon>0$. (A10) holds at each $\rho^*_{I,j}$ uniformly in $j$: $H_{I,j}:=\frac{\partial^2}{\partial\rho\,\partial\rho'}\bigl[\frac I2\sigma^2_I(\rho^*_{I,j})\bigr]$ is finite and positive definite uniformly in $I$ and $j\le J_I$.
\item[(A11)] $\sup_{\rho\in\mathcal P}\bigl|I\widehat\sigma^2_I(\rho) -I\{\sigma^2_I(\rho)+\sigma^2_{\tau,I}\}\bigr| = O_P(a_I)$ for a sequence $a_I\to 0$.
\end{itemize}

\noindent \textbf{Remark:} (A9') with $J_I=1$ is (A9) together with (A10). (A11) strengthens Proposition \ref{prop:cons_var} from consistency to a rate and holds with $a_I=I^{-1/2}(\log I)^{1/2}$ under (A5$'$), (A8$'$), (P6$'$) and (A12) of Appendix~\ref{app:hobias-lemmas}. 

Write $\sigma^{*2}_I:=\min_{\mathcal P}\sigma^2_I(\rho)$ and $Z_j:=\{\widehat\tau_I(\rho^*_{I,j})-\tau_w\}/\sigma^*_I$, so that $V(Z_j)=1$ for \emph{every} $j$ --- this is the tie itself. Selecting $\widehat\rho^*_I$ therefore does not
choose among estimators of different precision; it chooses among equally precise ones using
an event that is correlated with them, and it is that correlation, not any difference in
variance, that distorts the limit. 

To correct for this distortion, we propose a randomized version of TFGLS that replaces the exact minimizer $\widehat{\rho}^*_I$ by a draw from a vanishing band as follows. Let $\varsigma_I \to 0$ with $\varsigma_I/a_I\to\infty$. Define:
\[
\widehat{\mathcal P}_I:=\Bigl\{\rho\in\mathcal P:\;I\widehat\sigma^2_I(\rho)
\le \min_{\mathcal P} I\widehat\sigma^2_I+\varsigma_I\Bigr\},
\]
and let $\widetilde\rho^*_I$ be drawn from the normalized Lebesgue measure on $\widehat{\mathcal P}_I$, using auxiliary randomness independent of $U$. Then the randomized TFGLS estimator is $\widehat\tau_I(\widetilde\rho^*_I)$. 

$\varsigma_I=0$, or more generally any realization in which $\widehat{\mathcal P}_I$ has zero Lebesgue measure, gives TFGLS (with a slight abuse by which the then undefined normalized Lebesgue measure is replaced by the point mass at $\widehat\rho^*_I$). So this is a generalization of \eqref{eq:tfgls}--\eqref{eq:estrho} rather than a substitute. The two rate conditions do separate jobs. $\varsigma_I\to0$ makes every band point criterion-optimal. $\varsigma_I/a_I\to\infty$ makes the band wide enough that the conditional probability of landing in each basin is nonrandom up to $o_P(1)$ making the selection asymptotically uninformative about $U$ and thereby removing the limit-distorting correlation. This gives Proposition \ref{prop:bandTFGLS}(ii) below.

\begin{proposition}
\label{prop:bandTFGLS}
Consider model \eqref{eq:model}.
\begin{enumerate}
\item[(i)] Let $\mathcal P$ be compact. Let (A1)--(A8) hold. Let $\varsigma_I\to0$. Then
$\sigma^2_I(\widetilde\rho^*_I)/\sigma^{*2}_I\overset{p}{\to}1$: every point of the band, and hence
the draw, attains the oracle precision $\sigma^{*2}_I:= \min_{\rho \in \mathcal{P}} \sigma^2_I(\rho)$ to first order. No condition on $\mathcal{P}^*_I$
is used, so this holds also when $\mathcal{P}^*_I$ is a continuum.
\item[(ii)] If in addition (P6'), (A9'), (A11) and (A12) of Appendix~\ref{app:hobias-lemmas} hold, each $\rho^*_{I,j}$ is interior uniformly in $I$, i.e., $B(\rho^*_{I,j},r)\subseteq\mathcal P$ for all $j\le J_I$ and some fixed $r>0$, and $\varsigma_I/a_I\to\infty$, then
\[
\frac{\widehat\tau_I(\widetilde\rho^*_I)-\tau_w}{\sigma^*_I}\overset{d}{\longrightarrow}\mathrm{Normal}(0,1),
\qquad
\frac{\widehat\tau_I(\widetilde\rho^*_I)-\tau_w}{\widehat\sigma_I(\widetilde\rho^*_I)}
\;\overset{d}{\preceq}\;\mathrm{Normal}(0,1),
\]
with no restriction on the error law or on the treatment-effect heterogeneity.
\item[(iii)] If $J_I\equiv 1$, (ii) agrees with Proposition \ref{prop:tfgls}.
\end{enumerate}
\end{proposition}

\noindent \textbf{Remark:} It is not clear to us if the non-uniqueness of $\rho^*_I$ is of first-order importance. Even in the carefully crafted data generating processes (available from us) that led to non-unique $\rho^*_I$, we observed that the small-sample performance of TFGLS was as good as that of FGLS (that faced no such problems). Nevertheless, randomized TFGLS with the theoretical guarantees presented in Proposition \ref{prop:bandTFGLS} provides a generalization that is useful at least for the sake of completeness. Since (A5'), (A8'), (P6') and (A12) imply (A11) with $a_I=I^{-1/2}(\log I)^{1/2}$, the user can pick any $\varsigma_I\to0$ with $\varsigma_I/a_I\to\infty$. 

A user who prefers not to pick $\varsigma_I$ when the sample objective function displays multiple (approximate) global minima can always fall back on OLS or FGLS or a variance-minimizing convex combination of OLS and FGLS.

\subsubsection*{Proofs:}

\noindent \textbf{Proof of Proposition \ref{prop:tfgls-not-unique-rho}:}
Write $Y_I(\rho):=\sqrt I\{\widehat\tau_I(\rho)-\tau_w\}=\sqrt I\sum_i\kappa'_{i,I}(\rho)u_i$. By (A1), $E[Y_I(\rho)]=0$ and $V(Y_I(\rho))=I\sigma^2_I(\rho)\le(\sup_i\|\Sigma_i\|)IK_I=O(1)$ by (A2). Hence 
$Y_I(\rho)=O_P(1)$. Now, let $A_I := (\sum_i c^2_{i,I})^{1/2}$ with $c_{i,I}$ from (A4), $D_{\mathcal P} := \diam(\mathcal P)$, and $\omega(\mu) := \mu\,(1+\log(D_{\mathcal P}/\mu))^{1/2}$. Lemma \ref{lemma:maxineq} (with its constant $C$) gives $E\bigl[\sup_{\|\rho-\rho'\|\le\mu}|Y_I(\rho)-Y_I(\rho')|\bigr]\le C\,\omega(\mu)\,(IA_I^2)^{1/2}\le C'\omega(\mu)$ for all large $I$ and all $\mu\in(0,D_{\mathcal P}]$, for a finite $C'$ because $IA_I^2=O(1)$ by (A4). Since $\omega(\mu)\to 0$ as $\mu\downarrow 0$, Markov's inequality gives: for every $\eta,\varepsilon>0$ there is a
$\mu >0$, \emph{not depending on $I$}, with
\begin{equation}\label{eq:equicont2}
\limsup_I P\Bigl(\sup_{\substack{\rho,\rho'\in\mathcal P\\ \|\rho-\rho'\|<\mu}}|Y_I(\rho)-Y_I(\rho')|>\eta\Bigr)<\varepsilon.
\end{equation}

$\mathcal P$ is compact. So for the same $\mu$ there exist finite $N(\mu)$ and fixed $\rho_1,\dots,\rho_{N(\mu)} \in \mathcal P$ whose $\mu$-balls cover $\mathcal P$. For any $\rho\in\mathcal P$, let $\rho_{k(\rho)}$ be its nearest net point, so $\|\rho-\rho_{k(\rho)}\|<\mu$. Then:
\[
|Y_I(\rho)|\le|Y_I(\rho)-Y_I(\rho_{k(\rho)})|+|Y_I(\rho_{k(\rho)})|
\le \sup_{\|\rho-\rho'\|<\mu}|Y_I(\rho)-Y_I(\rho')| + \max_{k\le N(\mu)}|Y_I(\rho_k)| = O_P(1) + O_P(1)
\]
where the first $O_P(1)$ follows for this fixed $\mu$ from the bound $E\bigl[\sup_{\|\rho-\rho'\|\le\mu}|Y_I(\rho)-Y_I(\rho')|\bigr]\le C'\omega(\mu)$ above and Markov's inequality, and the second $O_P(1)$ follows using the union bound since $N(\mu)$ is finite and $Y(\rho) = O_p(1)$ for any given $\rho \in \mathcal P$.

The final claim is pathwise: for every realization
$|\widehat\tau_I(\widehat\rho_I)-\tau_w|\le\sup_{\rho\in\mathcal P}|\widehat\tau_I(\rho)-\tau_w|$, an
inequality between numbers, so no independence between $\widehat\rho_I$ and $U$ is needed. \qed

\bigskip

\noindent \textbf{Proof of Proposition \ref{prop:bandTFGLS}:} Let $R_I:=\sup_{\rho \in \mathcal P}\bigl|I\widehat\sigma^2_I(\rho)-I\{\sigma^2_I(\rho)+\sigma^2_{\tau,I}\}\bigr|$. $R_I=o_P(1)$ by Proposition \ref{prop:cons_var} under (A1)--(A8), and $R_I=O_P(a_I)$ under (A11).

\noindent\emph{(i)} For $\rho \in \widehat{\mathcal P}_I \subseteq \mathcal{P}$, it follows by using $\min_{\mathcal P}I\widehat\sigma^2_I\le I\widehat\sigma^2_I(\rho^*_I)$ and the definition of $R_I$ that:
\vspace{-.1cm}
\begin{equation*}
I\sigma^2_I(\rho)+I\sigma^2_{\tau,I}-R_I\;\le\;I\widehat\sigma^2_I(\rho)
\;\le\;I\widehat\sigma^2_I(\rho^*_I)+\varsigma_I
\;\le\;I\sigma^{*2}_I+I\sigma^2_{\tau,I}+R_I+\varsigma_I.
\vspace{-.1cm}
\end{equation*}
Hence $I\sigma^2_I(\rho)-I\sigma^{*2}_I\le\varsigma_I+2R_I = o_P(1)$. Divide by $I\sigma^{*2}_I$, which is bounded away from zero by (A3), for the result.

\medskip

\noindent\emph{(ii)} \textbf{Step 1:} We know from above that every $\rho\in\widehat{\mathcal
P}_I$ satisfies $I\{\sigma^2_I(\rho)-\sigma^{*2}_I\}\le\varsigma_I+2R_I\overset{p}{\to}0$. By
(A9'), the within-basin (A9) implies that no such $\rho$ can lie outside fixed neighborhoods of $\mathcal{P}^*_I$ for $I$ large; and, since each $\rho^*_{I,j}$ is interior so that the first-order condition $\partial \sigma^2_I(\rho^*_{I,j})/\partial \rho =0$ holds, (A10) implies that in each basin $j = 1,\ldots, J_I$ we have $I\{\sigma^2_I(\rho)-\sigma^{*2}_I\}\ge c\|\rho-\rho^*_{I,j}\|^2$ on $B(\rho^*_{I,j},r_0)$ with $r_0>0$ and $c>0$ uniform in $I$ and $j$, because the Hessian of $I\sigma^2_I(\rho)$ is Lipschitz on $\mathcal P$ uniformly in $I$. To see the latter, note that $I\sigma^2_I(\rho)=I\sum_i\kappa'_{i,I}(\rho)\Sigma_i\kappa_{i,I}(\rho)$ with only $\kappa_{[0],i,I}(\rho)$ depending on $\rho$, so its Hessian is $I\sum_i$ of terms of the form $\dot\kappa'_{0,i}\Sigma_{[00],i}\dot\kappa_{0,i}$ and $\ddot\kappa'_{0,i}(\Sigma_i\kappa_{i,I})_{[0]}$; by (A12) and $\sup_i\|\Sigma_i\|<\infty$ from (A1), each is Lipschitz in $\rho$ with constant $Cb_i^2$, and $I\sum_ib_i^2\le I K_I+1=O(1)$ by (A2). This also gives the uniform cubic remainder used in Step 2. Hence, with probability approaching 1 (w.p.a.1) $\widehat{\mathcal P}_I\subseteq\bigcup_{j\le J_I}B\bigl(\rho^*_{I,j},C(\varsigma_I+2R_I)^{1/2}\bigr)$, the balls with centers $\rho^*_{I,j}$'s and radius $C(\varsigma_I+2R_I)^{1/2}$ for some finite $C \ge 0$ being pairwise disjoint for $I$ large by (A9'). Let $F_I$ be index of the ball containing $\widetilde\rho^*_I$, so that
\vspace{-.1cm}
\begin{equation}
\label{eq:localrate}
\|\widetilde\rho^*_I-\rho^*_{I,F_I}\|=O_P\bigl((\varsigma_I+2R_I)^{1/2}\bigr)=o_P(1).
\vspace{-.1cm}
\end{equation}
Define $Z_I := \frac{\widehat\tau_I(\widetilde\rho^*_I)-\tau_w}{\sigma^*_I}$ and note that $Z_I - \sum_{j=1}^{J_I} 1(F_I = j) Z_j = (Y_I(\widetilde \rho^*_I) - Y_I(\rho^*_{I,F_I}))/(\sqrt{I} \sigma^*_I)$. Recall from \eqref{eq:equicont2} that conditional on the event $\{\|\widetilde \rho^*_I - \rho^*_{I,F_I}\| < \mu\}$, we have:
\[
|Y_I(\widetilde\rho^*_I)-Y_I(\rho^*_{I,F_I})|\le\sup_{\|\rho-\rho'\|<\mu}|Y_I(\rho)-Y_I(\rho')|.
\]
Since $P(\|\widetilde\rho^*_I-\rho^*_{I,F_I}\|\ge\mu)\to0$ by \eqref{eq:localrate}, 
it follows from \eqref{eq:equicont2} that $\limsup_I P\bigl(|Y_I(\widetilde\rho^*_I)-Y_I(\rho^*_{I,F_I})|>\eta\bigr)<\varepsilon$. Since $\eta,\varepsilon>0$ are arbitrary, $|Y_I(\widetilde\rho^*_I)-Y_I(\rho^*_{I,F_I})| = o_P(1)$. Now since $\sqrt I\,\sigma^*_I$ is bounded and bounded away from $0$ for $I$ large by (A2)--(A3), so dividing the $o_P(1)$ numerator by it preserves $o_P(1)$. Hence, $Z_I=\sum_j 1(F_I=j)Z_j +o_P(1)$.

\textbf{Step 2:} Define $p_{I,j}(U):=P(F_I=j\mid U)$, the
fraction of the band's volume lying in basin $j$ given the realized data --- this is the auxiliary randomness due to our way of picking the $\widetilde \rho^*_I \in \widehat{\mathcal{P}}_I$. This is well defined w.p.a.1: by Step 1 and $C(\varsigma_I+2R_I)^{1/2}<r$ for $I$ large, $\widehat{\mathcal P}_I$ lies in the interior of $\mathcal P$ w.p.a.1, and it contains the relatively open set $\{\rho\in\mathcal P: I\widehat\sigma^2_I(\rho)<\min_{\mathcal P}I\widehat\sigma^2_I+\varsigma_I\}\ni\widehat\rho^*_I$ by the continuity of $\widehat\sigma^2_I(\cdot)$ implied by (A4); hence $\widehat{\mathcal P}_I$ has positive Lebesgue measure w.p.a.1. We show that $p_{I,j}=p^0_{I,j}+o_P(1)$ for some nonrandom $p^0_{I,j}$ such that $\sum_{j=1}^{J_I}p^0_{I,j}=1$.

Let $\widehat\Delta_j:=I\widehat\sigma^2_I(\rho^*_{I,j})-\min_{\rho\in\mathcal P}I\widehat\sigma^2_I(\rho)$.
Since $\sigma^2_I(\rho^*_{I,j})=\sigma^{*2}_I$, the minimized population criterion, we have $I\widehat\sigma^2_I(\rho^*_{I,j})=I\sigma^{*2}_I+I\sigma^2_{\tau,I}+O_P(a_I)$ by (A11). Since $|\min_\rho f-\min_\rho g|\le\sup_\rho|f-g|$, (A11) also gives: $ \min_{\rho\in\mathcal P}I\widehat\sigma^2_I(\rho)
=\min_{\rho\in\mathcal P}I\bigl\{\sigma^2_I(\rho)+\sigma^2_{\tau,I}\bigr\}+O_P(a_I)
=I\sigma^{*2}_I+I\sigma^2_{\tau,I}+O_P(a_I)$. Hence $\widehat\Delta_j=O_P(a_I)$, uniformly in $j\le J_I$. Now, by (A11), (A10) and the uniformly Lipschitz Hessian from Step 1, for $\rho$ in the ball around $\rho^*_{I,j}$, using the first-order condition $\partial \sigma^2_I(\rho^*_{I,j})/\partial \rho =0$:
\[
I\widehat\sigma^2_I(\rho)-\min_{\rho \in \mathcal P}I\widehat\sigma^2_I(\rho) = I\left(\widehat\sigma^2_I(\rho)-\widehat \sigma^2_I(\rho^*_{I,j})\right) + \widehat\Delta_j = (\rho-\rho^*_{I,j})'H_{I,j}(\rho-\rho^*_{I,j})+O\bigl(\|\rho-\rho^*_{I,j}\|^3\bigr)+O_P(a_I).
\]
Recall from Step 1 that w.p.a.1 $\widehat{\mathcal P}_I\subseteq\bigcup_{k\le J_I}B\bigl(\rho^*_{I,k},C(\varsigma_I+2R_I)^{1/2}\bigr)$ and $R_I = O_p(a_I)$ by (A11) while $a_I = o(\varsigma_I)$ under the conditions of part (ii). Hence for $\rho \in \widehat{\mathcal P}_I \cap B\bigl(\rho^*_{I,j},C(\varsigma_I+2R_I)^{1/2}\bigr)$ we have $\|\rho -\rho^*_{I,j}\|=O(\varsigma_I^{1/2})$. Hence, $I\widehat\sigma^2_I(\rho)-\min_{\rho \in \mathcal P}I\widehat\sigma^2_I(\rho) \le \varsigma_I$ implies that $(\rho-\rho^*_{I,j})'H_{I,j}(\rho-\rho^*_{I,j})+O\bigl(\|\rho-\rho^*_{I,j}\|^3\bigr) \le \varsigma_I(1 + \epsilon_{I,j})$ where $\epsilon_{I,j}=O_P(a_I/\varsigma_I)+O(\varsigma_I^{1/2})=o_P(1)$ under the conditions of the proposition. That is, this set is sandwiched between ellipsoids with quadratic form $H_{I,j}$ and level
$\varsigma_I(1\pm\epsilon_{I,j})$. Its volume is therefore $V_{I,j}=v_{I,j}\{1+O(\epsilon_{I,j})\}$, where $v_{I,j}$ is the corresponding leading-order volume at level $\varsigma_I$ and is nonrandom. For $I$ large these ellipsoids, whose radii are $O(\varsigma_I^{1/2})$, lie in $B(\rho^*_{I,j},r)\subseteq\mathcal P$, so no boundary truncation arises.

Set $p^0_{I,j}:=v_{I,j}/\sum_{k\le J_I}v_{I,k}$, which is nonrandom, nonnegative and sums to
one. Since $J_I\le\bar J$, $\bar\epsilon_I:=\max_{k\le J_I}|\epsilon_{I,k}|=o_P(1)$ --- the only
place finiteness of $\bar J$ is used --- and forming the ratio $p_{I,j}=V_{I,j}/\sum_kV_{I,k}$ gives $p_{I,j} = p^0_{I,j} + o_P(1)$.

\textbf{Step 3:} Proposition \ref{prop:clt} gives $Z_j\overset{d}{\to}\mathrm{Normal}(0,1)$ for
every $j$, since $\sigma^2_I(\rho^*_{I,j})=\sigma^{*2}_I$. Take a bounded Lipschitz function $f$, with
Lipschitz constant $L$, and write $\bar Z_I:=\sum_{j\le J_I} 1(F_I=j)Z_j$, so that
$Z_I=\bar Z_I+r_I$ with $r_I=o_P(1)$ by Step 1. $|f(Z_I)-f(\bar Z_I)|\le L|r_I|$ by Lipschitz
and $\le2\|f\|_\infty$ by boundedness, so $\bigl|f(Z_I)-f(\bar Z_I)\bigr|\;\le\;\min\bigl(L|r_I|,\,2\|f\|_\infty\bigr)=:D_I$ and $D_I\overset{p}{\to}0$ while $0\le D_I\le2\|f\|_\infty$. Therefore, bounded convergence gives $E[D_I]\to0$.

Now note that, (a) given $U$ the quantities $Z_1,\dots,Z_{J_I}$ are fixed numbers, (b) the $\rho^*_{I,j}$ being nonrandom points, and hence the only remaining randomness is the auxiliary draw determining $F_I$. Hence $E[f(\bar Z_I)\mid U]=\sum_{j\le J_I}p_{I,j}\,f(Z_j)$, and therefore:
\vspace{-.2cm}
\begin{equation*}
E[f(Z_I)]=\sum_{j\le J_I}E\bigl[p_{I,j}\,f(Z_j)\bigr]+o(1)
=\sum_{j\le J_I}p^0_{I,j}\,E[f(Z_j)]
+\sum_{j\le J_I}E\bigl[(p_{I,j}-p^0_{I,j})f(Z_j)\bigr]+o(1).
\vspace{-.2cm}
\end{equation*}
The second sum on the RHS is bounded by $\|f\|_\infty\sum_j E|p_{I,j}-p^0_{I,j}|\to0$ using Step 2, boundedness of probabilities and the condition that $J_I \le \bar{J} < \infty$. The first sum on the RHS is a convex combination of the $E[f(Z_j)]$'s that all converge to the same limit $E[f(Z)]$, $Z\sim\mathrm{Normal}(0,1)$; hence it converges to $E[f(Z)]$ without any convergence of the
weights themselves. Therefore $E[f(Z_I)]\to E[f(Z)]$ for every bounded Lipschitz $f$, and since the bounded Lipschitz functions are convergence-determining, we obtain: $Z_I\overset{d}{\to}\mathrm{Normal}(0,1)$.

The last part follows in the same way as in Proposition \ref{prop:tfgls} after using Part (i).

\noindent\emph{(iii)} If $J_I=1$ then $p_{I,1}\equiv1$, and the above steps gives the result. \qed

\subsection{Lemmas for the Proof of Proposition \ref{prop:ho-bias}(i)}\label{app:hobias-lemmas}

\noindent For notational simplicity, throughout Appendices~\ref{app:hobias-lemmas} and \ref{app:hobias-boundary}, $\widehat{\beta} \equiv \widehat{\beta}(0)$ and $\widehat{\lambda}_i \equiv \widehat{\lambda}_i(0)$ denote the OLS estimates, i.e., the components of $\widehat{\eta}(0) = \widehat{\eta}_{\ols}$.

We are concerned with the expectations in (notation same as in the proof of Proposition \ref{prop:ho-bias}):
\vspace{-.1cm}
\begin{equation*}
\begin{aligned}
  E[\tauhattfgls-\tauw] 
  ~=~& -E\left[\Ghat'\HI^{-1}\sum_i\dot{\kappa}'_{0,i}\,u_{0,i}\right] + E[\rem_3]\\
  ~=~& -IE\left[\sum_i\bigl(\kappa'_i u_i + \delta_i - \phi_i - \psi_i\bigr)'\bigl(\dot{\kappa}'_{0,i}u_{0,i} - \dot \phi_i\bigr)'\HI^{-1}\sum_j\dot{\kappa}'_{0,j}u_{0,j}\right] + E[\rem_3]
\end{aligned}
\vspace{-.1cm}
\end{equation*}
using, for the two components of $\Ghat$, the decomposition \eqref{eq:decomp} for $\kappa'_i\hat{u}_i$ and an analogous decomposition for $\dot{\kappa}'_{0,i}\hat{u}_{0,i}$ where $\delta_i := w'_{[1],i}(\tau_{[1],i}-1_{N_{[1],i}}\tauw)$, $\phi_i := \kappa'_i x_i(\bhat-\beta)$,  $\psi_i := w'_{[1],i}1_{N_{[1],i}}(\tauhat(0)-\tauw)$ and $\dot\phi_i := \dot\kappa'_{0,i} x_{0,i}(\bhat-\beta)$. Our interest is in the order of the two components of $E[\tauhattfgls-\tauw]$.

We also recall the following notation and observations. Let $\phiI := \frac{\max_i\|\kappa_i\|^2}{\KI}$ where $\KI := \sup_{\rho\in\calP}\sum_i\|\kappa_{i,I}(\rho)\|^2$. Under (A2) and (A8): $\phiI=o(1)$, $\KI=O(I^{-1})$,
$\max_i b_i=o(I^{-1/2})$. Under (A8'): $\phiI=O(I^{-1})$, $\max_i b_i=O(I^{-1})$. Let $\ddot\kappa_{0,i}:=\frac{\partial^2\kappa_{[0],i,I}(\rhostar)}{\partial\rho\,\partial\rho'}$, and write $\dot\kappa_{0,i}(\rho)$ and $\ddot\kappa_{0,i}(\rho)$ when the derivatives are evaluated at a generic $\rho\in\calP$ instead of $\rhostar$. 

Throughout, $C$ denotes generic positive constants. Unless confusing, for any generic $z$, we write $z_{[j],i}$ as $z_{j,i}$ for $j=0,1$ and $i=1,\ldots,I$. Finally, $b_i:=\|w_{1,i}\|\vee I^{-1}$.

\noindent\textbf{Assumption (A12)} (entity-level regularity){\bf :} $\kappa_{[0],i,I}(\rho)$ is twice continuously differentiable on $\calP$, and for $f_i(\rho)\in\{\kappa_{[0],i,I}(\rho),\dot\kappa_{0,i}(\rho),\ddot\kappa_{0,i}(\rho)\}$ there is a finite $C$ such that $\|f_i(\rho)\|\le Cb_i$ and $\|f_i(\rho)-f_i(\rho')\|\le Cb_i\|\rho-\rho'\|$ for all $i$, $I$ and $\rho,\rho'\in\calP$.

\noindent\textbf{Remark:} (A12) is an entity-level version of (A2) and (A4) (recall $\kappa_{[1],i,I}=w_{[1],i}$). It holds, for instance, if $T$ is fixed, $a_{i,t}=1$ (entity intercepts) with $N_{[0],i}\ge1$ for all $i$, $\calP$ is convex, and (P2)--(P5) of Appendix~\ref{app:primitive} and (P6') hold. To see this, use \eqref{eq:kappa0}, \eqref{eq:beta_comp} and \eqref{eq:lambda_comp} to write $\kappa_{[0],i,I}(\rho)=-\bSig^{-1}_{[00],i}(\rho_0)\bigl\{\bSig_{[01],i}(\rho_1)w_{[1],i}+a_{[0],i}[\Pi^{-1}_I(\rho)v_I(\rho)]_{\lambda_i}+x_{[0],i}[\Pi^{-1}_I(\rho)v_I(\rho)]_\beta\bigr\}$, where $[\Pi^{-1}_I(\rho)v_I(\rho)]_\beta=S^{-1}_I(\rho)g_I(\rho)$. Then: (i) $a'_{[0],i}\bSig^{-1}_{[00],i}(\rho_0)a_{[0],i}=1'_{N_{[0],i}}\bSig^{-1}_{[00],i}(\rho_0)1_{N_{[0],i}}\ge N_{[0],i}/C\ge 1/C$ by (P2); (ii) by \eqref{eq:schur}, for every vector $\ell$, $\ell'S_I(\rho)\ell=\sum_i\min_{\gamma}(x_{[0],i}\ell-a_{[0],i}\gamma)'\bSig^{-1}_{[00],i}(\rho_0)(x_{[0],i}\ell-a_{[0],i}\gamma)\ge C^{-1}\ell'\bigl(\sum_i\tilde x'_{[0],i}\tilde x_{[0],i}\bigr)\ell$, so $\ell'S_I(\rho)\ell\ge cI\|\ell\|^2$ by (P2)--(P3); (iii) $\|v_{I,\lambda_i}(\rho)\|\le C\|w_{[1],i}\|$ and $\|g_I(\rho)\|=O(1)$ by \eqref{eq:vlambda_bound}, \eqref{eq:vbeta_bound}, (i), (P2) and (P4). Hence $[\Pi^{-1}_I(\rho)v_I(\rho)]_\beta=O(I^{-1})$ and $[\Pi^{-1}_I(\rho)v_I(\rho)]_{\lambda_i}=O(\|w_{[1],i}\|+I^{-1})=O(b_i)$, so $\|\kappa_{[0],i,I}(\rho)\|\le Cb_i$ by (P2) and (P4). By (P6') every factor in these expressions is three times continuously differentiable with derivatives of the same order as the factor; for the two inverses this uses (i)--(ii), e.g., $\partial S^{-1}_I=-S^{-1}_I(\partial S_I)S^{-1}_I$ with $\|\partial S_I\|\le CI$. So the first three derivatives of $\kappa_{[0],i,I}(\rho)$ are also bounded by $Cb_i$, and, $\calP$ being convex, the mean value inequality ($\|f(\rho)-f(\rho')\|\le\sup_{\bar\rho\in\calP}\|\partial f(\bar\rho)/\partial\rho'\|\,\|\rho-\rho'\|$) gives the Lipschitz bounds in (A12).

\begin{lemma}
\label{lemma:wtsum-order}
The following bounds hold for the sums involving the weights:
\begin{itemize}
  \item[(i)] $\sum_i\|\kappa_i\|^2\le K_I = O(I^{-1})$ and
        $\max_i\|\kappa_i\|= O(\Phi^{1/2}_IK^{1/2}_I) = o(I^{-1/2})$.
  \item[(ii)] $\sum_i\|\dot\kappa_{0,i}\|^2= O(K_I) = O(I^{-1})$ and
        $\max_i\|\dot\kappa_{0,i}\|=O(\Phi^{1/2}_IK^{1/2}_I) = o(I^{-1/2})$.
  \item[(iii)] $\sum_i\|w_{1,i}\|^2 = O(K_I) =O(I^{-1})$ and
        $\max_i\|w_{1,i}\|=O(\Phi^{1/2}_IK^{1/2}_I) = o(I^{-1/2})$.
  \item[(iv)] $\max_i|\delta_i|=o(I^{-1/2})$,
        where $\delta_i:=w'_{1,i}(\tau_{1,i}-1_{N_{1,i}}\tauw)$.
  \item[(v)] $\sum_i \|c_i\|^2\|d_i\| = O(\phiI^{1/2}K^{3/2}_I) = o(I^{-3/2})$ and  $\sum_i\|c_i\|^2\|d_i\|^2=O(\phiI K^{2}_I) = o(I^{-2})$ for $c_i,d_i= \kappa_i,\dot\kappa_{0,i},w_{1,i},\delta_i$.
  \item[(vi)] Under (A8'), every $o(\cdot)$ above is replaced by
$O(\cdot)$ with an additional factor of $I^{-1/2}$.  
\end{itemize}
\end{lemma}

\medskip 

\noindent \textbf{Proof of Lemma \ref{lemma:wtsum-order}:} (i) Follows from (A2).

(ii) By (A12), $\|\dot\kappa_{0,i}\|\leq Cb_i$ uniformly in $i$, $I$, and $\rho\in\calP$. Hence, $\sum_i\|\dot\kappa_{0,i}\|^2\leq C^2\sum_i b_i^2=O(I^{-1})$ and $\max_i\|\dot\kappa_{0,i}\|\leq C\max_i b_i=O(\Phi^{1/2}_IK^{1/2}_I)=o(I^{-1/2})$, where $\max_i b_i\le C\max_i\|\kappa_i\|=C\Phi^{1/2}_IK^{1/2}_I$ because $\kappa_{1,i}=w_{1,i}$ and, by (A3), $\max_i\|\kappa_i\|^2\ge I^{-1}\sum_i\|\kappa_i\|^2\ge cI^{-2}$.

(iii) This is assumption (A8).

(iv) $|\delta_i|\leq\|\tau_{1,i}-\tauw\|\cdot\|w_{1,i}\|
\leq C\max_i\|w_{1,i}\|=o(I^{-1/2})$, using (A7) and~(iii).

(v) $\sum_i\|\dot\kappa_{0,i}\|^2\|\kappa_i\|
\leq\max_i\|\kappa_i\|\cdot\sum_i\|\dot\kappa_{0,i}\|^2
=O(\sqrt{\phiI}\sqrt{\KI}K_I)= o(I^{-3/2})$, $\sum_i\|\dot\kappa_{0,i}\|^2\|\kappa_i\|^2
\leq\max_i\|\kappa_i\|^2\cdot\sum_i\|\dot\kappa_{0,i}\|^2=O(\phiI\KI K_I)= o(I^{-2})$,
using $\KI=O(I^{-1})$ from (A2), and the above results. The other choices for $c_i$ and $d_i$ give the same order analogously.

(vi) The result follows by noting from (A12) that (A8') gives $\max_i\|\kappa_i\| \le C(\max_i\|w_{1,i}\| + I^{-1}) = O(I^{-1})$ instead of $o(I^{-1/2})$.\qed

\medskip

\begin{lemma}
\label{lemma:esterror-order}
The following bounds hold for the estimation errors:
\begin{itemize}
  \item[(i)] $\bhat-\beta=\LambdaI^{-1}(1/I)\sum_i\tilde{x}'_{0,i}u_{0,i}$,
    where $\LambdaI:=(1/I)\sum_i\tilde{x}'_{0,i}\tilde{x}_{0,i}$
    is uniformly invertible by \textup{(A6)}, and
    $E[\|\bhat-\beta\|^2]=O(I^{-1})$ and $E[\|\bhat-\beta\|^4]=O(I^{-2})$.
    \item[(ii)] $\tauhat(0)-\tauw = \sum_i \kappa'_i(0)u_i$ satisfies $E[\tauhat(0)-\tauw]^2 = O(I^{-1})$ and $E[\tauhat(0)-\tauw]^4 = O(I^{-2})$.
  \item[(iii)] $\DI:=\|\bhat-\beta\|+|\tauhat(0)-\tauw|$ satisfies
    $E[\DI^2]=O(I^{-1})$ and $E[\DI^4]=O(I^{-2})$.
\end{itemize}
\end{lemma}

\medskip

\noindent \textbf{Proof of Lemma \ref{lemma:esterror-order}:} (i) $\bhat-\beta=(\sum_i\tilde{x}'_{0,i}\tilde{x}_{0,i})^{-1}
\sum_i\tilde{x}'_{0,i}u_{0,i} = \LambdaI^{-1}(1/I)\sum_i\tilde{x}'_{0,i}u_{0,i}$ by definition. $\|\LambdaI^{-1}\|=O(1)$ holds by (A6). (A1) implies $\tilde{x}'_{0,i}u_{0,i}$ are independent with mean
zero for $i$. (A6) implies$\|\tilde{x}_{0,i}\|\leq C$ uniformly. Hence using $\sup_iE\|u_i\|^2<\infty$ from (A1):
\[
  E\!\left[\|\bhat-\beta\|^2\right]
  \leq C\cdot\frac{1}{I^2}\sum_i\|\tilde{x}_{0,i}\|^2E\|u_{0,i}\|^2
  = O(I^{-1}).
\]
Similarly, now additionally using (A5) that $\sup_iE\|u_i\|^4<\infty$ it follows: 
\[
  E\!\left[\|\bhat-\beta\|^4\right]
  \leq C\cdot\frac{1}{I^4}\left[\sum_i\|\tilde{x}_{0,i}\|^4E\|u_{0,i}\|^4 + \left(\sum_i\|\tilde{x}_{0,i}\|^2E\|u_{0,i}\|^2\right)^2\right]
  = O(I^{-2}).
\]

(ii) Since $\tauhat(0)-\tauw = \sum_i \kappa'_i(0)u_i$, the results follow in the same way as part (i), but now using Lemma \ref{lemma:wtsum-order} (i) and (v) for the second and fourth moments respectively. (Replacing $\rho^*_I$ by 0 for the arguments in the terms in Lemma \ref{lemma:wtsum-order} does not invalidate the order of the terms.)

(iii) $E[\DI^2]=O(I^{-1})$ and $E[\DI^4]=O(I^{-2})$ follow from (i)--(ii) and, respectively, $(a+b)^2\leq 2(a^2+b^2)$ and $(a+b)^4\leq 8(a^4+b^4)$. \qed

\medskip

\noindent \textbf{Remark:} Lemma \ref{lem:bias-order} identifies the leading contribution to the higher-order bias, and Lemma \ref{lemma:rem3bound} shows that the remaining terms are asymptotically negligible.

\medskip

\begin{lemma}
\label{lem:bias-order}
(i) Under the maintained assumptions, the eight terms in the expansion of the first term of~\eqref{eq:bias-expand} (display at the start of this appendix) satisfy the following orders under
\textup{(A8)}: $\mathcal{BA}_1= O(\Phi^{1/2}_IK^{1/2}_I) = o(I^{-1/2})$, $\mathcal{BA}_2= O(\Phi^{1/2}_IK^{1/2}_I) = o(I^{-1/2})$, $\mathcal{BA}_3=O(\Phi^{1/2}_IK^{3/2}_I) = o(I^{-3/2})$, $\mathcal{BA}_4=O(\Phi_I K_I) = o(I^{-1})$, $\mathcal{BB}_1=O(\Phi^{1/2}_IK^{3/2}_I)= o(I^{-3/2})$, $\mathcal{BB}_2 = O(I^{-1})$, $\mathcal{BB}_3=O(I^{-2})$ and $\mathcal{BB}_4= O(I^{-2})$. 

(ii) If, additionally (A8') holds then $\Phi_I = O(I^{-1})$ and, consequently, $\mathcal{BA}_1,\mathcal{BA}_2=O(I^{-1})$, i.e., the leading bias terms (already $\mathcal{BB}_2 = O(I^{-1})$ regardless of (A8')), now have the standard order found in bias expansions, while $\mathcal{BA}_3,\mathcal{BA}_4,\mathcal{BB}_1,\mathcal{BB}_3,\mathcal{BB}_4 = O(I^{-2})$.

(iii) If there is no heterogeneity in the sense that $\delta_i = 0$ for all $i$ then $\mathcal{BA}_2 = \mathcal{BB}_2 = 0$. If $u_i \sim N(0,\Sigma_i)$ for all $i$ then $\mathcal{BA}_1 = \mathcal{BA}_3 =\mathcal{BA}_4 =\mathcal{BB}_1 = \mathcal{BB}_3 = \mathcal{BB}_4 =0$.

\end{lemma}

\medskip

\noindent \textbf{Proof of Lemma \ref{lem:bias-order}:} (i) The eight terms that we analyze here are:
\begin{eqnarray*}
\mathcal{BA}_1 &:=& -I \cdot E\left[\sum_i u'_i\kappa_i u'_{0,i}\dot\kappa_{0,i}H^{-1}_I\sum_j \dot\kappa'_{0,j}u_{0,j}\right],\\
\mathcal{BA}_2 &:=& -I \cdot E\left[\sum_i \delta_i u'_{0,i}\dot\kappa_{0,i}H^{-1}_I\sum_j \dot\kappa'_{0,j}u_{0,j}\right],\\
\mathcal{BA}_3 &:=& I \cdot E\left[(\widehat{\beta} - \beta)'\sum_i x'_{i}\kappa_{i}u'_{0,i}\dot\kappa_{0,i}H^{-1}_I\sum_j \dot\kappa'_{0,j}u_{0,j}\right],\\
\mathcal{BA}_4 &:=& I \cdot E\left[\sum_i w'_{1,i}1_{N_{1,i}}(\widehat{\tau}_I(0) - \tau_w)u'_{0,i}\dot\kappa_{0,i}H^{-1}_I\sum_j \dot\kappa'_{0,j}u_{0,j}\right],\\
\mathcal{BB}_1 &:=& I \cdot E\left[\sum_i u'_i\kappa_i (\widehat{\beta} - \beta)'x'_{0,i}\dot\kappa_{0,i}H^{-1}_I\sum_j \dot\kappa'_{0,j}u_{0,j}\right],\\
\mathcal{BB}_2 &:=& I \cdot E\left[\sum_i \delta_i (\widehat{\beta} - \beta)'x'_{0,i}\dot\kappa_{0,i}H^{-1}_I\sum_j \dot\kappa'_{0,j}u_{0,j}\right],\\
\mathcal{BB}_3 &:=& -I \cdot E\left[(\widehat{\beta} - \beta)'\sum_i x'_{i}\kappa_{i}(\widehat{\beta} - \beta)'x'_{0,i}\dot\kappa_{0,i}H^{-1}_I\sum_j \dot\kappa'_{0,j}u_{0,j}\right],\\
\mathcal{BB}_4 &:=& -I \cdot E\left[\sum_i w'_{1,i}1_{N_{1,i}}(\widehat{\tau}_I(0) - \tau_w)(\widehat{\beta} - \beta)'x'_{0,i}\dot\kappa_{0,i}H^{-1}_I\sum_j \dot\kappa'_{0,j}u_{0,j}\right]
\end{eqnarray*}
Using (A1), (A6), Lemma \ref{lemma:wtsum-order} (v):
\begin{eqnarray*}
\mathcal{BA}_1  &=& -I\sum_i E[u'_i\kappa_iu'_{0,i}\dot\kappa_{0,i}H^{-1}_I\dot\kappa'_{0,i}u_{0,i}] \le I \sup_iE\|u_i\|^3 \|H^{-1}_I\|\sum_{i}\|\kappa_i\| \|\dot\kappa_{0,i}\|^2 \\
&=& I O(1) O(\Phi^{1/2}_IK^{3/2}_I) = O(\Phi^{1/2}_IK^{1/2}_I) = o(I^{-1/2}).
\end{eqnarray*}
The other terms follow by the same steps, using (A1), (A6) and Lemma \ref{lemma:wtsum-order} (v), so we abbreviate them: the orders are determined by the factors involving $\kappa$, $\dot{\kappa}$, $w_{1,i}$ and $\delta_i$ (the others are $O(1)$ by assumption), and the orders in Lemma \ref{lemma:wtsum-order} are unchanged if $\rho^*_I$ is replaced by 0.
\begin{eqnarray*}
\mathcal{BA}_2  &=& -I\sum_i \delta_i E[u'_{0,i}\dot\kappa_{0,i}H^{-1}_I\dot\kappa'_{0,i}u_{0,i}] \le I \sup_iE\|u_i\|^2 \|H^{-1}_I\|\sum_{i}|\delta_i| \|\dot\kappa_{0,i}\|^2\\
&=& I  O(1) O(\Phi^{1/2}_IK^{3/2}_I) = O(\Phi^{1/2}_IK^{1/2}_I) = o(I^{-1/2}).
\end{eqnarray*} 
Additionally using Lemma \ref{lemma:esterror-order} (i) and (ii) respectively:
\begin{eqnarray*}
\mathcal{BA}_3  &=& I\sum_i E[u'_{0,i}\widetilde{x}_{0,i}\Lambda^{-1}_II^{-1}x'_{i}\kappa_{i}u'_{0,i}\dot\kappa_{0,i}H^{-1}_I\dot\kappa'_{0,i}u_{0,i}] \\
&\le& O(1)\sup_i E\|u_i\|^3 \sup_i \|\widetilde{x}_{0,i}\| \sup_i \|x_{i}\| \sum_{i}\|\kappa_{i}\|\|\dot\kappa_{0,i}\|^2 = O(1)O(\Phi^{1/2}_IK^{3/2}_I) = o(I^{-3/2}),\\
\mathcal{BA}_{4} &:=& I \sum_iw'_{1,i}1_{N_{1,i}} E\left[\kappa'_{i}(0)u_{i}u'_{0,i}\dot\kappa_{0,i}H^{-1}_I \dot\kappa'_{0,i}u_{0,i}\right] \\
&\le &
I  T^{1/2} \sup_iE\|u_i\|^3 \|H^{-1}_I\| \sum_i \|w_{1,i}\|\|\kappa_{i}(0)\|\|\dot\kappa_{0,i}\|^2 = I O(1) O(\Phi_I K^{2}_I) = O(\Phi_I K_I) = o(I^{-1}). 
\end{eqnarray*}
Now consider the $\mathcal{BB}$ terms.
\begin{eqnarray*}
\mathcal{BB}_1 &:=& I\sum_i E[u'_i\kappa_i u'_{0,i}\widetilde{x}_{0,i}I^{-1}\Lambda^{-1}_Ix'_{0,i}\dot \kappa_{0,i}H^{-1}_I\dot\kappa'_{0,i}u_{0,i}] \\
&\le& O(1) \sup_i E\|u_i\|^3 \sup_i \|\widetilde{x}_{0,i}\|\sum_i \|\kappa_i\|\|\dot\kappa_{0,i}\|^2 = O(1) O(\Phi^{1/2}_IK^{3/2}_I)= o(I^{-3/2}),\\
\mathcal{BB}_2 &:=& I\sum_i \sum_j \delta_i E[u'_{0,j}\widetilde{x}_{0,j}I^{-1}\Lambda^{-1}_I x'_{0,i}\dot \kappa_{0,i}H^{-1}_I\dot\kappa'_{0,j}u_{0,j}],\\
&\le& O(1)\cdot \sup_i E\|u_i\|^2 \sup_i \|\widetilde{x}_{0,i}\| \sup_i \|x_{0,i}\|\sum_i \|\delta_i\|\|\dot\kappa_{0,i}\|\sum_i \|\dot\kappa_{0,i}\|= O(1) O(1)O(I^{-1})= O(I^{-1})\\ 
\mathcal{BB}_3 &:=& -I\sum_i \sum_j E[u'_{0,j}\widetilde{x}_{0,j}I^{-1}\Lambda^{-1}_I x'_{i}\kappa_{i} \,u'_{0,j}\widetilde{x}_{0,j}I^{-1}\Lambda^{-1}_I x'_{0,i}\dot \kappa_{0,i}  H^{-1}_I\dot\kappa'_{0,j}u_{0,j}]\\
&\le& O(1)I^{-1} \sup_i E\|u_i\|^3 \sup_i \|\widetilde{x}_{0,i}\|^2\sup_i \|x_{i}\|^2\sum_i \|\kappa_{i}\|\|\dot\kappa_{0,i}\|\sum_j\|\dot\kappa_{0,j}\| =  O(I^{-1})\,O(I^{-1}) = O(I^{-2}),\\
\mathcal{BB}_{4} &:=& -I \sum_i\sum_j E\left[w'_{1,i}1_{N_{1,i}}\kappa'_j(0)u_{j}u'_{0,j}\widetilde{x}_{0,j}\Lambda^{-1}_II^{-1}x'_{0,i}\dot\kappa_{0,i}H^{-1}_I \dot\kappa'_{0,j}u_{0,j}\right]\\
&&\hspace{-2.7cm}\le O(1)  T^{1/2} \sup_iE\|u_i\|^3 \|H^{-1}_I\|\sup_i \|\widetilde{x}_{0,i}\| \sup_i \|x_{0,i}\|\sum_i \|w_{1,i}\|\|\dot\kappa_{0,i}\|\sum_j \|\kappa_{j}(0)\|\|\dot\kappa_{0,j}\|=O(I^{-1})\,O(I^{-1}) = O(I^{-2}).
\end{eqnarray*}

(ii) The conclusion for the other terms under (A8') is obvious. 

(iii) The proof is obvious and hence omitted.\footnote{The result for normality follows by noting that if, e.g., $v_j \sim N(0, \sigma_{jj})$ for $j=1,2,3$ with $Cov(v_j,v_k) = \sigma_{jk}$ for $j\ne k$ then: (i) $E[v^3_1] =0$, (ii) $E[v^2_1v_2] = E[v^2_1(v_2 - \frac{\sigma_{12}}{\sigma_{11}}v_1)] + E[v^3_1]\frac{\sigma_{12}}{\sigma_{11}} = 0 + 0 = 0$ using that zero covariance implies independence under normality for the first 0, and (i) for the second 0; and (iii) $E[v_1v_2v_3]=0$ because:
  \[E[v_1v_2v_3] = E\left[v_1v_2\left(v_3 - \left[\sigma_{32}, \sigma_{31}\right]\left[\begin{array}{cc} \sigma_{22} 
  & \sigma_{21}\\ \sigma_{21} & \sigma_{11}\end{array}\right]^{-1}\left[\begin{array}{c} v_2\\v_1\end{array}\right]  \right)\right] +  \left[\sigma_{32}, \sigma_{31}\right]\left[\begin{array}{cc} \sigma_{22} 
  & \sigma_{21}\\ \sigma_{21} & \sigma_{11}\end{array}\right]^{-1}E\left[\left[\begin{array}{c} v^2_2v_1\\v^2_1v_2\end{array}\right]\right] = 0 + 0 = 0
    \]
using that zero covariance implies independence under normality for the first 0, and (ii) for the second 0.} \qed

\begin{lemma}
\label{lemma:rem3bound}
The remainder term $\rem_3 := \rem_1'\bigl(\sum_j\dot\kappa'_{0,j}u_{0,j}\bigr) + \rem_2$
from the proof of Proposition \ref{prop:ho-bias} satisfies $E[\rem_3] = O(\Phi_I\,I^{-1/2})$ under the conditions of Proposition \ref{prop:ho-bias}, where $\Phi_I := \max_i\|\kappa_i\|^2/K_I$. In particular, under (A2) we have $\Phi_I=o(1)$ so $E[\rem_3] = o(\Phi_I^{1/2}I^{-1/2})$, and under additionally (A8') we have $\Phi_I=O(I^{-1})$ so $E[\rem_3] = O(I^{-3/2})$.
\end{lemma}

\medskip

The proof of Lemma \ref{lemma:rem3bound} uses three auxiliary lemmas, given first. We write $\|X\|_p:=(E\|X\|^p)^{1/p}$ for the $L^p$ norm and $\|\cdot\|$ for the Euclidean norm. Recall $b_i$ and the orders of $b_i$ and $\Phi_I$ from above; also $\sum_i b_i^2 = O(K_I)=O(I^{-1})$, $\|\kappa_i\|, \|\dot\kappa_{0,i}\|, \|\ddot\kappa_{0,i}\| \le Cb_i$ uniformly in $i,I,\rho\in\mathcal{P}$ by (A12), and $\Phi_I\ge cI^{-1}$ by (A2)--(A3). Also, let $q := (d_\rho\vee 4)+\varepsilon/2$, so that $2q = 2(d_\rho\vee 4)+\varepsilon$ is the same moment order in (A5'). $q>d_\rho$ (so packing integrals over $\mathcal{P}\subset\mathbb R^{d_\rho}$ converge) and $q\ge 4$ (so Markov at $L^q$ gives $O(\Phi_I^2)$). The moment requirements used below are: a summand \emph{linear} in $u_i$ at $L^p$ requires $E\|u_i\|^p<\infty$; a \emph{product of two terms linear} in $u_i$ at $L^p$ requires $E\|u_i\|^{2p}<\infty$. The largest demand below is $E\|u_i\|^{2q}$ — from linear summands at $p=2q$ and quadratic summands at $p=q$ — which is exactly (A5'). Throughout, the exponent $p$ is chosen per step: each invocation of Lemma \ref{aux:A} or H\"{o}lder below names the specific $p$ it uses, with $p\in\{2,4,8,q,2q\}$.

\newtheorem{auxlemma}{Lemma}
\renewcommand{\theauxlemma}{\Alph{auxlemma}}

\begin{auxlemma}\label{aux:A}
Let $\{Z_i(\rho)\}_{i=1}^I$, indexed by $\rho$ in a bounded set $\mathcal{P}\subset\R^{d_\rho}$, be independent with $\sum_i E[Z_i(\rho)]=0$ for every
$\rho\in \mathcal P$. Fix $p\ge2$ and write $S(\rho):=\sum_i Z_i(\rho)$.
\begin{itemize}
\item[(i)] \emph{Pointwise:} if $\|Z_i(\rho)\|_p\le Cb_i$ for all $i$ at a given $\rho$, then $\|S(\rho)\|_p=O(K_I^{1/2})=O(I^{-1/2})$. \emph{Uniform:} if also this holds for all $\rho$, $\|Z_i(\rho)-Z_i(\rho')\|_p\le Cb_i\|\rho-\rho'\|$, $p>d_\rho$, and each $\rho\mapsto Z_i(\rho)$ is continuous, then $\big\|\sup_{\rho}\|S(\rho)\|\big\|_p=O(K_I^{1/2})=O(I^{-1/2})$.
\item[(ii)] If instead these hold with $b_i$ replaced by $b_i^2$, then the same pointwise and uniform conclusions hold for $I\,S(\rho)$ with $O(\Phi_I^{1/2})$ in place of $O(K_I^{1/2})$; this is $O(I^{-1/2})$ under \emph{(A8')}.
\end{itemize}
\end{auxlemma}

\noindent\textbf{Proof of Lemma \ref{aux:A}:} Let $\bar Z_i(\rho):=Z_i(\rho)-E[Z_i(\rho)]$. These are independent and
mean-zero, with $\|\bar Z_i(\rho)\|_p\le 2\|Z_i(\rho)\|_p$ and $\|\bar Z_i(\rho)-\bar Z_i(\rho')\|_p\le 2\|Z_i(\rho)-Z_i(\rho')\|_p$ (as
$\|X-E[X]\|_p\le 2\|X\|_p$). Since $\sum_i E[Z_i(\rho)]=0$, we have $S(\rho)=\sum_i\bar Z_i(\rho)$, so Rosenthal's inequality for independent mean-zero summands gives, at a fixed $\rho$,
\[
\|S(\rho)\|_p\le C\Big[\big(\textstyle\sum_i a_i^{2}\big)^{1/2}
+\big(\textstyle\sum_i a_i^{p}\big)^{1/p}\Big],
\qquad a_i\in\{b_i,b_i^2\}.
\]
\emph{(i) $a_i=b_i$.} (A8) gives: the first term is $O(K_I^{1/2})$ and the second is $(\sum_i b_i^p)^{1/p}\le(\max_i b_i)^{(p-2)/p}(\sum_i b_i^2)^{1/p}= O(K_I^{1/2})$. \emph{(ii) $a_i=b_i^2$.} Follows similarly as $(i)$. These pointwise bounds need only $p\ge2$. For the suprema, the Lipschitz condition gives the same orders for $\|S(\rho)-S(\rho')\|_p$ with an extra factor $\|\rho-\rho'\|$. Since $\mathcal{P}$ is a bounded subset of $\mathbb{R}^{d_\rho}$, the packing numbers in Euclidean distance satisfy $\mathcal{D}(\epsilon,\mathcal{P},\|\cdot\|)\le(C_\mathcal{P}/\epsilon)^{d_\rho}$ for a generic constant $C_\mathcal{P}$, so $\psi_p^{-1}(\mathcal{D}(\epsilon,\mathcal{P},\|\cdot\|))\le (C_\mathcal{P}/\epsilon)^{d_\rho/p}$ for $\psi_p(x):=x^p$, which is integrable on $(0,\mathrm{diam}(\mathcal{P})]$ since $d_\rho/p<1$  ($\mathrm{diam}(\mathcal{P}) := \sup_{\rho,\rho' \in \mathcal{P}}\|\rho - \rho'\|$). Since each $Z_i(\cdot)$ is continuous, so is $S(\cdot)$, and hence $S$ is separable: suprema over $\mathcal{P}$ equal suprema over a countable dense subset, as required by Theorem 2.2.4 of \citet{VW96}. The increment bound above reads $\|S(\rho)-S(\rho')\|_p\le d(\rho,\rho')$ for the metric $d(\rho,\rho'):=L_I\|\rho-\rho'\|$ with $L_I=O(K_I^{1/2})$, so the constant in that theorem depends only on $\psi_p$ (i.e., on $p$), not on $I$. Under $d$, the diameter of $\mathcal{P}$ is $L_I\,\mathrm{diam}(\mathcal{P})$ and $\mathcal{D}(\epsilon,\mathcal{P},d)=\mathcal{D}(\epsilon/L_I,\mathcal{P},\|\cdot\|)$; changing variables $\epsilon\mapsto\epsilon/L_I$ in the entropy integral pulls out the factor $L_I$. Then, Corollary 2.2.5 of \citet{VW96}, applied componentwise to $S(\rho)$ with the Orlicz function $\psi_p$ and metric $d$, gives (with $C$ depending only on $p$ and $d_\rho$):
\[
  \Bigl\|\sup_{\rho,\rho'\in\mathcal{P}}\|S(\rho)-S(\rho')\|\Bigr\|_p \le C\cdot O(K_I^{1/2})\cdot\int_0^{\mathrm{diam}(\mathcal{P})}(C_\mathcal{P}/\epsilon)^{d_\rho/p}\,d\epsilon = O(K_I^{1/2}).
\]
Hence, for any fixed $\rho_0\in\mathcal P$, $\big\|\sup_{\rho}\|S(\rho)\|\big\|_p\le\|S(\rho_0)\|_p+\big\|\sup_{\rho}\|S(\rho)-S(\rho_0)\|\big\|_p=O(K_I^{1/2})$ by the pointwise bound; (ii) is analogous with $IS(\rho)$. \qed

\medskip

\begin{auxlemma}[extends Lemma~\ref{lemma:esterror-order}]\label{aux:B}
For every $p\ge2$ with $\sup_i E\|u_i\|^p<\infty$: $\|\bhat-\beta\|_p=O(I^{-1/2})$, $\|\that-\tau_w\|_p=O(I^{-1/2})$, and $\|D_I\|_p=O(I^{-1/2})$.
\end{auxlemma}

\noindent\textbf{Proof of Lemma \ref{aux:B}:} By Lemma \ref{lemma:esterror-order} (i), $\bhat-\beta=\Lambda_I^{-1}I^{-1}\sum_i\xt'u_{0,i}$ with
$\|\Lambda_I^{-1}\|=O(1)$ by (A6); the summands are independent, mean-zero,
$\|\xt'u_{0,i}\|_p\le C$ by (A6), so Rosenthal gives $\|\sum_i\xt'u_{0,i}\|_p=O(I^{1/2})$ and hence $\|\bhat-\beta\|_p=O(I^{-1/2})$.
For $\that-\tau_w=\sum_i\kappa_i'(0)u_i$, the summands are linear in $u_i$'s and $\|\kappa_i(0)\|\le Cb_i$, so the pointwise bound of Lemma \ref{aux:A}(i) at $\rho=0$ gives $O(K_I^{1/2})=O(I^{-1/2})$. The bound on $\|D_I\|_p$ follows by the triangle inequality. \qed

\medskip

\begin{auxlemma}\label{aux:C}
(i) For all $i$ and $\rho\in \mathcal P$, $\kappa_{i,I}'(\rho)\,a_i=0$; equivalently
$\kappa_{[0],i,I}'(\rho)\,a_{[0],i}=-\,w_{[1],i}'\,a_{[1],i}$, which does not
depend on $\rho$. Hence $\kdot'a_{[0],i}=0$ and $\kddot\!\cdot a_{[0],i}=0$.

(ii) For $f_i\in\{\kdot,\kddot\}$,
$\;f_i'\widehat u_{0,i}=f_i'u_{0,i}-f_i'\xt(\bhat-\beta)$; and
$\kappa_i'\widehat u_i=\kappa_i'u_i+c_i$ with $c_i:=\delta_i-\phi_i-\psi_i$.

(iii) For $f_i$ and $c_i$ from (ii), $\|f_i'\xt(\bhat-\beta)\|\le Cb_i\|\bhat-\beta\|$ and
$|c_i|\le Cb_i(1+D_I)$.
\end{auxlemma}

\noindent\textbf{Proof of Lemma \ref{aux:C}:} $(i)$ Follows from Footnote \ref{fn:fe_inv}. 

$(ii)$ With $\widehat u_{0,i}=u_{0,i}-a_{[0],i}(\widehat\lambda_i-\lambda_i)
-x_{0,i}(\bhat-\beta)$, the middle term drops by $(i)$, and
$f_i'x_{0,i}=f_i'\xt$ since $x_{0,i}-\xt\in\mathrm{col}(a_{[0],i})$ and
$f_i'a_{[0],i}=0$. The identity for $\kappa_i'\widehat u_i$ is \eqref{eq:decomp}.

$(iii)$ $\|f_i'\xt\|\le\|f_i\|\,\|\xt\|\le Cb_i$ by the order of $f_i$ and
(A6). $|c_i|\le|\delta_i|+|\phi_i|+|\psi_i|\le
Cb_i+Cb_i\|\bhat-\beta\|+Cb_i|\that-\tau_w|=Cb_i(1+D_I)$ by (A6)--(A7) and
$\|\kappa_i\|\le Cb_i$. \qed

\medskip

\noindent\textbf{Proof of Lemma \ref{lemma:rem3bound}:} Each sum below is handled by Lemma \ref{aux:A} after classifying its summand as either linear in $u_i$ with $\|\cdot\|_p\le Cb_i$ (Lemma \ref{aux:A}(i)), or a product of two terms linear in $u_i$ with $\|\cdot\|_p\le Cb_i^2$ (by Lemma \ref{aux:A}(ii)), the exponent $p$ is named at each step. Part A, except (A.iv), evaluates sums at $\rstar$ and needs only the pointwise bounds (any $p\ge2$); all uniform bounds use $p=q$ or, for sums linear in $u_i$ (via H\"{o}lder in (III)), $p=2q$; both exceed $d_\rho$ and need only $E\|u_i\|^{2q}<\infty$. We omit details that are similar to those in the other proofs. Estimation-error corrections are attached by H\"{o}lder using $\|\bhat-\beta\|_p,\|\that-\tau_w\|_p,\|D_I\|_p = O(I^{-1/2})$ from Lemma \ref{aux:B}.

\medskip

\noindent\underline{Part A: $L^p$ bounds for the basic quantities}

\smallskip
\emph{(A.i) $\|\Ghat\|_4=O(\Phi_I^{1/2})$.} Write $\Ghat=\Gtil+R_I$ with $\Gtil:=I\sum_i(\kdot'u_{0,i})(\kappa_i'u_i)$. $E[\Gtil]=0$ is the first order condition for \eqref{eq:optrho}; the summand for each $i$, while not mean-zero, has $\|\cdot\|_4 \le Cb_i^2$, so Lemma~\ref{aux:A}(ii) at $p=4$ gives $\|\Gtil\|_4=O(\Phi_I^{1/2})$. By Lemma~\ref{aux:C}(ii), $\kdot'\widehat u_{0,i}=\kdot'u_{0,i}+d_i$ with $d_i=-\kdot'\xt(\bhat-\beta)$, $\|d_i\|\le Cb_i\|\bhat-\beta\|$, and $\kappa_i'\widehat u_i=\kappa_i'u_i+c_i$. On the other hand,
\[
R_I=I\sum_i(\kdot'u_{0,i})c_i+I\sum_i d_i(\kappa_i'u_i)+I\sum_i d_i c_i
=:R_{I,1}+R_{I,2}+R_{I,3}.
\]

\emph{$R_{I,1}$} (split $R_{I,1}$ into $R^{(\delta)}_{I,1}-R^{(\phi)}_{I,1}-R^{(\psi)}_{I,1}$ along $c_i=\delta_i-\phi_i-\psi_i$). $R^{(\delta)}_{I,1}:= I\sum_i \delta_i (\dot \kappa'_{0i}u_{0i})$ has summands satisfying $\|\cdot\|_4 \le Cb_i^2$, and hence is $O(\Phi_I^{1/2})$ by Lemma~\ref{aux:A}(ii) at $p=4$. $R^{(\phi)}_{I,1} := I\sum_i (\dot \kappa'_{0i}u_{0i})\kappa'_ix_i(\widehat{\beta} - \beta) \equiv A_I(\bhat-\beta)$ where $A_I:=I\sum_i(\kdot'u_{0,i})\kappa_i'x_i$ has summands satisfying $\|\cdot\|_8 \le Cb_i^2$. So
$\|A_I\|_8=O(\Phi_I^{1/2})$ (Lemma~\ref{aux:A}(ii), $p=8$), and H\"{o}lder with Lemma~\ref{aux:B} gives $R^{(\phi)}_{I,1} = O(\Phi_I^{1/2}I^{-1/2})$. $R^{(\psi)}_{I,1}:= I\sum_i (\dot \kappa'_{0i}u_{0i})w'_{1i}\one_{N_{[1],i}}(\widehat{\tau}_I(0)-\tau_w)=O(\Phi_I^{1/2}I^{-1/2})$ gets the same treatment as the previous term (with summands $\|\cdot\|_8\le Cb_i^2$, so $p=8$), with $\kappa_i'x_i$ replaced by $w_{1,i}'\one_{N_{[1],i}}$ ($\le\sqrt T\,b_i$) and
$\bhat-\beta$ by $\that-\tau_w$. 

\emph{$R_{I,2}=-F_I(\bhat-\beta)$} where $F_I:=I\sum_i(\kdot'\xt)(\kappa_i'u_i)$ with $\|(\kdot'\xt)(\kappa_i'u_i)\|_8 \le Cb_i^2$, so
$\|F_I\|_8=O(\Phi_I^{1/2})$ (Lemma~\ref{aux:A}(ii), $p=8$) and H\"{o}lder with Lemma~\ref{aux:B} gives $R_{I,2} = O(\Phi_I^{1/2}I^{-1/2})$.

\emph{$R_{I,3}$:} by the two results in Lemma~\ref{aux:C}(iii), $|R_{I,3}|\le
CI\|\bhat-\beta\|(1+D_I)\sum_i b_i^2\le C\|\bhat-\beta\|(1+D_I)$, so by H\"{o}lder
and Lemma~\ref{aux:B}, $\|R_{I,3}\|_4=O(I^{-1/2})=O(\Phi_I^{1/2})$ (as $\Phi_I^{1/2}\ge cI^{-1/2}$).

Using $\|\Gtil\|_4=O(\Phi_I^{1/2})$ with $ R_{I,j}=O(\Phi_I^{1/2})$ for $j=1,2,3$, we obtain $\|\Ghat\|_4=O(\Phi_I^{1/2})$.

\smallskip
\emph{(A.ii) $\|\Hhat(\rstar)-H_I\|_4=O(\Phi_I^{1/2})$.} The sample Hessian is, like $\Ghat$, a sum of terms bilinear in $\widehat u$
in which one or both $\kappa$-factors are replaced by $\rho$-derivatives of $\kappa_{[0],i,I}$ (using Lemma \ref{aux:C}(ii) for $\kddot'\widehat u_{0,i}$). Every (deterministic) coefficient satisfies $\|\cdot\|\le Cb_i$, so every summand satisfies $\|\cdot\|_4\le Cb_i^2$, and the exact argument of (A.i) applies at $p=4$.

\smallskip
\emph{(A.iii) $\|S\|_p=O(I^{-1/2})$ for $p\in\{2,4\}$, where $S:=\sum_i\dot\kappa_{0,i}'u_{0,i}$.} This follows by Lemma \ref{aux:A}(i) since the summand $\dot\kappa_{0,i}'u_{0,i}$ is linear in $u_{0,i}$ with $\|\dot\kappa_{0,i}'u_{0,i}\|_p \le Cb_i$.

\smallskip
\emph{(A.iv) $\big\|\sup_\rho\|W(\rho)\|\big\|_2=O(I^{-1/2})$ for $W(\rho):=\sum_i\ddot\kappa_{0,i}(\rho)u_{0,i}$.} The summand is linear in $u_{0,i}$ with deterministic coefficient $\sup_\rho\|\ddot\kappa_{0,i}(\rho)\| \le C b_i$ as before, and Lipschitz in $\rho$ with constant $Cb_i$ by (A12). Then Lemma \ref{aux:A}(i) at $p=q>d_\rho$ gives $\big\|\sup_\rho\|W\|\big\|_q=O(I^{-1/2})$, and $\|\cdot\|_2\le\|\cdot\|_q$.

\medskip
\noindent\underline{Part B: $L^4$ bound for $\|\rhat-\rstar\|$}

Let $\mu_{\min}(\cdot)$ denote smallest eigenvalue. Note that $\mu_0:=\inf_I\mu_{\min}(H_I)>0$ by (A10). The population Hessian $H_I(\rho):=\frac{\partial^2}{\partial\rho\,\partial\rho'} \tfrac I2 \sigma_I^2(\rho)$ (with $H_I(\rstar)=H_I$) is Lipschitz with constant $L_{H}=O(1)$ (its derivative has terms $\le Cb_i^2$ and $I\sum_i b_i^2=O(1)$). Fix $\zeta>0$ with $L_{H}\zeta\le\tfrac16\mu_0$ and $\zeta<r_0:=\liminf_I \mathrm{dist}(\rstar,\partial\mathcal{P})$, which is positive because every limit point of $\{\rstar\}$ lies in the interior of the compact $\mathcal{P}$ (Proposition~\ref{prop:ho-bias}). Set $B:=\{\|\rho-\rstar\|\le\zeta\}$, so that $B\subset\mathrm{interior}(\mathcal{P})$ for all large $I$, and $A_I:=\{\rhat\in B\}\cap\{\sup_{\rho \in B}\|\Hhat(\rho)-H_I\|\le\tfrac12\mu_{\min}(H_I)\}$.

\emph{(B.i) On $A_I$:} Let $\bar H:=\int_0^1\Hhat(\rstar+t(\rhat-\rstar))\,dt$ be the sample Hessian averaged over the segment joining $\rstar$ and $\rhat$, which lies in the convex set $B$ (a single intermediate point need not exist when $d_\rho>1$). As $\rhat\in B\subset\mathrm{interior}(\mathcal{P})$, the first-order condition holds and gives $\bar H(\rhat-\rstar)=-\Ghat$. On $A_I$ the segment lies in $B$, so by Weyl's inequality and convexity of the norm, $\mu_{\min}(\bar H)\ge\mu_{\min}(H_I)-\sup_{\rho \in B} \|\Hhat(\rho) - H_I\| \ge \frac{1}{2}\mu_{\min}(H_I) > 0$. Hence, $\|\bar H^{-1}\| \le \sqrt{d_\rho}\,\|\bar H^{-1}\|_{\mathrm{op}} = \frac{\sqrt{d_\rho}}{\mu_{\min}(\bar H)} \le \frac{2\sqrt{d_\rho}}{\mu_{\min}(H_I)} = 2\sqrt{d_\rho}\,\|H_I^{-1}\|_{\mathrm{op}} \le 2\sqrt{d_\rho}\,\|H_I^{-1}\|$ (using the relation between the Frobenius norm $\|.\|$ and the operator norm $\|.\|_{\mathrm{op}}$). Since $\bar H(\rhat-\rstar)=-\Ghat$ we have, on $A_I$, $\|\rhat-\rstar\| \le \|\bar H^{-1}\|\,\|\Ghat\|$, i.e., $\|\rhat-\rstar\|\one(A_I)\le 2\sqrt{d_\rho}\,\|H_I^{-1}\|\,\|\Ghat\|$.

\emph{(B.ii) Off $A_I$:} $\|\rhat-\rstar\|\one(A_I^c)\le \mathrm{diam}(\mathcal{P})\,\one(A_I^c)$.

\emph{(B.iii) $P(A_I^c)=O(\Phi_I^2)$.} By the definitions of $A_I$ and $\mu_0$, $A^c_I \subseteq \{\rhat\notin B\}\cup\{\sup_{\rho \in B}\|\Hhat(\rho)-H_I\|\ge\tfrac12\mu_{0}\}$. Now, note that: $\sup_{\rho \in B}\|\Hhat(\rho) -H_I\|\le \sup_{\rho \in B}\|H_I(\rho) - H_I\| + \sup_{\rho \in B}\|\Hhat^{u}(\rho)- H_I(\rho)\| +\sup_{\rho \in B}\|\Hhat(\rho)-\Hhat^{u}(\rho)\| = (I) + (II) + (III)$, say, where we have introduced an intermediate term $\Hhat^{u}(\cdot)$ as the infeasible sample Hessian that uses true $u_i$'s instead of $\widehat{u}_i$'s. Note that $(I):= \sup_{\rho \in B}\|H_I(\rho) - H_I\| \le L_{H}\sup_{\rho \in B}\|\rho - \rho^*_I\|\le L_{H}\zeta \le \mu_0/6$ where the first inequality follows by the Lipschitz property, the second by definition of $B$ and the third by definition of $\zeta$. Hence $\sup_{\rho \in B}\|\Hhat(\rho)-H_I\|\ge\tfrac12\mu_0$ implies $(II)+(III)\ge\tfrac13\mu_0$, so $A^c_I\subseteq\{\rhat\notin B\}\cup\{(II)\ge\tfrac16\mu_0\}\cup\{(III)\ge\tfrac16\mu_0\}$. Now, we bound the probability of each of these three events by $O(\Phi_I^2)$.

\emph{Bound on $P((\mathrm{II})\ge\tfrac16\mu_0)$.} $\widehat H^u(\rho)-H_I(\rho)=I\sum_i(\vartheta_i(\rho)-E[\vartheta_i(\rho)])$ with $\vartheta_i(\rho) = (\dot \kappa_{0i}(\rho)u_{0i})(\dot \kappa_{0i}(\rho)u_{0i})' + (\kappa'_i(\rho)u_i)(\ddot\kappa_{0i}(\rho)u_{0i})$, independent and satisfying the pathwise bound $\|\vartheta_i(\rho)\|\le Cb_i^2\|u_i\|^2$ (and the analogous Lipschitz bound in $\rho$); hence $\|\vartheta_i(\rho)\|_q\le Cb_i^2\big\|\,\|u_i\|^2\big\|_q = Cb_i^2(E\|u_i\|^{2q})^{1/q}\le Cb_i^2$ using (A5'). By Lemma \ref{aux:A}(ii) at $p=q$, $\big\|(\mathrm{II})\big\|_q=O(\Phi_I^{1/2})$, so by Markov, $P\big((\mathrm{II})\ge\tfrac16\mu_0\big)\le(6/\mu_0)^q\,E\big[(\mathrm{II})^q\big]
=O(\Phi_I^{q/2})=O(\Phi_I^2)$ since $q\ge4$.

\emph{Bound on $P((\mathrm{III})\ge\tfrac16\mu_0)$.} Write $\widehat H(\rho)=I\sum_i \hat{\vartheta}_i(\rho)$ with $\hat{\vartheta}_i(\rho)$ same as $\vartheta_i(\rho)$ except that the errors are replaced by the residuals. Substitute the identities of Lemma \ref{aux:C}(ii)
for $\dot\kappa_{0,i}'\widehat u_{0,i}$, $\ddot\kappa_{0,i}\cdot\widehat u_{0,i}$, and $\kappa_i'\widehat u_i$. The terms quadratic in $(u_{0,i},u_i)$ alone reproduce $\vartheta_i(\rho)$ and cancel in $\widehat\vartheta_i(\rho)-\vartheta_i(\rho)$,
leaving terms of three types, each with $\|\cdot\|_q \le Cb_i^2$: (i) a linear in $u$ factor ($\|\cdot\|_{2q} \le Cb_i$) times an estimation-error factor ($\|\cdot\|_{2q} \le Cb_i\|\widehat\beta-\beta\|$ or $Cb_i|\widehat\tau_I(0)-\tau_w|$, or, for the
$\delta_i$-component of $c_i$, the deterministic $\delta_i\le Cb_i$), (ii) terms that are deterministic times squared estimation error (e.g. $\delta_i(\hat\tau_I(0)-\tau_w)^2$), and (iii) terms that are deterministic times a single estimation error (e.g. $-\delta_i\,\ddot\kappa_{0,i}(\rho)\widetilde x_{0,i}(\widehat\beta-\beta)$). Type (i) is $O(\Phi_I^{1/2})$
in $L^q$ by Lemma \ref{aux:A}(ii) (directly when the second factor is $\delta_i$, and via H\"{o}lder at conjugates $(2q,2q)$ with Lemma~\ref{aux:B} otherwise); type (ii) is $O(\Phi_I)=o(\Phi_I^{1/2})$ by Lemma \ref{aux:B}; type (iii) is $O(I^{-1/2})=O(\Phi_I^{1/2})$ uniformly in $\rho$ by Lemma \ref{aux:B}, since $I\sum_i|\delta_i|b_i\le CI\sum_ib_i^2=O(1)$ and $\Phi_I\ge cI^{-1}$. Hence $\big\|(\mathrm{III})\big\|_q=O(\Phi_I^{1/2})$, and so by Markov, $P\big((\mathrm{III})\ge\tfrac16\mu_0\big)=O(\Phi_I^{q/2})=O(\Phi_I^2)$ since $q\ge4$.

\emph{Bound on $P(\widehat{\rho}^*_I\notin B)$.} By (A9), there is $\Delta(\zeta)>0$ with $\liminf_I\inf_{\rho\in \mathcal{P}:\|\rho-\rho^*_I\|>\zeta}I(\sigma_I^2(\rho) -\sigma_I^2(\rho^*_I))\ge\Delta(\zeta)$. Since $\widehat{\rho}^*_I$ minimizes
$I\widehat\sigma_I^2(\rho)$ and $\sigma_{\tau,I}^2$ does not depend on $\rho$, $\{\widehat{\rho}^*_I\notin B\}\subseteq\{2R_I\ge\Delta(\zeta)\}$ with $R_I:=\sup_\rho|I\widehat\sigma_I^2(\rho)-I(\sigma_I^2(\rho)
+\sigma_{\tau,I}^2)|$. The dominant term of $R_I$ is quadratic in $u_i$ with summands satisfying $\|\cdot\|_q \le Cb_i^2$. Since its remaining estimation-error terms can be handled as in ($A.i$), we obtain $\|R_I\|_q=O(\Phi_I^{1/2})$ by Lemma \ref{aux:A}(ii) at $p=q$. Then, by Markov, $P(\widehat{\rho}^*_I\notin B)\le P(2R_I\ge\Delta(\zeta)) =O(\Phi_I^{q/2})=O(\Phi_I^2)$.

Combining the three bounds derived above, $P(A_I^c)=O(\Phi_I^2)$.

\emph{(B.iv)} $E\left[\|\widehat{\rho}^*_I - \rho^*_I\|^4\right] = E\left[\|\widehat{\rho}^*_I - \rho^*_I\|^4\one(A_I) \right] + E\left[\|\widehat{\rho}^*_I - \rho^*_I\|^4\one(A^c_I) \right]$. Note that, ($B.i$) gives:
$E\left[\|\widehat{\rho}^*_I - \rho^*_I\|^4\one(A_I) \right] \le 16 d_\rho^2 \|H_I^{-1}\|^4E[\|\Ghat\|^4]=O(\Phi_I^2)$ by ($A.i$) and (A10). On the other hand, by definition $E\left[\|\widehat{\rho}^*_I - \rho^*_I\|^4\one(A^c_I) \right] \le \mathrm{diam}^4(\mathcal{P})P(A^c_I) = O(\Phi_I^2)$ since $\mathcal{P}$ is bounded and (B.iii) gave $P(A_I^c)=O(\Phi_I^2)$. Combining them: $E\left[\|\widehat{\rho}^*_I - \rho^*_I\|^4\right] =O(\Phi_I^2)$.

\medskip
\noindent\underline{Part C: bounding $E[\mathrm{rem}_3]=E[\mathrm{rem}_2]+E[\mathrm{rem}_1'S]$}

\emph{(C.i) $E[\mathrm{rem}_2]=O(\Phi_I I^{-1/2})$.} $\mathrm{rem}_2 :=\tfrac12(\rhat-\rstar)'\big[\sum_i\ddot\kappa_{0,i}(\bar\rho)\widehat u_{0,i}\big](\rhat-\rstar)$, where $\bar\rho$ is an intermediate point on the segment joining $\rstar$ and $\rhat$ (a single such point exists since $\widehat\tau_I(\cdot)$ is scalar). By Lemma~\ref{aux:C}(ii), $\sum_i\ddot\kappa_{0,i}(\bar\rho)\widehat u_{0,i}= W(\bar\rho) -\sum_i\ddot\kappa_{0,i}(\bar\rho)\xt(\bhat-\beta)$, and $\sum_i\|\ddot\kappa_{0,i}\xt\|\le C\sum_i b_i=O(1)$. So its norm is
$\le\sup_\rho\|W(\rho)\|+C\|\bhat-\beta\|$. By Cauchy-Schwarz ($E[\|\rhat-\rstar\|^2 X]\le\|\rhat-\rstar\|_4^2\,\|X\|_2$) with ($A.iv$), ($B.iv$), Lemma \ref{aux:B}: 
\begin{equation*}
|E[\mathrm{rem}_2]|\le C\|\rhat-\rstar\|_4^2
\big(\big\|\sup_\rho\|W\|\big\|_2+C\|\bhat-\beta\|_2\big)
=O(\Phi_I)\,O(I^{-1/2})=O(\Phi_I I^{-1/2}).
\end{equation*}

\emph{(C.ii) $E[\rem_1'S] = O(\Phi_I I^{-1/2})$.}\; 

\emph{On $A_I$.} $\rem_1 := (\widehat{\rho}^*_I - \rho^*_I) + H_I^{-1}\hat G$ and
$\widehat{\rho}^*_I - \rho^*_I = -\bar H^{-1}\hat G$ on $A_I$. Hence,
$\rem_1\cdot\mathbf 1(A_I) = \bigl[-\bar H^{-1}\hat G + H_I^{-1}\hat G\bigr]\mathbf 1(A_I)
= H_I^{-1}\bigl(\bar H-H_I\bigr)\bar H^{-1}\hat G\cdot\mathbf 1(A_I)$. So
$\|\rem_1\,\mathbf 1(A_I)\| \le C\,\|\bar H-H_I\|\,\|\hat G\|$ on $A_I$, the constant
absorbing $\|H_I^{-1}\|$ and $\|\bar H^{-1}\|\le 2\sqrt{d_\rho}\,\|H_I^{-1}\|$. On $A_I$ the segment joining $\widehat{\rho}^*_I$ and $\rho^*_I$ lies in $B$; writing $\rho_t:=\rstar+t(\rhat-\rstar)$,
splitting $\bar H-H_I=\int_0^1\{\Hhat(\rho_t)-H_I(\rho_t)\}\,dt+\int_0^1\{H_I(\rho_t)-H_I\}\,dt$ and using the Lipschitz property of the \emph{nonrandom} map
$\rho\mapsto H_I(\rho)$ gives, on $A_I$,
\[
  \|\bar H-H_I\| ~\le~ \sup_{\rho\in B}\|\Hhat(\rho)-H_I(\rho)\|
  ~+~ L_H\,\|\widehat{\rho}^*_I - \rho^*_I\|.
\]
Both terms are $O(\Phi_I^{1/2})$ in $L^4$. For the first term, the computation in ($A.ii$) applies in the same way to $\Hhat(\rho)-H_I(\rho)$ at each $\rho$, its summands being bilinear in $\widehat u$ with coefficients bounded by $Cb_i$ uniformly in $\rho$ and uniformly Lipschitz in $\rho$ by (P6'), so Lemma \ref{aux:A}(ii) at $p=q>d_\rho$ delivers
the bound uniformly over $B$ exactly as it does for $W(\rho)$ in ($A.iv$). For the second term, use
($B.iv$). Finally, by H\"{o}lder with conjugate exponents $(4,4,2)$, and using ($A.i$) and ($A.iii$):
\begin{equation*}
  |E[\rem_1'S\,\mathbf 1_{A_I}]| \le C\,\bigl\|\,\|\bar H-H_I\|\,\bigr\|_4\,\|\hat G\|_4\,\|S\|_2
  = O(\Phi_I^{1/2})\cdot O(\Phi_I^{1/2})\cdot O(I^{-1/2}) = O(\Phi_I I^{-1/2}).
\end{equation*}

\emph{Off $A_I$.} From $\rem_1 = (\widehat{\rho}^*_I - \rho^*_I) + H_I^{-1}\hat G$ and boundedness of $\mathcal{P}$, $\|\rem_1\|\le\|\widehat{\rho}^*_I - \rho^*_I\|+C\|\hat G\| \le \mathrm{diam}(\mathcal{P}) + C\|\hat G\|$. So $\|\rem_1\,\mathbf 1(A_I^c)\|_4 \le \mathrm{diam}(\mathcal{P}) + C\|\hat G\|_4 = O(1)$ using ($A.i$). By H\"{o}lder with conjugate exponents $(4,4,2)$, and using the above, ($A.iii$) at $p=4$ and ($B.iii$):
\begin{equation*}
  |E[\rem_1'S\,\mathbf 1(A_I^c)]| \le \|\rem_1\,\mathbf 1(A_I^c)\|_4\cdot\|S\|_4\cdot P(A_I^c)^{1/2} = O(1)\cdot O(I^{-1/2})\cdot O(\Phi_I) = O(\Phi_I I^{-1/2}).
\end{equation*}
\emph{Combining on and off $A_I$:} $|E[\rem_1'S]| \le O(\Phi_I I^{-1/2}) + O(\Phi_I I^{-1/2}) = O(\Phi_I I^{-1/2})$.

\smallskip
\emph{(C.iii) Conclusion.}\; $|E[\rem_3]| \le |E[\rem_1'S]| + |E[\rem_2]| = O(\Phi_I I^{-1/2})$ by ($C.i$) and ($C.ii$). Under (A8), $\Phi_I=o(1)$, so $\Phi_I I^{-1/2} = o(\Phi_I^{1/2} I^{-1/2})$, i.e., the bound is of smaller order than $\Phi_I^{1/2} I^{-1/2}$, the order of the upper bound on $\Blead$. Under (A8'), $\Phi_I=O(I^{-1})$, so the bound is $O(I^{-3/2})$. \qed

\subsection{Higher-order bias when $\rho^*_I$ is affected by boundary-constraint}\label{app:hobias-boundary}

\begin{lemma}
\label{lemma:hobiasboundary}
Consider the scalar AR(1) working model, $d_\rho=1$, with a fixed compact parameter space
$\mathcal{P}=[\rho_{\min},\rho_{\max}]$, and let the conditions of Proposition \ref{prop:tfgls} hold together with assumptions (A5'), (P6') and (A12). Suppose that for all large $I$ the optimizer $\rhostar$ \eqref{eq:optrho} be a boundary point of $\mathcal{P}$, and that the corresponding one-sided derivative $g_I:=\frac{\partial}{\partial\rho}\bigl[I\sigma^2_I(\rhostar)\bigr]$ 
be non-degenerate ---  there exists $\underline c>0$ with $\liminf_I|g_I|\ge\underline c$, and its sign makes
$\rhostar$ a constrained minimizer, i.e.\ $g_I\le-\underline c$ if $\rhostar=\rho_{\max}$ and
$g_I\ge\underline c$ if $\rhostar=\rho_{\min}$. Let $\Phi_I:=\max_i\|\kappa_{i,I}(\rhostar)\|^2/K_I$. Then
\vspace{-.1cm}
\begin{equation*}
  \text{(i)}\quad P\bigl(\rhohat\ne\rhostar\bigr)=O(\Phi_I^2),
  \qquad\qquad
  \text{(ii)}\quad E[\widehat\tau_{w,\mathrm{tfgls}}-\tau_w]=O\bigl(\Phi_I\,I^{-1/2}\bigr).
\vspace{-.1cm}
\end{equation*}
Under (A8) the bias in (ii) is $o\bigl(\Phi_I^{1/2}I^{-1/2}\bigr)$, and under (A8') it is
$O(I^{-3/2})$. In both cases it is of smaller order than the upper bounds on $\Blead$ in Proposition \ref{prop:ho-bias}, i.e., $\Phi_I^{1/2}I^{-1/2}$ under (A8) and $I^{-1}$ under (A8$'$).
\end{lemma}

\medskip

\noindent \textbf{Proof of Lemma \ref{lemma:hobiasboundary}:} Without loss of generality we take $\rstar=\rho_{\max}$ with $g_I\le-\underline c<0$; the case $\rstar=\rho_{\min}$ (with $g_I\ge\underline c>0$) can be handled similarly, and if $\rstar$ alternates between the endpoints along $I$ the bound below applies along each subsequence.\footnote{We continue with the same notation and devices from before: $\kappa_i:=\kappa_{i,I}(\rstar)$; $\kdot(\rho):=\partial\kappa_{[0],i,I}(\rho)/\partial\rho$ and $\kddot(\rho):=\partial^2\kappa_{[0],i,I}(\rho)/\partial\rho^2$ are the scalar first and second derivatives ($d_\rho=1$), with $\kdot:=\kdot(\rstar)$; $b_i:=\|w_{[1],i}\|\vee I^{-1}$; $\|X\|_p:=(E\|X\|^p)^{1/p}$; and $q:=(d_\rho\vee4)+\varepsilon/2$, so that $d_\rho=1$ makes (A5') supply moments of $u_i$ up to order $2q\ge8$ and $q\ge4$. We use from Lemma \ref{lemma:wtsum-order} that $\|\kappa_{i,I}(\rho)\|,\|\kdot(\rho)\|,\|\kddot(\rho)\|\le Cb_i$ uniformly in $i,I,\rho$ (by (A12)), $\sum_i b_i^2=O(K_I)=O(I^{-1})$, and $IK_I=O(1)$; recall also $\Phi_I\ge cI^{-1}$.} Recall that the sample objective function and its probability limit (see Proposition \ref{prop:cons_var}) are $ \widehat M_I(\rho) =I\,\widehat\sigma^2_I(\rho)$ and $M_I(\rho):=I\sigma^2_I(\rho)+I\sigma^2_{\tau,I}$. Recall that (half times the) derivative of the sample objective function and the infeasible version that replaces the $\widehat{u}_i$'s by the $u_i$'s are $\Ghat(\rho) =I\sum_i\bigl(\kdot'(\rho)\widehat u_{0,i}\bigr)\bigl(\kappa_{i,I}'(\rho)\widehat u_i\bigr)$ and $\Gtil(\rho) =I\sum_i\bigl(\kdot'(\rho)u_{0,i}\bigr)\bigl(\kappa_{i,I}'(\rho)u_i\bigr)$. That is, $\tfrac12\partial \widehat M_I(\rho)/ \partial \rho=\Ghat(\rho)$ while:
\vspace{-.1cm}
\begin{equation}
\label{eq:hbb-center}
  \tfrac12 \frac{\partial M_I(\rho)}{\partial \rho}=E[\Gtil(\rho)],
  \text{~~so~~}
  g_I= \frac{\partial M_I(\rstar)}{\partial \rho}=2\,E[\Gtil(\rstar)].
\vspace{-.1cm}
\end{equation}
Since $g_I\ne0$ at the boundary, the interior first-order condition $E[\Gtil(\rstar)]=0$ underlying Proposition \ref{prop:ho-bias} is unavailable.

\smallskip

As noted in the third remark following Proposition \ref{prop:ho-bias}, we know by Cauchy-Schwarz that:
\begin{equation}\label{eq:hbb-red}
  \bigl|E[\widehat\tau_{w,\tfgls}-\tau_w]\bigr|
  \le 2\,\Bigl\|\sup_{\rho\in\mathcal P}|\widehat\tau_I(\rho)-\tau_w|\Bigr\|_2\;P(\rhat\ne\rstar)^{1/2}.
\end{equation}
The first factor is $O(K_I^{1/2})=O(I^{-1/2})$: the summand $\kappa_{i,I}'(\rho)u_i$ is linear in $u_i$ with $\|\kappa_{i,I}'(\rho)u_i\|_2\le Cb_i$, sums to mean zero, and satisfies $\|(\kappa_{i,I}(\rho) - \kappa_{i,I}(\rho'))'u_i\|_2 \le Cb_i\|\rho - \rho'\|$ by (A12), so Lemma~\ref{aux:A}(i) at $p=2$ applies. It therefore remains to prove (i): $P(\rhat\ne\rstar)=O(\Phi_I^2)$. We do it in the following steps.

\smallskip
\noindent\underline{Step 1: a fixed neighborhood on which $ M_I$ decreases.} Differentiating $\tfrac12 M_I(\rho)=\tfrac12 I\sigma^2_I(\rho)+\text{const}$ twice (allowed by (A12)) and bounding $\|\kddot(\rho)\|,\|\kdot(\rho)\|\le Cb_i$ by (A12) and $\|\Sigma_i\|\le C$:
\begin{equation}\label{eq:hbb-curv}
  | \partial^2 M_I(\rho)/\partial \rho^2|\le C\,I\sum_i b_i^2=O(1)=:H_{\max},
\end{equation}
uniformly in $\rho$. Fix the \emph{$I$-free} constants $\delta:=\underline c/(2H_{\max})>0$ and $N_\delta:=[\rho_{\max}-\delta,\rho_{\max}]\cap\mathcal P$. For $\rho\in N_\delta$ we have $\rho-\rstar\in[-\delta,0]$, so by \eqref{eq:hbb-curv} and $g_I\le-\underline c$, a mean-value expansion gives $ \partial M_I(\rho)/\partial \rho =g_I+ \left[\partial^2 M_I(\xi)/\partial \rho^2\right](\rho-\rstar)\le-\underline c+H_{\max}\delta=-\tfrac12\underline c<0$. Thus $ M_I$ is strictly decreasing on $N_\delta$, and $\argmin_{\rho \in N_\delta} M_I(\rho)=\rho_{\max}=\rstar$.

\smallskip
\noindent\underline{Step 2: Rate of convergence of first derivative in the neighborhood.}\;
We show
\begin{equation}\label{eq:hbb-grad}
  P\Bigl(\sup_{\rho\in N_\delta}\left|\frac{\partial}{\partial \rho} \left[\widehat M_I(\rho)- M_I(\rho)\right]\right|\ge\tfrac12\underline c\Bigr)=O(\Phi_I^2).
\end{equation}
By \eqref{eq:hbb-center}, $\tfrac12\partial (\widehat M_I(\rho)- M_I(\rho))/\partial \rho=\Ghat(\rho)-E[\Gtil(\rho)]=\bigl(\Gtil(\rho)-E[\Gtil(\rho)]\bigr)+R_I(\rho)$, where $R_I(\rho):=\Ghat(\rho)-\Gtil(\rho)$ is the residual-substitution term of Lemma \ref{lemma:rem3bound} --- we cannot directly use that proof for bounding because of the non-zero derivative at the boundary. 

\emph{Centered gradient:} Recall that $\widetilde G_i(\rho):=(\kdot'(\rho)u_{0,i})(\kappa_{i,I}'(\rho)u_i)$ is independent across $i$ and $\|\widetilde G_i(\rho)\|_q\le Cb_i^2(E\|u_i\|^{2q})^{1/q}\le Cb_i^2$ by (A5'), with the analogous Lipschitz bound in $\rho$ by (A12). This is exactly the quadratic-in-$u$ input to the maximal inequality Lemma \ref{aux:A}(ii) at $p=q$, which gives $\bigl\|\sup_{\rho\in N_\delta}|\Gtil(\rho)-E\Gtil(\rho)|\bigr\|_q=O(\Phi_I^{1/2})$.

\emph{Residual term:} $\bigl\|\sup_{\rho\in N_\delta}|R_I(\rho)|\bigr\|_q=O(\Phi_I^{1/2})$ by Lemma \ref{aux:C}(ii): writing $\kdot'(\rho)\widehat u_{0,i}=\kdot'(\rho)u_{0,i}-\kdot'(\rho)\xt(\bhat-\beta)$ and $\kappa_{i,I}'(\rho)\widehat u_i=\kappa_{i,I}'(\rho)u_i+c_i$ with $|c_i|\le Cb_i(1+D_I)$, the purely quadratic-in-$u$ part cancels against $\Gtil$, leaving the three types of terms in the bound on $(\mathrm{III})$ in the proof of Lemma \ref{lemma:rem3bound}, controlled in the same way using Lemma \ref{aux:B} ($\|\bhat-\beta\|_p,\|D_I\|_p=O(I^{-1/2})$) and Lemma \ref{aux:A}(ii).

Hence $\bigl\|\sup_{N_\delta}|\Ghat-E[\Gtil]|\bigr\|_q=O(\Phi_I^{1/2})$, and since $\partial [\widehat M_I- M_I]/\ \partial \rho=2(\Ghat-E\Gtil)$, Markov at $L^q$ gives \eqref{eq:hbb-grad}. On its complementary event, $\sup_{\rho \in N_\delta}|\partial [\widehat M_I(\rho)- M_I(\rho)]/ \partial \rho|<\tfrac12\underline c$, so together with $\partial M_I(\rho)/\partial \rho \le-\tfrac12\underline c$ from Step 1, $\partial \widehat M_I(\rho)/ \partial \rho \le \partial M_I(\rho)/ \partial \rho+\sup_{N_\delta}|\partial [\widehat M_I(\rho)- M_I(\rho)]/ \partial \rho|<-\tfrac12\underline c+\tfrac12\underline c=0$ for all $\rho\in N_\delta$. Thus $\widehat M_I(\rho)$ is strictly decreasing in $\rho \in N_\delta$, and $\argmin_{\rho \in N_\delta}\widehat M_I(\rho)=\rho_{\max}$.

\smallskip
\noindent\underline{Step 3: Convergence of $\rhat$.}\; We show $P(|\rhat-\rstar|>\delta)=O(\Phi_I^2)$, a rate that Lemma \ref{lemma:rhocons} (which gives only $o_P(1)$) does not by itself supply, via the $L^q$ strengthening of the objective function:
\begin{equation}\label{eq:hbb-unif}
  \Bigl\|\sup_{\rho\in\mathcal P}\bigl|\widehat M_I(\rho)- M_I(\rho)\bigr|\Bigr\|_q=O(\Phi_I^{1/2}).
\end{equation}
The decomposition \eqref{eq:decomp} and the steps in the proof of Proposition \ref{prop:cons_var} give that the leading term in $\widehat M_I(\rho)- M_I(\rho)=I[\widehat\sigma^2_I(\rho)-(\sigma^2_I(\rho)+\sigma^2_{\tau,I})]$ is $I[\sum_i\xi_i^2(\rho)-\sigma^2_I(\rho)]+2I\sum_i\xi_i(\rho)\delta_i$. The remaining estimation-of-$\eta$/$\tau_w$ terms (the $\phi_i,\psi_i$ terms of that proof) are treated below. The centered sum $\sum_i(\xi_i^2(\rho)-E\xi_i^2(\rho))$ has independent mean-zero summands with $\|\xi_i^2(\rho)-E\xi_i^2(\rho)\|_q\le 2\|\xi_i^2(\rho)\|_q\le Cb_i^2(E\|u_i\|^{2q})^{1/q}\le Cb_i^2$ by (A5$'$), so Lemma \ref{aux:A}(ii) at $p=q$ gives $\bigl\|I[\sum_i\xi_i^2-\sigma^2_I]\bigr\|_q=O(\Phi_I^{1/2})$. The cross term $2I\sum_i\xi_i\delta_i$ is linear in $u_i$ with coefficient $\le Cb_i^2$ (as $|\delta_i|\le C\|w_{[1],i}\|\le Cb_i$ by (A7) and Lemma \ref{lemma:wtsum-order}) and is bounded identically. The $\phi_i,\psi_i$ terms are $O(\Phi_I^{1/2})$ in $L^q$ as in Proposition \ref{prop:cons_var}. Two of them are not of smaller order: $-2I\sum_i\delta_i\psi_i=-2I(\widehat\tau_I(0)-\tau_w)\sum_i\delta_iw'_{[1],i}1_{N_{[1],i}}$, which does not depend on $\rho$, and $-2I\sum_i\delta_i\phi_i=-2I\bigl(\sum_i\delta_i\kappa_{i,I}'(\rho)x_i\bigr)(\bhat-\beta)$. As $|\delta_i|,|w'_{[1],i}1_{N_{[1],i}}|,\sup_\rho\|\kappa_{i,I}'(\rho)x_i\|\le Cb_i$ by (A6)--(A7) and (A12), both are bounded, uniformly in $\rho$, by $CI\sum_ib_i^2(|\widehat\tau_I(0)-\tau_w|+\|\bhat-\beta\|)=O(I^{-1/2})=O(\Phi_I^{1/2})$ in $L^q$ (Lemma \ref{aux:B}). The others are $o(\Phi_I^{1/2})$; e.g., $\|I\sum_i\phi_i^2\|_q\le I\|\bhat-\beta\|_{2q}^2\sum_i\|\kappa_{i,I}'x_i\|^2\le C\,I\cdot O(I^{-1})\cdot O(K_I)=O(I^{-1})=o(\Phi_I^{1/2})$ by Lemma~\ref{aux:C}(iii) and $\sum_i\|\kappa_{i,I}'x_i\|^2=O(K_I)$. Uniformity over $\rho\in\mathcal P$ follows from the stochastic-Lipschitz bound of Proposition \ref{prop:cons_var} (its $|a^2-b^2|=|a-b||a+b|$ argument, here in $L^q$ via (A12)) and Corollary 2.2.5 of \citet{VW96}, applied as in the proof of Lemma \ref{aux:A}: the summands are continuous in $\rho$ by (A4), so the process is separable, and the Lipschitz constant $O(\Phi_I^{1/2})$ enters the bound only as a multiplicative factor, proving \eqref{eq:hbb-unif}.

Now, by the well-separation condition (A9), for the fixed $\delta$ of Step 1, $\Delta_I(\delta):=\inf_{\rho\in\mathcal P:\,|\rho-\rstar|>\delta}[ M_I(\rho)- M_I(\rstar)]=\inf_{|\rho-\rstar|>\delta}I(\sigma^2_I(\rho)-\sigma^2_I(\rstar))\ge\Delta_0>0$ for all large $I$. As in the proof of Lemma \ref{lemma:rhocons}, $\rhat$ minimizing $\widehat M_I$ implies $ M_I(\rhat)- M_I(\rstar)\le2\sup_\rho|\widehat M_I(\rho)- M_I(\rho)|$, so $\{|\rhat-\rstar|>\delta\}\subseteq\{2\sup_\rho|\widehat M_I(\rho)- M_I(\rho)|\ge\Delta_I(\delta)\}$, and Markov at $L^q$ with \eqref{eq:hbb-unif} gives:
\begin{equation}\label{eq:hbb-esc}
  P(|\rhat-\rstar|>\delta)\le P\left(\sup_\rho|\widehat M_I(\rho)- M_I(\rho)|\ge\frac{\Delta_0}{2}\right)
  \le\left(\frac{2}{\Delta_0}\right)^q\,\bigl\|\sup_\rho|\widehat M_I(\rho)- M_I(\rho)|\bigr\|_q^{\,q}=O(\Phi_I^{q/2})=O(\Phi_I^2).
\end{equation}

By Step 2 (see its end), on the complement of \eqref{eq:hbb-grad} $\widehat M_I(\rho)$ is strictly decreasing on $N_\delta$, with unique minimizer $\rho_{\max}$ over $N_\delta$. Since $\rhat\in\mathcal P$ always, on the complement of \eqref{eq:hbb-esc} the bound $|\rhat-\rstar|\le\delta$ combined with $\rhat\le\rho_{\max}$ gives $\rhat\in N_\delta$, hence $\rhat=\rho_{\max}=\rstar$. Therefore, $\{\rhat\ne\rstar\}\subseteq\{\sup_{N_\delta}|\widehat M_I'- M_I'|\ge\tfrac12\underline c\}\cup\{|\rhat-\rstar|>\delta\}$, and \eqref{eq:hbb-grad}--\eqref{eq:hbb-esc} give $P(\rhat\ne\rstar)=O(\Phi_I^2)$, which is (i). Then (ii) follows by substituting into \eqref{eq:hbb-red} with the first factor $O(I^{-1/2})$: $|E[\widehat\tau_{w,\tfgls}-\tau_w]|=O(I^{-1/2})\cdot O((\Phi_I^2)^{1/2})=O(\Phi_I I^{-1/2})$. \qed

\subsection{Auxiliary material for the simulation experiment}\label{app:mcaux}

This appendix collects the simulation results referred to in Section~\ref{sec:mc-results} and, in Appendix~\ref{app:csxdes}, those for the original design of \citet{CSX2025} referred to in Section~\ref{sec:mc-design}. The residuals used by KFGLS$^*$ are those of the imputation fit with cohort and period effects in the untreated cells, and deviations from cohort-period means in the treated cells.

\subsubsection{Standard deviation relative to standard error}\label{app:sdse}

Tables \ref{tab:des1stdbse}--\ref{tab:morecohsdse} report the ratio of the Monte Carlo standard deviation of each estimator to the square root of the average of its estimated asymptotic variance. A ratio larger than one means that the asymptotic variance formula underestimates the variability of the estimator on average.

\begin{table}[!htbp]
\centering
\footnotesize
\renewcommand{\arraystretch}{1.0}
\setlength{\tabcolsep}{4pt}
\newcolumntype{C}{>{\centering\arraybackslash}p{1.2cm}}
\captionsetup{font=small,skip=4pt}
\begin{tabular}{rr@{\hspace{12pt}}|CCCCC@{\hspace{10pt}}|CCCCC}
\hline
\multicolumn{2}{c|}{Specifications} & \multicolumn{5}{c|}{$T = 10$} & \multicolumn{5}{c}{$T = 20$} \\
\hline
$I$ & $\nu_1$ & TFGLS & OLS & FGLS$^*$ & KFGLS$^*$ & CSX & TFGLS & OLS & FGLS$^*$ & KFGLS$^*$ & CSX \\
\hline
&&&&&&&&&&&\\[-1.6ex]
400 & $-1$ & 1.00 & 1.01 & 1.00 & 1.02 & 1.03 & 0.99 & 1.00 & 0.99 & 1.00 & 1.02 \\
400 & $-.5$ & 1.01 & 1.01 & 1.01 & 1.02 & 1.02 & 0.99 & 0.99 & 0.99 & 1.00 & 1.02 \\
400 & 0 & 1.01 & 1.00 & 1.01 & 1.02 & 1.03 & 0.99 & 0.99 & 0.99 & 1.02 & 1.04 \\
400 & .5 & 1.01 & 1.00 & 1.00 & 1.02 & 1.03 & 1.00 & 1.00 & 0.99 & 1.03 & 1.05 \\
400 & 1 & 1.01 & 1.01 & 1.01 & 1.02 & 1.03 & 1.01 & 1.00 & 1.01 & 1.04 & 1.08 \\[0.8ex]
\hline
&&&&&&&&&&&\\[-1.6ex]
100 & $-1$ & 0.98 & 1.00 & 0.99 & 1.03 & 1.08 & 0.99 & 1.00 & 0.98 & 1.10 & 1.28 \\
100 & $-.5$ & 1.00 & 1.00 & 0.99 & 1.04 & 1.09 & 0.99 & 0.99 & 0.99 & 1.10 & 1.29 \\
100 & 0 & 1.02 & 1.01 & 1.00 & 1.05 & 1.10 & 1.01 & 1.00 & 1.00 & 1.09 & 1.28 \\
100 & .5 & 1.02 & 1.01 & 1.01 & 1.05 & 1.11 & 1.02 & 1.01 & 1.01 & 1.12 & 1.30 \\
100 & 1 & 1.01 & 1.01 & 1.00 & 1.04 & 1.12 & 1.00 & 1.00 & 1.00 & 1.10 & 1.30 \\[0.8ex]
\hline
&&&&&&&&&&&\\[-1.6ex]
50 & $-1$ & 1.01 & 1.03 & 1.01 & 1.12 & 1.29 & 1.04 & 1.02 & 1.02 & 1.29 & 1.95 \\
50 & $-.5$ & 1.04 & 1.02 & 1.02 & 1.11 & 1.28 & 1.06 & 1.03 & 1.04 & 1.28 & 1.96 \\
50 & 0 & 1.04 & 1.02 & 1.02 & 1.12 & 1.28 & 1.06 & 1.03 & 1.03 & 1.27 & 1.96 \\
50 & .5 & 1.04 & 1.03 & 1.03 & 1.13 & 1.29 & 1.05 & 1.03 & 1.03 & 1.24 & 1.93 \\
50 & 1 & 1.04 & 1.04 & 1.03 & 1.11 & 1.27 & 1.03 & 1.03 & 1.02 & 1.25 & 1.91 \\[0.8ex]
\hline
\end{tabular}
\caption{Design 1: ratio of the Monte Carlo standard deviation of each estimator to the square root of the average of its estimated asymptotic variance, 10000 Monte Carlo trials. FGLS$^*$: the idealized infeasible FGLS (Section~\ref{sec:mc-estimators}). (Each trial draws its own block of years, with its own error scale; averaging variances rather than standard errors avoids distortion from this mixing.) Because $\widehat{\rho}^*_I$ minimizes the estimated variance, $\widehat{\sigma}^2_I(\widehat{\rho}^*_I)\le\widehat{\sigma}^2_I(\rho^*_I)$, so this selection should make the TFGLS standard error too small in finite samples (cf.\ \citet{WestgateBurchett2016} on selecting the working correlation in generalized estimating equations). In these designs the effect is small: the TFGLS ratio exceeds the larger of the OLS and FGLS$^*$ ratios (neither of which chooses $\rho$ by minimizing the estimated variance) by at most .03 at $I = 50$ and .01 at $I \ge 100$.}
\label{tab:des1stdbse}
\end{table}

\begin{table}[p]
\centering
\footnotesize
\renewcommand{\arraystretch}{1.05}
\setlength{\tabcolsep}{4pt}
\newcolumntype{C}{>{\centering\arraybackslash}p{1.2cm}}
\captionsetup{font=small,skip=4pt}
\begin{tabular}{rc@{\hspace{12pt}}|CCCCC@{\hspace{10pt}}|CCCCC}
\hline
\multicolumn{2}{c|}{Specifications} & \multicolumn{5}{c|}{$T = 10$} & \multicolumn{5}{c}{$T = 20$} \\
\hline
$I$ & Design & TFGLS & OLS & FGLS$^*$ & KFGLS$^*$ & CSX & TFGLS & OLS & FGLS$^*$ & KFGLS$^*$ & CSX \\
\hline
&&&&&&&&&&&\\[-1.6ex]
400 & (a) & 1.01 & 1.01 & 1.01 & 1.02 & 1.03 & 1.01 & 1.00 & 1.00 & 1.03 & 1.05 \\
400 & (b) & 1.01 & 1.01 & 1.01 & 1.02 & 1.03 & 1.01 & 1.00 & 1.01 & 1.03 & 1.06 \\
400 & (c) & 1.00 & 1.00 & 1.00 & 1.02 & 1.03 & 1.00 & 1.00 & 1.00 & 1.02 & 1.03 \\
400 & (d) & 1.01 & 1.01 & 1.00 & 1.02 & 1.03 & 1.01 & 1.00 & 1.01 & 1.03 & 1.05 \\
400 & (e) & 1.01 & 1.01 & 1.01 & 1.02 & 1.03 & 1.00 & 1.01 & 1.00 & 1.03 & 1.01 \\
400 & (f) & 1.01 & 1.00 & 1.00 & 1.02 & 1.03 & 1.00 & 1.00 & 1.00 & 1.03 & 1.06 \\
400 & (g) & 1.00 & 0.99 & 1.00 & 1.01 & 1.01 & 1.01 & 1.00 & 1.00 & 1.03 & 0.99 \\
400 & (h) & 1.01 & 1.01 & 1.01 & 1.02 & 1.03 & 1.00 & 1.00 & 1.00 & 1.02 & 1.06 \\
400 & (i) & 1.01 & 1.01 & 1.01 & 1.02 & 1.03 & 1.00 & 1.00 & 1.00 & 1.02 & 1.05 \\
400 & (j) & 1.01 & 1.00 & 1.01 & 1.02 & 1.03 & 1.01 & 1.01 & 1.01 & 1.03 & 1.06 \\
400 & (k) & 1.01 & 1.01 & 1.01 & 1.02 & 1.03 & 1.01 & 1.01 & 1.01 & 1.03 & 1.06 \\[0.8ex]
\hline
&&&&&&&&&&&\\[-1.6ex]
100 & (a) & 1.02 & 1.01 & 1.01 & 1.07 & 1.11 & 1.03 & 1.02 & 1.02 & 1.13 & 1.32 \\
100 & (b) & 1.03 & 1.02 & 1.02 & 1.07 & 1.13 & 1.04 & 1.03 & 1.03 & 1.13 & 1.33 \\
100 & (c) & 1.04 & 1.02 & 1.02 & 1.06 & 1.11 & 1.00 & 1.00 & 1.00 & 1.12 & 1.29 \\
100 & (d) & 1.03 & 1.02 & 1.02 & 1.07 & 1.12 & 1.03 & 1.03 & 1.02 & 1.13 & 1.33 \\
100 & (e) & 1.02 & 1.01 & 1.01 & 1.07 & 1.11 & 1.03 & 1.01 & 1.02 & 1.13 & 1.26 \\
100 & (f) & 1.02 & 1.01 & 1.01 & 1.06 & 1.11 & 1.02 & 1.02 & 1.01 & 1.12 & 1.32 \\
100 & (g) & 1.03 & 1.02 & 1.02 & 1.08 & 1.12 & 1.01 & 1.02 & 1.00 & 1.13 & 1.22 \\
100 & (h) & 1.02 & 1.02 & 1.02 & 1.06 & 1.12 & 1.01 & 1.01 & 1.00 & 1.12 & 1.33 \\
100 & (i) & 1.02 & 1.02 & 1.02 & 1.07 & 1.12 & 1.01 & 1.01 & 1.01 & 1.12 & 1.33 \\
100 & (j) & 1.00 & 1.02 & 1.00 & 1.03 & 1.10 & 1.01 & 1.01 & 1.00 & 1.10 & 1.31 \\
100 & (k) & 1.01 & 1.02 & 1.01 & 1.03 & 1.10 & 1.02 & 1.01 & 1.01 & 1.10 & 1.31 \\[0.8ex]
\hline
&&&&&&&&&&&\\[-1.6ex]
50 & (a) & 1.06 & 1.03 & 1.03 & 1.14 & 1.31 & 1.05 & 1.03 & 1.03 & 1.24 & 2.01 \\
50 & (b) & 1.06 & 1.03 & 1.03 & 1.13 & 1.31 & 1.05 & 1.03 & 1.03 & 1.23 & 2.05 \\
50 & (c) & 1.08 & 1.04 & 1.04 & 1.13 & 1.30 & 1.04 & 1.03 & 1.03 & 1.27 & 1.99 \\
50 & (d) & 1.06 & 1.03 & 1.03 & 1.14 & 1.31 & 1.05 & 1.03 & 1.03 & 1.24 & 2.03 \\
50 & (e) & 1.06 & 1.03 & 1.03 & 1.14 & 1.31 & 1.05 & 1.03 & 1.03 & 1.25 & 1.90 \\
50 & (f) & 1.05 & 1.03 & 1.03 & 1.12 & 1.29 & 1.04 & 1.02 & 1.02 & 1.23 & 2.06 \\
50 & (g) & 1.04 & 1.03 & 1.01 & 1.14 & 1.30 & 1.04 & 1.03 & 1.01 & 1.29 & 1.88 \\
50 & (h) & 1.03 & 1.03 & 1.02 & 1.13 & 1.31 & 1.06 & 1.05 & 1.05 & 1.28 & 2.10 \\
50 & (i) & 1.04 & 1.03 & 1.03 & 1.13 & 1.31 & 1.07 & 1.05 & 1.05 & 1.29 & 2.11 \\
50 & (j) & 1.03 & 1.04 & 1.03 & 1.10 & 1.29 & 1.05 & 1.05 & 1.03 & 1.24 & 2.05 \\
50 & (k) & 1.04 & 1.04 & 1.03 & 1.10 & 1.29 & 1.06 & 1.05 & 1.04 & 1.24 & 2.05 \\[0.8ex]
\hline
\end{tabular}
\caption{Design 2: ratios as defined in Table \ref{tab:des1stdbse}, 10000 Monte Carlo trials. Designs (a): i.i.d.; (b): AR(1) with coefficient .6; (c): AR(2) with $(\nu_1,\nu_2) = (0,-.9025)$; (d)--(e): MA(1) with $\theta_1 = .6, -.6$; (f)--(g): ARMA(1,1) with $(\nu_1,\theta_1) = (.6,.6), (-.6,-.6)$; (h)--(k): ARMA(1,1) $e_{i,t}$ with $(\nu_1,\theta_1) = (.85,.2)$ in (h), (j) and $(.85,-.2)$ in (i), (k), with a U-shaped (h, i) or falling (j, k) variance over time. As discussed in the note to Table \ref{tab:des1stdbse}, the selection of $\widehat{\rho}^*_I$ should make the TFGLS standard error too small in finite samples; here the TFGLS ratio exceeds the larger of the OLS and FGLS$^*$ ratios by at most .04 at $I = 50$, .02 at $I = 100$ and .01 at $I = 400$.}
\label{tab:des2stdbse}
\end{table}

\begin{table}[!htbp]
\centering
\footnotesize
\renewcommand{\arraystretch}{1.2}
\setlength{\tabcolsep}{4pt}
\newcolumntype{C}{>{\centering\arraybackslash}p{1.2cm}}
\captionsetup{font=small,skip=4pt}
\begin{tabular}{rc@{\hspace{12pt}}|CCCCC}
\hline
$I$ & Design & TFGLS & OLS & FGLS$^*$ & KFGLS$^*$ & CSX \\
\hline
&&&&&&\\[-1.6ex]
400 & 1, $\nu_1 = -1$ & 1.02 & 1.03 & 1.04 & 1.05 & 1.04 \\
400 & 1, $\nu_1 = -.5$ & 1.03 & 1.03 & 1.02 & 1.09 & 1.08 \\
400 & 1, $\nu_1 = 0$ & 0.99 & 0.99 & 0.99 & 1.03 & 1.04 \\
400 & 1, $\nu_1 = .5$ & 1.00 & 1.00 & 1.00 & 1.02 & 1.05 \\
400 & 1, $\nu_1 = 1$ & 1.00 & 1.01 & 1.00 & 1.03 & 1.08 \\[0.8ex]
\hline
&&&&&&\\[-1.6ex]
100 & 1, $\nu_1 = -1$ & 1.08 & 1.06 & 1.07 & 1.13 & 1.38 \\
100 & 1, $\nu_1 = -.5$ & 1.03 & 1.01 & 1.01 & 1.09 & 1.28 \\
100 & 1, $\nu_1 = 0$ & 1.05 & 1.04 & 1.04 & 1.11 & 1.29 \\
100 & 1, $\nu_1 = .5$ & 1.01 & 1.00 & 0.99 & 1.05 & 1.25 \\
100 & 1, $\nu_1 = 1$ & 1.04 & 1.04 & 1.04 & 1.09 & 1.31 \\[0.8ex]
\hline
&&&&&&\\[-1.6ex]
400 & 2(a) & 0.99 & 0.99 & 0.99 & 1.03 & 1.03 \\
400 & 2(b) & 1.03 & 1.03 & 1.02 & 1.05 & 1.07 \\
400 & 2(c) & 1.01 & 1.00 & 1.00 & 1.02 & 1.04 \\
400 & 2(d) & 1.01 & 1.00 & 1.00 & 1.04 & 1.05 \\
400 & 2(e) & 1.01 & 1.00 & 1.00 & 1.04 & 1.04 \\
400 & 2(f) & 1.02 & 1.01 & 1.01 & 1.04 & 1.07 \\
400 & 2(g) & 1.01 & 1.01 & 1.00 & 1.05 & 1.04 \\
400 & 2(h) & 1.01 & 1.02 & 1.01 & 1.05 & 1.07 \\
400 & 2(i) & 1.01 & 1.00 & 1.00 & 1.04 & 1.06 \\
400 & 2(j) & 1.02 & 1.02 & 1.02 & 1.03 & 1.07 \\
400 & 2(k) & 1.02 & 1.01 & 1.02 & 1.03 & 1.07 \\[0.8ex]
\hline
&&&&&&\\[-1.6ex]
100 & 2(a) & 1.04 & 1.03 & 1.03 & 1.15 & 1.31 \\
100 & 2(b) & 1.05 & 1.03 & 1.03 & 1.14 & 1.31 \\
100 & 2(c) & 1.03 & 1.02 & 1.02 & 1.13 & 1.29 \\
100 & 2(d) & 1.05 & 1.03 & 1.02 & 1.15 & 1.32 \\
100 & 2(e) & 1.04 & 1.03 & 1.03 & 1.16 & 1.30 \\
100 & 2(f) & 1.05 & 1.04 & 1.03 & 1.14 & 1.32 \\
100 & 2(g) & 1.03 & 1.02 & 1.01 & 1.18 & 1.30 \\
100 & 2(h) & 1.03 & 1.03 & 1.02 & 1.15 & 1.34 \\
100 & 2(i) & 1.04 & 1.03 & 1.03 & 1.16 & 1.32 \\
100 & 2(j) & 1.02 & 1.02 & 1.02 & 1.08 & 1.30 \\
100 & 2(k) & 1.01 & 1.01 & 1.01 & 1.08 & 1.29 \\[0.8ex]
\hline
\end{tabular}
\caption{More cohorts (Table \ref{tab:morecoh}), $T = 10$: ratios as defined in Table \ref{tab:des1stdbse}, 10000 Monte Carlo trials. As discussed in the note to Table \ref{tab:des1stdbse}, the selection of $\widehat{\rho}^*_I$ should make the TFGLS standard error too small in finite samples; here the TFGLS ratio exceeds the larger of the OLS and FGLS$^*$ ratios by at most .02 at $I = 100$ and .01 at $I = 400$.}
\label{tab:morecohsdse}
\end{table}

\FloatBarrier
\subsubsection{CSX empirical size based on clustered bootstrap standard errors}\label{app:boot}

Following Table 2 of \citet{CSX2025}, for $I=100$ and $I=50$ we compute a bootstrap standard error (BS-SE) of CSX, the standard deviation of 300 clustered bootstrap replications that resample firms with replacement and fully re-estimate CSX, on the Monte Carlo trials of Tables \ref{tab:des1T10}--\ref{tab:des2T20}. Tables \ref{tab:des1boot} and \ref{tab:des2boot} compare the empirical size of the nominal 5\% test based on it with that based on the asymptotic standard error (AS-SE). Pooling Designs 1 and 2, the bootstrap changes the rejection rate from 7.3--8.4\% to 4.6--6.4\% for $(I,T)=(100,10)$, from 11.2--13.7\% to 3.3--4.8\% for $(50,10)$, from 10.6--14.3\% to 2.5--4.9\% for $(100,20)$, and from 23.7--34.6\% to 0.0\% for $(50,20)$. The last case is a breakdown of the bootstrap, not mere conservatism: the BS-SE exceeds three times the AS-SE in 96.6\% of the trials and ten times in 27.0\% (median ratio 5.8), whereas the corresponding ratio for CSX in Tables \ref{tab:des1stdbse} and \ref{tab:des2stdbse} is only 1.9--2.1. A likely reason is that, with about 17 firms per cohort and $T=20$, the cohort-specific covariance matrices estimated by CSX are (nearly) singular, and resampling firms with replacement makes them more so, so that the re-estimated CSX is erratic.

\begin{table}[!hbp]
\centering
\footnotesize
\renewcommand{\arraystretch}{1.1}
\setlength{\tabcolsep}{4pt}
\newcolumntype{C}{>{\centering\arraybackslash}p{1.8cm}}
\begin{tabular}{rr|CC@{\hspace{10pt}}|CC}
\hline
\multicolumn{2}{c|}{Specifications} & \multicolumn{2}{c|}{$T = 10$} & \multicolumn{2}{c}{$T = 20$} \\
\hline
 & & \multicolumn{2}{c|}{Empirical size using} & \multicolumn{2}{c}{Empirical size using} \\
$I$ & $\nu_1$ & AS-SE & BS-SE & AS-SE & BS-SE \\
\hline
&&&&&\\[-1.4ex]
100 & $-1$ & 7.3 & 5.2 & 10.6 & 3.2 \\
100 & $-.5$ & 7.6 & 5.7 & 11.8 & 3.9 \\
100 & $0$ & 8.0 & 6.2 & 12.9 & 4.4 \\
100 & $.5$ & 8.4 & 6.2 & 13.9 & 4.8 \\
100 & $1$ & 7.8 & 6.4 & 14.3 & 4.9 \\[1.4ex]
\hline
&&&&&\\[-1.4ex]
50 & $-1$ & 11.2 & 3.5 & 23.7 & 0.0 \\
50 & $-.5$ & 11.6 & 3.8 & 25.8 & 0.0 \\
50 & $0$ & 11.9 & 4.0 & 28.5 & 0.0 \\
50 & $.5$ & 12.4 & 4.5 & 30.3 & 0.0 \\
50 & $1$ & 12.7 & 4.8 & 30.9 & 0.0 \\[1.4ex]
\hline
\end{tabular}
\caption{Design 1: Rejection probability (\%) of the truth by nominal 5\% tests based on CSX, for $T=10$ (left panel) and $T=20$ (right panel), 10000 Monte Carlo trials. AS-SE / BS-SE: tests based on the t-ratio using the ASymptotic Standard Error (the CSX column of Tables \ref{tab:des1T10} and \ref{tab:des1T20}) / the BootStrap Standard Error of Appendix~\ref{app:boot}, as in Table 2 of \citet{CSX2025}.}
\label{tab:des1boot}
\end{table}

\begin{table}[!tbp]
\centering
\footnotesize
\renewcommand{\arraystretch}{1.1}
\setlength{\tabcolsep}{4pt}
\newcolumntype{C}{>{\centering\arraybackslash}p{1.8cm}}
\begin{tabular}{rr|CC@{\hspace{10pt}}|CC}
\hline
\multicolumn{2}{c|}{Specifications} & \multicolumn{2}{c|}{$T = 10$} & \multicolumn{2}{c}{$T = 20$} \\
\hline
 & & \multicolumn{2}{c|}{Empirical size using} & \multicolumn{2}{c}{Empirical size using} \\
$I$ & Design & AS-SE & BS-SE & AS-SE & BS-SE \\
\hline
&&&&&\\[-1.4ex]
100 & (a) & 8.4 & 5.2 & 13.6 & 3.3 \\
100 & (b) & 8.4 & 5.2 & 13.7 & 3.5 \\
100 & (c) & 7.8 & 4.6 & 12.4 & 3.0 \\
100 & (d) & 8.3 & 5.3 & 14.0 & 3.3 \\
100 & (e) & 7.8 & 5.0 & 12.0 & 2.8 \\
100 & (f) & 8.0 & 4.7 & 14.0 & 3.2 \\
100 & (g) & 8.2 & 4.7 & 10.9 & 2.5 \\
100 & (h) & 8.2 & 5.0 & 14.2 & 3.4 \\
100 & (i) & 8.3 & 5.1 & 14.0 & 3.4 \\
100 & (j) & 7.8 & 5.0 & 13.7 & 3.3 \\
100 & (k) & 7.9 & 4.9 & 13.9 & 3.3 \\[1.4ex]
\hline
&&&&&\\[-1.4ex]
50 & (a) & 13.4 & 3.5 & 31.7 & 0.0 \\
50 & (b) & 13.7 & 3.5 & 32.6 & 0.0 \\
50 & (c) & 12.8 & 3.3 & 31.4 & 0.0 \\
50 & (d) & 13.4 & 3.5 & 32.2 & 0.0 \\
50 & (e) & 13.6 & 3.4 & 29.3 & 0.0 \\
50 & (f) & 12.8 & 3.4 & 33.5 & 0.0 \\
50 & (g) & 13.0 & 3.4 & 29.0 & 0.0 \\
50 & (h) & 13.3 & 3.7 & 34.6 & 0.0 \\
50 & (i) & 13.3 & 3.9 & 34.6 & 0.0 \\
50 & (j) & 13.0 & 3.7 & 33.1 & 0.0 \\
50 & (k) & 13.2 & 3.7 & 33.1 & 0.0 \\[1.4ex]
\hline
\end{tabular}
\caption{Design 2: rejection probabilities as in Table \ref{tab:des1boot}, with AS-SE from the CSX column of Tables \ref{tab:des2T10} and \ref{tab:des2T20}. Designs (a): i.i.d.; (b): AR(1) with coefficient .6; (c): AR(2) with $(\nu_1,\nu_2) = (0,-.9025)$; (d)--(e): MA(1) with $\theta_1 = .6, -.6$; (f)--(g): ARMA(1,1) with $(\nu_1,\theta_1) = (.6,.6), (-.6,-.6)$; (h)--(k): ARMA(1,1) $e_{i,t}$ with $(\nu_1,\theta_1) = (.85,.2)$ in (h), (j) and $(.85,-.2)$ in (i), (k), with a U-shaped (h, i) or falling (j, k) variance over time.}
\label{tab:des2boot}
\end{table}

\FloatBarrier
\subsubsection{The original design of \citet{CSX2025}}\label{app:csxdes}

Tables \ref{tab:desAT10}--\ref{tab:desAsdse} report the results for the design of Section 5.2 of \citet{CSX2025} (see our footnote \ref{fn:CSXdes}) in which the FE, $\tilde{u}_{i,0}$ and all the innovations $\tilde{\epsilon}_{i,t}$ are drawn independently from the pools of Compustat estimates; everything else is as in Sections~\ref{sec:mc-design}--\ref{sec:mc-estimators}.

The results are similar to those for Design 1 but more favorable to TFGLS. OLS and FGLS$^*$ are never more precise than TFGLS. CSX is up to 7\% more precise than TFGLS when $I=400$, but 2--8\% less precise when $I=50$ and $T=10$, 4--10\% when $I=100$ and $T=20$, and 25--34\% when $I=50$ and $T=20$; KFGLS$^*$ is up to 6\% more precise than TFGLS when $I=400$, but up to 14\% less precise when $I=50$ and $T=20$. TFGLS rejects the truth in at most 6.2\% of the trials, CSX in up to 33\%, and KFGLS$^*$ in up to 14\%; as in Designs 1 and 2, the standard errors of KFGLS$^*$ and CSX understate their variability when $I$ is small relative to $T$ (Table \ref{tab:desAsdse}).

\begin{table}[!hbp]
\centering
\footnotesize
\renewcommand{\arraystretch}{1.1}
\setlength{\tabcolsep}{4pt}
\newcolumntype{C}{>{\centering\arraybackslash}p{1.2cm}}
\captionsetup{font=small,skip=4pt}
\begin{tabular}{rr@{\hspace{12pt}}|CCCC@{\hspace{10pt}}|CCCCC}
\hline
\multicolumn{2}{c|}{Specifications}
& \multicolumn{4}{c|}{\begin{tabular}[c]{@{}c@{}}RMSE relative\\to TFGLS\end{tabular}}
& \multicolumn{5}{c}{\begin{tabular}[c]{@{}c@{}}Rejection probability (\%) of truth\\by 5\% Wald test\end{tabular}} \\
\hline
$I$ & $\nu_1$ & OLS & FGLS$^*$ & KFGLS$^*$ & CSX & TFGLS & OLS & FGLS$^*$ & KFGLS$^*$ & CSX \\
\hline
&&&&&&&&&&\\[-1.6ex]
400 & $-1$ & 1.77 & 1.03 & 0.97 & 0.96 & 4.5 & 4.6 & 4.6 & 5.0 & 5.4 \\
400 & $-.5$ & 1.08 & 1.03 & 0.94 & 0.93 & 4.6 & 4.8 & 4.6 & 5.0 & 5.0 \\
400 & 0 & 1.04 & 1.04 & 0.95 & 0.94 & 5.2 & 4.9 & 4.9 & 5.3 & 5.7 \\
400 & .5 & 1.07 & 1.03 & 0.96 & 0.95 & 4.8 & 4.3 & 4.5 & 4.8 & 5.1 \\
400 & 1 & 1.23 & 1.01 & 1.01 & 1.00 & 4.1 & 4.5 & 4.1 & 4.3 & 5.0 \\[0.8ex]
\hline
&&&&&&&&&&\\[-1.6ex]
100 & $-1$ & 1.77 & 1.04 & 0.99 & 0.99 & 4.5 & 4.8 & 4.2 & 6.1 & 7.5 \\
100 & $-.5$ & 1.08 & 1.04 & 0.95 & 0.96 & 5.6 & 5.0 & 5.1 & 6.5 & 7.8 \\
100 & 0 & 1.07 & 1.07 & 0.96 & 0.97 & 5.7 & 5.1 & 5.0 & 6.4 & 7.5 \\
100 & .5 & 1.10 & 1.05 & 0.97 & 0.97 & 5.0 & 4.5 & 4.3 & 5.7 & 7.6 \\
100 & 1 & 1.23 & 1.02 & 1.03 & 1.02 & 4.7 & 4.9 & 4.5 & 5.7 & 7.7 \\[0.8ex]
\hline
&&&&&&&&&&\\[-1.6ex]
50 & $-1$ & 1.78 & 1.04 & 1.02 & 1.02 & 5.4 & 5.1 & 4.8 & 8.1 & 12.0 \\
50 & $-.5$ & 1.11 & 1.05 & 0.97 & 1.02 & 5.8 & 5.3 & 5.2 & 7.8 & 12.2 \\
50 & 0 & 1.08 & 1.08 & 0.98 & 1.03 & 6.2 & 5.6 & 5.6 & 7.8 & 13.0 \\
50 & .5 & 1.10 & 1.04 & 1.00 & 1.06 & 5.8 & 5.3 & 5.2 & 7.7 & 12.6 \\
50 & 1 & 1.24 & 1.01 & 1.04 & 1.08 & 5.4 & 5.3 & 4.9 & 7.4 & 12.5 \\[0.8ex]
\hline
\end{tabular}
\caption{Original design of \citet{CSX2025}, $\mathbf{T = 10}$, 10000 Monte Carlo trials. Left panel: RMSE of estimators relative to that of TFGLS ($> 1$ means estimator is less precise than TFGLS). Right panel: rejection probability (\%) of the truth by the nominal 5\% Wald test.}
\label{tab:desAT10}
\end{table}

\begin{table}[!tbp]
\centering
\footnotesize
\renewcommand{\arraystretch}{1.02}
\setlength{\tabcolsep}{4pt}
\newcolumntype{C}{>{\centering\arraybackslash}p{1.2cm}}
\captionsetup{font=small,skip=4pt}
\begin{tabular}{rr@{\hspace{12pt}}|CCCC@{\hspace{10pt}}|CCCCC}
\hline
\multicolumn{2}{c|}{Specifications}
& \multicolumn{4}{c|}{\begin{tabular}[c]{@{}c@{}}RMSE relative\\to TFGLS\end{tabular}}
& \multicolumn{5}{c}{\begin{tabular}[c]{@{}c@{}}Rejection probability (\%) of truth\\by 5\% Wald test\end{tabular}} \\
\hline
$I$ & $\nu_1$ & OLS & FGLS$^*$ & KFGLS$^*$ & CSX & TFGLS & OLS & FGLS$^*$ & KFGLS$^*$ & CSX \\
\hline
&&&&&&&&&&\\[-1.6ex]
400 & $-1$ & 2.18 & 1.03 & 0.96 & 0.97 & 4.7 & 4.7 & 4.7 & 5.2 & 5.3 \\
400 & $-.5$ & 1.05 & 1.02 & 0.94 & 0.94 & 5.1 & 4.6 & 4.9 & 5.5 & 5.7 \\
400 & 0 & 1.02 & 1.02 & 0.95 & 0.95 & 4.8 & 4.6 & 4.5 & 5.2 & 5.8 \\
400 & .5 & 1.05 & 1.02 & 0.96 & 0.95 & 4.9 & 4.8 & 4.9 & 5.3 & 6.0 \\
400 & 1 & 1.32 & 1.01 & 1.01 & 1.01 & 4.4 & 4.8 & 4.6 & 5.0 & 6.6 \\[0.8ex]
\hline
&&&&&&&&&&\\[-1.6ex]
100 & $-1$ & 2.20 & 1.03 & 1.01 & 1.05 & 4.3 & 4.3 & 4.1 & 7.7 & 11.4 \\
100 & $-.5$ & 1.06 & 1.03 & 0.97 & 1.04 & 5.2 & 4.7 & 5.0 & 8.1 & 11.6 \\
100 & 0 & 1.03 & 1.04 & 0.97 & 1.04 & 5.3 & 5.1 & 5.0 & 7.8 & 12.7 \\
100 & .5 & 1.07 & 1.04 & 0.99 & 1.05 & 5.5 & 4.8 & 5.0 & 8.0 & 13.4 \\
100 & 1 & 1.34 & 1.02 & 1.05 & 1.10 & 4.8 & 4.9 & 4.7 & 7.0 & 13.1 \\[0.8ex]
\hline
&&&&&&&&&&\\[-1.6ex]
50 & $-1$ & 2.21 & 1.02 & 1.14 & 1.25 & 4.5 & 5.1 & 4.0 & 14.4 & 28.2 \\
50 & $-.5$ & 1.07 & 1.04 & 0.99 & 1.28 & 5.8 & 5.0 & 5.3 & 11.9 & 28.7 \\
50 & 0 & 1.05 & 1.05 & 0.99 & 1.28 & 6.2 & 5.2 & 5.2 & 11.5 & 31.3 \\
50 & .5 & 1.08 & 1.05 & 1.01 & 1.30 & 5.7 & 5.0 & 5.1 & 11.1 & 31.9 \\
50 & 1 & 1.32 & 1.01 & 1.10 & 1.34 & 5.3 & 5.3 & 4.7 & 11.1 & 32.8 \\[0.8ex]
\hline
\end{tabular}
\caption{Original design of \citet{CSX2025}, $\mathbf{T = 20}$, 10000 Monte Carlo trials. Left panel: RMSE of estimators relative to that of TFGLS ($> 1$ means estimator is less precise than TFGLS). Right panel: rejection probability (\%) of the truth by the nominal 5\% Wald test.}
\label{tab:desAT20}
\par\bigskip 

\centering
\footnotesize
\renewcommand{\arraystretch}{1.02}
\setlength{\tabcolsep}{4pt}
\newcolumntype{C}{>{\centering\arraybackslash}p{1.2cm}}
\captionsetup{font=small,skip=4pt}
\begin{tabular}{rr@{\hspace{12pt}}|CCCCC@{\hspace{10pt}}|CCCCC}
\hline
\multicolumn{2}{c|}{Specifications} & \multicolumn{5}{c|}{$T = 10$} & \multicolumn{5}{c}{$T = 20$} \\
\hline
$I$ & $\nu_1$ & TFGLS & OLS & FGLS$^*$ & KFGLS$^*$ & CSX & TFGLS & OLS & FGLS$^*$ & KFGLS$^*$ & CSX \\
\hline
&&&&&&&&&&&\\[-1.6ex]
400 & $-1$ & 1.01 & 1.01 & 1.01 & 1.02 & 1.03 & 1.00 & 1.00 & 1.00 & 1.03 & 1.03 \\
400 & $-.5$ & 1.01 & 1.01 & 1.01 & 1.02 & 1.02 & 1.00 & 1.00 & 1.00 & 1.02 & 1.02 \\
400 & 0 & 1.02 & 1.01 & 1.01 & 1.02 & 1.03 & 0.99 & 0.99 & 0.99 & 1.02 & 1.04 \\
400 & .5 & 0.99 & 0.99 & 0.99 & 1.00 & 1.01 & 1.00 & 0.99 & 0.99 & 1.02 & 1.05 \\
400 & 1 & 0.99 & 1.00 & 0.99 & 1.00 & 1.02 & 1.01 & 1.01 & 1.01 & 1.03 & 1.06 \\[0.8ex]
\hline
&&&&&&&&&&&\\[-1.6ex]
100 & $-1$ & 1.00 & 1.00 & 0.99 & 1.05 & 1.12 & 0.99 & 1.02 & 0.99 & 1.13 & 1.27 \\
100 & $-.5$ & 1.04 & 1.02 & 1.03 & 1.08 & 1.13 & 1.03 & 1.02 & 1.02 & 1.13 & 1.29 \\
100 & 0 & 1.03 & 1.02 & 1.02 & 1.06 & 1.12 & 1.03 & 1.01 & 1.01 & 1.12 & 1.30 \\
100 & .5 & 1.01 & 1.01 & 1.01 & 1.05 & 1.10 & 1.03 & 1.02 & 1.02 & 1.13 & 1.32 \\
100 & 1 & 1.02 & 1.02 & 1.01 & 1.06 & 1.12 & 1.01 & 1.01 & 1.00 & 1.11 & 1.29 \\[0.8ex]
\hline
&&&&&&&&&&&\\[-1.6ex]
50 & $-1$ & 1.02 & 1.02 & 1.00 & 1.12 & 1.27 & 1.00 & 1.04 & 0.98 & 1.36 & 1.90 \\
50 & $-.5$ & 1.03 & 1.03 & 1.02 & 1.12 & 1.29 & 1.05 & 1.03 & 1.04 & 1.26 & 1.95 \\
50 & 0 & 1.06 & 1.03 & 1.03 & 1.13 & 1.30 & 1.06 & 1.04 & 1.04 & 1.26 & 2.00 \\
50 & .5 & 1.04 & 1.03 & 1.03 & 1.13 & 1.31 & 1.05 & 1.02 & 1.03 & 1.24 & 2.03 \\
50 & 1 & 1.04 & 1.03 & 1.02 & 1.11 & 1.28 & 1.04 & 1.04 & 1.02 & 1.24 & 1.98 \\[0.8ex]
\hline
\end{tabular}
\caption{Original design of \citet{CSX2025}: ratios as defined in Table \ref{tab:des1stdbse}, 10000 Monte Carlo trials. As discussed in the note to Table \ref{tab:des1stdbse}, the selection of $\widehat{\rho}^*_I$ should make the TFGLS standard error too small in finite samples; here the TFGLS ratio exceeds the larger of the OLS and FGLS$^*$ ratios by at most .03 at $I = 50$, .02 at $I = 100$ and .01 at $I = 400$.}
\label{tab:desAsdse}
\end{table}

\FloatBarrier

\subsubsection{Bias, empirical size and standard errors under entity-level heterogeneity}\label{app:hetaux}

\begin{table}[!htbp]
\centering
\footnotesize
\renewcommand{\arraystretch}{1.07}
\setlength{\tabcolsep}{3.3pt}
\captionsetup{font=small,skip=4pt}
\begin{tabular}{c@{\hspace{8pt}}|cccccc@{\hspace{7pt}}|cccccc}
\hline
& \multicolumn{12}{c}{Bias relative to the Monte Carlo standard deviation} \\
\hline
Design & TFGLS & TFGLS$^*$ & OLS & FGLS & KFGLS$^*$ & CSX & TFGLS & TFGLS$^*$ & OLS & FGLS & KFGLS$^*$ & CSX \\
\hline
& \multicolumn{6}{c|}{$I=4000$} & \multicolumn{6}{c}{$I=400$} \\[0.3ex]
(a) & -0.01 & 0.00 & 0.00 & 0.00 & 0.00 & 0.00 & 0.01 & 0.01 & 0.01 & 0.00 & 0.00 & -0.01 \\
(b) & 0.00 & 0.00 & -0.01 & 0.00 & 0.00 & 0.00 & -0.01 & -0.01 & -0.01 & -0.01 & -0.02 & -0.02 \\
(c) & 0.02 & 0.01 & 0.01 & 0.01 & 0.03 & 0.03 & 0.00 & 0.00 & 0.00 & 0.00 & -0.02 & -0.01 \\
(d) & -0.01 & -0.01 & -0.01 & -0.01 & -0.01 & -0.01 & 0.00 & 0.01 & 0.00 & 0.01 & 0.00 & 0.00 \\
(e) & 0.01 & 0.01 & 0.00 & 0.01 & 0.00 & 0.00 & -0.01 & 0.00 & 0.00 & 0.00 & 0.00 & 0.00 \\
(f) & -0.01 & -0.01 & -0.01 & -0.01 & -0.01 & -0.01 & 0.01 & 0.01 & 0.01 & 0.01 & 0.01 & 0.00 \\
(g) & -0.01 & -0.01 & 0.01 & -0.01 & -0.01 & 0.00 & 0.00 & 0.00 & 0.00 & 0.00 & 0.01 & 0.00 \\
(h) & -0.01 & -0.01 & 0.00 & 0.00 & -0.01 & -0.01 & 0.01 & 0.01 & 0.01 & 0.01 & 0.00 & 0.01 \\
(i) & 0.00 & 0.00 & -0.01 & 0.00 & 0.00 & 0.00 & 0.00 & 0.00 & -0.01 & -0.01 & 0.00 & 0.00 \\
(j) & -0.02 & -0.02 & -0.01 & -0.01 & -0.01 & -0.01 & 0.00 & -0.01 & 0.00 & 0.00 & 0.00 & 0.00 \\
(k) & 0.00 & 0.00 & 0.01 & 0.01 & 0.00 & 0.00 & 0.00 & 0.00 & 0.00 & 0.00 & 0.00 & 0.00 \\[0.8ex]
\hline
& \multicolumn{6}{c|}{$I=2000$} & \multicolumn{6}{c}{$I=100$} \\[0.3ex]
(a) & 0.01 & 0.01 & 0.01 & 0.01 & 0.00 & 0.00 & 0.01 & 0.02 & 0.02 & 0.02 & 0.01 & 0.00 \\
(b) & 0.01 & 0.01 & 0.02 & 0.02 & 0.02 & 0.02 & -0.03 & -0.03 & -0.03 & -0.03 & -0.01 & -0.01 \\
(c) & 0.01 & 0.00 & 0.00 & 0.00 & 0.01 & 0.02 & 0.01 & 0.01 & 0.01 & 0.01 & -0.02 & -0.02 \\
(d) & -0.01 & -0.01 & -0.01 & -0.01 & -0.01 & 0.00 & 0.00 & 0.00 & 0.00 & 0.00 & -0.01 & 0.00 \\
(e) & 0.01 & 0.01 & 0.01 & 0.01 & 0.00 & 0.00 & 0.00 & 0.00 & 0.00 & 0.00 & -0.01 & 0.00 \\
(f) & 0.00 & 0.00 & 0.00 & 0.00 & 0.00 & -0.01 & 0.01 & 0.01 & 0.01 & 0.01 & 0.02 & 0.02 \\
(g) & 0.01 & 0.01 & 0.01 & 0.01 & 0.00 & 0.00 & 0.01 & 0.01 & 0.00 & 0.01 & 0.01 & 0.01 \\
(h) & 0.00 & 0.00 & 0.00 & 0.00 & 0.00 & 0.00 & 0.01 & 0.00 & 0.01 & 0.01 & 0.00 & 0.00 \\
(i) & -0.01 & -0.01 & -0.01 & -0.01 & -0.01 & -0.01 & -0.01 & -0.01 & -0.02 & -0.02 & -0.01 & -0.01 \\
(j) & 0.00 & 0.00 & 0.00 & 0.00 & 0.00 & 0.00 & 0.00 & 0.00 & -0.02 & -0.01 & -0.01 & -0.01 \\
(k) & 0.01 & 0.01 & 0.00 & 0.00 & 0.01 & 0.00 & -0.02 & -0.01 & -0.01 & -0.01 & -0.01 & -0.01 \\[0.8ex]
\hline
& \multicolumn{6}{c|}{$I=1000$} & \multicolumn{6}{c}{$I=50$} \\[0.3ex]
(a) & -0.02 & -0.02 & -0.02 & -0.02 & -0.02 & -0.02 & 0.03 & 0.02 & 0.02 & 0.02 & 0.00 & 0.00 \\
(b) & 0.00 & 0.00 & 0.01 & 0.01 & 0.00 & 0.00 & -0.01 & -0.01 & -0.02 & -0.01 & -0.01 & 0.01 \\
(c) & 0.00 & 0.00 & 0.00 & 0.00 & 0.02 & 0.02 & 0.00 & -0.01 & 0.00 & 0.00 & -0.01 & -0.01 \\
(d) & 0.00 & 0.00 & 0.00 & 0.00 & 0.00 & -0.01 & 0.00 & 0.00 & 0.00 & 0.00 & -0.01 & 0.00 \\
(e) & 0.01 & 0.01 & 0.00 & 0.01 & 0.01 & 0.01 & 0.00 & 0.01 & 0.01 & 0.01 & 0.02 & 0.00 \\
(f) & -0.01 & -0.01 & 0.00 & -0.01 & -0.01 & -0.01 & -0.01 & -0.01 & 0.00 & -0.01 & -0.01 & 0.00 \\
(g) & 0.00 & 0.00 & 0.01 & 0.00 & 0.00 & -0.01 & 0.00 & -0.01 & 0.00 & -0.01 & 0.01 & 0.00 \\
(h) & -0.01 & -0.01 & 0.00 & -0.01 & -0.01 & -0.01 & 0.01 & 0.00 & 0.00 & 0.00 & 0.01 & 0.01 \\
(i) & 0.01 & 0.01 & 0.00 & 0.00 & 0.01 & 0.01 & -0.01 & -0.01 & -0.02 & -0.02 & -0.01 & -0.01 \\
(j) & 0.01 & 0.01 & 0.01 & 0.01 & 0.01 & 0.01 & -0.01 & -0.01 & 0.00 & 0.00 & 0.00 & -0.01 \\
(k) & 0.00 & 0.00 & 0.01 & 0.01 & 0.00 & 0.00 & 0.00 & 0.00 & 0.01 & 0.00 & 0.00 & 0.00 \\[0.8ex]
\hline
\end{tabular}
\caption{Design 2 with entity-level heterogeneity $\tau_{i,t}=\tau_t(E_i)+\sqrt{T-t}\,z_i$ (Section~\ref{sec:mc-hetit}), $\mathbf{T = 20}$, 10000 Monte Carlo trials; estimators as in Table~\ref{tab:hetT20}. Bias of each estimator for the realized target $ES_{\text{avg}}$ divided by its Monte Carlo standard deviation.}
\label{tab:hetbias}
\end{table}

\begin{table}[!htbp]
\centering
\footnotesize
\renewcommand{\arraystretch}{1.07}
\setlength{\tabcolsep}{3.3pt}
\captionsetup{font=small,skip=4pt}
\begin{tabular}{c@{\hspace{8pt}}|cccccc@{\hspace{7pt}}|cccccc}
\hline
& \multicolumn{12}{c}{Empirical size (\%)} \\
\hline
Design & TFGLS & TFGLS$^*$ & OLS & FGLS & KFGLS$^*$ & CSX & TFGLS & TFGLS$^*$ & OLS & FGLS & KFGLS$^*$ & CSX \\
\hline
& \multicolumn{6}{c|}{$I=4000$} & \multicolumn{6}{c}{$I=400$} \\[0.3ex]
(a) & 0.0 & 0.0 & 0.0 & 0.0 & 0.0 & 0.0 & 0.0 & 0.0 & 0.0 & 0.0 & 0.0 & 0.0 \\
(b) & 0.0 & 0.0 & 0.0 & 0.0 & 0.0 & 0.0 & 0.0 & 0.0 & 0.0 & 0.0 & 0.0 & 0.0 \\
(c) & 0.0 & 0.0 & 0.0 & 0.0 & 0.0 & 0.0 & 0.0 & 0.0 & 0.0 & 0.0 & 0.0 & 0.0 \\
(d) & 0.0 & 0.0 & 0.0 & 0.0 & 0.0 & 0.0 & 0.0 & 0.0 & 0.0 & 0.0 & 0.0 & 0.0 \\
(e) & 0.0 & 0.0 & 0.0 & 0.0 & 0.0 & 0.0 & 0.0 & 0.0 & 0.0 & 0.0 & 0.0 & 0.0 \\
(f) & 0.0 & 0.0 & 0.0 & 0.0 & 0.0 & 0.0 & 0.0 & 0.0 & 0.0 & 0.0 & 0.0 & 0.0 \\
(g) & 0.0 & 0.0 & 0.0 & 0.0 & 0.0 & 0.0 & 0.0 & 0.0 & 0.0 & 0.0 & 0.0 & 0.0 \\
(h) & 0.3 & 0.3 & 1.0 & 0.7 & 0.3 & 0.3 & 0.4 & 0.5 & 1.3 & 1.0 & 0.6 & 0.8 \\
(i) & 0.1 & 0.1 & 0.2 & 0.1 & 0.1 & 0.0 & 0.1 & 0.0 & 0.2 & 0.1 & 0.1 & 0.1 \\
(j) & 0.3 & 0.3 & 1.7 & 0.9 & 0.3 & 0.3 & 0.4 & 0.4 & 1.4 & 1.0 & 0.6 & 0.7 \\
(k) & 0.0 & 0.0 & 0.2 & 0.1 & 0.0 & 0.0 & 0.1 & 0.1 & 0.4 & 0.2 & 0.0 & 0.1 \\[0.8ex]
\hline
& \multicolumn{6}{c|}{$I=2000$} & \multicolumn{6}{c}{$I=100$} \\[0.3ex]
(a) & 0.0 & 0.0 & 0.0 & 0.0 & 0.0 & 0.0 & 0.0 & 0.0 & 0.0 & 0.0 & 0.0 & 0.3 \\
(b) & 0.0 & 0.0 & 0.0 & 0.0 & 0.0 & 0.0 & 0.0 & 0.0 & 0.0 & 0.0 & 0.0 & 0.5 \\
(c) & 0.0 & 0.0 & 0.0 & 0.0 & 0.0 & 0.0 & 0.0 & 0.0 & 0.0 & 0.0 & 0.0 & 0.3 \\
(d) & 0.0 & 0.0 & 0.0 & 0.0 & 0.0 & 0.0 & 0.0 & 0.0 & 0.0 & 0.0 & 0.0 & 0.5 \\
(e) & 0.0 & 0.0 & 0.0 & 0.0 & 0.0 & 0.0 & 0.0 & 0.0 & 0.0 & 0.0 & 0.0 & 0.3 \\
(f) & 0.0 & 0.0 & 0.0 & 0.0 & 0.0 & 0.0 & 0.0 & 0.0 & 0.0 & 0.0 & 0.0 & 0.9 \\
(g) & 0.0 & 0.0 & 0.0 & 0.0 & 0.0 & 0.0 & 0.0 & 0.0 & 0.0 & 0.0 & 0.0 & 0.3 \\
(h) & 0.3 & 0.3 & 0.9 & 0.6 & 0.4 & 0.4 & 0.7 & 0.7 & 1.5 & 1.1 & 1.6 & 4.9 \\
(i) & 0.1 & 0.1 & 0.2 & 0.2 & 0.1 & 0.1 & 0.1 & 0.0 & 0.3 & 0.2 & 0.4 & 2.9 \\
(j) & 0.3 & 0.3 & 1.5 & 1.0 & 0.3 & 0.4 & 0.6 & 0.5 & 1.9 & 1.2 & 1.4 & 4.5 \\
(k) & 0.0 & 0.0 & 0.2 & 0.1 & 0.0 & 0.0 & 0.2 & 0.1 & 0.6 & 0.4 & 0.4 & 2.4 \\[0.8ex]
\hline
& \multicolumn{6}{c|}{$I=1000$} & \multicolumn{6}{c}{$I=50$} \\[0.3ex]
(a) & 0.0 & 0.0 & 0.0 & 0.0 & 0.0 & 0.0 & 0.0 & 0.0 & 0.0 & 0.0 & 0.0 & 13.5 \\
(b) & 0.0 & 0.0 & 0.0 & 0.0 & 0.0 & 0.0 & 0.0 & 0.0 & 0.0 & 0.0 & 0.1 & 14.9 \\
(c) & 0.0 & 0.0 & 0.0 & 0.0 & 0.0 & 0.0 & 0.0 & 0.0 & 0.0 & 0.0 & 0.0 & 12.3 \\
(d) & 0.0 & 0.0 & 0.0 & 0.0 & 0.0 & 0.0 & 0.0 & 0.0 & 0.0 & 0.0 & 0.0 & 14.1 \\
(e) & 0.0 & 0.0 & 0.0 & 0.0 & 0.0 & 0.0 & 0.0 & 0.0 & 0.0 & 0.0 & 0.0 & 12.6 \\
(f) & 0.0 & 0.0 & 0.0 & 0.0 & 0.0 & 0.0 & 0.0 & 0.0 & 0.0 & 0.0 & 0.3 & 16.4 \\
(g) & 0.0 & 0.0 & 0.0 & 0.0 & 0.0 & 0.0 & 0.0 & 0.0 & 0.0 & 0.0 & 0.0 & 13.8 \\
(h) & 0.3 & 0.3 & 1.0 & 0.8 & 0.4 & 0.5 & 1.0 & 0.8 & 1.7 & 1.3 & 3.9 & 23.1 \\
(i) & 0.0 & 0.0 & 0.2 & 0.1 & 0.1 & 0.1 & 0.2 & 0.1 & 0.5 & 0.4 & 1.5 & 20.3 \\
(j) & 0.4 & 0.4 & 1.6 & 1.0 & 0.4 & 0.4 & 1.1 & 0.8 & 2.3 & 1.7 & 3.6 & 23.8 \\
(k) & 0.1 & 0.1 & 0.3 & 0.2 & 0.0 & 0.0 & 0.4 & 0.2 & 0.8 & 0.6 & 1.5 & 19.4 \\[0.8ex]
\hline
\end{tabular}
\caption{Design 2 with entity-level heterogeneity $\tau_{i,t}=\tau_t(E_i)+\sqrt{T-t}\,z_i$ (Section~\ref{sec:mc-hetit}), $\mathbf{T = 20}$, 10000 Monte Carlo trials; estimators as in Table~\ref{tab:hetT20}. Empirical size (in \%) of the nominal 5\% two-sided $t$-test for the realized target $ES_{\text{avg}}$.}
\label{tab:hetsize}
\end{table}

\begin{table}[!htbp]
\centering
\footnotesize
\renewcommand{\arraystretch}{1.07}
\setlength{\tabcolsep}{3.3pt}
\captionsetup{font=small,skip=4pt}
\begin{tabular}{c@{\hspace{8pt}}|cccccc@{\hspace{7pt}}|cccccc}
\hline
& \multicolumn{12}{c}{Standard deviation relative to standard error} \\
\hline
Design & TFGLS & TFGLS$^*$ & OLS & FGLS & KFGLS$^*$ & CSX & TFGLS & TFGLS$^*$ & OLS & FGLS & KFGLS$^*$ & CSX \\
\hline
& \multicolumn{6}{c|}{$I=4000$} & \multicolumn{6}{c}{$I=400$} \\[0.3ex]
(a) & 0.13 & 0.13 & 0.13 & 0.13 & 0.14 & 0.15 & 0.14 & 0.13 & 0.13 & 0.13 & 0.18 & 0.26 \\
(b) & 0.25 & 0.25 & 0.26 & 0.25 & 0.26 & 0.26 & 0.26 & 0.25 & 0.26 & 0.25 & 0.28 & 0.34 \\
(c) & 0.13 & 0.13 & 0.15 & 0.15 & 0.10 & 0.12 & 0.14 & 0.13 & 0.15 & 0.15 & 0.15 & 0.24 \\
(d) & 0.19 & 0.19 & 0.20 & 0.19 & 0.20 & 0.20 & 0.20 & 0.20 & 0.20 & 0.20 & 0.23 & 0.29 \\
(e) & 0.07 & 0.07 & 0.08 & 0.07 & 0.08 & 0.10 & 0.08 & 0.07 & 0.08 & 0.07 & 0.14 & 0.23 \\
(f) & 0.38 & 0.38 & 0.40 & 0.38 & 0.38 & 0.38 & 0.38 & 0.37 & 0.40 & 0.37 & 0.39 & 0.43 \\
(g) & 0.06 & 0.06 & 0.08 & 0.06 & 0.07 & 0.09 & 0.07 & 0.06 & 0.08 & 0.06 & 0.14 & 0.23 \\
(h) & 0.67 & 0.67 & 0.76 & 0.73 & 0.67 & 0.67 & 0.69 & 0.68 & 0.77 & 0.74 & 0.70 & 0.74 \\
(i) & 0.54 & 0.54 & 0.62 & 0.60 & 0.54 & 0.54 & 0.55 & 0.55 & 0.63 & 0.61 & 0.57 & 0.60 \\
(j) & 0.66 & 0.66 & 0.80 & 0.75 & 0.66 & 0.67 & 0.67 & 0.67 & 0.79 & 0.74 & 0.69 & 0.72 \\
(k) & 0.54 & 0.54 & 0.66 & 0.63 & 0.53 & 0.53 & 0.55 & 0.54 & 0.66 & 0.63 & 0.56 & 0.59 \\[0.8ex]
\hline
& \multicolumn{6}{c|}{$I=2000$} & \multicolumn{6}{c}{$I=100$} \\[0.3ex]
(a) & 0.13 & 0.13 & 0.13 & 0.13 & 0.14 & 0.16 & 0.16 & 0.13 & 0.13 & 0.13 & 0.28 & 0.57 \\
(b) & 0.25 & 0.25 & 0.27 & 0.25 & 0.26 & 0.27 & 0.28 & 0.26 & 0.27 & 0.26 & 0.37 & 0.62 \\
(c) & 0.14 & 0.14 & 0.15 & 0.15 & 0.11 & 0.13 & 0.16 & 0.14 & 0.16 & 0.16 & 0.27 & 0.57 \\
(d) & 0.20 & 0.19 & 0.20 & 0.19 & 0.20 & 0.21 & 0.22 & 0.20 & 0.20 & 0.20 & 0.32 & 0.60 \\
(e) & 0.08 & 0.07 & 0.08 & 0.07 & 0.09 & 0.12 & 0.12 & 0.07 & 0.08 & 0.07 & 0.26 & 0.56 \\
(f) & 0.37 & 0.37 & 0.40 & 0.37 & 0.38 & 0.38 & 0.40 & 0.38 & 0.41 & 0.38 & 0.48 & 0.70 \\
(g) & 0.07 & 0.06 & 0.08 & 0.06 & 0.08 & 0.11 & 0.10 & 0.06 & 0.08 & 0.06 & 0.25 & 0.55 \\
(h) & 0.67 & 0.67 & 0.76 & 0.72 & 0.67 & 0.68 & 0.69 & 0.68 & 0.78 & 0.74 & 0.78 & 0.97 \\
(i) & 0.54 & 0.54 & 0.62 & 0.60 & 0.54 & 0.54 & 0.56 & 0.54 & 0.63 & 0.61 & 0.64 & 0.85 \\
(j) & 0.66 & 0.66 & 0.80 & 0.76 & 0.67 & 0.67 & 0.69 & 0.67 & 0.81 & 0.76 & 0.76 & 0.95 \\
(k) & 0.54 & 0.54 & 0.66 & 0.63 & 0.54 & 0.54 & 0.57 & 0.55 & 0.67 & 0.64 & 0.63 & 0.83 \\[0.8ex]
\hline
& \multicolumn{6}{c|}{$I=1000$} & \multicolumn{6}{c}{$I=50$} \\[0.3ex]
(a) & 0.13 & 0.13 & 0.13 & 0.13 & 0.15 & 0.19 & 0.18 & 0.14 & 0.14 & 0.14 & 0.39 & 1.27 \\
(b) & 0.26 & 0.25 & 0.26 & 0.25 & 0.27 & 0.29 & 0.30 & 0.26 & 0.28 & 0.27 & 0.47 & 1.31 \\
(c) & 0.14 & 0.13 & 0.15 & 0.15 & 0.12 & 0.16 & 0.19 & 0.14 & 0.16 & 0.16 & 0.37 & 1.24 \\
(d) & 0.20 & 0.20 & 0.20 & 0.20 & 0.21 & 0.24 & 0.24 & 0.20 & 0.21 & 0.20 & 0.42 & 1.30 \\
(e) & 0.08 & 0.07 & 0.08 & 0.07 & 0.11 & 0.15 & 0.14 & 0.08 & 0.08 & 0.08 & 0.36 & 1.24 \\
(f) & 0.37 & 0.37 & 0.39 & 0.37 & 0.38 & 0.39 & 0.42 & 0.39 & 0.41 & 0.39 & 0.57 & 1.38 \\
(g) & 0.07 & 0.06 & 0.08 & 0.06 & 0.10 & 0.14 & 0.14 & 0.07 & 0.09 & 0.07 & 0.36 & 1.29 \\
(h) & 0.67 & 0.67 & 0.76 & 0.73 & 0.68 & 0.69 & 0.72 & 0.70 & 0.79 & 0.75 & 0.89 & 1.63 \\
(i) & 0.54 & 0.54 & 0.62 & 0.60 & 0.55 & 0.56 & 0.59 & 0.56 & 0.65 & 0.63 & 0.75 & 1.51 \\
(j) & 0.67 & 0.66 & 0.80 & 0.75 & 0.67 & 0.68 & 0.71 & 0.68 & 0.82 & 0.77 & 0.87 & 1.64 \\
(k) & 0.55 & 0.54 & 0.65 & 0.62 & 0.55 & 0.56 & 0.60 & 0.56 & 0.68 & 0.65 & 0.74 & 1.48 \\[0.8ex]
\hline
\end{tabular}
\caption{Design 2 with entity-level heterogeneity $\tau_{i,t}=\tau_t(E_i)+\sqrt{T-t}\,z_i$ (Section~\ref{sec:mc-hetit}), $\mathbf{T = 20}$, 10000 Monte Carlo trials; estimators as in Table~\ref{tab:hetT20}. Ratio of the Monte Carlo standard deviation of each estimator to the square root of the average of its estimated variance; a ratio below 1 means a conservative standard error.}
\label{tab:hetsdse}
\end{table}

\subsection{Bootstrap standard errors in the hospital admissions example}\label{app:hrsboot}

Table~\ref{tab:hrsboot} presents the bootstrap version of the right panel of Table~\ref{tab:hrs} in Section~\ref{sec:emp-hrs}. The bootstrap standard error (BS-SE) is the standard deviation of 1000 bootstrap replications that resample individuals with replacement, as in \citet{CSX2025}. Every estimator is recomputed in every replication, including the choice of $\rho$ by TFGLS and TFGLS$(\rho_0,\rho_1)$, the estimate of $\rho$ by FGLS and the choice between the two versions of KFGLS$^*$.

\begin{table}[H]
\centering
\captionsetup{skip=3pt}
\caption{Hospital admissions: ratio with respect to TFGLS of the length of 95\% confidence interval based on bootstrap standard errors (left panel), and ratio of bootstrap to analytic standard error (right panel)}
\label{tab:hrsboot}
\small
\begin{tabular}{lrrrrcrrrr}
\toprule
& \multicolumn{4}{c}{Ratio of interval length} & & \multicolumn{4}{c}{ratio: BS-SE / AS-SE} \\
\cmidrule(lr){2-5}\cmidrule(l){7-10}
Estimator & $ES(0)$ & $ES(1)$ & $ES(2)$ & $ES_{\text{avg}}$ & & $ES(0)$ & $ES(1)$ & $ES(2)$ & $ES_{\text{avg}}$ \\
\midrule
TWFE & \textbf{0.96} & 1.22 & 1.15 & 1.11 & & 1.00 & 1.01 & 1.00 & 1.01 \\
CS-SA & 1.05 & 1.16 & 1.25 & 1.18 & & 1.01 & 1.00 & 1.01 & 1.01 \\
CS-dCDH & 1.01 & 1.12 & 1.25 & 1.02 & & 0.98 & 1.00 & 1.01 & 1.00 \\
BJS(OLS) & 1.05 & 1.16 & 1.02 & 1.04 & & 0.99 & 1.00 & 0.98 & 0.99 \\
FGLS & 1.21 & 1.12 & 1.07 & 1.08 & & 1.01 & 1.01 & 1.00 & 1.01 \\
KFGLS$^*$ & 1.03 & 1.18 & 1.19 & 1.16 & & 1.01 & 1.08 & 1.14 & 1.17 \\
CSX & 1.01 & 1.03 & 1.10 & 1.05 & & 1.08 & 1.10 & 1.10 & 1.12 \\
\midrule
TFGLS & 1.00 & 1.00 & 1.00 & 1.00 & & 0.99 & 0.98 & 0.96 & 0.96 \\
TFGLS ($\rho_0,\rho_1$) & \textbf{0.94} & \textbf{0.97} & 1.00 & \textbf{0.93} & & 1.00 & 1.01 & 0.96 & 0.99 \\
\bottomrule
\end{tabular}
\vspace{0.2em}
\begin{minipage}{\textwidth}
\footnotesize \textit{Notes.} 1000 bootstrap replications resampling individuals. Bold: interval shorter than that of TFGLS. Monte Carlo standard errors of the left-panel ratios (standard deviations over 400 resamples of the 1000 replications) are at most .03; they reflect the finite number of replications, not sampling uncertainty.
\end{minipage}
\end{table}

\FloatBarrier

\putbib\enlargethispage{\baselineskip}
\end{bibunit}

\end{document}